\documentclass[11pt,a4paper]{article}
\pdfoutput=1
\usepackage[T1]{fontenc}
\usepackage[english]{babel}
\usepackage[utf8]{inputenc}
\usepackage[normalem]{ulem}
\usepackage[dvipsnames]{xcolor}
\usepackage{tikz}
\usetikzlibrary{decorations.pathmorphing, decorations.markings, calc, arrows.meta, patterns}
\usepackage{simpler-wick}

\renewcommand{\baselinestretch}{1.1}
\usepackage[textwidth=16cm, textheight=22cm]{geometry}

\usepackage{tocloft} 
\usepackage[colorlinks=true, linkcolor=black, citecolor=blue, urlcolor=blue, linktoc=all, pagebackref]{hyperref}

\usepackage{amsmath}
\usepackage{amsfonts}
\usepackage{amssymb}
\usepackage{mathtools}
\usepackage{latexsym}
\usepackage{mathrsfs}
\usepackage{braket}
\usepackage{slashed}

\usepackage{enumitem}
\usepackage{subcaption}
\usepackage{cite}

\newcommand{\CI}{{\cal I}}
\newcommand{\cI}{\mathcal{I}}
\newcommand{\cL}{\mathcal{L}}
\newcommand{\cH}{\mathcal{H}}
\newcommand{\cT}{\mathcal{T}}
\newcommand{\bp}{\mathbf{p}}
\newcommand{\bq}{\mathbf{q}}
\newcommand{\bx}{\mathbf{x}}

\newcommand{\parslash}{\partial\kern-0.52em\raisebox{0.09em}{\slash}\kern0.08em}

\numberwithin{equation}{section}

\newcommand\be{\begin{equation}}
\newcommand\ee{\end{equation}}

\newcommand\nn{\nonumber}

\let\originalleft\left
\let\originalright\right
\renewcommand{\left}{\mathopen{}\mathclose\bgroup\originalleft}
\renewcommand{\right}{\aftergroup\egroup\originalright}

\newcommand{\hypgeo}[2]{%
  \operatorname{%
    {\vphantom{\mathnormal{F}}}_{#1}%
    \kern-\scriptspace
    \mathnormal{F}_{#2}%
  }%
}

\newcommand{\re}{\operatorname{Re}}

\newcommand\neutralColor{BurntOrange}
\newcommand\chargedColor{RoyalBlue}

\DeclareMathOperator{\sn}{sn}
\DeclareMathOperator{\cn}{cn}
\DeclareMathOperator{\dn}{dn}
\DeclareMathOperator{\zn}{zn}

\DeclareMathOperator{\Tr}{Tr}
\DeclareMathOperator{\tr}{tr}

\DeclareMathOperator{\arctanh}{arctanh}
\DeclareMathOperator{\arccoth}{arccoth}
\newcommand{\smat}[1]{\bigl( \begin{smallmatrix} #1 \end{smallmatrix} \bigr)}
\newcommand{\parfrac}[2]{\frac{\partial #1}{\partial #2}}
\newcommand{\wb}{\overline}
\newcommand{\tps}[2]{\texorpdfstring{\ensuremath{#1}}{#2}}
\newcommand{\tpsb}[2]{\texorpdfstring{\ensuremath{\boldsymbol{#1}}}{#2}}

\begin{document}

\setcounter{page}{0}
\thispagestyle{empty}

\parskip 3pt

\font\mini=cmr10 at 2pt

\hypersetup{pageanchor=false}
\begin{titlepage}
\begin{flushright}
CERN-TH-2026-102, SISSA 05/2026/FISI
\end{flushright}
    \vspace{1cm}
    \begin{center}
	\vspace*{-.6cm}
	\begin{center}
		\vspace*{1.1cm}
		{\centering \Large\textbf{Perturbative, Nonperturbative and Exact Aspects \\[0.5em] of Crystalline Phases in the Gross--Neveu Model}}
	\end{center}
	\vspace{0.8cm}
	{\bf Francesco Benini$^{a,b}$, Ohad Mamroud$^{a,b}$, Tomas Reis$^{c}$, and Marco Serone$^{a,b}$}
	\vspace{1.cm}
    
  ${}^a\!\!$
	{\em SISSA, Via Bonomea 265, I-34136 Trieste, Italy} \\

 	\vspace{.3cm}
 ${}^b\!\!$
	{\em INFN, Sezione di Trieste, Via Valerio 2, I-34127 Trieste, Italy}\\
	\vspace{.3cm}
 ${}^c\!\!$
    {\em Theoretical Physics Department, CERN, 1211 Geneva 23, Switzerland}

	        \vspace{.8cm}

        {\small{\texttt{fbenini@sissa.it,omamroud@sissa.it,tomas.reis@cern.ch, serone@sissa.it}}}
    \end{center}
    \vskip 15pt
    
    \begin{abstract}
        We study the crystalline phase of the $O(2N)$ Gross--Neveu model with a chemical potential for $a \leq N-2$ of the fermions. We analyze the problem in three independent ways: using perturbative QFT methods, a semiclassical large $N$ analysis, and integrability techniques (both at finite and large $N$). The resulting picture is consistent across all three approaches: at sufficiently large chemical potential $h$, an inhomogeneous phase emerges in which $a$-particle bound states condense and which, at large $N$, corresponds to a periodically oscillating chiral condensate. In this phase, the usual dynamically generated scale $\Lambda$ is replaced by two new dynamically generated scales $\Lambda_{\rm n}$ and $\Lambda_{\rm c}$. These two scales govern the multiple nonperturbative effects in the theory, corresponding in particular to the mass gaps of neutral and charged excitations on top of the inhomogeneous vacuum, respectively. They also control the nonperturbative corrections to observables such as the free energy and provide the parameters characterizing the oscillatory profile of the mean field at large $N$. In this paper, we provide the necessary details of each of the three methods, thereby complementing the results announced in a previous, shorter publication.
    \end{abstract}
    
\end{titlepage}
\hypersetup{pageanchor=true}

{%
  \renewcommand{\baselinestretch}{.88}
  \parskip=0pt
  \setcounter{tocdepth}{2}
  \tableofcontents
}

\section{Introduction}

Characterizing the phase diagram of strongly interacting matter at finite temperature and density is a central problem, relevant to phenomena such as superconductivity and the physics of compact astrophysical objects. These phases may exhibit spontaneous breaking of various symmetries, including spacetime symmetries through the formation of crystals. Whereas finite-temperature effects can at least be treated numerically using Euclidean lattice methods, finite-density effects remain more challenging because of the notorious sign problem.

For this reason, it is useful to focus on simpler theories that qualitatively resemble the four-dimensional theories of interest, such as quantum chromodynamics (QCD). A renowned example is the $O(2N)$-symmetric Gross--Neveu (GN) model: a theory of $N$ massless Dirac fermions $\psi^i$ in two spacetime dimensions (2d) coupled through four-fermion interactions \cite{Gross:1974jv}:
\begin{equation}
\label{eq:UV1}
\mathscr{L}_0 = \sum_{j=1}^N \bar\psi^{j} i \slashed{\partial}\psi^{j} + \frac{g^2}2 \biggl( \sum_{j=1}^N \bar\psi^{j} \psi^{j} \!\biggr)^{\!\!2} \,.
\end{equation}
Notably, like QCD, this theory is asymptotically free and undergoes dynamical mass generation in the infrared (IR). When the Lagrangian is defined at some scale $\Lambda_\text{UV}$, the dynamical mass scale at 1-loop order is
\begin{equation}
\label{Lambda}
\Lambda \approx \Lambda_\text{UV} \; e^{ - \tfrac{ \pi }{ (N-1) \, g^2(\Lambda_\text{UV}) } } \,.
\end{equation} 
The vacuum is gapped, with spontaneous breaking of a $\mathbb{Z}_2$ chiral symmetry. The GN model is also of intrinsic interest, with applications in condensed-matter systems such as conducting polymers \cite{Campbell:1981dc} and inhomogeneous superconductors \cite{Thies:2006ti}.

Thanks to integrability, the full spectrum of asymptotic states above the vacuum, as well as their $S$-matrix and scattering amplitudes, are known for any $N$ \cite{Zamolodchikov:1978xm, Karowski:1980kq}. The most physically relevant states for our purposes belong to the rank-$r$ completely antisymmetric representation of $O(2N)$, with mass
\begin{equation}
\label{eq:spectrum}
m_r = m \, \frac{\sin\bigl( \tfrac{\pi r}{2N-2} \bigr)}{ \sin \bigl( \tfrac{\pi}{2N-2} \bigr)} \qquad\text{for}\qquad  1 \leq r \leq N-2 \,,
\end{equation}
where the nonperturbative mass $m$ is proportional to the UV scale \eqref{Lambda} through a known $N$-dependent constant \cite{fnw1}. In addition, there are states in the two conjugate spinor representations of $O(2N)$, the kink and the anti-kink, with mass
\begin{equation}
\label{eq:skinkmass}
m_\text{k} = m_{\bar{\text{k}}} = \frac{m}{ 2\sin \bigl( \tfrac{\pi}{2N-2} \bigr)} \,.
\end{equation}
Furthermore, there are states with mass $m_r$ that belong to rank-$r'$ representations with $0 \leq r' < r$ and $r'  \text{ mod } 2 = r \text{ mod } 2$.

The analytical tractability of the GN model has motivated studies of its behavior at finite density. This can be achieved by adding to the Lagrangian a chemical potential $h$ for the common $U(1)$ particle-number symmetry of the first $a\leq N$ fermions:%
\footnote{As we will see below, when $h$ exceeds a critical value $h_\text{crit}$, the ground state develops a finite charge density.}
\begin{equation}
\label{eq:UV}
\mathscr{L}_h = \sum_{j=1}^N \bar\psi^{j} i \slashed{\partial}\psi^{j} 
+ \frac{g^2}2 \biggl( \sum_{j=1}^N \bar\psi^{j} \psi^{j} \!\biggr)^{\!\!2} + h \! \sum_{m=1}^a \bar\psi^m \gamma^0 \psi^{m}  \,.
\end{equation}
Such a chemical potential breaks the global symmetry $O(2N)$ down to $U(a) \times O(2N-2a)$.%
\footnote{Note that $h$ breaks charge conjugation. That is why the unbroken symmetry is $U(a)$ rather than $O(2a)$.}
The $N$ Dirac fermions then split into charged fermions $\psi^i$ ($i=1,\ldots, a$) and neutral fermions $\psi^i$ ($i=a+1,\ldots,N$).

Past work focused on the large $N$ limit with a chemical potential for the global $U(1)$ fermion-number symmetry, namely $a=N$. While early studies assumed translational invariance \cite{Wolff:1985av, Barducci:1994cb}, later work established that the true phase diagram includes inhomogeneous crystalline phases at low temperature $T$ and sufficiently large $h$ \cite{Schon:2000he, Thies:2003kk, Schnetz:2004vr, Basar:2009fg, Melin:2024oee}.%
\footnote{Inhomogeneous phases have also been studied in other variants of the GN model \cite{Basar:2009fg, Ciccone:2022zkg, Ciccone:2023pdk} and, more recently, in \cite{Thies:2025mro}. They have also been investigated in two-dimensional QCD at finite temperature and chemical potential; see, e.g., \cite{Azaria:2016mqb, Lajer:2021kcz}.}
These phases arise because kinks condense in the vacuum, and a Peierls-like instability is responsible for the appearance of vacuum inhomogeneities. The results of \cite{Thies:2003kk, Schnetz:2004vr, Schon:2000he, Basar:2009fg} are based on the solution of certain semiclassical self-consistency equations based on ingenious ans\"atze, and they apply in the strict large $N$ limit. 

In a largely independent line of investigation, the GN model \eqref{eq:UV} with $a=1$, defined at a scale $\Lambda_\text{UV} \propto h$, was analyzed with distinct motivations. In this context, the primary observable is the relative free energy $\mathcal{F} = F(h)-F(0)$, which can be determined exactly, both at finite and at large $N$, via Bethe ansatz methods \cite{Polyakov:1983tt}. Early studies were motivated by the search for the exact relation between the mass gap $m$ in the vacuum and the dynamically generated scale $\Lambda$ of the theory in the modified minimal subtraction scheme $\overline{\rm MS}$  \cite{fnw1, fnw2}. More recently, technical advances in the analytic treatment of Bethe ansatz integral equations \cite{Volin:2009wr, Marino:2019eym} have enabled a systematic weak-coupling expansion of $\mathcal{F}$, both for the GN model and for other 2d integrable theories. This development has unveiled the resurgent structure of the perturbative series and clarified its connection to renormalon phenomena \cite{Marino:2019eym, DiPietro:2021yxb, Marino:2021dzn, Bajnok:2022xgx}. If renormalons originate from condensates \cite{Shifman:1978bx}, they should appear in $\mathcal{F}$ in the form $(\Lambda/h)^{n \Delta}$, where $n$ is a positive integer and $\Delta$ is the classical dimension of the condensing operator. Instead, in the case $a=1$ of the GN model, Ref.~\cite{Marino:2021dzn} unexpectedly found renormalons with fractional powers, $(\Lambda/h)^{(N-1)/(N-2)}$, which do not fit the usual pattern.

The aim of this work is to study the GN model at $T=0$ and finite density, in the presence of a chemical potential for a generic number $1\leq a\leq N-2$ of fermions as in \eqref{eq:UV}. As we will see, this corresponds to studying the theory in the presence of different kinds of ``matter''. We consider both finite and large $N$. A particularly noteworthy byproduct --- which in fact constituted the primary motivation for this study --- is the resolution of the renormalon puzzle mentioned above and the establishment of a connection between the two previously disparate lines of research.

In order to obtain a compelling picture of the IR physics of the Gross--Neveu model for generic values of $a$ and $h$, and to confirm the expected relation between renormalons and condensates, we use three complementary techniques.
The first method is based on a suitable generalization of the semiclassical techniques used in \cite{Schon:2000he, Thies:2003kk, Schnetz:2004vr, Basar:2009fg}. For brevity, we refer to this method as the ``semiclassical analysis''. The physics is particularly transparent in this approach, but it applies only in the large $N$ limit.
The second method is the Bethe ansatz (BA), as in \cite{fnw1, fnw2} and subsequent works, properly extended to generic $a$ and generalized to study the dispersion relations of different particles. Extracting physical results from the BA is less straightforward; notably, however, this method also works at finite $N$.
The third method is standard perturbation theory in QFT. Since the model is at nonzero chemical potential, the perturbative description involves excitations around a Fermi surface. It is well known that dynamically generated scales near Fermi surfaces can be determined from the RG behavior of low-energy couplings using EFT techniques. These are typically four-fermion interactions obtained by integrating out other degrees of freedom; see, e.g., \cite{Benfatto:1990zz, Polchinski:1992ed} for superconductivity, or \cite{Evans:1998ek, Schafer:1998na} for color superconductivity in QCD. In contrast to these cases, the GN model contains four-fermion interactions already in the UV, so we can adopt a microscopic description and determine such dynamically generated scales directly from the UV theory, paying careful attention to the effect of the chemical potential. The physics is clear in this method, and it applies at finite $N$, but the final results are based on perturbative arguments and some observables, such as the IR spectrum, remain out of reach.

\subsection{Summary of results}

Let us summarize the main results of the paper. First, when $h$ is large one can define the model at a high scale $\Lambda_\text{UV} = 2h$, where it is weakly coupled, and then use perturbation theory. The $O(2N)$-symmetric GN model \eqref{eq:UV1} develops a dynamically generated scale $\Lambda$ which at 1-loop order reads
\begin{equation}
\Lambda \approx 2h \, e^{ - \tfrac{ \pi }{ (N-1) \, g^2(2h) } } \,.
\end{equation} 
However, when coupled to a chemical potential, the theory in \eqref{eq:UV} develops {\it two} dynamically generated scales. At 1-loop order, they read
\begin{equation}\begin{aligned}
\label{eq:lambda-12}
\Lambda_\mathrm{n} &= 2h \, \Bigl( \frac{\Lambda}{2h} \Bigr) \rule{0pt}{1em}^{\frac{N-1}{N-a-1}} \,\approx\, 2h \, e^{ - \tfrac{\pi}{(N-a-1) \, g^2(2h) } } \,, \\
\Lambda_\mathrm{c} &= 2 h \, \Bigl( \frac{\Lambda}{2h} \Bigr) \rule{0pt}{1em}^{\frac{2(N-1)}{a}} \:\approx\, 2h \, e^{ - \tfrac{2\pi}{ a\, g^2(2h) } } \,,
\end{aligned}\end{equation}
for $2\leq a \leq N-2$ and $N>2$. The formula for $\Lambda_\text{c}$ also applies for $a=N$, where it reproduces the scale found in \cite{Melin:2024oee}. For $a=1$, the scale $\Lambda_\text{n}$ is given by \eqref{eq:lambda-12}, while $\Lambda_\text{c}$ turns out not to  exist. 
For $a=N-1$, the scale $\Lambda_\text{n}$ vanishes and clearly it does not exist for $a=N$. In perturbation theory at large densities, $h \gg \Lambda$, the scale $\Lambda_\mathrm{n}$ is the scale at which the perturbative interactions between light neutral-fermion excitations blow up, while $\Lambda_\mathrm{c}$ is the scale at which the interactions between light charged-fermion excitations close to the Fermi surface blow up. The identification of these scales leads to an immediate resolution of the renormalon puzzle: the fractional powers found in \cite{Marino:2021dzn} disappear when the results are expressed in terms of the correct dynamical scale. Indeed, for $a=1$ and up to numerical factors we have 
\begin{equation}
\Big(\frac{\Lambda}{h}\Big)^{\frac{N-1}{N-2}} \to \Big(\frac{\Lambda_\text{n}}{h}\Big) \,, 
\end{equation}
with no fractional powers. For $a>1$, in addition, we predict new renormalons associated with the scale $\Lambda_\text{c}$.

To determine the properties of the theory at nonvanishing density, we turn to a large $N$ semiclassical analysis, where 
$N\to \infty$, $a\to \infty$, with $y\equiv a/N$ kept fixed. We show that at positive density the homogeneous solutions for the Hubbard--Stratonovich field $\sigma(x) = -g^2 \sum_{j=1}^N \bar\psi^j(x) \, \psi^j(x)$ are unstable for any $a$, as found for $a=N$ in \cite{gn-inst}, indicating that the vacuum configuration must be inhomogeneous (see Fig.~\ref{fig-gamma}). To identify such a configuration we solve the saddle-point equation for $\sigma(x)$, which involves the Green function of the Dirac operator with potential $\sigma(x)$. The solution is obtained by extending to generic $a$ the methods employed in \cite{Schon:2000he, Thies:2003kk, Schnetz:2004vr}. We find an inhomogeneous time-independent crystal-like condensate, which at high densities simplifies to
\begin{equation}
\label{eq:intro-sigma-osc}
\bigl\langle  \sigma(x) \bigr\rangle \,\approx\, \Lambda_\mathrm{n} +  \Lambda_\mathrm{c} \sin(2 h x^1) \,,
\end{equation}
where $\Lambda_\mathrm{n}$ and $\Lambda_\mathrm{c}$ are as in \eqref{eq:lambda-12}, whereas $x^1$ parametrizes the spatial direction. The transition between the homogeneous and inhomogeneous solutions occurs at a critical value $h_\text{crit}$ and, at infinite volume, is characterized by a second-order phase transition.%
\footnote{At finite volume and large $N$ there is a series of first-order phase transitions, where the number of particles populating the ground state grows in integer jumps as $h$ is increased. In the infinite-volume limit this becomes a second-order phase transition where the density changes continuously.}
At low densities just above the phase transition, $\bigl\langle  \sigma(x) \bigr\rangle$ approaches the inhomogeneous semiclassical configuration associated with a single bound state in the rank-$a$ completely antisymmetric representation \cite{Dashen:1975xh}, reported in \eqref{eq:DHN}. This agrees with the fact that at $T=0$ the point $h_\text{crit}$ is where the ground state starts being populated by the particles with the highest charge-to-mass ratio. At fixed $a\leq N-2$, the particles with the highest ratio correspond to a specific nondegenerate weight state in the rank-$a$ completely antisymmetric representation of $O(2N)$. For $a=N-1$, instead, they correspond to the kink state, and the analysis is expected to be similar to the one for $a=N$. This explains why we focus on the range $1\leq a \leq N-2$. The large $N$ analysis also shows that the very same scales $\Lambda_\mathrm{n}$ and $\Lambda_\mathrm{c}$ control the dispersion relations of fermion modes around (\ref{eq:intro-sigma-osc}), with $\Lambda_\mathrm{n}$ equal to the mass gap of the neutral fermions and $\frac{1}{2}\Lambda_\mathrm{c}$ equal to the mass gap of the charged ones; see \eqref{eq:neutralGap} and \eqref{eq:chargedGap}.

Further properties of the model at both finite and large $N$, as well as extensive checks of the above findings, are derived using integrability and BA techniques. Thanks to a proper identification of the neutral and charged fermion excitations in terms of Bethe ansatz variables, we confirm that $\Lambda_\mathrm{n}$ and $\Lambda_\mathrm{c}$, respectively, govern their mass gaps also at finite $N$ (see Fig.~\ref{fig:mass-gaps}). At large $N$ the dispersion relations found using integrability are in complete agreement with those obtained from the semiclassical analysis (see Fig.~\ref{fig:disp-colors}).

Besides, we find that the GN model exhibits a \emph{massless} mode for all values of $a$. At infinite $N$ this mode is the Goldstone boson for broken translations (the phonon) in the crystal phase, as discussed in \cite{Melin:2024oee} for $a=N$. The phonon is nonrelativistic at low densities and approaches the speed of light at high densities. Strict long-range order is forbidden at finite $N$ \cite{Mermin1966, Coleman1973}, and quantum corrections are expected to melt any crystalline structure. Nevertheless, gapless excitations persist and give rise to a quasi-long-range ordered phase.

Using the Bethe ansatz we can also compute the free energy $\mathcal{F}$, as in early work. At large $N$ the result agrees with the expression computed in the semiclassical analysis, for any $a$ (see Fig.~\ref{fig:free-energy}). For the specific case $a=N/2$, the free energy and the mass gap of charged fermions can be determined analytically; see \eqref{eq:nH5} and \eqref{eq:nH16}. 
Building on the methodologies developed in \cite{Marino:2021dzn}, we rewrite the exact expression for $\mathcal{F}$ as a transseries expansion, which again reveals the scales $\Lambda_\mathrm{n}$ and $\Lambda_\mathrm{c}$.%
\footnote{In fact, we first determined these scales from \eqref{eq:ts-120}.} 
Schematically, we find
\begin{equation}
\label{eq:ts-120}
    \mathcal{F}(h) \sim h^2  \sum_{m,n \,\geq\, 0} \frac{\Lambda_\mathrm{n}^{2m} \, \Lambda_\mathrm{c}^{2n} }{ h^{2(m+n)}} \, \sum_{k \,\geq\, 0} \, a^{[m,n]}_{k} \, g^{2k} \,-\, c_0 \, \Lambda^2
\end{equation}
where $a^{[m,n]}_{k}$ are coefficients that can be determined recursively.

\subsection{Structure of the paper}

The paper is relatively long and technically involved, as it relies on several distinct methodologies. To facilitate reading, each section has been made as self-contained as possible, so that readers primarily interested in a specific part can jump to it without going through the others.

We begin in sec.~\ref{sec:pert theory} with the perturbative QFT computation. We show how a theory with a single marginal coupling in the UV can exhibit 1-loop amplitudes that diverge at different energy and momentum scales, which are identified with the dynamically generated scales $\Lambda_\text{n}$ and $\Lambda_\text{c}$. Sec.~\ref{sec:large-N} is devoted to the semiclassical analysis at large $N$. After establishing the instability of the homogeneous large $N$ solution, we review the techniques of \cite{Thies:2003br, Schnetz:2005ih} used to construct inhomogeneous solutions. We then derive and analyze the saddle-point equations in the high- and low-density limits, and obtain the fermion dispersion relations as well. The Bethe ansatz method is discussed in sec.~\ref{sec:tba-all}. We briefly review the method and discuss how to identify the physical states. We then determine the transseries expansion \eqref{eq:ts-120}, derive several analytic results at large $N$, and proceed to an extensive numerical analysis of mass gaps, dispersion relations, and comparisons with the semiclassical analysis.  We conclude in sec.~\ref{sec:conclusions} by discussing open questions and possible directions for future investigation. Several appendices complement the paper.

Beyond presenting new results, this paper provides the necessary background and details for deriving the results presented in \cite{Benini:2025riz}, which contains a short version of some of our key findings.

\section{RG flow at finite density}
\label{sec:pert theory}

\begin{figure}
    \centering
    \begin{center}
        \begin{tikzpicture}
        \draw[->,very thick] (-5.2,0)--(5.2,0) node[right]{$q_1$};
        \draw[->,very thick] (0,-1/2)--(0,3) node[left]{$q_0$};
        \filldraw[\neutralColor, fill=\neutralColor] (0,0) circle (2pt) node[below left]{$0$};
        \draw[dashed,\neutralColor] (0,0) circle (1.25);
        \draw[<->,\neutralColor] ({cos(-25)*0.1},{sin(-25)*0.1}) -- ({cos(-25)*1.25},{sin(-25)*1.25}) node[midway, xshift=-1pt, yshift=-5pt] {$\Lambda_{\rm n}$};
        \draw[-{Latex[length=8pt, width=6pt]}, thick, \neutralColor] (1,2) to[out=-120, in=30] (0,0);
        \filldraw[\chargedColor, fill=\chargedColor] (3.5,0) circle (2pt) node[below]{$2h$};
        \draw[dashed,\chargedColor] (3.5,0) circle (0.75);
        \draw[<->,\chargedColor] ({3.5+cos(30)*0.1},{sin(30)*0.1}) -- ({3.5+cos(30)*0.75},{sin(30)*0.75}) node[midway, xshift=-5pt, yshift=7pt] {$\Lambda_{\rm c}$};
        \draw[-{Latex[length=8pt, width=6pt]}, thick, \chargedColor] (1,2) to[out=-30, in=120] (3.5,0);
        \filldraw[\chargedColor, fill=\chargedColor] (-3.5,0) circle (2pt) node[below]{$-2h$};
        \draw[dashed,\chargedColor] (-3.5,0) circle (0.75);
        \draw[<->,\chargedColor] ({-3.5+cos(150)*0.1},{sin(150)*0.1}) -- ({-3.5+cos(150)*0.75},{sin(150)*0.75}) node[midway,  xshift=5pt, yshift=7pt] {$\Lambda_{\rm c}$};
        \draw[-{Latex[length=8pt, width=6pt]}, thick, \chargedColor] (1,2) to[out=-150, in=60] (-3.5,0);
        \filldraw[black, fill=black] (1,2) circle (2pt) node[above]{$\Lambda_{UV}$};
        \end{tikzpicture}
        \end{center}
    \caption{\label{fig:RG flow}%
    Schematic depiction of the RG flow with chemical potential. We start with a small coupling at some UV scale $\Lambda_\text{UV}$. At some radius around zero momentum for the neutral fermions and around the Fermi surface for the charged fermions, the coupling explodes and perturbation theory breaks down.}    
\end{figure}
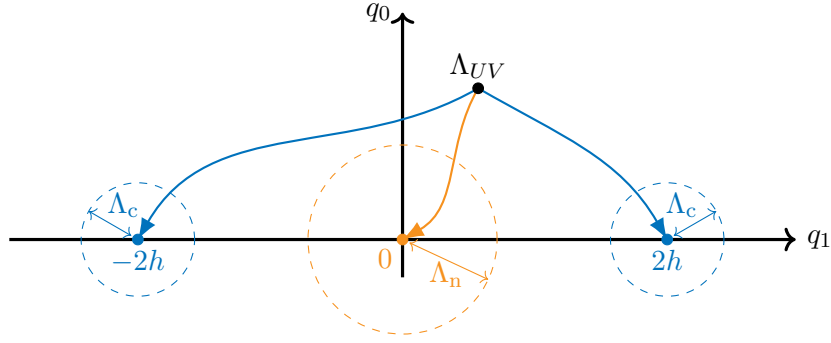

In this section we show that the two scales $\Lambda_{\rm n, c}$ naturally arise in perturbation theory from the model with a chemical potential \eqref{eq:UV}. To do so, we track the RG flow of the coupling $g^2(\mu)$ starting from some UV scale $\Lambda_\text{UV}$ where the coupling $g^2(\Lambda_\text{UV})$ is small and perturbation theory applies. When approaching a scale at which light particles can run in the loop, large logs are generated that cause perturbation theory to break down. There are three such regions, sketched in Fig.~\ref{fig:RG flow}: one at vanishing energy and momentum where the light particles are neutral fermions, and two near the Fermi surface where the light particles are charged fermions. In order to compare the dynamically generated scales with BA results, we set $\Lambda_\text{UV} \propto h$ at the end of the computation, and show that the breakdown happens at radii $\Lambda_{\rm n}$ and $\Lambda_{\rm c}$ around the points where the light particles running in the loop go on-shell. 

\subsection{Setting up the computation}
\label{sec:Feynman rules}

The chemical potential $h$ is a dimensionful quantity with dimension of mass, thus it does not affect the renormalization properties of the UV theory, which remains the same as in the case $h=0$. This can also be understood by treating the chemical potential as a background gauge field for the $SO(2N)$ global symmetry, and then using spurion analysis to argue that there are no new relevant gauge-invariant terms that can be generated. In addition, $h$ does not renormalize because it is the coupling associated with the time component of a conserved current. 
This guaranties that the renormalized theory does not need new counterterms, and we use:
\begin{equation}
\label{eq:renormalized Lag}
    \mathscr{L} = \sum_{j=1}^N i \bar\psi^{j}\bigl( \slashed{\partial} - ih_j\gamma^0\bigr)\psi^{j} 
    + \frac{g^2}2 \biggl( \sum_{j=1}^N \bar\psi^{j} \psi^{j} \!\biggr)^{\!\!2}
    + \delta_Z \sum_{j=1}^N i \bar\psi^{j} \bigl( \slashed{\partial} - ih_j\gamma^0\bigr)\psi^{j} 
    + \frac{\delta_g}2 \biggl( \sum_{j=1}^N \bar\psi^{j} \psi^{j} \!\biggr)^{\!\!2} ,
\end{equation}
where 
\begin{equation}
    h_j = \begin{cases}
        h & \text{for } 1 \le j \le a \\
        0 & \text{otherwise,}
    \end{cases}
\end{equation}
while $\delta_Z$, $\delta_g$ control the counterterms. The Minkowski metric is $\eta_{00}=1$, $\eta_{11} = -1$. Spinor indices $\alpha,\beta \in \{+,-\}$ are raised and lowered using the Levi--Civita tensor $\epsilon_{\alpha\beta}$ with $\epsilon_{+-} = \epsilon^{-+} = 1$:%
\footnote{\label{foo:sec2conv}Our conventions for gamma matrices in this section are $(\gamma^0)^\alpha{}_\beta = (\sigma^1)^\alpha{}_\beta =\begin{psmallmatrix} 0 & 1\\ 1 & 0 \end{psmallmatrix}$,
$(\gamma^1)^\alpha{}_\beta = (i\sigma^2)^\alpha{}_\beta =\begin{psmallmatrix} 0 & 1 \\ -1 & 0 \end{psmallmatrix}$,
with chirality matrix $(\gamma^*)^\alpha{}_\beta = (\gamma^1\gamma^0)^\alpha{}_\beta = (\sigma^3)^\alpha{}_\beta = \begin{psmallmatrix} 1 & 0 \\ 0 & -1 \end{psmallmatrix}$.
Whenever spinor indices are left implicit, it is understood that $\bar\psi =\bar\psi_\alpha$ and $\psi = \psi^\alpha$. We will use different conventions in section~\ref{sec:large-N}.} 
\be
    \psi_\alpha = \epsilon_{\alpha\beta}\psi^\beta\,, \qquad  \psi^\alpha = \epsilon^{\alpha\beta} \psi_\beta \,,
\ee
and similarly for $\bar\psi = \psi^\dagger \gamma^0$. 

The tree-level fermion propagator is
\be
\label{eq:FermProp}
\Bigl\langle \psi^{\alpha j}(p) \, \bar\psi^j_\beta (-p) \Bigr\rangle = \;\;
\begin{tikzpicture}
	\draw [-{Latex[length=8]}, line width=0.6] (-1, 0) -- (0.2, 0);
	\draw [line width=0.6] (-1, 0) -- (1, 0);
	\draw [-{Stealth[length=5]}] (-0.6, 0.25) -- (0.6, 0.25) node[above, pos=0.5] {\small$j,p$};
\end{tikzpicture}
\;\; = \biggl( \frac{i}{\slashed{p} + h_j \gamma^0} \biggr)_{\!\raisebox{0.7em}[0pt][0pt]{\scriptsize$\alpha$}\beta} = \frac{i \, ( \slashed{p} + h_j \gamma^0)^\alpha{}_\beta }{ (p_0+h_j)^2 - \bp^2} \;,
\ee
where $p = (p_0, \bp\equiv p_1)$. The presence of a chemical potential affects the Feynman $i\epsilon$ prescription, as reviewed in app.~\ref{app:Canonical Quantization Propagator}. In \eqref{eq:FermProp} there is no $i\epsilon$ term, but the integration contour for $p_0$  is deformed away from the real axis in a way that depends on both $\bp$ and the Fermi momentum $p_F = |h_j|$, as depicted in \eqref{eq:appContour}.

In order to track how particle species flow in diagrams, it is useful to split the Feynman rule for the tree-level quartic vertex as follows:
\begin{equation}
\begin{aligned}
\label{eq:tree level vertex}
& \bigl\langle \bar\psi_\alpha^i \, \psi_\beta^j \, \bar\psi_\gamma^k \, \psi_\delta^l \bigr\rangle = {} \\
& \raisebox{-3.3em}{\begin{tikzpicture}[line width=0.6]
	\draw (-0.6, 0.6) node [above left] {\small$\bar\psi^i_\alpha$} -- (0.6, -0.6) node [below right] {\small$\bar\psi^k_\gamma$};
	\draw (-0.6, -0.6) node [below left] {\small$\psi^j_\beta$} -- (0.6, 0.6) node [above right] {\small$\psi^l_\delta$};
	\draw [-{Latex[length=7]}] (-0.6, 0.6) -- +(-45: 0.6);
	\draw [-{Latex[length=7]}] (0.6, -0.6) -- +(135: 0.6);
	\draw [-{Latex[length=7, reversed]}] (-0.6, -0.6) -- +(45: 0.5);
	\draw [-{Latex[length=7, reversed]}] (0.6, 0.6) -- +(-135: 0.5);
    \filldraw (0,0) circle [radius = 0.05];
	\node at (1.5, 0) {$=$};
\begin{scope}[shift={(3.0, 0)}]
	\draw [gray!40!white] (0.05, 0) circle [radius = 0.15];
	\draw (-0.6, 0.6) node [above left] {\small$\bar\psi^i_\alpha$} -- (0, 0) -- (-0.6, -0.6) node [below left] {\small$\psi^j_\beta$};
	\draw (0.7, -0.6) node [below right] {\small$\bar\psi^k_\gamma$} -- (0.1, 0) -- (0.7, 0.6) node [above right] {\small$\psi^l_\delta$};
	\draw [-{Latex[length=7]}] (-0.6, 0.6) -- +(-45: 0.6);
	\draw [-{Latex[length=7]}] (0.7, -0.6) -- +(135: 0.6);
	\draw [-{Latex[length=7, reversed]}] (-0.6, -0.6) -- +(45: 0.5);
	\draw [-{Latex[length=7, reversed]}] (0.7, 0.6) -- +(-135: 0.5);
	\node at (1.5, 0) {$+$};
\end{scope}
\begin{scope}[shift={(6.0, 0)}]
	\draw [gray!40!white] (0, 0) circle [radius = 0.15];
	\draw (-0.55, 0.6) node [above left] {\small$\bar\psi^i_\alpha$} -- (0, 0.05) --  (0.55, 0.6) node [above right] {\small$\psi^l_\delta$};
	\draw (0.55, -0.6) node [below right] {\small$\bar\psi^k_\gamma$} -- (0, -0.05) -- (-0.55, -0.6) node [below left] {\small$\psi^j_\beta$};
	\draw [-{Latex[length=7]}] (-0.55, 0.6) -- +(-45: 0.55);
	\draw [-{Latex[length=7]}] (0.55, -0.6) -- +(135: 0.55);
	\draw [-{Latex[length=7, reversed]}] (-0.55, -0.6) -- +(45: 0.45);
	\draw [-{Latex[length=7, reversed]}] (0.55, 0.6) -- +(-135: 0.45);
	\node [right] at (1.3, 0) {$= \; ig^2 \bigl(\delta^{ij}\delta^{kl}\epsilon_{\beta\alpha}\epsilon_{\delta\gamma} - \delta^{il}\delta^{kj}\epsilon_{\delta\alpha}\epsilon_{\beta\gamma}\bigr)$\,.};
\end{scope}
\end{tikzpicture}}
\end{aligned}
\end{equation}

\subsection{The 1-loop diagrams} 
\label{sec:1 loop diagrams}

Let us consider the renormalization of two-point and four-point functions. Up to 1-loop order, the propagator is given by
\begin{equation}
\bigl\langle \psi^j(q) \, \bar\psi^j(-q) \bigr\rangle = \;\;
\raisebox{-0.5em}{\begin{tikzpicture}[line width=0.6]
	\draw (-0.7, 0) -- (0.7, 0); \draw [-{Latex[length=7]}] (-0.5, 0) -- +(0.65, 0) node[shift = {(-0.16, 0.3)}] {\small$q$};
	\node at (1.2, 0) {$+$};
\begin{scope}[shift={(2.8, 0)}]
	\draw [gray!40!white] (0, 0) circle [radius = 0.15];
	\draw [looseness = 1.5] (-1, 0) to[out = 0, in = 180] (-0.15, 0) to[out = 165, in = 180] (0, 0.9) to[out = 0, in = 15] (0.15, 0) to[out = 0, in = 180] (1, 0);
	\draw [-{Latex[length=7]}] (-1, 0) -- +(0.5, 0) node[shift = {(-0.16, 0.3)}] {\small$q$};
	\draw [-{Latex[length=7]}] (0.2, 0) -- (0.8, 0) node[shift = {(-0.16, 0.3)}] {\small$q$};
	\draw [-{Latex[length=7]}] (0, 0.89) -- +(0.14, 0) node[shift = {(-0.14, -0.25)}] {\small$p$};
	\node at (1.5, 0) {$+$};
\end{scope}
\begin{scope}[shift={(5.8, 0)}]
	\draw [gray!40!white] (0, 0.025) circle [radius = 0.15];
	\draw (-1, 0) -- (1, 0);
	\draw [looseness = 1.7] (0, 0.05) to[out = 165, in = 180] (0, 0.9) to[out = 0, in = 15] (0, 0.05);
	\draw [-{Latex[length=7]}] (-1, 0) -- +(0.5, 0) node[shift = {(-0.16, 0.3)}] {\small$q$};
	\draw [-{Latex[length=7]}] (0.2, 0) -- (0.8, 0) node[shift = {(-0.16, 0.3)}] {\small$q$};
	\draw [-{Latex[length=7]}] (0, 0.89) -- +(0.14, 0) node[shift = {(-0.14, -0.25)}] {\small$p$};
	\node at (1.5, 0) {$+$};
\end{scope}
\begin{scope}[shift={(8.8, 0)}]
	\draw (-1, 0) -- (1, 0);
	\draw [-{Latex[length=7]}] (-1, 0) -- +(0.5, 0) node[shift = {(-0.16, 0.3)}] {\small$q$};
	\draw [-{Latex[length=7]}] (0.2, 0) -- (0.8, 0) node[shift = {(-0.16, 0.3)}] {\small$q$};
	\draw [fill = white] (0, 0) circle [radius = 0.17];
	\draw (135: 0.17) -- (-45: 0.17);
	\draw (-135: 0.17) -- (45: 0.17);
	\node at (0, 0.45) {\small$\delta_Z$};
\end{scope}
\end{tikzpicture}}
\;\;.
\end{equation}
Since the loop diagrams do not depend on the external momentum, no divergences can appear. In fact, both loop diagrams vanish.%
\footnote{Both loop diagrams vanish because their integrands are the sum of functions odd either in $p_1$ or $p_0 + h_j$. The ``disconnected'' diagram, in addition, is proportional to the vanishing traces of gamma matrices.}
We can thus set $\delta_Z=0$.

\begin{figure}[t]
\centering
\begin{subfigure}{0.18\textwidth}
\centering
\begin{tikzpicture}[line width=0.6]
	\draw [gray!40!white] (-0.57, 0) circle [radius = 0.13];
	\draw [gray!40!white] (0.57, 0) circle [radius = 0.13];
	\draw (-1.3, 0.7) -- (-0.6, 0) -- (-1.3, -0.7);
	\draw (-0.54, 0) to[out = 90, in = 90, looseness=1.7] (0.54, 0) to[out = -90, in = -90, looseness=1.7] (-0.54, 0);
	\draw (1.3, 0.7) -- (0.6, 0) -- (1.3, -0.7);
	\draw [-{Latex[length=7]}] (-1.3, 0.7) -- +(-45: 0.6);
	\draw [-{Latex[length=7]}] (1.3, -0.7) -- +(135: 0.6);
	\draw [-{Latex[length=7, reversed]}] (-1.3, -0.7) -- +(45: 0.55);
	\draw [-{Latex[length=7, reversed]}] (1.3, 0.7) -- +(-135: 0.55);
	\draw [-{Latex[length=7]}] (-0.05, 0.53) -- +(0: 0.18) node[above, pos=0.2] {\small$p$};
	\draw [-{Latex[length=7]}] (0.05, -0.53) -- +(180: 0.18) node[below, pos=0.2] {\small$p-q$};
	\end{tikzpicture}
\caption{\label{fig:1-loop-1}}
\end{subfigure}
\begin{subfigure}{0.18\textwidth}
\centering
\begin{tikzpicture}[line width=0.6]
	\draw [gray!40!white] (-0.5, 0) circle [radius = 0.13];
	\draw [gray!40!white] (0.57, 0) circle [radius = 0.13];
	\draw (-1.15, 0.7) -- (-0.5, 0.05) to[out = 90, in = 90, looseness=1.7] (0.54, 0) to[out = -90, in = -90, looseness=1.7] (-0.5, -0.05) -- (-1.15, -0.7);
	\draw (1.3, 0.7) -- (0.6, 0) -- (1.3, -0.7);
	\draw [-{Latex[length=7]}] (-1.15, 0.7) -- +(-45: 0.6);
	\draw [-{Latex[length=7]}] (1.3, -0.7) -- +(135: 0.6);
	\draw [-{Latex[length=7, reversed]}] (-1.15, -0.7) -- +(45: 0.55);
	\draw [-{Latex[length=7, reversed]}] (1.3, 0.7) -- +(-135: 0.55);
	\draw [-{Latex[length=7]}] (-0.05, 0.54) -- +(0: 0.18) node[above, pos=0.2] {\small$p+q$};
	\draw [-{Latex[length=7]}] (0.05, -0.54) -- +(180: 0.18) node[below, pos=0.2] {\small$p$};
\end{tikzpicture}
\caption{\label{fig:1-loop-2}}
\end{subfigure}
\begin{subfigure}{0.18\textwidth}
\centering
\begin{tikzpicture}[line width=0.6]
	\draw [gray!40!white] (-0.57, 0) circle [radius = 0.13];
	\draw [gray!40!white] (0.5, 0) circle [radius = 0.13];
	\draw (-1.3, 0.7) -- (-0.6, 0) -- (-1.3, -0.7);
	\draw (1.15, 0.7) -- (0.5, 0.05) to[out = 90, in = 90, looseness=1.7] (-0.54, 0) to[out = -90, in = -90, looseness=1.7] (0.5, -0.05) -- (1.15, -0.7);
	\draw [-{Latex[length=7]}] (-1.3, 0.7) -- +(-45: 0.6);
	\draw [-{Latex[length=7]}] (1.15, -0.7) -- +(135: 0.6);
	\draw [-{Latex[length=7, reversed]}] (-1.3, -0.7) -- +(45: 0.55);
	\draw [-{Latex[length=7, reversed]}] (1.15, 0.7) -- +(-135: 0.55);
	\draw [-{Latex[length=7]}] (-0.05, 0.54) -- +(0: 0.18) node[above, pos=0.2] {\small$p$};
	\draw [-{Latex[length=7]}] (0.05, -0.54) -- +(180: 0.18) node[below, pos=0.2] {\small$p-q$};
\end{tikzpicture}
\caption{\label{fig:1-loop-3}}
\end{subfigure}
\begin{subfigure}{0.2\textwidth}
\centering
\hfill
\begin{tikzpicture}[line width=0.6]
	\draw [gray!40!white] (0.01, 0.5) circle [radius = 0.13];
	\draw [gray!40!white] (0.01, -0.5) circle [radius = 0.13];
	\draw (-0.5, 1.0) -- (-0.03, 0.5) to[out = 180, in = 180, looseness=1.0] (-0.03, -0.5) -- (-0.5, -1.0);
	\draw (0.9, 1.0) -- (0.05, -0.5);
	\fill [white] (0.33, 0) circle [radius = 0.05]; \fill [white] (0.53, 0.35) circle [radius = 0.05];
	\draw (0.05, -0.5) to[out = 0, in = 0, looseness=2.2] (0.05, 0.5);
	\fill [white] (0.53, -0.35) circle [radius = 0.05];
	\draw (0.05, 0.5) -- (0.9, -1.0);
	\draw [-{Latex[length=7]}] (-0.5, 1) -- +(-47: 0.5);
	\draw [-{Latex[length=7]}] (0.9, -1) -- +(119.5: 0.5);
	\draw [-{Latex[length=7, reversed]}] (-0.5, -1) -- +(47: 0.4);
	\draw [-{Latex[length=7, reversed]}] (0.9, 1) -- +(-119.5: 0.4);
	\draw [-{Latex[length=7]}] (-0.32, 0.05) -- +(-90: 0.18) node[left, pos=0.2] {\small$p$};
	\draw [-{Latex[length=7]}] (0.685, 0.05) -- +(-90: 0.18) node[right, pos=0.2] {\small$q_{13}{-}p$};
\end{tikzpicture}
\caption{\label{fig:1-loop-4}}
\end{subfigure}
\begin{subfigure}{0.2\textwidth}
\centering
\hfill
\begin{tikzpicture}[line width=0.6]
	\draw [gray!40!white] (0, 0.5) circle [radius = 0.13];
	\draw [gray!40!white] (0, -0.5) circle [radius = 0.13];
	\draw (-0.7, 1.0) -- (-0.05, 0.5) to[out = 180, in = 180, looseness=1.7] (-0.05, -0.5) -- (-0.7, -1.0);
	\draw (0.7, 1.0) -- (0.05, 0.5) to[out = 0, in = 0, looseness=1.7] (0.05, -0.5) -- (0.7, -1.0);
	\draw [-{Latex[length=7]}] (-0.7, 1) -- +(-37.5: 0.6);
	\draw [-{Latex[length=7]}] (0.7, -1) -- +(142.5: 0.6);
	\draw [-{Latex[length=7, reversed]}] (-0.7, -1) -- +(37.5: 0.5);
	\draw [-{Latex[length=7, reversed]}] (0.7, 1) -- +(-142.5: 0.5);
	\draw [-{Latex[length=7]}] (-0.53, 0.05) -- +(-90: 0.20) node[left, pos=0.2] {\small$p$};
	\draw [-{Latex[length=7]}] (0.53, -0.05) -- +(90: 0.20) node[right, pos=0.2] {\small$p{-}q_{14}$};
\end{tikzpicture}
\caption{\label{fig:1-loop-5}}
\end{subfigure}
\caption{\label{fig:1-loop-correlator}%
The five diagrams that contribute to the four-point function at 1-loop.}
\end{figure}
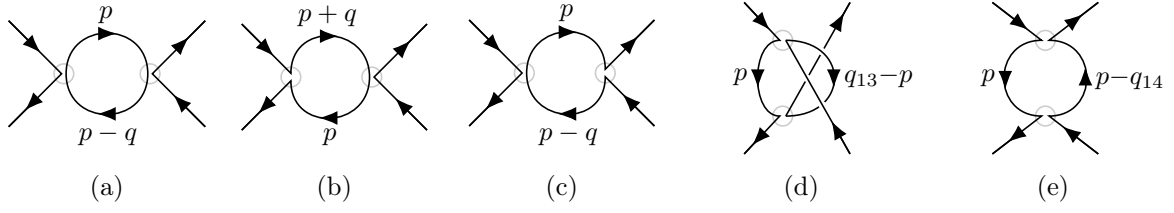

Consider then the connected (amputated) four point function:
\begin{equation}
\label{eq:corr def}
    \Gamma^{i j}_{\alpha\beta\gamma\delta}(q_1,q_2,q_3,q_4) \equiv \left\langle  \bar\psi_\alpha^{i}(q_1) \, \psi_\beta^{i}(q_2) \, \bar\psi_\gamma^{j}(q_3) \, \psi_\delta^{j}(q_4) \right\rangle \,,
\end{equation}
for $i\neq j$ and with spinor indices $\alpha, \beta, \gamma, \delta \in \{+,-\}$. For convenience, we denote the center-of-mass 2-momentum by $q \equiv q_1 + q_2 = -(q_3+q_4)$, and also use the variables $q_{13} \equiv q_1 + q_3$ and $q_{14} \equiv q_1 + q_4$. Diagrammatically, the four-point function can be written as
\begin{equation}
\label{eq:1loopGamma}
\begin{tikzpicture}[baseline=(current bounding box.center)]
	\draw [line width=0.6] (-0.9, 0.9) node [left] {\small$\bar\psi^i_\alpha$} -- (0.9, -0.9) node [right] {\small$\bar\psi^j_\gamma$};
	\draw [line width=0.6] (-0.9, -0.9) node [left] {\small$\psi^i_\beta$} -- (0.9, 0.9) node [right] {\small$\psi^j_\delta$};
	\draw [-{Latex[length=7]}] (-0.9, 0.9) -- +(-45: 0.6);
	\draw [-{Latex[length=7]}] (0.9, -0.9) -- +(135: 0.6);
	\draw [-{Latex[length=7, reversed]}] (-0.9, -0.9) -- +(45: 0.55);
	\draw [-{Latex[length=7, reversed]}] (0.9, 0.9) -- +(-135: 0.55);
	\draw [-{Stealth[length=5]}] (-0.95, 0.55) node[shift={(-0.1, -0.25)}] {\small$q_1$}-- +(-45: 0.6);
	\draw [-{Stealth[length=5]}] (-0.55, -0.95) node[shift={(0.4, 0)}] {\small$q_2$}-- +(45: 0.6);
	\draw [-{Stealth[length=5]}] (0.95, -0.55) node[shift={(0.1, 0.3)}] {\small$q_3$}-- +(135: 0.6);
	\draw [-{Stealth[length=5]}] (0.55, 0.95) node[shift={(-0.4, 0)}] {\small$q_4$}-- +(-135: 0.6);
	\fill [white] (0,0) circle [radius = 0.35];
	\draw [pattern = north west lines, line width = 0.6] (0,0) circle [radius = 0.35];
	\node at (2.2, 0) {$=$};
\begin{scope}[shift={(4,0)}]
	\draw [gray!40!white] (0.05, 0) circle [radius = 0.15];
	\draw [line width=0.6] (-0.9, 0.9) -- (0, 0) -- (-0.9, -0.9);
	\draw [line width=0.6]  (1.0, 0.9) -- (0.1, 0) -- (1.0, -0.9);
	\draw [-{Latex[length=7]}] (-0.9, 0.9) -- +(-45: 0.6);
	\draw [-{Latex[length=7]}] (1.0, -0.9) -- +(135: 0.6);
	\draw [-{Latex[length=7, reversed]}] (-0.9, -0.9) -- +(45: 0.55);
	\draw [-{Latex[length=7, reversed]}] (1.0, 0.9) -- +(-135: 0.55);
	\node at (1.7, 0) {$+$};
\end{scope}
\begin{scope}[shift={(7.5,0)}]
	\draw [line width=0.6] (-0.9, 0.9) -- (0.9, -0.9);
	\draw [line width=0.6] (-0.9, -0.9) -- (0.9, 0.9);
	\draw [-{Latex[length=7]}] (-0.9, 0.9) -- +(-45: 0.6);
	\draw [-{Latex[length=7]}] (0.9, -0.9) -- +(135: 0.6);
	\draw [-{Latex[length=7, reversed]}] (-0.9, -0.9) -- +(45: 0.55);
	\draw [-{Latex[length=7, reversed]}] (0.9, 0.9) -- +(-135: 0.55);
	\draw [fill=white, line width=0.6] (0, 0) circle [x radius = 0.6, y radius = 0.4] node {\small 1-loop};
	\node at (1.7, 0) {$+$};
\end{scope}
\begin{scope}[shift={(11,0)}]
	\draw [line width=0.6] (-0.9, 0.9) -- (0.9, -0.9);
	\draw [line width=0.6] (-0.9, -0.9) -- (0.9, 0.9);
	\draw [-{Latex[length=7]}] (-0.9, 0.9) -- +(-45: 0.6);
	\draw [-{Latex[length=7]}] (0.9, -0.9) -- +(135: 0.6);
	\draw [-{Latex[length=7, reversed]}] (-0.9, -0.9) -- +(45: 0.55);
	\draw [-{Latex[length=7, reversed]}] (0.9, 0.9) -- +(-135: 0.55);
	\draw [fill=white, line width=0.6] (0, 0) circle [radius = 0.17];
	\draw [line width=0.6] (150: 0.17) -- (-30: 0.17);
	\draw [line width=0.6] (60: 0.17) -- (-120: 0.17);
	\node at (-0.4, 0.05) {\small$\delta_g$};
	\node at (1.2, 0) {.};
\end{scope}
\end{tikzpicture}
\end{equation}
The first term on the right-hand side is the tree-level vertex \eqref{eq:tree level vertex} to which only the first tensor structure contributes because we chose $i \neq j$; the second term is the 1-loop contribution, which consists of the five diagrams in Fig.~\ref{fig:1-loop-correlator}; the third term is the counterterm $\delta_g$ which has the same structure as the first term.
The 1-loop integral in all five diagrams can be expressed through the function
\begin{equation}
    \label{eq:cI def}
    \cI_{\alpha\beta\gamma\delta}(q,h_1,h_2) = \int\! \frac{d^2p}{(2\pi)^2} \left(\frac{i}{\slashed{p} + h_1\gamma^0}\right)_{\!\alpha\beta} \left(\frac{i}{\slashed{p} - \slashed{q} + h_2\gamma^0}\right)_{\!\gamma\delta} \,,
\end{equation}
which we need to regulate. As a regularization scheme we introduce a cutoff $M$ on the spatial momentum only, keeping undeformed the $p_0$ integration, so $\int\!\frac{d^2p}{(2\pi)^2} \to\int_{-M}^{M} \frac{dp_1}{2\pi} \int \frac{dp_0}{2\pi}$. To perform the integrals, we rotate the $p_0$ integration contour to Euclidean signature, $p_0 = i p_2$, and then close it in the complex plane. The full details are given in app.~\ref{app:cI computation}. After rotating also the external momentum to Euclidean signature, $q_0 = i q_2$, the result is
\begin{equation}\begin{aligned}
\label{eq:cI result}
    \cI_{\alpha\beta\gamma\delta}(q,h_1,h_2) &= \frac{\gamma_{d} - \gamma_a}{2} \, B(q,h_1,h_2) + \frac{\gamma_{c}-\gamma_{b}}{2} \, A(q,h_1,h_2) \\
    &\quad - \frac{i}{4\pi} \frac{\slashed{q}_{\alpha\beta}\slashed{q}_{\gamma\delta} + i (h_1 - h_2) \bigl( q_2 \gamma_a + i(h_1 - h_2 + iq_2) \gamma_d \bigr)}{\left(h_1-h_2-q_1+iq_2\right)\left(h_1-h_2+q_1+iq_2\right)} + \frac{i\gamma_a}{4\pi} \,,
\end{aligned}\end{equation}
where $\gamma_{a}=\gamma^{0}_{\alpha\beta}\gamma^0_{\gamma\delta}$, $\gamma_{b}=\gamma^{0}_{\alpha\beta}\gamma^1_{\gamma\delta}$, $\gamma_{c}=\gamma^{1}_{\alpha\beta}\gamma^0_{\gamma\delta}$, $\gamma_{d}=\gamma^{1}_{\alpha\beta}\gamma^1_{\gamma\delta}$, so
\begin{equation}
    \gamma_a - \gamma_d = \eta_{\mu\nu}\gamma^\mu_{\alpha\beta}\gamma^\nu_{\gamma\delta} \,, \qquad \gamma_b - \gamma_c = \epsilon_{\mu\nu}\gamma^\mu_{\alpha\beta}\gamma^\nu_{\gamma\delta} \,,
\end{equation}
and 
\begin{subequations}
\label{eq:def A B}
\begin{align}
    A(q,h_1,h_2) &= \frac{i}{8\pi} \log\Biggl[ \frac{\left(h_1+h_2+q_{1}\right)^2+q_2^2}{\left(h_1+h_2-q_{1}\right)^2+q_2^2} \Biggr] \,, \\
    B(q,h_1,h_2) &= \frac{i}{8\pi} \log \Biggl[ \frac{ \left(( h_1 + h_2 - q_1)^2 + q_2^2 \right) \left( (h_1 + h_2 + q_1)^2 + q_2^2 \right) }{16M^4} \Biggr] \,.
\end{align}
\end{subequations}
In cases where $h_1 = h_2=h$, we will sometimes write only $A(q,h)$, $B(q,h)$ for brevity's sake.

Since the diagrams involve fermions, one needs to be careful in deriving the sign with which each diagram contributes, which we do in app.~\ref{app:1-loop diagrams}. Overall, the purely 1-loop contribution to $\Gamma^{i j}_{\alpha\beta\gamma\delta}$ (second term in \eqref{eq:1loopGamma}) equals
\begin{multline}
    (ig^2)^2 \biggl[ - \epsilon_{\alpha\beta} \, \epsilon_{\gamma\delta} \, \sum_{k=1}^N \, \cI^{\lambda \phantom{\sigma} \sigma}_{\phantom{\lambda} \sigma\phantom{\sigma}\lambda}(q,h_k,h_k) + \cI^{\phantom{\beta} \phantom{\lambda} \lambda}_{\beta\lambda\phantom{\lambda}\alpha}(-q, h_i, h_i) \, \epsilon_{\delta\gamma} \\[-0.5em]
    {} + \epsilon_{\beta\alpha} \, \cI^{\phantom{\delta} \phantom{\lambda} \lambda}_{\delta \lambda \phantom{\lambda} \gamma}(q,h_j,h_j) - \cI_{\beta\alpha\delta\gamma}(q_{13},h_i,-h_j) + \cI_{\beta\alpha\delta\gamma}(q_{14},h_i,h_j) \biggr] \,.
\end{multline}
The last two diagrams of Fig.~\ref{fig:1-loop-correlator} are responsible for the last two terms in the sum above:
\begin{align}
\label{eq:last two diagrams}
    (\text{\ref{fig:1-loop-4}}) + (\text{\ref{fig:1-loop-5}}) = \text{non-logarithmic terms}
    &+ g^4 \eta_{\mu\nu} \gamma^\mu_{\beta\alpha} \gamma^\nu_{\delta\gamma} \bigl[B(q_{14},h_i,h_j) - B(q_{13},h_i,-h_j)\bigr] \nn \\
    &+ g^4 \epsilon_{\mu\nu} \gamma^\mu_{\beta\alpha} \gamma^\nu_{\delta\gamma} \bigl[A(q_{14},h_i,h_j) - A(q_{13},h_i,-h_j)\bigr] \,,
\end{align}
where the non-logarithmic terms come from the second line of \eqref{eq:cI result}. Note that these diagrams do not depend on the cutoff $M$.

Using the explicit form of $\cI$ in \eqref{eq:cI result} we finally find$\,$%
\footnote{There was a typo in the sign of the $\epsilon_{\alpha\beta} \, \gamma^*_{\gamma\delta}$ term in equation (9) of \cite{Benini:2025riz} (v1), whereas it should read as in \eqref{eq:four point function explicit} here. This typo affects the correct components of $\Gamma$ that appear later in  \cite{Benini:2025riz} (v1), such that the LHS of (13) there should read $\Gamma^{mn}_{+--+}$, as in \eqref{eq:renorm cond charged} here, while the LHS of (15) should read $\Gamma^{mn}_{-++-}$. This does not affect the results of \cite{Benini:2025riz}.} 
\begin{equation}
\label{eq:four point function explicit}
\begin{aligned}
    \Gamma&^{ij}_{\alpha\beta\gamma\delta}(q_1,q_2,q_3,q_4) =  g^4 \, \epsilon_{\alpha\beta} \, \gamma^*_{\gamma\delta} \, A(q,h_j) - g^4 \, \gamma^*_{\alpha\beta} \, \epsilon_{\gamma\delta} \, A(q,h_i) + (\text{\ref{fig:1-loop-4}}) + (\text{\ref{fig:1-loop-5}}) \\
    &\quad + \epsilon_{\alpha\beta} \, \epsilon_{\gamma\delta} \biggl[ ig^2 + i\delta_g - 2(N-a) \, g^4 B(q,0) - 2a \, g^4B(q,h) + g^4 B(q,h_i) + g^4 B(q,h_j)\biggr] \,,
\end{aligned}
\end{equation}
where $h_i$ and $h_j$ are the chemical potentials of the external states of type $i$ and $j$, respectively, and we used $A(-q,h) = -A(q,h)$.
For our considerations the two relevant classes of correlators in \eqref{eq:four point function explicit} are the neutral ones, denoted by $\Gamma^{\rm neut}_{\alpha\beta\gamma\delta}$, in which the external particles have $h_i = h_j = 0$, and the charged ones, denoted by $\Gamma^{\rm ch}_{\alpha\beta\gamma\delta}$, in which both external particles have $h_i = h_j = h$.

The neutral correlator is quite simple, as $A(q,0) = 0$, and is given by
\begin{equation}
\label{eq:Gamma4Neut}
    \Gamma^{\rm neut}_{\alpha\beta\gamma\delta} = \epsilon_{\alpha\beta} \, \epsilon_{\gamma\delta} \biggl[ig^2 + i\delta_g - 2(N-a-1) \, g^4 B(q,0) - 2a \, g^4B(q,h)\biggr] + (\text{\ref{fig:1-loop-4}}) + (\text{\ref{fig:1-loop-5}}) \,. 
\end{equation}
The components of the charged-charged correlator that have a leading order contribution are$\,$%
\footnote{To compute it we used the fact that, with two lower indices, $\gamma^*_{\alpha\beta} = \begin{psmallmatrix} 0 & -1 \\ -1 & 0 \end{psmallmatrix}$. Moreover, the components shown in \eqref{eq:charged correlators components} all have $\alpha\neq\beta$ and $\gamma\neq\delta$, and those components vanish for all the matrices contributing to \eqref{eq:last two diagrams}, as can be seen by lowering the indices for the $\gamma$ matrices written in footnote~\ref{foo:sec2conv}.}
\begin{subequations}
\label{eq:charged correlators components}
    \begin{align}
        \label{eq:charged correlators -+-+}
        \Gamma_{-+-+}^{\rm ch} &= \phantom{-} ig^2 + i\delta_{g} - 2(N-a) \, g^4 B(q,0) - 2(a-1) \, g^4 B(q,h) \,, \\
        \label{eq:charged correlators +-+-}
        \Gamma_{+-+-}^{\rm ch} & = \phantom{-} ig^2 + i\delta_{g} - 2(N-a) \, g^4 B(q,0) - 2(a-1) \, g^4 B(q,h) \,, \\
        \label{eq:charged correlators -++-}
        \Gamma_{-++-}^{\rm ch} &= -ig^2 - i\delta_{g} + 2(N-a) \, g^4 B(q,0) +2(a-1) \, g^4 B(q,h) + 2 \, g^4 A(q,h) \,, \\
        \label{eq:charged correlators +--+}
        \Gamma_{+--+}^{\rm ch} &= -ig^2 - i\delta_{g} + 2(N-a) \, g^4 B(q,0) + 2(a-1) \, g^4 B(q,h) - 2 \, g^4 A(q,h) \,.
    \end{align}
\end{subequations}
All other components of $\Gamma^\text{ch}$ have either the first two or the last two spinor indices equal to each other (for example as in $\Gamma^{\rm ch}_{++--}$). The second row in \eqref{eq:four point function explicit} vanishes for these components, which are not associated to vertices of the Lagrangian \eqref{eq:renormalized Lag}. Therefore they cannot give rise to UV divergences and dynamically generated scales, and can be neglected for our purposes.

In order to gain some physical intuition for these different components of the amplitude, we show in app.~\ref{app:Physical amplitudes} that they are related to different physical scattering processes: for example, for $h>0$, $\Gamma^{\rm ch}_{+--+}$ is responsible for the scattering of a particle and a hole of one type to a particle and a hole of another type close to the upper part of the Fermi surface, where $q = (q_1, q_2) \to (2h,0)$.%
\footnote{In higher dimensions, such particle-hole pairs have been identified as the Goldstone bosons associated with the breaking of boosts due to the Fermi surface \cite{Alberte:2020eil}.} 
The component $\Gamma^{\rm ch}_{-++-}$ governs the same kind of scattering close to the lower part of the Fermi surface, where $q \to (-2h,0)$. For massless fermions we couldn't find a scattering process close to the Fermi surface that involves the other two polarizations $\Gamma^{\rm ch}_{+-+-}$ and $\Gamma^{\rm ch}_{-+-+}$, but for massive fermions they can also contribute.

We see that there are four interesting choices of momenta where the logarithms in $A(q)$ or $B(q)$ could become large and the correlators diverge. The first possibility is when $|q| \to \infty$, which turns out to be harmless because the theory is asymptotically free. The second one is when $q \to 0$, in which case the contributions from loops containing neutral fermions grow larger and larger and $B(q,0)$ diverges. The other two cases are $q_1 \to \pm 2h$, $q_2 \to 0$, so that the scattering involves fermions near the Fermi surface and loops containing the charged fermions diverge.

\subsection{The beta function}
\label{sec:beta func}

We now compute the $\beta$-function close to the points of interest mentioned above, showing that indeed perturbation theory breaks down there and new scales emerge.

First, we choose a renormalization condition in order to precisely define our coupling, which we do at the (Euclidean) scale $\vec q = \bar\mu + \vec \mu$, where $\bar\mu$ is the point we would like to approach (for instance zero momentum or the Fermi surface) whereas $\vec\mu$ is the sliding scale.%
\footnote{Here $\mu_2$ is the Wick rotated energy scale: $\mu_0 = i\mu_2$.}
Here we explicitly distinguish the 2-vector $\vec\mu = (\mu_1,\mu_2)$ from its magnitude $\mu = |\vec\mu| = \sqrt{\mu_1^2 + \mu_2^2}$. We also find it convenient to work with the kinematics 
\begin{equation}
    q_1 = q_2 = \frac{\bar \mu + \vec \mu}{2} \,,\qquad\quad q_3 = -\frac{\bar\mu}{2} - (\mu_1, 0) \,,\qquad\quad q_4 = -\frac{\bar\mu}{2} - (0,\mu_2) \,,
\end{equation}
which implies
\begin{equation}
\label{eq:renorm cond kinematics}
    q \equiv q_1 + q_2 = \bar\mu + \vec \mu \,, \qquad q_{13} \equiv q_1 + q_3 = \Bigl(-\frac{\mu_1}{2},\frac{\mu_2}{2}\Bigr) \,, \qquad q_{14} \equiv q_1 + q_4 = \Bigl(\frac{\mu_1}{2},-\frac{\mu_2}{2}\Bigr) \,. 
\end{equation}
In our conventions all the $q_i$'s are incoming.

In the UV, where $|q| \gg h$, the couplings between different species of fermions are all the same and are Lorentz covariant, so it does not matter which scattering amplitude we pick to define the coupling. At lower scales, where $h$ becomes more important, instead, the different correlators are no longer related to each other by symmetry. We then define the coupling based on the interactions between the degrees of freedom that remain light around $\bar\mu$. In order to flow towards $\bar\mu = (0,0)$ we define $g^2$ through a correlator of the neutral fermions; in order to flow towards the Fermi surface we define $g^2$ through a correlator of the charged fermions, as we detail below.%
\footnote{\label{footnote:1}%
Strictly speaking, for $\mu < h$ the couplings governing neutral and charged fermion interactions should be denoted differently, say $g_\text{n}^2$ and $g_\text{c}^2$, as they run with different beta functions. In what follows we will nevertheless denote both of them simply as $g^2$ to contrast our microscopic analysis from one based on an EFT approach, where we systematically write down all possible four-fermion couplings and match them to the UV theory. The basic physics is of course the same. Neglecting threshold corrections, the two couplings $g_\text{n}^2$ and $g_\text{c}^2$ run independently for $\mu < h$ according to \eqref{eq:beta func neutral IR} and \eqref{eq:beta func charged IR}. Their relative strength is fixed by demanding $g_\text{n}^2(h) = g_\text{c}^2(h) = g^2(h)$. For $\mu > h$ the two couplings unify and run together according to \eqref{eq:BetaUV}.}

\paragraph{The UV at \tpsb{\mu \to \infty}{mu->infinity}.}
We begin with the first point where \eqref{eq:four point function explicit} might diverge: the UV where $q = \mu \gg h$. The chemical potential can be neglected to leading order, and
\begin{equation}
    \Gamma^{ij}_{\alpha\beta\gamma\delta}(\mu) = 
     \epsilon_{\alpha\beta} \, \epsilon_{\gamma\delta} \biggl[ig^2 + i\delta_g - i \, \frac{N-1}{\pi} \, g^4 \log\left(\frac{\mu}{2M}\right) + O(h/\mu)\biggr] \,. 
\end{equation}
As anticipated, this amplitude respects the $O(2N)$ symmetry and is Lorentz covariant. We choose the physical renormalization condition
\begin{equation}
    \Gamma^{ij}_{\alpha\beta\gamma\delta}(\vec\mu) = ig^2 \, \epsilon_{\alpha\beta} \, \epsilon_{\gamma\delta} \qquad \qquad \text{(for $\mu \gg h$)}\,,
\end{equation}
which fixes the 1-loop counterterm $\delta_g = \frac{N-1}{\pi} \, g^4 \log\left(\frac{\mu}{2M}\right) + O(h/\mu)$. We note that at leading order in $h/\mu$, the counterterm (and thus the amplitude) depends on $\vec \mu$ only through its magnitude $\mu$, despite the fact that the theory breaks Lorentz invariance. As we will see below, this is also the case as we approach other points of interest.

We now denote the amplitude at some arbitrary momenta $\vec q$, given the renormalization condition above at the scale $\vec\mu$, by $\Gamma^{ij}_{\alpha\beta\gamma\delta}(\vec q,\vec\mu)$. The $\beta$-function $\beta(g^2) = \mu \frac{\partial g^2}{\partial \mu}$ is determined through the Callan--Symanzik equation, which in the absence of an anomalous dimension (as in our case) reads
\begin{equation}
    \left(\mu\frac{\partial}{\partial\mu} + \beta(g^2) \frac{\partial}{\partial g^2} \right) \, \Gamma^{ij}_{\alpha\beta\gamma\delta}(\vec q, \vec\mu) = 0  \,.
\end{equation}
The only dependence on $\mu$ in $\Gamma(\vec p,\vec \mu)$ comes from the counterterm, and so at one-loop level
\begin{equation}\label{eq:BetaUV}
    \beta(g^2) = -\mu \frac{\partial}{\partial\mu} \delta_g = -\frac{N-1}{\pi} g^4 + O(h/\mu) \qquad \qquad \text{(approaching the UV)}\,.
\end{equation}
This $\beta$ function is asymptotically free, meaning that we can increase $\mu$ as much as we want and perturbation theory holds. This is of no surprise, as in this regime the theory is practically the Gross--Neveu model without chemical potential. 

In the theory without a chemical potential, this $\beta$-function is exact at any $\mu$ and we can use it all the way to the IR, \textit{i.e.}\ scales $\mu \ll \Lambda_\text{UV}$. Perturbation theory breaks down when the coupling explodes, which at 1-loop happens at the dynamically generated scale
\begin{equation}
\label{eq:Lambda UV}
    \Lambda \approx \Lambda_\text{UV} \; e^{-\frac{\pi}{(N-1) \, g^2(\Lambda_\text{UV})}} \,.
\end{equation}

\paragraph{IR at $\boldsymbol{\bar\mu = (0,0)}$.}
Suppose we want to look at processes where the center of momentum is $q = \mu \ll h$. The light degrees of freedom there are the neutral fermions, and so we would like to define our coupling through their scattering, and our renormalization condition would read
\begin{equation}
    \Gamma^{\rm neut}_{\alpha\beta\gamma\delta}(\vec\mu) = i g^2 \, \epsilon_{\alpha\beta}\epsilon_{\gamma\delta} \,, \qquad \qquad \text{(when $\bar\mu=0$ and $\mu \ll h$)}\,.
\end{equation}
Since we are working with $\vec q_{13} = -\vec q_{14}$, and since $A(\vec \mu,0) = 0$, then only the second line in \eqref{eq:four point function explicit} does not vanish. This renormalization condition implies 
\begin{align}
    \delta_g & = - 2i(N-a-1)g^4B(\vec \mu,0) - 2iag^4 B(\vec \mu,h) \nn \\
    & = \frac{N-a-1}{\pi}g^4\log\left(\frac{\mu}{2M}\right) + \frac{a}{\pi}g^4 \log\left(\frac{h}{M}\right) + O(\mu/h) \,.
\end{align}
Repeating the same steps as above, we find that the one-loop beta function is
\begin{equation}
\label{eq:beta func neutral IR}
    \beta(g^2) = -\frac{N-a-1}{\pi} \, g^4 + O(\mu/h) \,, \qquad\qquad \text{(approaching $(0,0)$)}\,.
\end{equation}
The coupling diverges as we go to lower energies. Our original theory was defined in some UV scale $\Lambda_{UV}$ with some coupling $g^2(\Lambda_{UV})$ such that, without chemical potential, it dynamically generated the scale $\Lambda$ in \eqref{eq:Lambda UV}. Starting from the same UV Lagrangian, in the presence of chemical potential we now need to use \eqref{eq:beta func neutral IR}, and the new dynamically generated scale is
\begin{equation}
    \Lambda_1 = \Lambda_\text{UV} \; e^{-\frac{\pi}{(N-a-1) \, g^2(\Lambda_\text{UV})}} = \Lambda^{\frac{N-1}{N-a-1}} \, \Lambda_\text{UV}^{-\frac{a}{N-a-1}} \,.
\end{equation}
In order to relate to the BA results we pick the UV scale $\Lambda_\text{UV} = 2h$. Note that for perturbation theory to be consistent we have to take $h \gg \Lambda$. This allows us to identify the scale $\Lambda_1$ with the one found using the other methods,
\begin{equation}
    \Lambda_{\rm n} = \Lambda_1 = \Lambda^{\frac{N-1}{N-a-1}} \, (2h)^{\frac{a}{N-a-1}} \,.
\end{equation}
Since the loops that govern the $\beta$-function in this scale are those containing the light neutral fermions, we associate this scale with them.

The same scale $\Lambda_{\rm n}$ would have emerged had we simply given a large mass $h$ to the charged fermions, instead of turning on a chemical potential, making them decouple from the $\beta$ function in the IR. However, unlike with a mass term, adding a chemical potential leaves some charged excitations light. These are particles and holes around the Fermi surface, which we study next.

\paragraph{IR at the Fermi surface.}
Excitations near the Fermi surfaces have a momentum $|q|\approx h$, positive (negative) near the upper (lower) Fermi ``surface'' (two points in $d=2$). Without loss of generality, we focus on the upper Fermi surface. The total spatial momentum of two excitations is around $2h$, so we take $q = \bar \mu + \vec \mu$, with $\bar\mu = (2h,0)$ and $\mu \ll h$. The precise kinematics are as in \eqref{eq:renorm cond kinematics}. The light modes are charged particle and hole excitations. We define the coupling through a correlator that involves these modes. As is evident from \eqref{eq:charged correlators components}, in this regime the correlator $\Gamma$ is not Lorentz covariant. We choose the renormalization condition by demanding that a particular spinorial component of the charged-charged correlator is given purely by its tree-level contribution. We pick this condition to be
\begin{equation}
\label{eq:renorm cond charged}
    \Gamma^{\rm ch}_{+--+}(\bar \mu + \vec \mu) = ig^2 \, \epsilon_{+-} \epsilon_{-+} = -ig^2 \,, \qquad\qquad \text{when } \bar \mu = (2h,0) \text{ and }\mu \ll h.
\end{equation}
This is exactly the component that contributes to the scattering of particles and holes around the Fermi-surface, as detailed in the appendix~\ref{app:freefermions}, see \eqref{eq:particle hole different flavors}.%
\footnote{Note that the opposite convention for $\pm$ signs in the polarization is used in \cite{Benini:2025riz}.}
We comment below about other choices of spinorial indices for the amplitude. 

At 1-loop order, the correlator is \eqref{eq:charged correlators -++-} where, unlike in the neutral scattering, it includes a non-trivial contribution from the terms proportional to $\epsilon \gamma^*$. Enforcing the renormalization condition we can determine 
\begin{align}
    \label{eq:charged counterterm}
    \delta_g & = i g^4 \bigl[2(N-a) B(\bar\mu + \vec \mu,0) - 2(a-1) B(\bar\mu + \vec \mu,h) + 2A(\bar\mu + \vec \mu,h) \bigr] \nn  \\
    & = -g^4\left[ \frac{N-a}{\pi}\log\left(\frac{h}{M}\right) - \frac{a-1}{2\pi}\log\left(\frac{\mu h}{M^2}\right) + \frac{1}{2\pi}\log\left(\frac{4h}{\mu}\right) + O(\mu/h)\right] \,.
\end{align}
Once again we find that to leading order the counterterm (and thus the amplitude) depends only on the magnitude $\mu$. The one-loop $\beta$-function can be computed as before through the Callan--Symanzik equation, and reads
\begin{equation}\label{eq:beta func charged IR}
    \beta(g^2) = -\mu \frac{\partial}{\partial\mu} \delta_g = -\frac{a}{2\pi} g^4 \,.
\end{equation}
As we get closer to the Fermi surface the coupling grows, until it diverges at the dynamically generated scale
\begin{equation}
    \Lambda_2 = \Lambda_\text{UV} \; e^{-\frac{2\pi}{a \, g^2(\Lambda_\text{UV})}} = \Lambda^{\frac{2(N-1)}{a}} \, \Lambda_\text{UV}^{1-\frac{2(N-1)}{a}} \,, 
\end{equation}
which when setting $\Lambda_\text{UV} = 2h$  becomes 
\begin{equation}
    \Lambda_\text{c} = \Lambda_2 = \Lambda^{\frac{2(N-1)}{a}} \, (2h)^{1-\frac{2(N-1)}{a}} \,,
\end{equation}
which arises from the divergence of loops containing charged fermions.

The choice of polarization in $\Gamma^{\rm ch}_{+--+}$ is specific but not arbitrary, and it gives the correct scale for the breakdown of perturbation theory when approaching the Fermi surface. To see why, we consider 
a scattering amplitude with other spinorial indices, such as $\Gamma^{\rm ch}_{+-+-}(\bar\mu  + \vec \mu)$ given in \eqref{eq:charged correlators +-+-}. The major difference is that the contributions from the two $\epsilon \gamma^*$ terms in \eqref{eq:four point function explicit} now cancel each other, and so after using \eqref{eq:charged counterterm} we have
\begin{equation}
    \Gamma^{\rm ch}_{+-+-}(\bar\mu + \vec \mu) =  ig^2\left(1 + 2ig^2A(\bar\mu + \vec \mu,h) \right) \,,
\end{equation}
where we plugged in the value of the counterterm that we determined above. If we inspect the family of multi-loop diagrams coming from iterating the diagrams above, we see that each additional loop will add a factor of $2ig A(\bar\mu + \vec\mu,h)$. Ultimately, their contribution to the four point function will be, suppressing the factors coming from the external propagators,
\begin{equation}
    ig^2\sum_{\ell = 0}^\infty \bigl( 2ig^2 A(\bar\mu + \vec \mu,h) \bigr)^\ell  = \frac{ig^2}{1 + \frac{g^2}{4\pi} \log\left(\frac{\left(4h+\mu_1\right)^2+\mu_2^2}{\mu_1^2+\mu_2^2}\right)}\,.
    \label{eq:iterated loops}
\end{equation}
The denominator is always positive (and larger than 1), and therefore the correlator is always perturbatively under control as long as $g$ is under control, and will break down only at the scale $\Lambda_{\rm c}$ found above.%
\footnote{\label{foo:other schemes}%
Had we used this second correlator, $\Gamma^\text{ch}_{+-+-}$, in our renormalization condition we would have run into a pathological situation instead. Due to the lack of the additional term $A(q,h)$ in \eqref{eq:charged correlators +-+-}, the $\beta$-function would have looked like $\beta(g^2) = - \frac{a-1}{2\pi} \, g^4$, (wrongly) implying a smaller dynamically generated scale, $\tilde \Lambda_2 \approx \exp\left(-\frac{2\pi}{(a-1) \, g^2(\Lambda_\text{UV})}\right) \ll \Lambda_2$. However, had we then tried to calculate the scattering of particles and holes in $\Gamma_{+--+}^\text{ch}$, we would have found that the iterated loops contribute through a formula analogous to \eqref{eq:iterated loops} but with an opposite sign compared to before,
\begin{equation}
    -ig^2\sum_{\ell = 0}^\infty \bigl(-2ig^2 A(\bar\mu + \vec \mu,h) \bigr)^\ell  = \frac{-ig^2}{1 - \frac{g^2}{4\pi} \log\left(\frac{\left(4h+\mu_1\right)^2+\mu_2^2}{\mu_1^2+\mu_2^2}\right)}\,.
\end{equation}
The minus sign in the denominator will cause it to diverge at scales where $\mu \ll h$, hinting that we can no longer trust perturbation theory there. This divergence occurs irrespective of the choice of $\Lambda_\text{UV}$, and therefore of $\tilde \Lambda_2$, meaning that some amplitude could diverge while the coupling is still small. This signals that defining the physical coupling as that component is not a good scheme.

Near the other component of the Fermi surface, $\bar\mu = (-2h,0)$, the consistent choice requires defining the coupling through another spinorial structure, $\Gamma^\text{ch}_{-++-}$, resulting in the same dynamically generated scale $\Lambda_\text{c}$.} 

As mentioned in footnote \ref{footnote:1}, we could have alternatively used an EFT approach near the Fermi surface, writing down all possible four-fermion couplings allowed by the symmetries at the scales below $h$. One would then find different beta functions for the different couplings, naively resulting in various dynamically generated scales. But perturbation theory will break down at the largest among those, which would be $\Lambda_{\rm c}$.

\paragraph{Comments on $\boldsymbol{a = 1,\; N-1,\; N}$.}
When $a = 1$, or $a=N-1,N$, the above computation clearly does not hold, as it assumes $i\neq j$, and we do not have two distinct charged or neutral flavors $i$ and $j$ available. When one scatters fermions of the same flavor, $i = j$, there are additional diagrams, identical to those in Fig.~\ref{fig:1-loop-correlator} up to exchanging $(2\leftrightarrow4, \beta \leftrightarrow\delta)$ and an additional overall minus sign for all diagrams, coming from exchanging two fermions. The details are deferred to appendix~\ref{app:Identical Fermions}. Ultimately, one finds that near the Fermi surface $\beta = -\frac{a-1}{2\pi} \, g^4$, which for $a > 1$ will not properly regulate other amplitudes, as in footnote~\ref{foo:other schemes}. Near the neutral IR, the beta function remains the same, $\beta = -\frac{N-a-1}{\pi} \, g^4$. 

For $a=1$, however, there is no other amplitude that could diverge near the Fermi surface, so the $\beta$-function vanishes and the scale $\Lambda_\text{c}$ does not appear. This should not come as a surprise. Near the Fermi surface the contribution of the neutral fermions is negligible, and we effectively have a theory of a single massless Dirac Fermion with $g^2(\bar\psi\psi)^2$ interaction, notably dual to a compact free scalar (with chemical potential for the winding mode). The radius of the scalar is related to the coupling $g^2$ and does not run. The theory is gapless.

For $a=N-1$, the $\beta$-function for the neutral fermions vanishes and so does $\Lambda_\text{n}$. Lastly, for $a=N$, all fermions are charged and $\Lambda_\text{n}$ does not exist.

Observe that for any value of $a$, both neutral and charged fermions are lighter than the gapped fermions at $h=0$, as $\Lambda_{\text{n},\text{c}} < \Lambda \approx 2h \exp\left(-\frac{1}{ (N-1)\pi g^2}\right)$. 
Between the two, the lightest are the charged ones for $a < \frac{2(N-1)}{3}$ and the neutral ones otherwise.

We thus find from perturbative arguments alone that the physically relevant dynamically generated scale at which perturbation theory breaks down are $\Lambda_\text{n,c}$ instead of the naive scale $\Lambda$.

\section{Large \tpsb{N}{N} semiclassical analysis}
\label{sec:large-N}

In this section we perform a large $N$ analysis of the model in the limit
\begin{equation}
    g^2 \to 0 \,,\qquad N \to \infty \,,\qquad a \to \infty \,,\qquad \text{keeping} \qquad g^2 N = \lambda \,,\quad \frac{a}{N} = y \qquad \text{fixed.}
\end{equation}
As is typical in vectorial large $N$ limits, the theory classicalizes around tractable saddle points. We first analyze saddle points where the Hubbard--Stratonovich field is constant, which are a generalization of the textbook calculation of the large $N$ mass gap of the Gross--Neveu model in the absence of charges, $y=0$. We show that these solutions are unstable when $y \neq 0$, motivating us to consider more general inhomogeneous configurations.
We then show how to determine saddle points when the Hubbard--Stratonovich field is not constant, and analyze the inhomogeneous solutions which permit us to capture the physics of the theory in both high- and low-density limits.

\subsection{Setup}

We start from the Lagrangian \eqref{eq:UV} and rewrite it as
\be
\label{Lagrangian with sigma}
    \cL = \sum_{j=1}^a \bar\psi^{(j)} \bigl( i \gamma^\mu \partial_\mu - \sigma + \gamma^0 h \bigr) \psi^{(j)}
    + \sum_{j=a+1}^N \bar\psi^{(j)} \bigl( i \gamma^\mu \partial_\mu - \sigma \bigr) \psi^{(j)}
    - \frac1{2g^2} \, \sigma^2 \,,
\ee
where $\sigma = -g^2 \sum \bar\psi \psi$ is the Hubbard--Stratonovich field. Integrating it out would bring us back to the original Lagrangian. We will work in Euclidean signature. We thus set $t \equiv x^0 = -i\tau$, $d^2\vec x = -i \, d^2\vec x_\text{E}$, $\partial_0 = \parfrac{}{t} = i \, \partial_{0\text{E}} = i \parfrac{}{\tau}$, $\gamma^1 = i \gamma^1_\text{E}$. In this section we explicitly denote spacetime 2-vectors as $\vec x$, and reserve $x \equiv x^1$ for the spatial component. The Euclidean action is 
\be
    S_\text{E} = \int\! d^2\vec x_\text{E} \Biggl[ \, \sum_{j=1}^a \psi^{(j) \dag} \gamma^0 \bigl(  \parslash_\text{E} +\sigma - \gamma^0h \bigr) \psi^{(j)} + \sum_{j=a+1}^N \psi^{(j) \dag} \gamma^0 \bigl( \parslash_\text{E} +\sigma \bigr) \psi^{(j)} + \frac1{2g^2} \, \sigma^2 \Biggr]
\ee
where $\parslash_\text{E} = \gamma^0 \partial_{0\text{E}} + \gamma^1_E \partial_1$, and the path integral measure is $e^{-S_\text{E}}$.
In order to evaluate the action it is convenient to compactify the thermal direction so that the inverse temperature is $\beta$, and in order to keep the action free of IR divergences we compactify also the spatial direction to length $L$. The fermions obey antiperiodic boundary conditions along the thermal direction and periodic around the spatial one, as appropriate for a thermal computation. Ultimately, we are interested in studying the ground state of the system at large volume, therefore we take the limit $\beta \to \infty$, $L \to \infty$ and consider the \emph{grand-canonical potential density}
\begin{equation}
    \Phi = \frac{S_\text{E}}{N\beta L} \,.
\end{equation}
When evaluated on the solution that minimizes the action, this observable $\Phi_\text{on-shell}$ is the large $N$ free energy $\mathcal{F}/N$ which can be compared with the results from integrability.

The fermions can be integrated out. The resulting action is nonlocal:
\begin{equation}
\label{nonlocal action}
    \frac{S_\text{E}}N = - y \Tr \log \Bigl[\gamma^0 \bigl( \parslash_\text{E} + \sigma(\vec x) - \gamma^0 h \bigr)\Bigr] - (1{-}y) \Tr \log \Bigl[\gamma^0 \bigl( \parslash_\text{E} + \sigma(\vec x) \bigr) \Bigr] + \frac1{2\lambda} \!\int\! d^2\vec x_\text{E} \, \sigma(\vec x)^2
\end{equation}
where $\Tr$ denotes the trace of the appropriate differential operator. In the 't~Hooft limit of large $N$ with fixed $\lambda = g^2N$, the action becomes semiclassical. The equation of motion derived from this action is
\be
\label{saddle point equation}
y \tr \left[\gamma^0 \, \frac1{\gamma^0 \bigl( \parslash_\text{E} + \sigma(\vec x) - \gamma^0 h \bigr)} \bigg|_{(\vec x,\vec x)} \right]
+
(1-y) \tr \left[\gamma^0 \, \frac1{\gamma^0 \bigl( \parslash_\text{E} + \sigma(\vec x)  \bigr)} \bigg|_{(\vec x,\vec x)} \right]
= \frac1\lambda \, \sigma(\vec x) \,.
\ee
Here $\tr$ is a trace over spinor indices only, and inside the trace there is the diagonal part $G_h(\vec x, \vec x)$ of the Green function of the Dirac operator, denoted by $G_h(\vec x, \vec y) \equiv \bigl(\gamma^0 \bigl( \parslash_\text{E} + \sigma(\vec x) - \gamma^0 h \bigr) \bigr)^{-1} (\vec x,\vec y)$. The equation of motion is thus an inverse problem: we need to find a function $\sigma(\vec x)$ such the Green function $G_h(\vec x,\vec y)$ for a differential operator that depends on $\sigma(\vec x)$ solves this equation. Below we will do so by making the ansatz that $\sigma(\vec x)$ belongs to a certain family of functions, diagonalize the differential operator, compute the Green function, and solve the saddle-point equation.

In the large $N$ limit, the field $\sigma$ behaves semiclassically. In this section we will study its saddle points and find the one with minimal action density $\Phi$, which describes the phase of the system. We will first present homogeneous solutions in which $\sigma(x)$ is constant, as solving the saddle-point equation is simpler. However, these do not provide a good characterization of the dense phase since they turn out to be unstable. We will then study inhomogeneous periodic configurations belonging to the family of reflectionless ansatze \cite{Schnetz:2005ih, Schnetz:2005vh}.
Such inhomogeneous solutions describe a crystal of bound states; they capture the expected physics and match the numerical analysis. We will see that in these solutions the two scales $\Lambda_{\rm n}$ and $\Lambda_{\rm c}$ play a natural role in the high-density limit, characterizing key aspects of the solution.

\subsection[Instability of large \tps{N}{N} homogeneous configurations]{Instability of large \tpsb{N}{N} homogeneous configurations}
\label{sec:hom-instability}

We now approach the task of finding saddle-point solutions at large $N$. At zero chemical potential, the solution is a well-known constant profile for the Hubbard--Stratonovich field $\sigma(x) = \Lambda$ \cite{Gross:1974jv}. Let us start by finding the possible homogeneous solutions at finite $h \geq 0$ and $0 < y < 1$.

When $\sigma(x) = \sigma$ is a constant, we can rewrite the action in Fourier space. The large $N$ potential density for constant solutions is given by:
\be
\Phi(\sigma) = \frac{\sigma^2}{2\lambda} - y \!\int\!\! \frac{d^2p}{(2\pi)^2} \, \log \det \bigl( i \slashed{p}_E - h \gamma^0 + \sigma \bigr) - (1-y) \!\int\!\! \frac{d^2p}{(2\pi)^2} \, \log \det \bigl( i \slashed{p}_E + \sigma \bigr) \, + \, C \,,
\ee
where $\det$ is a determinant over spinor indices, we choose $\gamma^0 = - \sigma_1 = \smat{ 0 & -1 \\ -1 & 0 }$, $\gamma^1 = i \sigma_3$, $\gamma^1_\text{E} = \sigma_3 = \smat{ 1 & 0 \\ 0 & -1}$, $\gamma^0 \gamma^1_\text{E} = i \sigma_2 = \smat{0 & 1 \\ -1 & 0}$,%
\footnote{Note that this basis differs from the one used in sec.~\ref{sec:pert theory}. The physics does not depend on the choice of basis, but the computation in the next section significantly simplifies using this choice.}
and $C$ is a constant affected by the choice of counterterms, \textit{a priori} containing divergences. The equation of motion comes from extremizing the action, $\Phi'(\sigma) = 0$. We thus consider the derivative of the potential,
\be
\label{eq:Vprime-def}
\Phi'(\sigma) =  \frac{\sigma}{\lambda} - y \, \sigma \, J(h,\sigma) - (1-y) \, \sigma \, J(0,\sigma) \,,
\ee
where, as before, we put a cutoff $M$ in $p_1$ and no cutoff in $p_2$:
\be
\begin{aligned}
J(h,\sigma) 
&\equiv \int_{-M}^M \frac{dp_1}{2\pi} \int_{-\infty}^\infty \frac{dp_2}{2\pi} \; \frac{2}{ p_1^2 + (p_2+ih)^2 + \sigma^2} \\
&= \frac{1}{2\pi} \log \left( \frac{4M^2}{\sigma^2} \right) + \Theta(h-\sigma) \, \frac{1}{\pi}
\log \biggl( \frac{h - \sqrt{ h^2 - \sigma^2}}\sigma \,\biggr) + \mathcal{O}\bigl( M^{-2} \bigr) \,.
\end{aligned}
\ee
For $y=0$, we recover the well known relation between the 't~Hooft coupling $\lambda$ defined at the cutoff $M$ and the expectation value $\sigma_0$ of $\sigma$:
\be
\label{thooft-to-cutoff}
\frac{1}{\lambda}  = \frac{1}{2\pi} \log\left(\frac{4M^2}{\sigma_0^2}\right) \,,
\ee
as the cutoff $M$ is sent to infinity. This can also be expressed in terms of the dynamically generated scale $\Lambda$:
\be
\label{eq2}
\sigma_0 = 2M \, e^{-\frac{\pi}{\lambda}} \equiv \Lambda \,.
\ee
Such a scale determines the mass gap at leading order at large $N$, as apparent from \eqref{Lagrangian with sigma}:
\begin{equation}
\label{eq:gap-largeN}
\frac{m}{\Lambda} = 1 + \mathcal{O}\Bigl( \frac{1}{N} \Bigr) \,,
\end{equation}
therefore in the large $N$ analysis we can use $m$ and $\Lambda$ interchangeably (see footnote~\ref{foo: m/Lambda finite N} for the finite $N$ version of this equation). The scale $\Lambda$ can be used to rewrite \eqref{eq:Vprime-def} in a cutoff independent way as
\begin{equation}
    \Phi'(\sigma) = \frac{\sigma}{\pi} \, \Biggl[ \frac12 \log\frac{\sigma^2}{\Lambda^2} - y \, \Theta(h-\sigma) \, \log \biggl( \frac{h - \sqrt{ h^2 - \sigma^2}}\sigma \,\biggr) \Biggr] \,.
\end{equation}
In order to find the regularized action (or free energy), we integrate this expression. We determine the $h$-dependent integration constant as follows. At $\sigma = 0$, \textit{i.e.} around the naive perturbative vacuum, the free energy should match the one for massless fermions which, at leading order in $N$, are free. We then have 
\be
\Phi\bigl( \sigma = 0; h \bigr) = c - \frac{y \, h^2}{2\pi} \,.
\ee
Here $c$ is an $h$-independent constant which in turn is fixed by requiring that the free energy vanishes for $h=0$ at the nonperturbative vacuum $\sigma = m$, hence $\Phi\bigl( \sigma = m = \Lambda; h=0 \bigr) = 0$. In this way we find $c= \Lambda^2/4\pi$.
The free energy is then uniquely fixed to be
\be
\label{eq:Phi-hom}
\Phi(\sigma; h)= \frac{\sigma^2}{4\pi} \log \frac{\sigma^2 }{e \, \Lambda^2} - \frac{y}{2\pi} \, \Theta(h-\sigma ) \biggl[ h \sqrt{h^2 - \sigma^2} + \sigma^2 \log\biggl( \frac{h - \sqrt{h^2 - \sigma^2}}\sigma \,\biggr) \biggr] + \frac{\Lambda^2}{4\pi} \,.
\ee
This matches the expression we will obtain with a more careful treatment of counterterms in the case of inhomogeneous $\sigma$.%
\footnote{App.~\ref{app:homLimit} shows how \eqref{eq:Phi-hom} can be reproduced as a limit of the analysis used for inhomogeneous configurations, discussed from section~\ref{subsec:InhoSol} onward.}

Let us inspect the constant solutions to the equation of motion $\Phi'(\sigma)=0$ when $h>0$.
First, there is the trivial solution $\sigma=0$ with $\Phi\bigl( \sigma = 0; h \bigr) = \frac{\Lambda^2}{4\pi} - y \frac{h^2}{2\pi}$.
Second, as long as $h \leq \Lambda$, there is the nonperturbative solution $\sigma = \sigma_0 = \Lambda$, whose action is
\begin{equation}
\label{eq:phi-m}
\Phi(\sigma_0) = 0 \,.
\end{equation}
Third, two new real saddles $\sigma_\pm$ appear when $h>h_y$, where
\begin{equation}
\label{h-y}
h_y = (1-y)^{-\frac{1}{2}(1-y)} \, (1+y)^{-\frac{1}{2}(1+y)} \, \Lambda \,.
\end{equation}
The values of $\sigma$ at these saddles are at the solutions of 
\begin{equation}
\label{hard-min}
\frac{\sigma}{\Lambda} = \biggl( \frac{h - \sqrt{h^2 - \sigma^2} }\sigma \,\biggr)^y \,.
\end{equation}
These solutions $\sigma_\pm(h)$ satisfy $h > \sigma_+ > \sigma_- > 0$. The higher solution, $\sigma_+$, increases with $h$ until it ceases to exist at $h = \Lambda$, where it collides with the saddle $\sigma = \Lambda$ and they disappear together. The lower solution, $\sigma_-$, continues to exist for all $h > h_y$ and monotonically decreases towards $0$. From \eqref{hard-min}, one can read that for $h\gg \Lambda$ it scales with the neutral fermion scale:
\be
\label{eq5}
\sigma_- \,\approx\, 2h \, \biggl( \frac{\Lambda}{2h} \biggr)^{\frac{1}{1-y}} = \Lambda_{\rm n} \,.
\ee

We are ready to determine which one of these solutions is the dominant one. The values $\Phi(\sigma = 0)$ and $\Phi(\sigma_+)$ are always local maxima, while $\Phi(\sigma_0)$ and $\Phi(\sigma_-)$ are always local minima. For $h < h_y$, the global minimum is at $\sigma_0$. At $h = h_y$, when the solutions $\sigma_\pm$ appear, $\sigma_0$ is still the global minimum. However, $\Phi(\sigma_-)$ decreases monotonically with $h$ and at some critical value $h_* < \Lambda$ the two saddles have an equal action: $\Phi(\sigma_0; h_*) = \Phi(\sigma_-;h_*)$. From that point on, when $h > h_*$, the global minimum among the homogeneous solutions switches to $\Phi(\sigma_-)$. However, there is a crucial feature that this analysis misses: at some $h_{\rm pt} < h_*$ the $\sigma_-$ solution becomes unstable with respect to certain inhomogeneous perturbations, and the true minimum of the system is described by a completely different profile for $\sigma$.

\begin{figure}[t]
\centering
\begin{subfigure}{0.495\textwidth}
\centering
\includegraphics[width=0.95\textwidth]{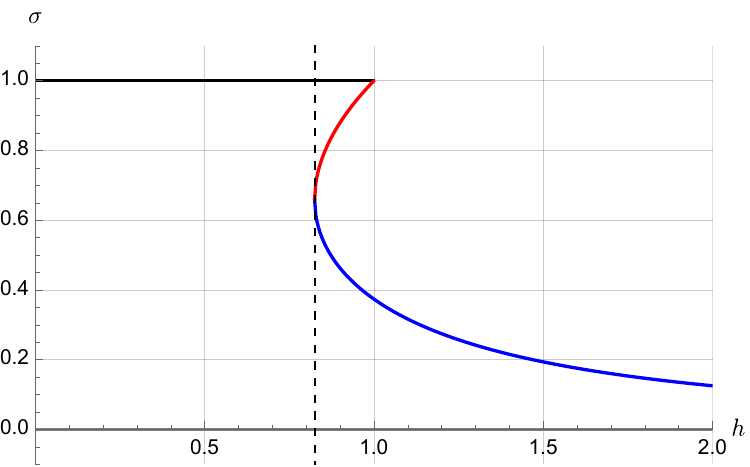}
\caption{\label{fig-saddles}%
$\sigma$}
\end{subfigure}
\hfill
\begin{subfigure}{0.495\textwidth}
\centering
\includegraphics[width=0.95\textwidth]{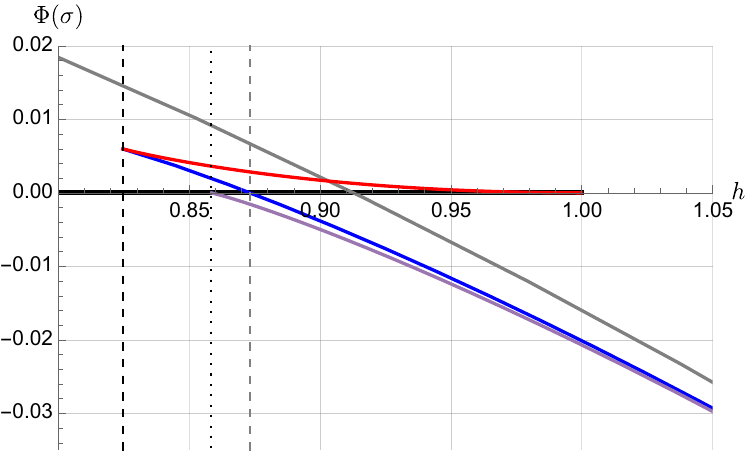}
\caption{\label{fig-action}%
$\Phi(\sigma)$}
\end{subfigure}
\caption{\label{fig-saddles-action}%
The different constant saddles (left) and their action (right) as a function of $h$ for $y=6/10$ in units where $\Lambda=1$. The gray and black solid lines are $\sigma=0$ and $\sigma=\Lambda$, respectively, whereas the red and blue lines correspond to $\sigma_+$ and $\sigma_-$, respectively. The vertical dashed black line marks $h_y$ as defined in \eqref{h-y}. For the plot on the right, $h_*$ is the point where the black and blue lines cross and is marked by the dashed gray line. Meanwhile, the dotted line marks $h_{\rm pt} = \frac{2}{\pi y}\sin\left(\tfrac{\pi y}{2}\right)$ where the phase transition happens, and the purple line is the action of the inhomogeneous solution.}
\end{figure}

In Figure~\ref{fig-saddles-action} we illustrate an example of the constant saddles and their action, including the contrast with the free energy of the inhomogeneous solution given in \eqref{grand potential on-shell y<1}.

\subsubsection{Instability analysis}

It was pointed out in \cite{gn-inst} that, in the case $y=1$, the homogeneous solutions becomes unstable. In this subsection we extend that analysis to general $y$ and find, again, that for $h$ higher than $h_*$ the least-action constant solution becomes unstable to spatially-modulated perturbations. This instability will also give a clue of the periodicity of the actual inhomogeneous solution.

To evaluate the stability of the solution we calculate the second derivative of the action with respect to $\sigma$. In order to simplify the calculation, we consider variations with only spatial momentum but no temporal momentum. This is physically motivated because we expect the crystalline phase to vary in space but not in time. The derivative has two components: a constant term coming from the term $\sigma^2$ in the action, and a bubble integral coming from the determinant:
\begin{equation}
    \Gamma^{(2)}(h,\sigma,q) = \frac{\delta^2 S}{\delta \sigma(q) \, \delta \sigma(-q)} = 
    \frac{1}{\lambda}
    \;-\;\;
    \begin{tikzpicture}[line width=0.6, baseline={(0,-0.08)}]
    \draw (0,0) circle [radius = 0.5];
    \draw [-{Latex[length=7]}] (0.05, 0.49) -- ++(-0.2, 0);
    \draw [-{Latex[length=7]}] (-0.05, -0.49) -- ++(0.2, 0);
    \draw [decorate, decoration=snake] (0.5, 0) -- (1.25, 0);
    \draw [decorate, decoration=snake] (-0.5, 0) -- (-1.25, 0);
    \end{tikzpicture}
    \; .
\end{equation}
The solution $\sigma$ is stable when $\Gamma^{(2)}(h,\sigma,q) > 0$, and unstable when $\Gamma^{(2)}(h,\sigma,q) < 0$.

The calculation of the bubble diagram in the presence of a constant $\sigma$ is similar to the calculation of Feynman diagrams in section~\ref{sec:pert theory}, but with an added mass. The integral itself is 
\begin{align}
& \CI(h, \sigma, q) = \\
& 2\int_{-M}^{M} \frac{dp_1}{2\pi} \int_{{\mathbb R}} \frac{dp_2}{2\pi} \,
\frac{\frac14 q^2 - p_1^2 - (p_2+ih)^2 + \sigma^2 }
{
\Bigl( \bigl( p_1 + \frac{q_1}2 \bigr)^2 + \bigl( p_2 + \frac{q_2}{2} + ih \bigr)^2 + \sigma^2 \Bigr)
\Bigl( \bigl( p_1 - \frac{q_1}{2} \bigr)^2 + \bigl( p_2 - \frac{q_2}{2} + ih \bigr)^2 + \sigma^2 \Bigr)
} \,. \nn
\end{align}
From now on we assume that $q_\mu$ is purely in the spatial direction, \textit{i.e.} $q_2=0$, and denote $q_1$ simply by $q$. This integral can be done by Jordan's lemma in $p_2$ followed by direct integration in $p_1$, and has been calculated in \cite{gn-inst}. In our case, it is useful to separate the $h$-dependent part of the integral as follows:
\begin{align}
\CI(h,\sigma,q) &= \CI(\sigma, q) + \mathcal{J} \bigl( h, \sigma, q \bigr) \,, \\
\CI(\sigma,q) &= \frac{1}{2\pi}\log\left( \frac{\sigma^2}{4M^2} \right) + \frac{1}{\pi} \sqrt{1+\frac{4\sigma^2}{q^2}} \, \arccoth\left( \sqrt{1+\frac{4\sigma^2}{q^2}}\right) \,, \\
\mathcal{J}(h,\sigma,q) &= - \Theta(h-\sigma) \, \frac{1}{\pi} \Biggl[ \log\biggl( \frac{ h - \sqrt{h^2 - \sigma^2}}\sigma \biggr) \nn \\
& \qquad\qquad\qquad\qquad\qquad
{} + \sqrt{1+\frac{4\sigma^2}{q^2}} \;\; \mathfrak{T} \left( \sqrt{1-\frac{\sigma^2}{h^2}} \; \sqrt{1+\frac{4\sigma^2}{q^2}} \right) 
\Bigg] \,, \label{eq:whereisA}
\end{align}
where the function $\mathfrak{T}$ appearing in \eqref{eq:whereisA} equals
\begin{equation}
\mathfrak{T}(x) = \Theta(x-1) \, \arccoth(x) + \Theta(1-x) \, \arctanh(x) = \frac{1}{2} \arctanh\left( \frac{2x}{1+x^2} \right)
\end{equation}
and, as before, we assume that $h > 0$. The bubble is then given by
\begin{equation}
    \begin{tikzpicture}[line width=0.6, baseline={(0,-0.09)}]
    \draw (0,0) circle [radius = 0.5];
    \draw [-{Latex[length=7]}] (0.05, 0.49) -- ++(-0.2, 0);
    \draw [-{Latex[length=7]}] (-0.05, -0.49) -- ++(0.2, 0);
    \draw [decorate, decoration=snake] (0.5, 0) -- (1.25, 0);
    \draw [decorate, decoration=snake] (-0.5, 0) -- (-1.25, 0);
    \end{tikzpicture}
\;=\, 
- (1-y) \; \CI(0,\sigma,q) - y \; \CI(h,\sigma,q) = - \; \CI(\sigma,q) - y \; \mathcal{J}(h,\sigma,q) \,.
\end{equation}
Using the gap equation \eqref{thooft-to-cutoff} we write the finite expression
\begin{equation}
\label{eq:gamma-two}
\begin{multlined}
\Gamma^{(2)}(h,\sigma,q) =
\frac{1}{2\pi} \log \left( \frac{\sigma^2}{\Lambda^2} \right) + \frac{1}{\pi} \sqrt{1 + \frac{4 \sigma^2}{q^2}} \, \arccoth\left( \sqrt{1+\frac{4 \sigma^2}{q^2}} \right) \\
- \Theta(h-\sigma) \, \frac{y}{\pi} \, \Biggl[ \log \left( \frac{h - \sqrt{h^2 - \sigma^2}}\sigma \right) + \sqrt{1 + \frac{4 \sigma^2}{q^2}} \;\; \mathfrak{T}\left( \sqrt{1-\frac{\sigma^2}{h^2}} \; \sqrt{1 + \frac{4 \sigma^2}{q^2}} \right) \Biggr] \,.
\end{multlined}
\end{equation}
For $y=1$ this reproduces the results of \cite{gn-inst}.

\begin{figure}[t]
\centering
\includegraphics[width=0.65\textwidth]{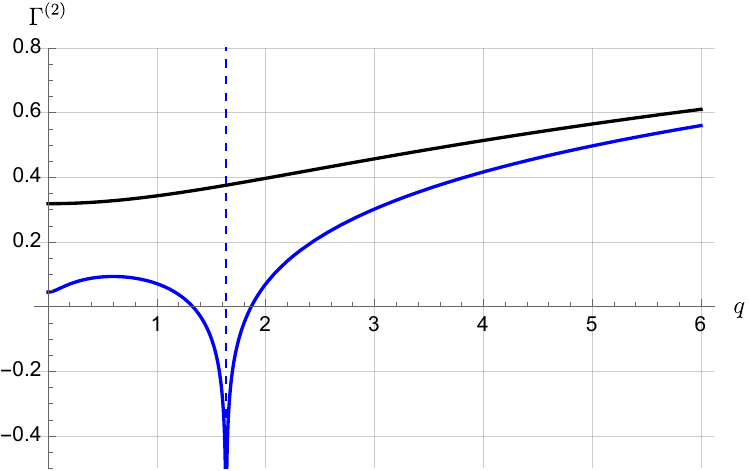}
\caption{\label{fig-gamma}%
$\Gamma^{(2)}(h, \sigma, q)$ for $y=6/7$, $m=1$ and $h=0.82$, which satisfies $h>h^*$ such that $\sigma_-$ is the global minimum. The black line is $\Gamma^{(2)}(h,\sigma_0, q)$ and the blue lines represent $\Gamma^{(2)}(h,\sigma_-, q)$. The dashed line marks $q=2\sqrt{h^2-\sigma_-^2}$, around which $\Gamma^{(2)}(h, \sigma_-, q)$ signals an instability.}
\end{figure}

At $\sigma=0$ we can simplify \eqref{eq:gamma-two} as
\begin{equation}
\Gamma^{(2)}(h,0,q) =\frac{1}{2\pi} \log\biggl( \frac{q^2}{\Lambda^2} \biggr) + \frac{y}{4 \pi } \log \biggl[ \biggl( 1 - \frac{4 h^2}{q^2} \biggr)^{\!2} \,\biggr] \,,
\end{equation}
which always has an instability at $q \to 0$, as expected since it is a local maximum.%
\footnote{The case $y=1$, analyzed in \cite{gn-inst}, is slightly different. In that case, there is an instability around $q \approx 0$ only for $h<\Lambda/2$, while the instability around $q \approx 2h$ is always present.}
Another useful case is $\sigma = m = \Lambda > h$,
\begin{equation}
\Gamma^{(2)}(h,m,q) = \frac{1}{\pi} \, \sqrt{1 + \frac{4 \Lambda^2}{q^2}} \; \arccoth \left( \sqrt{1+\frac{4 \Lambda^2}{q^2}} \right)>0 \,, \quad\text{for}\quad m > h \,,
\end{equation}
which is stable for all momenta $q$.

It is also useful to observe that for $h<\sigma$ the function is monotonic in $q \geq 0$ and 
\begin{equation}
\Gamma^{(2)}(h<\sigma, \sigma, 0) = \frac{ \log(\sigma/\Lambda) + 1}{\pi} \,.
\end{equation}
Thus if $h < \sigma < \Lambda/e$ there is an instability at $q=0$.
On the other hand, if $h>\sigma$ then at $q^2 = 4 (h^2 - \sigma^2)$ the function $\Gamma^{(2)}$ diverges to negative infinity since $\mathfrak{T}(1) = +\infty$, triggering an instability. Thus a saddle with constant $\sigma$ is only stable as long as $h<\sigma$ and $\sigma >\Lambda/e$. A plot of $\Gamma^{(2)}(q)$ around the two local minima is given in Figure~\ref{fig-gamma}.

\begin{figure}[t]
\centering
\includegraphics[width=0.65\textwidth]{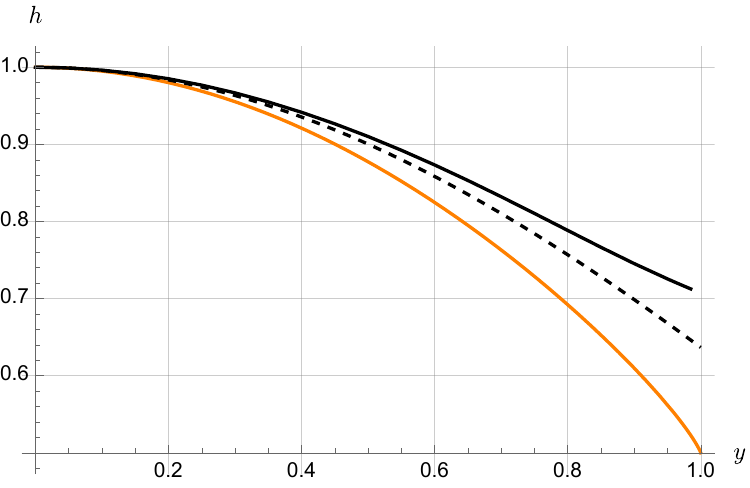}
\caption{\label{fig-h-y}%
Comparison between the functions $h_y$ (orange) where the constant saddles $\sigma_\pm$ appear, $h_\text{pt}$ (dashed) equal to the mass/charge ratio and where the actual phase transition occurs, and $h_*(y)$ (black) where the constant saddle $\sigma_-$ becomes subdominant with respect to $\sigma_0$.}
\end{figure}

Recalling the results of the previous section, we have that the dominant constant saddle is $\sigma_0$ for $0< h <h_*$ and $\sigma_-$ for $h > h_*$. According to the previous criterion, the saddle $\sigma_0$ is stable but the saddle $\sigma_-$ is not.
Naively, one could guess that the phase transition to the space-dependent solution occurs at $h_*$, when $\sigma_0$ becomes subdominant with respect to the unstable $\sigma_-$. However, like in the case $y=1$, this overestimates the transition point,  which actually takes place when the particle with the smallest mass-to-charge ratio starts to populate the vacuum, as we will later see from both semiclassical and integrability arguments. This bound state is the one with $a = yN$, whose mass/charge ratio at large $N$ is given by 
\begin{equation}
\frac{m_a}{q_a} = \frac{2 \Lambda}{\pi y}\sin \frac{\pi y}{2} + \mathcal{O} \biggl( \frac{1}{N} \biggr) \,.
\end{equation}
The actual phase transition occurs at $h_\text{pt} = \frac{m_a}{q_a}$.  In Figure~\ref{fig-h-y} we plot $h_*$ as a function of $y$ and compare it with $h_{\rm pt}$. Like in the case $y=1$ in \cite{gn-inst}, the guess for the first-order phase transition overestimates the phase transition, and $h_*\rightarrow\tfrac{1}{\sqrt{2}}$ as $y\rightarrow 1$.

The nature of the instability, however, does offer clues about the inhomogeneous solutions we will find in the next section. At large $h$, the unstable minimum is $\sigma_- \approx \Lambda_{\rm n}$ and the instability is at spatial momentum $q_1 \approx 2h$, thus anticipating the form \eqref{eq:intro-sigma-osc}.

\subsection{Inhomogeneous solutions}
\label{subsec:InhoSol}

As we established in the previous section, homogeneous solutions are not stable saddle points of the large $N$ semiclassical analysis. One must then inspect inhomogeneous solutions. This program has been carried out in the case $y=1$, where a crystal of kink-antikink bound states can be captured by analytical ansatze even in the presence of a mass term, see \textit{e.g.}\ \cite{Thies:2003br, Thies:2003kk, Thies:2005wv, Thies:2006ti, Thies:2024anc, Schnetz:2004vr, Schnetz:2005vh, Schnetz:2005ih}. In this section we generalize the analysis to $y<1$, though we restrict to zero temperature and no mass term. The analysis that follows makes heavy use of elliptic functions of various kinds. We refer the reader to appendix~\ref{app:ellipticFunbasics} for their definitions and main properties.

\subsubsection{Action and diagonalization problem}
\label{sec: saddle point analysis}

Let us go back to the saddle-point equation \eqref{saddle point equation} for $\sigma(x)$ with the goal of finding non-constant solutions. In order to compute the Green function and the action in that case, we need to diagonalize a Dirac operator. It is convenient to work at finite temperature $T = 1/\beta$, namely on a temporal circle of length $\beta$, and take the $T \rightarrow 0$ limit in the end. We want to solve the eigenvalue problem:
\be\label{eq:DopeDefi}
\mathcal{D} \Psi(\tau,x) = E \, \Psi(\tau,x) \qquad\text{where}\qquad \mathcal{D} \equiv \gamma^0 \bigl( \parslash_\text{E} + \sigma(x) - \gamma^0 h \bigr) \;.
\ee
We assume that $\sigma$ only depends on $x$ (space) and not on Euclidean time ($\tau$). A thermal computation requires imposing  antiperiodic boundary conditions for fermions, and so we decompose
\be
\Psi(\tau,x) = \beta^{-\frac12} \, e^{\frac{i\pi}\beta (2k+1) \tau} \, \psi_\omega(x) \;,
\ee
which gives the equation
\be
\label{diagonalized energies}
\biggl( \frac{i\pi}\beta (2k+1) - h + \gamma^0 \gamma^1_\text{E} \, \parfrac{}{x} + \gamma^0 \sigma(x) \biggr) \psi_\omega(x) = E_{k,\omega} \, \psi_\omega(x) \;.
\ee
We choose $\gamma^0 = - \sigma_1 = \smat{ 0 & -1 \\ -1 & 0 }$, $\gamma^1 = i \sigma_3$, $\gamma^1_\text{E} = \sigma_3 = \smat{ 1 & 0 \\ 0 & -1}$, $\gamma^0 \gamma^1_\text{E} = i \sigma_2 = \smat{0 & 1 \\ -1 & 0}$.%
\footnote{These are the same gamma matrices as in the case of constant $\sigma$, but they differ from the ones used in sec.~\ref{sec:pert theory}. The physics does not depend on this choice, but it significantly simplifies the following analysis.}
We use components $\psi = \smat{ \phi_+ \\ \phi_-}$. We also set the label $\omega$ such that
\be
    E_{k,\omega} = \frac{i\pi}\beta (2k+1) + \omega - h \;,
\ee
so that we are left with the differential equations:
\be
\label{system for eigenfunctions 1st order}
    \left( \begin{matrix} 0 & \parfrac{}{x} {-} \sigma \\ - \parfrac{}{x} {-} \sigma & 0 \end{matrix} \right) \psi_\omega = \omega \, \psi_\omega
    \qquad\text{that is}\qquad
    \left\{ \begin{aligned} \bigl[ - \tfrac{\partial}{\partial x} - \sigma(x) \bigr] \phi_+(x) &= \omega \, \phi_-(x) \;, \\
    \bigl[ \tfrac{\partial}{\partial x} - \sigma(x) \bigr] \phi_-(x) &= \omega \, \phi_+(x) \;.
    \end{aligned} \right.
\ee
An important but simple observation is that $h$ shifts only the eigenvalue $E_{k,\omega}$, and thus the eigensystem in the $\omega$ basis is the same for the charged ($h\neq 0$) and neutral ($h= 0$) sector. Notice as well that if $(\phi_+, \phi_-)$ solve the system with eigenvalue $\omega$, then $(\phi_+, - \phi_-)$ solve it with eigenvalue $-\omega$. From $\bar\psi_\omega \psi_\omega = - 2 \re (\phi_+^* \phi_-)$ it also follows that
\be
\bar\psi_\omega \psi_\omega = - \bar\psi_{-\omega} \psi_{-\omega} \;.
\ee
The two equations in (\ref{system for eigenfunctions 1st order}) can be decoupled to:%
\footnote{The system (\ref{system for eigenfunctions 1st order}) implies both equations in (\ref{eqn for eigenfunctions 2nd order}). If $\omega\neq0$, each one of the equations in (\ref{eqn for eigenfunctions 2nd order}) implies the system (\ref{system for eigenfunctions 1st order}). For instance, if $\phi_+(x)$ solves (\ref{eqn for eigenfunctions 2nd order}), then define $\phi_-(x) = \frac1\omega \bigl[ - \parfrac{}{x} - \sigma(x) \bigr] \phi_+(x)$ so that the pair $(\phi_+, \phi_-)$ solves the system (\ref{system for eigenfunctions 1st order}). If $\omega=0$ it is more subtle. Suppose that $\phi_+(x)$ solves (\ref{eqn for eigenfunctions 2nd order}). One possibility is that $\bigl[ - \parfrac{}{x} - \sigma(x) \bigr] \phi_+(x) = 0$. If not, define $\phi_- = \bigl[ - \parfrac{}{x} - \sigma(x) \bigr] \phi_+$ and then $\bigl[ \parfrac{}{x} - \sigma(x) \bigr] \phi_-(x) = 0$.}
\be
\label{eqn for eigenfunctions 2nd order}
\biggl( - \frac{\partial^2}{\partial x^2} \mp \sigma'(x) + \sigma(x)^2 \biggr) \phi_\pm(x) = \omega^2 \, \phi_\pm(x) \;.
\ee
Assuming that the eigenfunctions form a complete and orthonormal set of functions on the cylinder with respect to the Hermitian product defined by $\dagger$, the Green function is given by $G(\vec x, \vec y \mkern2mu ) = \sum_n \frac1{\lambda_n} \, \Psi_n(\vec x \mkern1mu ) \, \Psi^\dag_n(\vec y \mkern1mu )$, where $\vec x, \vec y$ are 2-dimensional coordinates, $n$ labels the eigenfunctions, and $\lambda_n$ are the eigenvalues. As we will discuss, it will be important to identify the space of functions that are part of the Hilbert space. The traces on the LHS of the saddle-point equation (\ref{saddle point equation}) then read:
\be
\label{trace of Green function}
\tr \bigl( G(\vec x, \vec x \mkern1mu ) \gamma^0 \bigr) = \sum_{k,\omega} \tr \left(\frac{ \Psi_{k,\omega}(\vec x \mkern1mu ) \, \wb\Psi_{k,\omega}( \vec x \mkern1mu ) }{E_{k, \omega}}\right) = \sum_{k,\omega} \frac{ \bar\psi_\omega(x) \, \psi_\omega(x) }{ \beta \, E_{k,\omega}} \;,
\ee
where the dependence on $h$ is only in $E_{k, \omega}$ and the dependence on $\tau$ drops.
We consider a periodic $\sigma(x)$ with period $\ell$.%
\footnote{The period of sigma can be much smaller than the volume of the space $L$ described before, and can remain finite when $L \to \infty$.}
Therefore the system \eqref{system for eigenfunctions 1st order} has a continuous spectrum of solutions by Bloch's theorem, with a density of states $\rho(\omega)$. These solutions are periodic up to a phase, such that
\begin{equation}
\label{eq:def Bloch momentum}
    \psi_\omega(x + \ell) = e^{-i \, \ell \, p(\omega)} \, \psi_\omega(x) \,,
\end{equation}
and $p(\omega)$ is called the Bloch momentum.
Using the symmetry of the spectrum, we write
\be
\tr \bigl( G(\vec x, \vec x \mkern1mu ) \gamma^0 \bigr) = \int_0^\infty \! d\omega\, \rho(\omega) \sum_{k=-\infty}^\infty \biggl( \frac1{ \beta \, E_{k,\omega}} - \frac1{\beta \, E_{k,-\omega}} \biggr) \, \bar\psi_\omega(x) \, \psi_\omega(x) \;.
\ee
Using $E_{k,\omega}$ in (\ref{diagonalized energies}), the sum over $k$ converges and the saddle-point equation (\ref{saddle point equation}) becomes
\begin{multline}
\label{thermal saddle point equation}
y\int_0^\infty \! d\omega \, \rho(\omega) \, \biggl( \frac1{1+ e^{\beta(h - \omega)}} - \frac1{ 1 + e^{\beta(h+\omega)}} \biggr) \, \bar\psi_\omega(x) \, \psi_\omega(x) \\
{} + (1-y) \int_0^\infty \! d\omega \, \rho(\omega) \, \biggl( \frac1{1+ e^{-\beta \omega}} - \frac1{ 1 + e^{\beta\omega}} \biggr) \, \bar\psi_\omega(x) \, \psi_\omega(x)
= \frac{\sigma(x)}\lambda \;.
\end{multline}
In the zero-temperature limit $\beta \to \infty$ (with $h \geq 0$) the saddle-point equation reduces to
\be
\label{EOM with partial chem pot}
(1-y) \int_0^h d\omega \, \rho(\omega) \, \bar\psi_\omega \psi_\omega(x) + \int_h^\infty d\omega \, \rho(\omega) \, \bar\psi_\omega \psi_\omega(x) = \frac1\lambda \, \sigma(x) \;.
\ee
Note that the integral  still requires regularization for large values of $\omega$, which we will address with a hard cutoff later. 

We can also use the diagonalized basis to evaluate the logarithmic terms in the action. Let
\be
I \bigl[ \sigma(x); h \bigr] = \Tr \log \mathcal{D} = \Tr \log \bigl[\gamma^0 \bigl( \parslash_\text{E} + \sigma(x) -\gamma^0 h\bigr)\bigr] \;.
\ee
For $h>0$ we find 
\begin{align}
I[\sigma;\mu] &= \int_\mathbb{R} d\omega \, \rho(\omega) \log \left[\, \prod_{k=-\infty}^{+\infty} \biggl( \frac{i\pi}\beta (2k+1) + \omega - h \biggr) \right] \\
&= \int_0^\infty d\omega \, \rho(\omega) \log \Bigl[ \bigl( 1 + e^{- \beta\omega - \beta h} \bigr) \bigl( 1 + e^{\beta \omega - \beta h} \bigr) \Bigr] + \text{$\sigma$-independent divergent terms} \,. \nn
\end{align}
Following \cite{Schnetz:2005ih}, we discard the $\sigma$-independent divergent terms which can be seen as proportional to the trace of the identity operator. In the $\beta \to \infty$ limit we get
\be
\label{nonlinear action at zero temp}
I[\sigma;\mu] = \beta \int_h^\infty d\omega \, \rho(\omega) \, (\omega - h) + \mathcal{O}(\beta^0) \;.
\ee
We recognize the leading term as $\beta$ times (minus) a free energy.%
\footnote{The reason why the integral is minus the free energy is that it should be regarded as an integral over the occupied states. These form the Dirac see and correspond to $\omega<0$.}
The normalized action, or density of grand-canonical potential, is then given by
\begin{equation}
\label{eq:def-Phi}
    \Phi = \frac{S_{\rm E}}{NL\beta} = - y \int_h^\infty \! d\omega \, \rho(\omega) \, (\omega - h) - (1-y) \int_0^\infty \! d\omega \, \rho(\omega) \, \omega + \frac{1}{2\lambda} \, \langle \sigma^2 \rangle - C_0 - y \, h \, C_1 \;,
\end{equation}
where $C_0$ and $C_1$ are counterterms for the background energy and charge, respectively, which will cancel polynomial divergences.%
\footnote{The counterterms can also be phrased as a counterterm for the background metric $\int\! \sqrt{g}$ and for the background gauge field $\int\!\sqrt{g}A_\mu J^\mu$, respectively.}

\subsubsection{Reflectionless ansatz}

In this subsection we review the ansatz proposed in \cite{Schnetz:2005ih,Schnetz:2005vh} and generalize it to the problem at hand. 
The ansatz is often called the \emph{reflectionless ansatz} in the literature, because if one solves the Dirac equation in the potential $\sigma(x)$, the reflection coefficient vanishes and the wavefunction is entirely transmitted up to a phase \cite{Dashen:1975xh,Feinberg:2003qz}.
We study the following family of profiles for $\sigma$:
\be
\label{eq:sigma-prof}
\sigma(x) = A \, \mathfrak{m} \sn(b| \mathfrak{m}) \sn(Ax| \mathfrak{m}) \sn(Ax+b| \mathfrak{m}) + A \, \frac{\cn(b| \mathfrak{m}) \dn(b| \mathfrak{m})}{\sn(b| \mathfrak{m})} \;,
\ee
where $\sn$, $\cn$, $\dn$ are Jacobi elliptic functions (see \eqref{eq:Ellipt1}--\eqref{eq:Ellipt5} for their definitions), whereas $A$, $\mathfrak{m}$, $b$ are parameters to be determined. When not specified, the parameter of the elliptic functions is $\mathfrak{m}$.
Note that $A$ has dimension of mass and the other two are dimensionless. The explicit eigenfunctions $\psi_\omega(x)$ that solve \eqref{diagonalized energies} with this ansatz can be found in appendix~\ref{app:eigenfuncs}.
The ansatz can be motivated from the fact that the resulting ``potentials'' $u_\pm$ defined in \eqref{potentials u+-} correspond to periodic sums of the potentials created by a $\sigma(x)$ with a single bound state, see \cite{Schnetz:2005ih, PhysRevA.44.R2251}. It can also be motivated through an inverse scattering problem, see \textit{e.g.}{} \cite{Dashen:1975xh, Schnetz:2005ih}. 

\begin{figure}[t]
\centering
\begin{subfigure}{0.495\textwidth}
\centering
\includegraphics[width=0.95\textwidth]{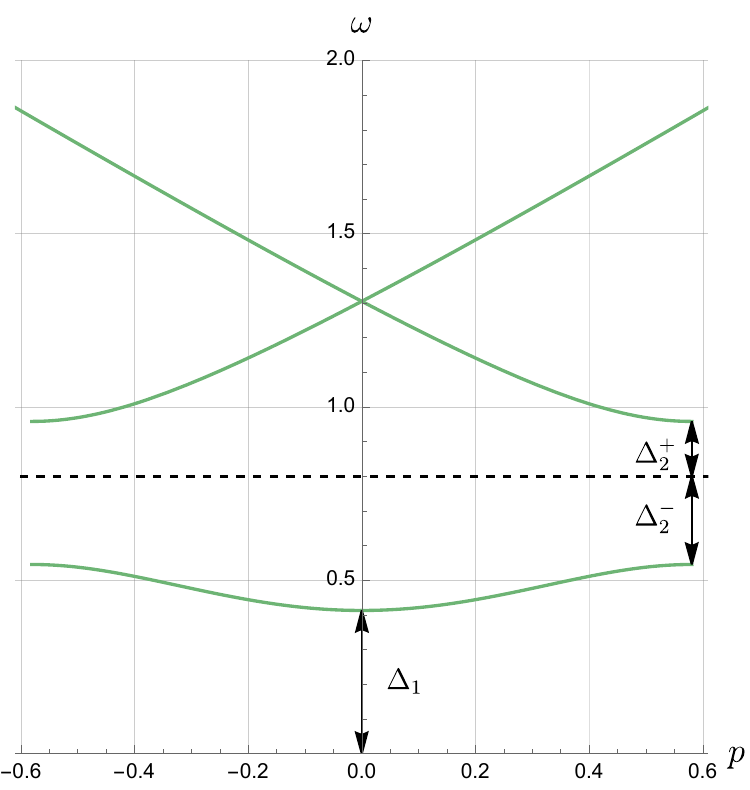}
\caption{$h = 0.8 \, \Lambda$, $\; y=2/3$}
\label{fig-disp-med}
\end{subfigure}
\hfill
\begin{subfigure}{0.495\textwidth}
\centering
\includegraphics[width=0.95\textwidth]{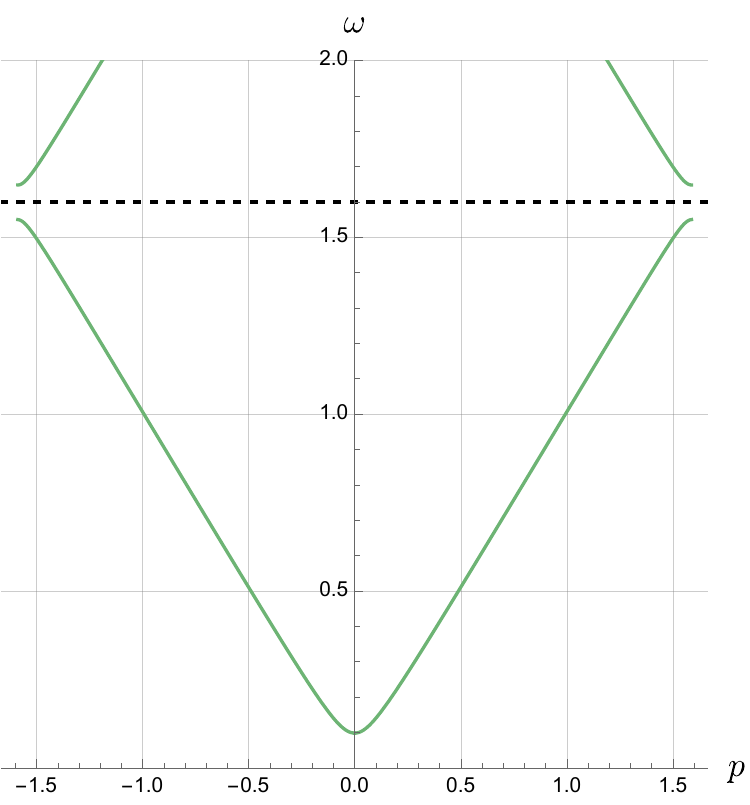}
\caption{$h = 1.6 \, \Lambda$, $\; y=2/3$}
\label{fig-disp-high}
\end{subfigure}
\caption{Dispersion relation for two values of $h$ and $y=2/3$, in units where $\Lambda=1$. Obtained from \eqref{eq:disp-analytic} using the parameters of the physical solutions.}
\label{fig-disp-rel}
\end{figure}

The function $\sigma(x)$ is periodic with period $2K(\mathfrak{m})/A$ in $x$, where $K$ is the elliptic integral of the first kind defined in \eqref{eq:Kdefinition}. The eigenfunctions $\psi_\omega(x)$ are then quasi-periodic, collecting an $\omega$-dependent phase under shift by a period. That phase is the Bloch momentum \eqref{eq:def Bloch momentum}, denoted by $p(\omega)$.  In appendix~\ref{app:eigenfuncs} we find it to be
\be
\label{eq:disp-analytic}
\tilde p(\tilde\omega) = \pm i \zn\biggl( F \Bigl( \arcsin \sqrt{ \tfrac{1}{\mathfrak{m}} \bigl( \sn(b|\mathfrak{m})^{-2} - \tilde\omega^2 \bigr) } \, \Big| \, \mathfrak{m} \Bigr) \, \bigg| \, \mathfrak{m} \biggr) + \frac\pi{2K(\mathfrak{m})} \;,
\ee
where $\tilde p = p/A$, $\tilde \omega = \omega/A$, $F$ is the incomplete elliptic integral of the first kind \eqref{eq:Ellipt1}, and $\zn$ is the Jacobi zeta function \eqref{eq:jacobi zeta}.
This relation is interpreted as a dispersion relation, relating the energy $\omega$ to the momentum $p$. Two examples of this dispersion relation are shown in Figure~\ref{fig-disp-rel}.

There is a small subtlety to mention: if the system is in some finite volume $L$, the solutions need to be exactly periodic in $x$ with period $L$. This, in principle, would introduce a constraint between the parameters of the ansatz:
\begin{equation}
\label{eq:periodicity constraint}
A = \frac{2 \, r \, K(\mathfrak{m})}{L} \qquad \text{for some} \quad r\in \mathbb{N}  \,.
\end{equation}
In the infinite-volume limit, however, $A$ becomes a continuous parameter. This apparently minor detail will become relevant later.

With this ansatz, the spectrum of the Dirac operator can be found analytically, as presented for example in \cite{Schnetz:2005ih, Dunne:1997ia}. We review the eigensystem in appendix~\ref{app:eigenfuncs}. The spectrum has two continuous bands. Introducing the label
\begin{equation}
\label{eq:vareps-def}
\varepsilon(\omega) = \frac{\omega^2}{A^2} + 1 + \mathfrak{m} - \frac1{\sn(b| \mathfrak{m})^2} \;,
\end{equation}
the bands are
\be
\label{eq:bands}
\varepsilon \;\in\; [\mathfrak{m},1] \,\cup\, [1+ \mathfrak{m}, \infty] \;.
\ee
Note that the gap between $\omega = 0$ and the bottom of the lower band vanishes when $b \to K(\mathfrak{m})$, and the gap between the two bands vanishes when $\mathfrak{m} \to 0$. The lower band shrinks to a point in the limit $\mathfrak{m} \to 1$. For later use, we define the gaps
\begin{equation}
\label{eq:three scales}
\Delta_1 \equiv A \, \sqrt{\frac{1}{\sn^2(b)} - 1} \,, \qquad \Delta_2^- \equiv h - A \, \sqrt{\frac{1}{\sn^2(b)} - \mathfrak{m}} \,, \qquad \Delta_2^+ \equiv \frac{A}{\sn(b)} - h \,.
\end{equation}
These three gaps are, respectively, the minimal energy $\omega$ in the lower band (representing the mass gap for neutral fermions), the distance between the chemical potential $h$ and the top of the lower band, and the distance between $h$ and the bottom of the upper band, as illustrated in Figure~\ref{fig-disp-med}. When $h$ is between the bands, $\Delta_2^-$ and $\Delta_2^+$ give the mass gap for charged fermions and holes. 

In order to compute the density of states, remember that our computation started from placing the system in a spatial box of volume $L$, where all fields obeyed periodic boundary conditions. In that setup, $L$ is a multiple of the period of $\sigma(x)$, and the spectrum is discrete and equally spaced in the Bloch momentum $p$ defined by \eqref{eq:def Bloch momentum}.  The density of states can then be computed by taking the limit of an infinite box. Rescaling $p = A \, \tilde{p}$, this gives 
\be
\label{sum to integral over states}
\sum\nolimits_n \;\to\; \int 2 \, \frac{d p}{2\pi} = \frac A\pi \int d\tilde p = \int \rho(\omega) \, d\omega \;.
\ee
The factor of $2$ comes from the two states with equal $\omega$ but opposite $p$, as explained in more detail in appendix~\ref{app:eigenfuncs}. The density of states in $\varepsilon$ is then shown to be
\be
\label{density of states}
\frac{d \tilde p}{d\varepsilon} = \frac{ \bigl\lvert \varepsilon - \mathfrak{m} - \frac{E(\mathfrak{m})}{K(\mathfrak{m})} \bigr\rvert }{ 2 \sqrt{ (\varepsilon - \mathfrak{m})(\varepsilon -1)(\varepsilon - 1 - \mathfrak{m}) }} \;,
\ee
where $E$ is the elliptic integral of the second kind, defined in \eqref{eq:Edefinition}.
The argument of the square root is always positive (for $\varepsilon$ within the bands), whereas the argument of the absolute value is positive on the upper band and negative on the lower band. We can now see that the integration measure is $\rho(\omega) \, d\omega = \frac{A}{\pi} \frac{d\tilde p}{d\varepsilon} \, d\varepsilon$.

From the spectral problem one derives some other useful formulas, also found in \cite{Schnetz:2005ih}. We define the reduced dimensionless potential
\be
\label{eq:sigma-tilde}
\tilde\sigma(\xi) = \mathfrak{m} \sn(b|\mathfrak{m}) \sn(\xi|\mathfrak{m}) \sn(\xi+b|\mathfrak{m}) + \frac{\cn(b|\mathfrak{m}) \dn(b|\mathfrak{m})}{\sn(b|\mathfrak{m})}
\ee
so that $\sigma(x) = A \, \tilde\sigma(Ax)$. We also define the parameters
\be
\label{eq:stu}
s = \frac1{\sn^2(b)} \;,\qquad t = \frac{\cn(b) \dn(b) }{ \sn^3(b) } \;,\qquad u = 1 - \frac EK \;, \qquad \eta = \sn(b| \mathfrak{m})^{-2} -1 - \mathfrak{m} \;,
\ee
that satisfy $t^2 = s(s-1)(s-\mathfrak{m})$. The product of eigenvectors that shows up in the saddle-point equation and in \eqref{trace of Green function} can be written as
\be
\label{psi-bar psi}
\bar\psi_\omega \psi_\omega(x) = \frac{(\omega/A) \, \tilde\sigma(\xi) - t A / \omega }{ (\omega/A)^2 -s + u} = \frac{(\omega/A) \, \tilde\sigma(\xi) - t A / \omega }{ \varepsilon(\omega) - \mathfrak{m} - E/K}
\ee
with $\xi = Ax$.
Notice that, as a function of $x$, $\bar\psi_\omega \psi_\omega$ has a piece proportional to $\sigma(x)$ and a constant, for all values of $\omega$. The proportionality constants are functions of $\omega$. Furthermore,  $\bar\psi_\omega \psi_\omega$ is odd in $\omega$. The spatial averages of the potentials can also be written simply in terms of these parameters,
\be
\label{spatial averages}
\langle \tilde\sigma \rangle = \zn(b) + t/s \;,\qquad\qquad \langle \tilde\sigma^2 \rangle = \eta + 2u \;.
\ee
With these data on the spectral problem, we can solve the equations of motion, find the saddle point, and evaluate the action.

\subsection{Saddle-point solution and its limits}
\label{sec: GN model part chem pot semiclassical}

\subsubsection{Saddle-point equations}

We start by solving the saddle-point equation \eqref{EOM with partial chem pot}.
We make the assumption that $\omega=h$ lies between the two bands in the spectrum. This is motivated by the physics of Peierls instability and from the solution at $y=1$. It has the simplifying effect of turning integrals in $\omega$ over $[0,h]$ into integrals over the lower band, and integrals over $[h,+\infty)$ into integrals over the upper band. This assumption can be checked to be consistent in the end of the computation, numerically for arbitrary $h$, and analytically in the low- and high-density limits.

The LHS of \eqref{EOM with partial chem pot} contains two integrals of $\bar\psi_\omega \psi_\omega(\xi)$ over $\omega$. The result can be rewritten as
\begin{equation}
\label{eq:lhs saddle point c def}
(1-y) \int_0^h d\omega \, \rho(\omega) \, \bar\psi_\omega \psi_\omega(x) + \int_h^\infty d\omega \, \rho(\omega) \, \bar\psi_\omega \psi_\omega(x) = c_1 \, \sigma(x) + c_2 \;,
\end{equation}
where $c_1$ and $c_2$ are $\xi$-independent. 
Matching with the RHS of (\ref{EOM with partial chem pot}) gives the two equations
\be
c_1 = \frac{1}{\lambda} \;, \qquad\qquad c_2=0 \;.
\ee
We can work out the integrals using the formulas in appendix~\ref{sec: useful integrals}.
Let us consider the pure constant term first:
\begin{equation}\begin{aligned}
c_2 &= \biggl[ (1-y) \!\int_\mathfrak{m}^1 + \int_{1 + \mathfrak{m}}^\infty \,\biggr] \, \frac A\pi \, \frac{d\tilde p}{d\varepsilon} \, d\varepsilon \; \frac{ -tA}{\omega \, (\varepsilon - \mathfrak{m} - E/K)} \\
&= \frac{tA}{2\pi} \, \biggl[ (1-y) \!\int_\mathfrak{m}^1 - \int_{1 + \mathfrak{m}}^\infty \,\biggr] \; \frac{d\varepsilon}{\sqrt{ (\varepsilon + \eta)( \varepsilon - \mathfrak{m}) (\varepsilon -1) (\varepsilon -1 - \mathfrak{m})}} \;.
\end{aligned}\end{equation}
The equation $c_2=0$ implies that the two integrals must cancel out, or alternatively that $t=0$. The latter does not lead to a full solution of the equations for generic $y$ and $h$, thus we need to cancel the integrals with each other. The integrals are convergent and one gets$\,$%
\footnote{Eqn.~\eqref{first saddle point equation} can also be written as $y \, K(q^2) = F\bigl( \arcsin(\sn b) \,\big|\, q^2 \bigr)$ or as $\dn^2 \bigl( (1-y) \, K(q^2) \,\big|\, q^2 \bigr) = \mathfrak{m}$.}
\be
\label{first saddle point equation}
(1-y) \, K(q^2) = F \bigl( \arcsin(\dn b) \,\big|\, q^2 \bigr)
\quad\qquad\text{with}\qquad q^2 = \frac{1-\mathfrak{m}}{\dn^2 b} \;,
\ee
where $\dn b \equiv \dn(b|\mathfrak{m})$. At fixed $0<y<1$ and $0<\mathfrak{m}<1$, \eqref{first saddle point equation} has a unique solution for $b$ in the range $b \in \,\bigr( 0, K(\mathfrak{m}) \bigr)$. The extreme values $b=0$ and $b= K(\mathfrak{m})$ correspond to $y=0$ and $y=1$, respectively.
Indeed, for $b\to 0$ we have $\arcsin(\dn b) \to \frac\pi2$ and $F\bigl( \frac\pi2 \big| \, q^2 \bigr) = K(q^2)$, so \eqref{first saddle point equation} is satisfied for $y\to 0$.
For $b \to K(\mathfrak{m})$, instead, we have $q^2 \to 1$ and $K(q^2) \to \infty$, for any $\mathfrak{m}$. Since $F\bigl( \arcsin\sqrt{1-\mathfrak{m}} \big| 1 \bigr)$ is finite, consistency with \eqref{first saddle point equation} requires $y\to 1$.

Let us now consider the term proportional to $\sigma(x)$ in \eqref{eq:lhs saddle point c def}:
\be
c_1=
- \frac 1{2\pi} \, \biggl[ (1-y) \!\int_\mathfrak{m}^1 - \int_{1 + \mathfrak{m}}^\infty \,\biggr] \; \frac{ \sqrt{\varepsilon + \eta} \quad d\varepsilon }{ \sqrt{(\varepsilon-\mathfrak{m}) (\varepsilon - 1) (\varepsilon -1 -\mathfrak{m}) }} \;.
\ee
The lower-band integral is convergent and its result is reported in \eqref{eq:UsefulIntD2}. 
The upper-band integral is instead divergent and requires a cutoff $\varepsilon_\text{max}$. 
The result is reported in \eqref{eq:UsefulIntD}. We can relate the energy cutoff $\varepsilon_\text{max}$ with the momentum cutoff $M = p_\text{max}$ appearing in \eqref{thooft-to-cutoff} by using the dispersion relation \eqref{density of states}. Expanding the dispersion relation for $\varepsilon \to \infty$, integrating it and then inverting it, we obtain
\begin{equation}
\label{eq:eps-max}
\varepsilon_\text{max} = \frac{M^2}{A^2} + 2 \, \left( 1 - \frac{E(\mathfrak{m})}{K(\mathfrak{m})} \right) + \mathcal{O}\left(M^{-2}\right) \;.
\end{equation}
The divergence coming from the first term in \eqref{eq:UsefulIntD} cancels out with the one in the 't~Hooft coupling through \eqref{thooft-to-cutoff}. The equation $c_1 = 1/\lambda$ reduces then to the relation
\be
\label{second saddle point equation bis}
\log \frac{\Lambda^2}{(2+\eta)A^2} + \frac{2}{\chi \sn^2 b} F\bigl( \check p \,\big|\, q^2 \bigr) - \frac{2\mathfrak{m}}{\chi} \, \Pi\bigl( \check p \,\big|\, 1-\mathfrak{m}, q^2 \bigr) = (1-y) \, \frac{2(\mathfrak{m} + \eta)}\chi \, \Pi \bigl( \sn^2(b) \, q^2, q^2 \bigr)
\ee
with
\begin{equation}
\chi = \sqrt{1+\eta} \;,\qquad\qquad
\check p  = \arcsin(\dn b) \;.
\end{equation}
This equation fixes the dimensionless ratio $A/\Lambda$ in terms of $\mathfrak{m}, b, y$. As a check, consider the case $y=1$ and $b = K(\mathfrak{m})$. After multiple simplifications it reduces to $A^2 = \Lambda^2 / \mathfrak{m}$ as found in \cite{Thies:2003br}.
For $y\neq 1$, we can use (\ref{first saddle point equation}) on the second term on the LHS, and the identity (\ref{identity on Pi}) on the term on the RHS. We obtain:
\be
\label{second saddle point equation}
\log \biggl( \sqrt{2+\eta} \; \frac A\Lambda \biggr) = \frac{\mathfrak{m}}{\chi} \, \Bigl[ (1-y) \, \Pi(1-\mathfrak{m}, q^2) - \Pi\bigl( \arcsin(\dn b) \,\big|\, 1-\mathfrak{m}, q^2 \bigr) \Bigr] \;.
\ee
This is the second saddle point equation.

Note that because $h$ lies between the bands, none of the equations depends on it. Furthermore, the equations of motion give two constraints, \eqref{first saddle point equation} and \eqref{second saddle point equation}, but our ansatz contains three parameters: $A$, $b$ (or $q$) and $\mathfrak{m}$. Thus, apparently, the equations of motion give a one-parameter family of saddle points without specifying the dominant saddle at a specific $h$.

To solve this riddle, consider first what happens when the system is placed in a finite volume. In any finite volume $L$, there is a third constraint on these parameters that comes from the periodicity of the solution, as described by eqn.~\eqref{eq:periodicity constraint}. This implies that one of the parameters, say $A$, is not continuous but rather discrete, with spacing $\delta A \equiv 2K(\mathfrak{m})/L$. Therefore, one of the parameters cannot be directly determined from the saddle-point equations that are only sensitive to infinitesimal deformations. 

For any fixed and finite $L$, then,  we need to find the dominant saddle. Formally, we can use \eqref{eq:periodicity constraint} to determine $A$ up to a discrete choice of $r \in \mathbb{Z}$. For each $r$ we have two free parameters, $\mathfrak{m}$ and $b$ (or $q$), and we can use the two equations \eqref{first saddle point equation} and \eqref{second saddle point equation} to determine them. After that, we need to find the $r = r^*$ that minimizes the on-shell action:
\begin{equation}
\Phi \bigl( A(r^*,\mathfrak{m}),\, \mathfrak{m},\, b \bigr) = \min_{r \,\in\, \mathbb{Z}} \; \Phi \bigl( A(r,\mathfrak{m}),\, \mathfrak{m},\, b \bigr) \;.
\end{equation}
This determines the value of the third parameter $A$ for the dominant solution. All other allowed values are still valid saddles, albeit subdominant.

Such a procedure would be cumbersome. However, it considerably simplifies when the periodicity of $\sigma(x)$ is much smaller then the spatial volume, $2K(\mathfrak{m})/A \ll L$, and thus the discrete values are very close to each other, $\frac{\delta A}{A} \ll 1$. Assuming that the action is a smooth function of $A$, we can determine the dominant solution by taking a derivative of $\Phi$ with respect to $A$. We can then find the dominant value of $A$ by setting
\begin{equation}
\frac{d}{d A} \, \Phi(A,\mathfrak{m},b) = \frac{\partial \Phi}{\partial A} + \frac{\partial \mathfrak{m}}{\partial A} \, \frac{\partial \Phi}{\partial \mathfrak{m}} + \frac{\partial b}{\partial A} \, \frac{\partial \Phi}{\partial b} = 0 \;, \qquad  \text{when} \quad 2K(\mathfrak{m})/A \ll L \;.
\end{equation}
Thus, when the equations of motion $\frac{\partial \Phi}{\partial \mathfrak{m}} = \frac{\partial \Phi}{\partial b} = 0$ are solved, we are left with one additional condition:
\begin{equation}
\label{eq:potential extremization}
\frac{\partial \Phi}{\partial A} = 0 \;, \qquad\qquad \text{when} \quad 2K(\mathfrak{m})/A \ll L \;.
\end{equation}
Note that when the periodicity of the solution is of the same order as $L$, we can no longer use this equation. This will be particularly important in the low-density limit.

\paragraph{Computing the grand-canonical density.}
The density of grand-canonical potential \eqref{eq:def-Phi} can be organized into the following terms:
\begin{equation}
\label{grand potential integrals}
\Phi = - h \, 
\underbrace{\rule[-1.4em]{0pt}{0em}
\bigl( \varrho_{\rm bare} + y \, C_1 \bigr)
}_{\varrho} \;
\underbrace{\rule[-1.4em]{0pt}{0em} 
- \int_h^\infty \! d\omega \, \rho(\omega) \, \omega 
}_{E_1} 
\; \underbrace{\rule[-1.4em]{0pt}{0em} - (1-y) \int_0^h \! d\omega \, \rho(\omega) \, \omega 
}_{ E_3} \;  + 
\underbrace{\rule[-1.4em]{0pt}{0em} \frac{A^2}{2\lambda} \, \langle \tilde\sigma^2 \rangle 
}_{ E_2} \;{} - C_0
\end{equation}
where 
\be
\varrho_{\rm bare} = - y \int_h^\infty \! d\omega \, \rho(\omega) \;.
\ee
In this expression $\varrho$ accounts for the charge density, while $E_1$, $E_2$, and $E_3$ capture the energy of the configuration. $E_1$ accounts for the vacuum fluctuations in the upper band, coming from both charged and neutral fermions, while $E_3$ accounts for the fluctuations in the lower band, which can only come from neutral excitations. $E_2$ comes from the potential energy of the configuration. Finally, $C_0$ and $C_1$ are counterterms.

The integral of $\varrho_\text{bare}$ is divergent and can be computed by introducing the momentum cutoff $M$ and using \eqref{eq:usefulD4}. One finds:
\be
\label{eq:bare-density}
\varrho_{\rm bare} = - y \int_h^M \! d\omega \, \rho(\omega) = - \frac{yA}{2\pi} \int_{1+\mathfrak{m}}^{\varepsilon_\text{max}} \! \frac{ \varepsilon - \mathfrak{m} - E/K }{ \sqrt{ (\varepsilon - \mathfrak{m})(\varepsilon - 1)(\varepsilon - 1 - \mathfrak{m}) }} \, d\varepsilon = \frac{yA}{2K(\mathfrak{m})} - \frac{yM}{\pi} \,.
\ee
We choose the divergent part of the counterterm $C_1$ to cancel the cutoff dependency. The finite part is zero so that the charge density vanishes in the low-density limit $\mathfrak{m}\rightarrow 1$, and thus 
\be
\label{eq:chargedensity}
\varrho = \frac{yA }{ 2K(\mathfrak{m})} \;,
\ee
which is independent of $b$. Note that this is $y$ times the inverse of the period, since each ``bump'' in $\sigma$ carries charge $yN$. This will be more evident when we study the low-density limit.

The term $E_1$ (upper band) in \eqref{grand potential integrals} is also divergent and reads
\be
E_1 = - \int_h^M \! d\omega \, \rho(\omega) \, \omega = - \frac{A^2}{2\pi} \int_{1+\mathfrak{m}}^{\varepsilon_\text{max}} \!\frac{ (\varepsilon - \mathfrak{m} - E/K) \, \sqrt{\varepsilon + \eta} \;\; d\varepsilon }{ \sqrt{ (\varepsilon - \mathfrak{m}) (\varepsilon -1) (\varepsilon - 1 - \mathfrak{m}) }} \;.
\ee
The result of the integral is reported in \eqref{eq:usedulD5}. It is both quadratically and logarithmically divergent due, respectively, to the first and second term on the right-hand side of \eqref{eq:usedulD5}, which are proportional to $\varepsilon_\text{max}$ and $\log \varepsilon_\text{max}$.
We choose the counterterm $C_0$ to be
\begin{equation}
C_0 = -\frac{M^2}{2\pi} - \frac{\Lambda^2}{4\pi}
\end{equation}
such that it cancels the $M^2$ divergence of $E_1$ and it ensures the correct normalization of the free energy. 
The term $E_2$, using \eqref{spatial averages}, equals
\be
E_2 = \frac{A^2}{2\lambda} \, \langle \tilde\sigma^2 \rangle = -\frac{A^2}{2\pi} \left( 2 \, \frac {E(\mathfrak{m})}{K(\mathfrak{m})} - \eta - 2 \right) \log \frac{2M}{\Lambda} \;,
\ee
and its logarithmic divergence precisely cancels the one of $E_1$.
The term $E_3$ (lower band) in \eqref{grand potential integrals} reads
\be
E_3 = {} - (1-y) \int_0^h \! d\omega \, \rho(\omega) \, \omega = - \frac{(1-y) \, A^2}{2\pi} \int_
\mathfrak{m}^1 \! \frac{ \bigl\lvert \varepsilon - \mathfrak{m} - E/K \bigr\rvert  \, \sqrt{\varepsilon + \eta} \;\; d\varepsilon }{ \sqrt{ (\varepsilon - \mathfrak{m}) (\varepsilon -1) (\varepsilon - 1 - \mathfrak{m}) }} \;.
\ee
The integral is convergent and is reported in \eqref{eq:usedulD7}.
Collecting all terms, we finally find:
\begin{align}
\label{eq:Phi intermediate}
\Phi &= \frac{A^2}{2\pi} \Biggl\{ \Bigl( 2 \tfrac EK - 2 + \mathfrak{m} - \frac\eta2 \Bigr) + \Bigl( 2 \tfrac EK - 2 - \eta \Bigr) \log \frac{\Lambda}{ A \sqrt{2+\eta}} + \chi \Bigl[ E(\check p \,|\, q^2) - (1-y) E(q^2) \Bigr] \nn \\
&\qquad\quad + \frac{\mathfrak{m}}{\chi} \Bigl( 2 \tfrac EK - 2 - \eta \Bigr) \biggl[ F(\check p \,|\, q^2) - \Pi (\check p \,|\, 1-\mathfrak{m}, q^2) - (1-y) \Bigl( K(q^2) - \Pi(1-\mathfrak{m}, q^2) \Bigr) \biggr] \nn\\
&\qquad\quad + \chi \Bigl( 2 \tfrac EK - 2 + \mathfrak{m} \Bigr) \Bigl[ F(\check p \,|\, q^2) - (1-y) K(q^2) \Bigr] \Biggr\} - \frac{y \, h \, A}{2K(\mathfrak{m})} +
\frac{\Lambda^2}{4\pi}\;.
\end{align}
By setting to zero the variation with respect to $A$ we obtain a relation between $A$ and the chemical potential $h$. Using the saddle-point equations (\ref{first saddle point equation}) and (\ref{second saddle point equation}) it simplifies to:
\be
\label{chemical potential}
h = \frac{2A}{y\, \pi} \biggl[ E(\mathfrak{m}) + \chi \, K(\mathfrak{m}) \Bigl( E(\check p \,|\, q^2) - (1-y) E(q^2) \Bigr)  - (1-\mathfrak{m}) \, K(\mathfrak{m}) \biggr] \;.
\ee
The potential on the solutions gives the free energy
\be
\label{grand potential on-shell y<1}
\frac{\mathcal{F}}{N} = \Phi_\text{on-shell} = \frac{A^2}{4\pi} \biggl( 2 \, \frac{E(\mathfrak{m})}{ K(\mathfrak{m})} - 2 - \eta \biggr) - \frac{y \, h \, A}{4K(\mathfrak{m})}+\frac{\Lambda^2}{4\pi} \;.
\ee
Note that setting $y=1$ and the parameters accordingly, these expressions reduce to their analogues in \cite{Thies:2003br}.

\begin{figure}[t!]
\centering
\begin{subfigure}{0.31\textwidth}
\includegraphics[width=\textwidth]{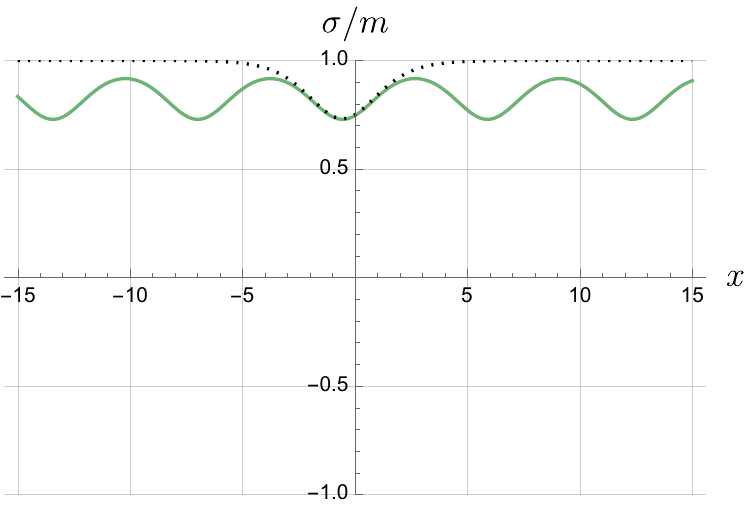}
\caption{$y=1/3$ ($\tilde\rho \approx 1.14$)}
\end{subfigure}
\hspace{0.01\textwidth}
\begin{subfigure}{0.31\textwidth}
\includegraphics[width=\textwidth]{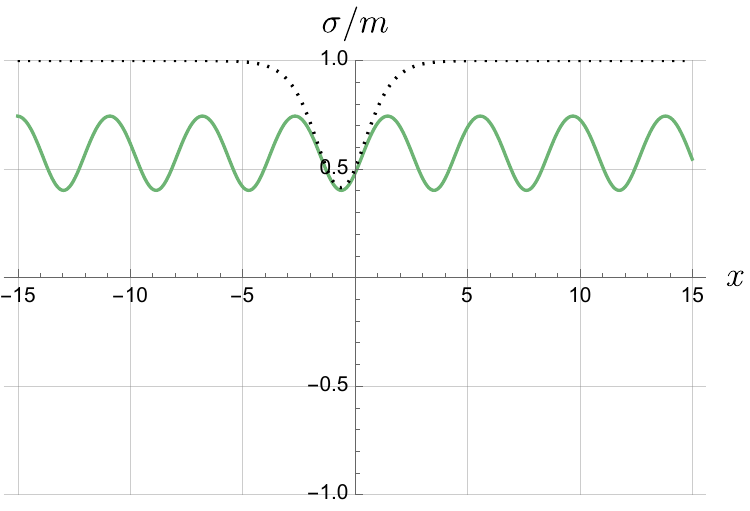}
\caption{$y=1/2$ ($\tilde\rho \approx 1.68$)}
\end{subfigure}
\hspace{0.01\textwidth}
\begin{subfigure}{0.31\textwidth}
\includegraphics[width=\textwidth]{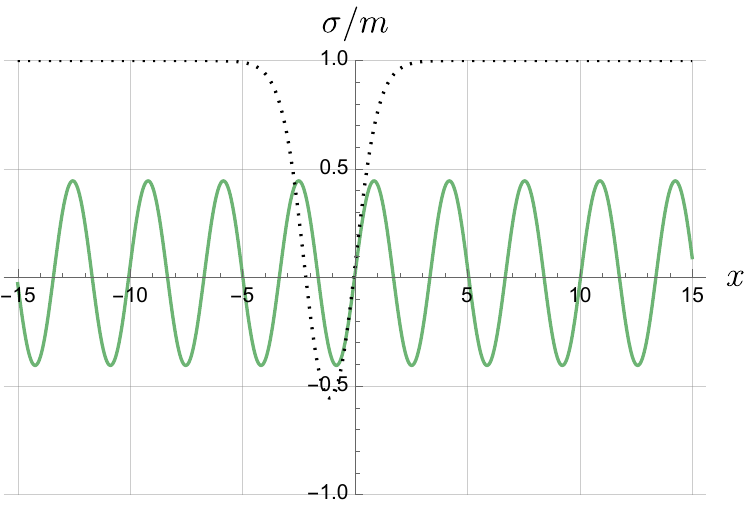}
\caption{$y=6/7$ ($\tilde\rho \approx 1.89$)}
\end{subfigure}
\caption{\label{fig:med-dens}%
Plot of $\sigma(x)$ with $h = 0.97 \, \Lambda$ for different values of $y$. The dotted line is the single rank-$a$ particle solution found in \cite{Dashen:1975xh}.}
\end{figure}

Inverting the parameters in terms of $h$ is analytically unfeasible. We can solve the equations \eqref{first saddle point equation}, \eqref{second saddle point equation} and \eqref{chemical potential} numerically and plot some examples such as those in Figure~\ref{fig:med-dens}. It is convenient to introduce the normalized density
\begin{equation}
\label{eq:norm-dens}
\tilde\varrho = \frac{A/\Lambda}{2 K(\mathfrak{m}) \sin\bigl( \tfrac{\pi y}{2}\bigr)} 
\, \Biggl[ 4 + \frac{1}{2} \log \Biggl( \frac{ 1 + \sin\bigl( \tfrac{\pi y}{2} \bigr) }{ 1 -\sin \bigl( \tfrac{\pi y}{2} \bigr)} \Biggr) \Biggr] \;,
\end{equation}
where we multiply the particle density by the typical width of a solution containing a single rank-$a$ particle \cite{Dashen:1975xh}. In this normalization, low density means $\tilde\varrho \ll 1$, medium density means $\tilde\varrho \sim 1$, and high density means $\tilde\varrho \gg 1$. Visually, at low density we observe well separated particle ``bumps'' (Fig.~\ref{fig:low-dens}), at medium density they somewhat overlap (Fig.~\ref{fig:med-dens}), whereas at high density they overlap so much that their tips form a sinusoidal profile (Fig.~\ref{fig:hi-dens}). 

We can gain some analytic control in the limits $\mathfrak{m} \rightarrow 1$ and $\mathfrak{m} \rightarrow 0$ which correspond, as we will see below, to the low- and high-density limits, respectively. The analysis can also be extended to the homogeneous case, which we do in appendix~\ref{app:homLimit}, re-deriving the results of section~\ref{sec:hom-instability}.

\subsubsection{Low-density limit}

Let us compute the solution in the limit $\mathfrak{m} \to 1$. We keep $0<y<1$ fixed and, as we will see, this means that also $b$ remains fixed. We use the following limits:
\begin{equation}\begin{aligned}
& \sn(b|\mathfrak{m}) \to \tanh b \;,\qquad \cn(b|\mathfrak{m}) \,,\, \dn(b|\mathfrak{m}) \to \cosh^{-1} b \;, \qquad q^2 = \frac{1-\mathfrak{m}}{\dn^2 b} \to 0 \\
& K(\mathfrak{m}) \to \infty \;,\qquad (1-\mathfrak{m}) \, K(\mathfrak{m}) \to 0 \;,\qquad E(\mathfrak{m}) \to 1 \;, \\[0.4em]
& \sqrt{2+\eta} \to \tanh^{-1} b \;,\quad \chi \to \sinh^{-1} b \;, \\[0.3em]
& F(x|q^2) \,,\, E(x|q^2) \,,\, \Pi(x|1-\mathfrak{m}, q^2) \to x \;,\qquad K(q^2) \,,\, E(q^2) \,,\, \Pi(1-\mathfrak{m},q^2) \to \frac\pi2 \;.
\end{aligned}\end{equation}
The first saddle point equation (\ref{first saddle point equation}) reduces to:
\be
(1-y) \, \frac\pi2 = \arcsin \frac1{\cosh b}
\qquad\Leftrightarrow\qquad \tanh b = \sin \bigl( y \, \tfrac\pi2 \bigr) \;.
\ee
The second saddle point equation (\ref{second saddle point equation}) reduces to:
\be
\log\biggl( \frac{A}{\Lambda \tanh b} \biggr) = \sinh(b) \biggl[ (1-y) \frac\pi2 - \arcsin \frac1{\cosh b} \biggr] = 0 \;.
\ee
This implies
\be
A = \tanh(b) \, \Lambda = \sin \bigl( y \, \tfrac\pi2 \bigr) \, \Lambda \;.
\ee
Plugging this value in \eqref{eq:chargedensity}, we see that the limit $\mathfrak{m}\to 1$ corresponds to low density, $\varrho\to 0$, as $K(\mathfrak{m})$ diverges in the limit. The divergence of $K(\mathfrak{m})$ makes the analysis of the third saddle-point equation (\ref{chemical potential}) more delicate.
Using
\be\begin{aligned}
F(x | q^2) &= x + \tfrac18 \cosh^2(b) \, \bigl( 2x - \sin(2x) \bigr) (1-\mathfrak{m}) + \mathcal{O}(1-\mathfrak{m})^2 \;, \\
E(x | q^2) &= x - \tfrac18 \cosh^2(b) \, \bigl( 2x - \sin(2x) \bigr) (1-\mathfrak{m}) + \mathcal{O}(1-\mathfrak{m})^2 \;, 
\end{aligned}\ee
combined with (\ref{first saddle point equation}), one can reduce $E(\check p|q^2) - (1-y) E(q^2)$  to $\mathcal{O}(1-\mathfrak{m})$ terms. Additionally, $(1-\mathfrak{m}) K(\mathfrak{m}) \to 0$. Then \eqref{chemical potential} simplifies to:
\be
\label{lower bound chem pot}
h_\text{pt} = \frac{2A}{y \,\pi } = \frac{2}{y\, \pi} \sin \biggl( \frac{y \, \pi}2 \biggr) \, \Lambda \;.
\ee
This is the minimal value of the chemical potential in the low-density limit of the condensed phase. This result is consistent with the condensation of the rank-$a$ bound state since it corresponds precisely to the mass-to-charge ratio (at large $N$), as follows from \eqref{eq:spectrum}, see the discussion in appendix~\ref{app:anti}.

\begin{figure}[t!]
\centering
\begin{subfigure}{0.31\textwidth}
\includegraphics[width=\textwidth]{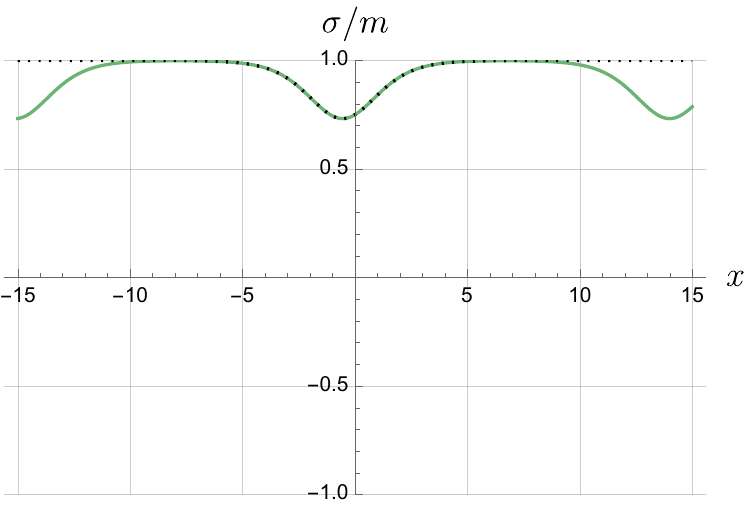}
\caption{$y=1/3$ ($\tilde\rho \approx 0.63$)}
\end{subfigure}
\hspace{0.01\textwidth}
\begin{subfigure}{0.31\textwidth}
\includegraphics[width=\textwidth]{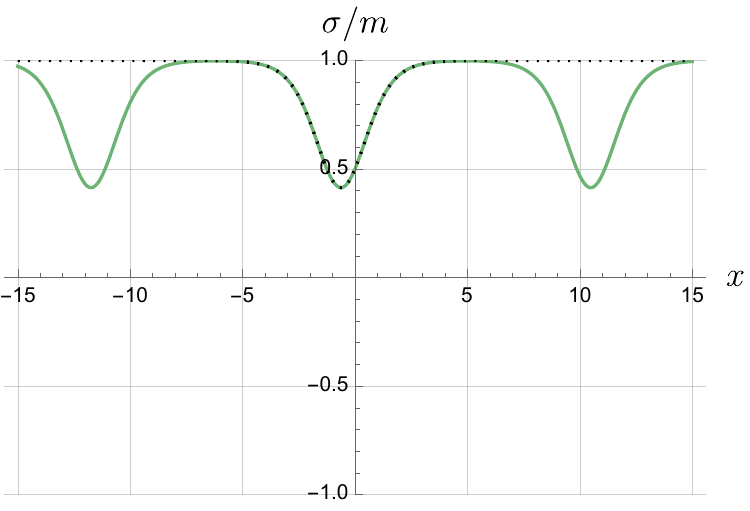}
\caption{$y=1/2$ ($\tilde\rho \approx 0.62$)}
\end{subfigure}
\hspace{0.01\textwidth}
\begin{subfigure}{0.31\textwidth}
\includegraphics[width=\textwidth]{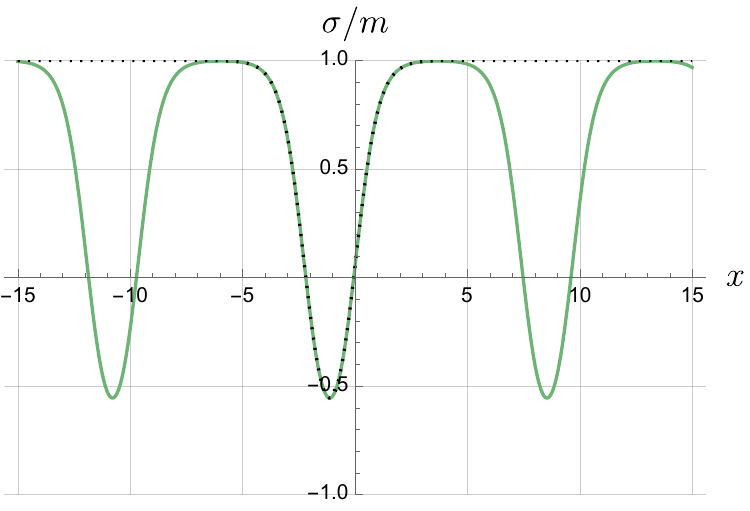}
\caption{$y=6/7$ ($\tilde\rho \approx 0.66$)}
\end{subfigure}
\caption{\label{fig:low-dens}%
Plot of $\sigma(x)$ with $h=1.0001 \, h_\text{pt}$ for different values of $y$, with $h_\text{pt}$ defined in \eqref{lower bound chem pot}. The dotted line is the single rank-$a$ particle solution found in \cite{Dashen:1975xh}.}
\end{figure}

The condensate is $\sigma(x) = A \, \tilde\sigma\bigl( Ax - \frac b2 \bigr)$, where we shifted the origin of the $x$ coordinates to center the soliton at $x=0$. Introducing the variable $y_\text{D} = \sin\bigl( \frac{y \, \pi}2 \bigr)$, in the limit $\mathfrak{m} \to 1$ we also have $y_D = \tanh(b)$ and can write the condensate as:
\be
\label{eq:DHN}
\sigma(x) = \Lambda \, \biggl[ 1 + y_\text{D} \tanh \biggl( \Lambda \, y_\text{D} \, x - \frac14 \log \frac{1 + y_\text{D}}{ 1 - y_\text{D}} \biggr) - y_D \tanh \biggl( \Lambda \, y_\text{D} \, x + \frac14 \log \frac{1 + y_\text{D}}{ 1 - y_\text{D}} \biggr) \biggr] \;.
\ee
This is precisely the single rank-$a$ soliton particle in eqn.~(3.28) of \cite{Dashen:1975xh}. 
Thus we show that the low-density limit is given by the condensation of a single bound state at the critical value of the chemical potential. In Figure~\ref{fig:low-dens} we show some numerical solutions of $\sigma(x)$ at low density, displaying increasingly separated single-particle solitons.

\begin{figure}[t!]
\centering
\begin{subfigure}{0.49\textwidth}
\centering
\includegraphics[width=\textwidth]{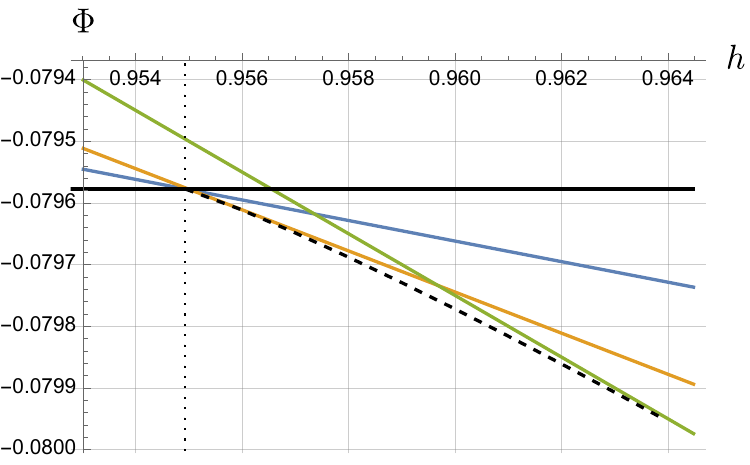}
\caption{$L=20$, $r = 1$ (blue), $2$ (orange) and $3$ (green)}
\label{fig:action-L20}
\end{subfigure}
\hspace{0.01\textwidth}
\begin{subfigure}{0.475\textwidth}
\centering
\includegraphics[width=\textwidth]{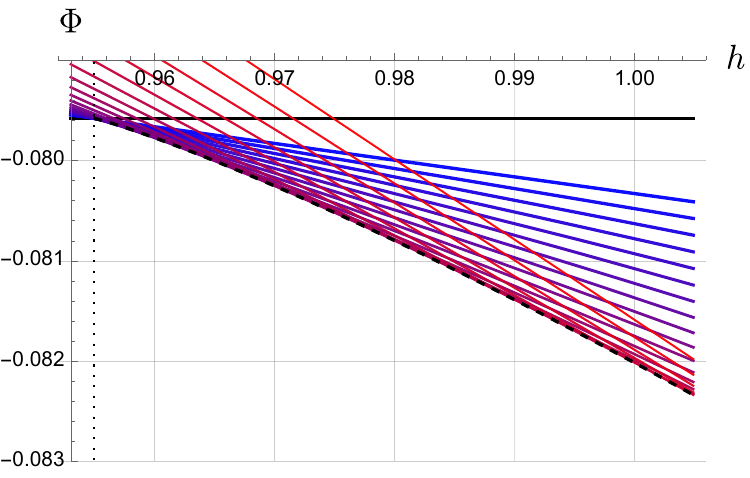}
\caption{$L=100$, $r$ from $5$ (blue) to $24$ (red).}
\label{fig:action-L100}
\end{subfigure}
\caption{\label{fig:Phi-finite-L}%
Plot of $\Phi(h)$ at finite $L$ for different values of $r$ (colored full lines) compared with $\Phi(h)$ for the infinite-volume solution (black dashed line) and with the action for $\sigma=m$ (black solid line). Plotted with $y=1/3$ and $\Lambda=1$.}
\end{figure}

At infinitesimally low density, it might make more physical sense to study the finite-volume solutions. As previously explained, in such case we do not extremize the grand potential with respect to $A$, but rather consider the family of solutions labeled by the integer $r$ in \eqref{eq:periodicity constraint}. As discussed below \eqref{second saddle point equation}, for any fixed $r$ the parameters $\mathfrak{m}$, $b$ for these solutions are independent of $h$, and so the potential $\Phi$ in \eqref{eq:Phi intermediate} has a linear dependence on $h$. Indeed, it be interpreted as the specific free energy $\Phi = \frac{1}{L}(E - h Q)$ of a configuration with energy $E$ and charge $Q = L \, \varrho = y \, r$. Close to the phase transition $h_\text{pt}$, the $r=1$ solution (approximately given by \eqref{eq:DHN} for large $L$) overtakes the homogeneous solution $\sigma=m$. Then, at a slightly higher value of $h$, the $r=2$ solution (approximately given by two solitons equally spaced in $L$) becomes the preferred solution. We exemplify this behavior in Figure~\ref{fig:action-L20}. At large $L$, these succession of discrete solutions approximates the curve generated by assuming infinite volume, as seen in Figure~\ref{fig:action-L100}. Thus at finite volume we have a \emph{first-order} phase transition, which only appears as second order in the infinite-volume limit.%
\footnote{In generating the plots of Figures~\ref{fig:action-L20} and \ref{fig:action-L100}, we solve \eqref{first saddle point equation} and \eqref{second saddle point equation} in conjunction with \eqref{eq:periodicity constraint}. This might neglect some $1/L$ effects because \eqref{first saddle point equation} and \eqref{second saddle point equation} are derived using a continuous spectrum, while at finite volume one should consider a discrete one. However, since $h$ is between the two bands, this will not change the fact that $\Phi = \frac1L(E - h Q)$ has linear dependence on $h$, nor the value of the charge $Q$, but will only correct the energy $E$. The picture of successive first-order transitions between these different states remains valid.}

\subsubsection{High-density limit}
\label{sec: high-density limit}

Next, consider the limit $\mathfrak{m} \to 0$ with $y$ fixed. In this limit, it is convenient to re-parametrize $b$ in terms of $\nu$ as follows:
\be
b = K(\mathfrak{m}) - \nu \;.
\ee
It is also useful to introduce the notation 
\begin{equation}\begin{aligned}
\label{eq:p-x}
q^2 &\equiv 1-p &&\quad\Rightarrow\quad& p &= \mathfrak{m} \; \frac{\cn^2(b|\mathfrak{m})}{\dn^2(b|\mathfrak{m})} = \sin^2(\nu) \, \mathfrak{m} + \mathcal{O}(\mathfrak{m}^2) \;, \\
\dn(b|\mathfrak{m}) &\equiv \sqrt{1-x^2} &&\quad\Rightarrow\quad& x &= \sqrt\mathfrak{m} \, \sn(b|\mathfrak{m}) = \mathfrak{m}^\frac12 \cos(\nu) + \mathcal{O}(\nu^2 \mathfrak{m}^\frac32) \;.
\end{aligned}\end{equation}
When handling the incomplete elliptic integrals, taking both $p\rightarrow 0$ and $x\rightarrow 0$ requires additional care since we obtain a singular limit in both the argument and the parameter.
An important consideration in this limit, which we derive below, is that the consistent solution requires also $\nu\rightarrow 0$. Then, from \eqref{eq:p-x}, we read that $p/x^2 \sim \nu^2 \rightarrow 0 $ and $p/\mathfrak{m} \sim \nu^2\rightarrow 0$, which allows us to use the expansions derived in appendix~\ref{app: series expansions}.

Using the leading order terms in \eqref{eq:K-1-p} and \eqref{eq:F-expasion}, we can expand the first saddle-point equation (\ref{first saddle point equation}) to find 
\be
\label{nu high density}
\nu \,\approx\, 2^{\frac{1-2y}{1-y}} \, \mathfrak{m}^{\frac{y}{2-2y}} \;,
\ee
which is consistent with $\nu \rightarrow 0$. The second saddle-point equation (\ref{second saddle point equation}) can be analyzed using the limits \eqref{eq:Pi-exp} and \eqref{eq:incomp-Pi-exp}. At leading order, the equation reduces to
\be
\log \biggl( \sqrt2 \; \frac A\Lambda \biggr) = \mathfrak{m} \, \biggl( (1-y) \, \frac{1}{2\mathfrak{m}} \log \frac{4}{\nu^2} - \frac{1}{2\mathfrak{m}} \log 2 \biggr) \;,
\ee
whose solution is
\be
A \,\approx\, 2^{-y} \, \nu^{-(1-y)} \, \Lambda \,\simeq\, 2^{y-1} \, \mathfrak{m}^{-y/2} \, \Lambda \;.
\ee
The second equality is obtained using (\ref{nu high density}). Plugging this value of $A$ in \eqref{eq:chargedensity} clearly shows that the limit $\mathfrak{m} \to 0$ corresponds to the high-density limit $\varrho\to \infty$.

After using the limits \eqref{eq:E-1-p} and \eqref{eq:incomp-E-exp}, the third saddle-point equation (\ref{chemical potential}) simply gives
\be
\label{relations for mu high density}
h \,\approx\, A \qquad\Rightarrow\qquad \frac h\Lambda \,\approx\, 2^{y-1} \, \mathfrak{m}^{-y/2} \,\approx\, 2^{-y} \, \nu^{y-1} \;.
\ee
This equation determines $\mathfrak{m}$ (or $\nu$) as a function of $h/\Lambda$.
Using the subleading terms in \eqref{eq:incomp-E-parms},
we can obtain the more refined expansion 
\be
\label{expansion A high density}
\frac{A}{h} = 1 + \frac{\mathfrak{m}}4 + \frac{11y-4}{64y} \, \mathfrak{m}^2 - \frac{\nu^2}2 + \mathcal{O}\Bigl( \mathfrak{m}^3 \,,\, \nu^2\mathfrak{m} \log \nu^2\mathfrak{m} \,,\, \nu^4 \Bigr)
\ee
that will be useful to evaluate the grand-canonical potential.

We can determine the fermion gaps using \eqref{eq:three scales}. The gap for the neutral fermions is
\be
\label{eq:neutralGap}
\Delta_1 = A \sqrt{ \sn^{-2}(b|\mathfrak{m}) - 1} \;\approx A\, \nu = 2^{- \frac{y}{1-y}} \, h \, \Bigl( \frac\Lambda h\Bigr)^{\frac1{1-y}} \;,
\ee
where we used $\sqrt{ \sn^{-2}(b) - 1} = \nu + \mathcal{O}(\nu\mathfrak{m}, \nu^3)$ and only kept the leading-order terms. The gaps for the charged fermions are
\be
\label{eq:chargedGap}
\left. \begin{aligned}
\Delta_2^+ &= A \sn^{-1}(b|\mathfrak{m}) - h \\
\Delta_2^- &= h - A \sqrt{ \sn^{-2}(b|\mathfrak{m}) - \mathfrak{m}}
\end{aligned} \quad \right\}
\approx \frac{A\, \mathfrak{m}}4 = 2^{-\frac2y} \, h \, \Bigl( \frac\Lambda h \Bigr)^\frac2y \;.
\ee
To compute the limits we expressed $h$ in terms of $A$ using (\ref{chemical potential}), and then expanded the expression that multiplies $A$ at leading order. The sum of the two scales can more easily be obtained using the expansion $\sn^{-1}(b) - \sqrt{ \sn^{-2}(b) - \mathfrak{m}} = \frac{\mathfrak{m}}2 + \mathcal{O}(\nu^2\mathfrak{m}, \mathfrak{m}^2)$. We can thus see that these scales match the ones defined in \eqref{eq:lambda-12}:
\begin{equation}
\Delta_1 \approx \Lambda_{\rm n} \;,\qquad\qquad \Delta_2^+ \approx \Delta_2^- \approx \frac{\Lambda_{\rm c}}{2} \;.
\end{equation}
We conclude that, at large $N$, the scales $\Lambda_\text{n}$ and $\Lambda_\text{c}$ correspond to the mass gaps of neutral and charged particles, respectively. We will show in section~\ref{sec:tba-all} that this also applies at finite $N$. Observe that at high density the gaps become small and $\Delta_2^\pm$ become identical, as visible in Figure~\ref{fig-disp-high}.

\begin{figure}[t!]
\centering
\begin{subfigure}{0.31\textwidth}
\includegraphics[width=\textwidth]{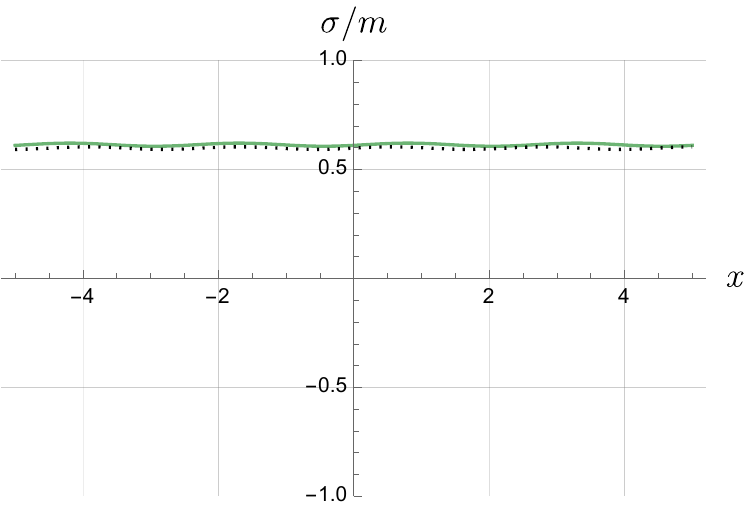}
\caption{$y=1/3$ ($\tilde\rho \approx 3.64$)}
\end{subfigure}
\hspace{0.01\textwidth}
\begin{subfigure}{0.31\textwidth}
\includegraphics[width=\textwidth]{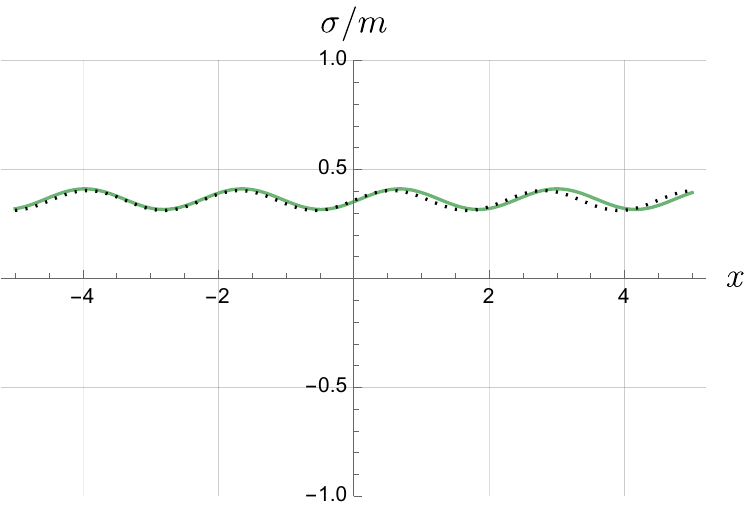}
\caption{$y=1/2$ ($\tilde\rho \approx 2.97$)}
\end{subfigure}
\hspace{0.01\textwidth}
\begin{subfigure}{0.31\textwidth}
\includegraphics[width=\textwidth]{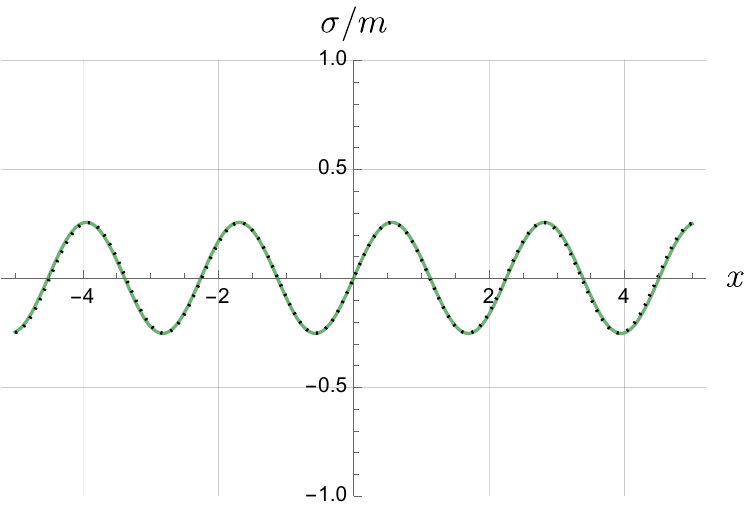}
\caption{$y=6/7$ ($\tilde\rho \approx 2.81$)}
\end{subfigure}
\caption{\label{fig:hi-dens}%
Plot of $\sigma(x)$ with $h = 1.4 \, \Lambda$ for different values of $y$. The dotted line is the limiting solution \eqref{eq:sigma-sine}.}
\end{figure}

The two scales also appear in the high-density limit of the condensate. Indeed
\begin{equation}\begin{aligned}
\tilde\sigma(\xi) &= \mathfrak{m} \; \frac{ \cn(\nu) \cn(\nu-\xi) \sn(\xi) }{ \dn(\nu) \dn(\nu-\xi) } + (1-\mathfrak{m}) \, \frac{\sn(\nu)}{\cn(\nu) \dn(\nu)} \\
&\approx \frac{\mathfrak{m}}2 \sin(2\xi) + \mathcal{O}(\nu^2 \mathfrak{m}, \mathfrak{m}^2) + \nu + \mathcal{O}(\nu^3, \nu \mathfrak{m}) \;,
\end{aligned}\end{equation}
so that
\be
\label{eq:sigma-sine}
\sigma(x) \,\approx\, \Lambda_{\rm n} + \Lambda_{\rm c} \sin(2h x) \;.
\ee
In Figure~\ref{fig:hi-dens} we plot the exact solution of $\sigma(x)$ at high $h$ and compare it with \eqref{eq:sigma-sine}. Finally, using the expansion (\ref{expansion A high density}), the free energy (\ref{grand potential on-shell y<1}) can be expanded as 
\be
\label{grand potential high density y less 1}
\frac{\mathcal{F}}{N} = - \frac{h^2}{2\pi} \, \biggl( y + \frac{(1-y)}2 \, \Bigl( \frac{\Lambda_{\rm n}}{h} \Bigr)^2
+ \frac{y}{8} \, \Bigl( \frac{\Lambda_{\rm c}}{h} \Bigr)^2
+ \mathcal{O}\bigl( \nu^4 \,,\, \nu^2\mathfrak{m} \,,\, \nu^2\mathfrak{m}^2 \bigr) \biggr) 
+ \frac{\Lambda^2}{4\pi} \;.
\ee
Thus, at large $N$ and high density, the scales $\Lambda_{\rm n}$ and $\Lambda_{\rm c}$ directly parameterize the profile of $\sigma$ and the nonperturbative terms in the free energy.

\section{Bethe ansatz and finite \tpsb{N}{N} results}
\label{sec:tba-all}

In this section we discuss BA techniques, the third approach used to study the GN model at finite temperature.
We start in section~\ref{sec:tba-general} with some general considerations. In section~\ref{sec:tba-finite} we analyze nonperturbative effects at finite $N$ with the Wiener--Hopf method. Analytic results at large $N$ are discussed in section~\ref{subsec:BAlargeN}, while numerical results at both large and finite $N$ are collected in section~\ref{sec:tba-num}.

\subsection{Integral equations and observables}
\label{sec:tba-general}

Finite-density states in integrable quantum field theories can be studied with the BA, thus providing an exact method to extract many observables, as pointed out in \cite{Polyakov:1983tt}. This requires the exact $S$-matrix amplitudes of particles in the vacuum, which are also known from integrability.
A particularly important observable is the zero-temperature relative free energy 
\be
\label{eq:FhPCF1}
\mathcal{F}(h) \equiv  F(h)-F(0) \,.
\ee
Here $h$ is the common fugacity for a number $a$ of $U(1)$ factors within $O(2N)$, as in \eqref{eq:UV}.
In the large $N$ limit, the free energy $F$ relates to the action potential $\Phi$ evaluated at the minimal saddle.
As mentioned in the introduction, at $T=0$ we expect that the vacuum will be populated only by the particles with the highest $q_a/m_a$
ratio, in order to minimize the shifted Hamiltonian $H- h \sum_i Q_i$. When $a$ fugacities are switched on, such particles
are specific polarizations of the rank-$a$ bound states of the model, see appendix~\ref{app:anti}. There is only a single polarization state
of this kind for any $a$, so the resulting scattering matrix is one-dimensional, like in the well-studied $a=1$ case.

The density of states in rapidity space $\chi_a(\theta)$ is specified by the BA integral equation
\be
\label{eq:FhPCF3}
\chi_a(\theta) - \int_{-B}^B  d \theta' \, K_a(\theta-\theta') \, \chi_a (\theta') = m_a \cosh \theta \,, 
\ee
where $\chi_a(\theta)$ is supported on the interval $[-B,B]$, with $B$ a ``Fermi rapidity'' yet to be determined, and $m_a$ is the mass of the rank-$a$ state as given in \eqref{eq:spectrum}. Note that while $\theta$ is the relative rapidity of the particles \emph{in the vacuum}, the actual momentum distribution $p_a$ of the particles \emph{in the medium} is governed by $\chi_a$:
\be
\label{eq:chi-dpdt}
\chi_a(\theta)  = \frac{d p_a}{d\theta} \,.
\ee
The integral kernel appearing in the BA equation is given by 
\be
\label{eq:FhPCF4}
K_a(\theta) = \frac{1}{2 \pi i} \, \frac{d  \log S_a(\theta)}{d\theta} \,,
\ee
where $\theta=\theta_1-\theta_2$ is the relative rapidity of the two scattering states, and $S_a$ is their corresponding $S$-matrix amplitude \cite{fendley-tba}.

The energy per unit length $e_a$ and the density $\rho_a$ are given by:
\be
\label{eq:FhPCF5}
e_a = \frac{m_a}{\pi} \int_{-B}^B d \theta \, \chi_a(\theta)  \cosh \theta \,, \qquad\qquad \rho_a = \frac1{2\pi} \int_{-B}^B d \theta\, \chi_a(\theta) \,.
\ee
The parameter $B$ is related to the Fermi momentum $p_F$ of particles and, at fixed density, its value is fixed by the density $\rho$.%
\footnote{The parameter $B$ is thus a function of $\rho$ or $h$ and it depends on $a$, but to avoid cluttering we simply write $B$, leaving any dependence implicit.}
Naturally, the Fermi momentum and the density are related by
\begin{equation}
\label{eq:pf-pirho}
p_F = \pi \, \rho_a \,.
\end{equation}
The free energy $\mathcal{F}(h)$ can be obtained by taking a Legendre transform of $e(\rho)$.

In an equivalent, more useful, formulation of the BA equations, the basic quantity is a function $\epsilon_a(\theta)$ which satisfies the integral equation 
\be
\label{eq:FhPCF7}
\epsilon_a(\theta) - \int_{-B}^{B} d\theta'\, K_a(\theta-\theta') \, \epsilon_a(\theta') = a \, h - m_a \cosh \theta
\ee
and the boundary conditions
\be
\label{eq:eps-bc}
\epsilon_a(\pm B) = 0 \,.
\ee
This boundary condition fixes $B$ for a given $h$.
In terms of $\epsilon_a$, the relative free energy is more directly given by
\be
\label{eq:FhPCF9}
\mathcal{F}_a(h)=  - \frac{m_a}{2 \pi} \int_{-B}^B \! d \theta \, \epsilon_a(\theta) \cosh\theta \,.
\ee
For $|\theta|< B$, we can interpret $\epsilon_a$ as the energy of hole excitations --- configurations where a particle of momentum $p_a$ is removed from the Dirac sea. As we also study probe particles on top of this state, we will refer to the $a$-particle bound states that populate the vacuum as the \textit{background} particles. At zero coupling, the relative free energy for $a$ charged fermions reads
\be
\label{eq:Freegeq0}
\mathcal{F}_\text{free}(h) = - \frac{a \, h^2}{2\pi} \,.
\ee

The BA equations \eqref{eq:FhPCF3} and \eqref{eq:FhPCF7} in the GN model have been the subject of an intense study since \cite{fnw1,fnw2}. A revival of interest occurred in the last years, see \cite{Serone:2024uwz} for an overview and \cite{Bajnok:2025mxi} for other recent developments since then. Such studies were focusing on more formal aspects of the BA equations, mostly for $a=1$ and for background particles only (see \cite{Melin:2024oee} for an exception), motivated by resurgence and renormalon considerations. Here we build on some of the previous developments, generalizing to arbitrary $a$ and finding the two scales $\Lambda_\text{n}$ and $\Lambda_\text{c}$.

In addition, we also consider more general BA equations in which probe particles are introduced in the medium. A probe rank-$b$ bound-state particle has momentum and hole energy distribution which we denote by $\overline\chi_b$ and $\overline\epsilon_b$. They are given by the combined probe BA equations:
\begin{align}
\overline\chi_b(\theta) &=  m_b \cosh \theta + \int_{-B}^B  d\theta' \, K_{ab}(\theta-\theta') \, \chi_a (\theta') \,, \label{eq:chiProbe} \\
-\overline\epsilon_b(\theta) &= m_b \cosh \theta - \int_{-B}^{B} d\theta' \, K_{ab}(\theta-\theta') \, \epsilon_a(\theta') - \alpha \, h \,, \label{eq:epsProbe}
\end{align}
where $\alpha = \min(a,b)$ as dictated by \eqref{app:qrelMax}, and $K_{ab}$ is the kernel associated to the vacuum scattering of rank-$b$ and rank-$a$ particles as in \eqref{eq:kernel-ab} below. Clearly, once $\chi_a$ and $\epsilon_a$ are known, \eqref{eq:chiProbe} and \eqref{eq:epsProbe} directly give, upon integration, the momentum and energy distributions of such particles. For neutral excitations, since $p_b(\theta)$ is an odd function of $\theta$ and the states are excitations on top of the vacuum, we choose the integration constant such that $p_b(0)=0$. Charged states arise as excitations of the Fermi surface, so it is convenient to choose the constant such that $p_b(0)=p_\text{F}$. 
Note that $p_b(\theta)$ is still an odd function of $\theta$, yet this choice gives rise to a continuous function, because the momentum of charged states are defined on a Brillouin zone $p\sim p+2p_\text{F}$. From \eqref{eq:chiProbe} and \eqref{eq:epsProbe} we can then also determine the dispersion relation $\epsilon_b = \omega_b(p_b)$ of the probe, or its inverse $p_b = p_b(\omega_b)$, from which we can in particular
extract the mass gap of the excitations. An easier way to determine the mass gap $M_b$ requires only knowledge of $\overline\epsilon_b$:
\be
\label{eq:mass-gap-eps}
M_b = \min_{\theta \,\in\, [-B,B]} \; \bigl\lvert \bar\epsilon_b(\theta) \bigr\rvert \,.
\ee
The relation \eqref{eq:mass-gap-eps} applies also to background excitations if we replace $\bar\epsilon_b$ with $\epsilon_a$.
For probe excitations the minimum in \eqref{eq:mass-gap-eps} occurs at $\theta=0$, while for background excitations, due to \eqref{eq:eps-bc}, the minimum is at $\theta=\pm B$ where it vanishes. So, \emph{background excitations are massless for any $N$}. We will come back to this important point later in this section.

In the following, we will focus on three kinds of probes: 
\begin{itemize}
    \item Charged fermion excitations, given by $b=1$.
    \item Charged fermion holes, given by $b=a-1$.
    \item Neutral fermions, given by $b=a+1$.
\end{itemize} 
The charged fermion is the simplest probe. Next is the charged fermion hole, which is also natural from the large $N$ perspective. From the BA perspective it corresponds to $b=a-1$ because, after removing a charged fermion from the background, one is left with a bound state with one fermion that behaves as a probe.%
\footnote{Note that we are working in the limit of finite density but infinite particle number and length. Thus the created lower bound state acts as a probe, but the fact that the background has one less particle is a negligible effect.}
Meanwhile, a neutral fermion probe, which is also a natural probe in the large $N$ analysis, immediately binds with a background particle to form the energetically favorable $a+1$ bound state which acts as a probe. This description will be checked and validated in section~\ref{sec:tba-num}.

One of the goals of introducing these probes, as we will see later, is to check the observation in the large $N$ limit that, at high densities, the mass gaps are ruled by two nonperturbative scales:
\be
\lim_{h\to \infty} M_1 = \lim_{h\to \infty} M_{a-1} \,\propto\, \Lambda_\text{c} \,, \qquad\qquad  \lim_{h\to \infty} M_{a+1} \,\propto\, \Lambda_\text{n} \,.
\ee
We will check that this holds at finite $N$.

\bigskip

We end by reporting the explicit form of the kernels $K_{ab}(\theta)$ and their Fourier transforms 
\be
\widetilde K_{ab}(\omega) = \int_{-\infty}^\infty \!d\theta\, e^{i \omega \theta} \, K_{ab}(\theta) \,,
\ee
which will be extensively used in the rest of this section. 
We have:
\begin{equation}
\label{eq:kernel-ab}
\widetilde K_{ab}(\omega) = \delta_{ab} - e^{\pi  \Delta  | \omega | } \; \frac{ \sinh (\pi  b \Delta  \omega ) \, \cosh \bigl(\frac{\pi \omega}{2} (2 a \Delta -1) \bigr) }{ \sinh(\pi  \Delta  \omega ) \, \cosh \bigl( \frac{\pi  \omega}{2} \bigr) } \,, \qquad\quad 1 \leq b \leq a \leq N-2 \,, 
\end{equation}
where
\begin{equation}
\Delta = \frac{1}{2N-2} \,.
\end{equation}
Moreover $K_a = K_{aa}$ and $K_{ba} \equiv K_{ab}$ for $a>b$. In rapidity space the kernel is given by
\begin{equation}
K_{ab}(\theta) = f_{2N,|a - b| - N + 1} (\theta) - f_{2N,(a + b) - N + 1} (\theta) \,,
\end{equation}
where $f$ are sums of polygamma functions (see \textit{e.g.} \cite{melin2023integrability} for a derivation):
\begin{align}
\label{fnm-aux}
& f_{n,m}(\theta) = - f_{n,-m}(\theta) = \sum _{r=0}^{m-1} \kappa _{\frac{m-2 r-2}{n-2}}(\theta) \,, \qquad m>0 \,, \qquad \kappa_{\frac{1}{2}} \equiv - \kappa_{-\frac{1}{2}} \,, \\
& \kappa_a(x) = \frac{1}{4\pi^2} \Biggl[
 \psi \biggl( \frac{a}{2} + \frac{3}{4} + \frac{i x}{2 \pi } \biggr)
+ \psi \biggl(\frac{a}{2}+\frac{3}{4}-\frac{i x}{2 \pi } \biggr)
-\psi \biggl( \frac{a}{2}+\frac{1}{4}+\frac{i x}{2 \pi } \biggr)
-\psi \biggl( \frac{a}{2}+\frac{1}{4}-\frac{i x}{2 \pi } \biggr)
\Biggr] \,. \nn 
\end{align}

\subsection[Integrability at finite \tps{N}{N} and trans-series]{Integrability at finite \tpsb{N}{N} and trans-series}
\label{sec:tba-finite}

Equations of the form of \eqref{eq:FhPCF7} are amenable to a numerical treatment at finite coupling, as we will discuss in section~\ref{sec:tba-num}. However, for analytical purposes, it is instructive to study them in the weak coupling limit. As worked out in  \cite{Zamolodchikov:1995xk, Volin:2009wr, Marino:2019eym, Marino:2021dzn, Bajnok:2022xgx}, one can in principle recursively extract the full analytical trans-series from the integral equation using the Wiener-Hopf method, first used in this kind of problems in \cite{hmn,fnw1}. In this section, we briefly review the key logical steps of the method and apply it to the case at hand. We refer to \cite{Reis:2022tni} for a more pedagogical review. See also \cite{Bajnok:2025mxi} for a very effective application of the 
method in 2d integrable theories.

In a nutshell, the Wiener--Hopf method to solve an integral equation like \eqref{eq:FhPCF7} consists in decomposing its Fourier transform into two equations, which are respectively analytic in the upper and lower half planes. Matching the two, and taking into account of the boundary conditions, will allow us to 
find an iterative solution, directly in terms of the free energy $\mathcal{F}$.
In order to take a Fourier transform in a convenient way, it is necessary to extend \eqref{eq:FhPCF7}, valid only in the range $\theta\in[-B,B]$, to the full real axis, keeping $\epsilon$ supported on the interval. This is achieved by writing
\begin{equation}
    \epsilon(\theta) - \int_{-B}^B K_a(\theta-\theta')\epsilon(\theta') d\theta' = g(\theta)+Y(\theta-B)+Y(-\theta-B),
    \label{eq:long-TBA}
\end{equation}
where $g$ extends the original r.h.s. to
\begin{equation}
    g(\theta) = ah\, \Theta(B^2-\theta^2)-\frac{ m_a e^{-\theta}}{2} \Theta(\theta+B)-\frac{ m_a e^{\theta}}{2} \Theta(-\theta+B)
\end{equation}
and $Y(\theta)$ is an unknown function (admitting a Fourier transform $\widetilde Y(\omega)$) supported on the positive reals, necessary for \eqref{eq:long-TBA} to hold for $|\theta|> B$. 
 We define two functions $G_\pm$ such that
\begin{equation}
    1-\widetilde K_a(\omega) = \frac{1}{G_+(\omega)G_-(\omega)},
\end{equation}
where $G_+(\omega)$ and $G_-(\omega)$ are respectively analytic in the upper and lower half planes.
Here and in what follows 
\begin{equation}
\widetilde f(\omega) = \frac{1}{2\pi} \int_{-\infty}^{+\infty}\!d\theta\, e^{i \omega \theta} f(\theta)
\end{equation}
denotes the Fourier transform of a generic function $f(\theta)$.
Due to the evenness of $K(\theta)$, the functions $G_\pm$ must be a reflection of each other: $G_\pm (\omega) = G_\mp(-\omega)$.

The Fourier transform of the extended integral equation can be arranged into
\begin{equation}\label{eq:WHready}
    \frac{e^{-iB\omega}\widetilde\epsilon(\omega)}{G_+(-\omega)}=e^{-iB\omega}G_+(\omega) \widetilde g(\omega)+G_+(\omega) \widetilde Y(\omega) + e^{-2i B\omega} G_+(\omega)\widetilde Y(-\omega).
\end{equation}
We apply the Wiener--Hopf decomposition to \eqref{eq:WHready}, splitting it into a sum of two functions, respectively analytic in the upper and lower half planes, obtaining in this way two equations.  We will not report the details of the decomposition, which follow the lines of the $a=1$ case, explicitly worked out e.g. in chapter 5.2 of \cite{Reis:2022tni}.  The result is conveniently written in terms of two functions defined as
\begin{equation}\label{eq:urDef}
\begin{split}
    u(\omega) & \equiv \frac{(\omega-i)G_+(\omega)}{i ah G_+(0)}\left(\widetilde Y(\omega)- \frac{iah}{\omega+i0} + \frac{i m_a e^B}{2(\omega-i)}\right) \,, \\
    f(\omega) & \equiv - \frac{\omega+i}{\omega-i}\frac{G_-(\omega)}{G_+(\omega)}.
\end{split}
\end{equation}
In terms of these functions, the two equations read:%
\footnote{In order to write the two eqations in the form \eqref{eq:u-eq-original}, we also used the boundary condition $\epsilon(\pm B) = 0$, which in momentum space reads $\lim_{\kappa\to\infty} \kappa e^{- \kappa B}\widetilde \epsilon(i \kappa) = 0$.}
\begin{align}
\label{eq:u-eq-original}
    u(\omega) & = \frac{i}{\omega}+\frac{1}{2\pi i}\int_\mathbb{R}\frac{e^{2iB\omega'}f(\omega')u(\omega')}{\omega'+\omega+i0}d\omega' \,, \\
      \frac{e^{i B\omega} \widetilde\epsilon(\omega)}{G_+(\omega)} & = \frac{iah G_+(0)}{\omega+i}\left(\frac{i}{\omega}-\frac{1}{2\pi i}\int_\mathbb{R}\frac{e^{2iB\omega'}f(\omega')u(\omega')}{\omega'-\omega-i0}d\omega'\right).  \label{eq:G-eps-original}
\end{align}
From the definition of $u(\omega)$ in \eqref{eq:urDef}, we have
\begin{equation}
    u(i) = \frac{m_ae^B}{2ah}\frac{G_+(i)}{G_+(0)}.
    \label{eq:u-bc}
\end{equation}
Using \eqref{eq:G-eps-original} and \eqref{eq:u-bc}, the free energy can be written
in terms of the unknown function $u(\omega)$:
\begin{equation}
 \mathcal{F}(h)= - \frac{m_a}{2\pi} \widetilde \epsilon(i) = - \frac{(ah)^2}{2\pi} u(i) G_+^2(0)\left(1-\frac{1}{2\pi i}\int_\mathbb{R}\frac{e^{2iB\omega'}f(\omega')u(\omega')}{\omega'-i}d\omega'\right).
    \label{eq:u-free-energy}
\end{equation}
The analysis so far applies for generic holomorphic functions $G_\pm(\omega)$. In the case at hand, for the kernel $K_{a}$ in \eqref{eq:kernel-ab}, we specifically have
\begin{equation}
    G_+(\omega) = e^{-\Delta \omega\log(-i\omega)} r(\omega)\,,
    \label{eq:G-template}
\end{equation}
where
\begin{equation}
r(\omega) = 
e^{\frac{i \omega \mathfrak{b}}{2} }
\frac{\Gamma (1-i a \Delta  \omega ) \Gamma \left(\frac{1}{2}- ( 1-2 a \Delta   )\frac{i\omega}{2} \right)}{\sqrt{a} \Gamma \left(\frac{1}{2}-\frac{i \omega }{2}\right) \Gamma (1-i \Delta  \omega )}
\label{eq:rDef-a}
\end{equation}
is a meromorphic function $r(\omega)$ with poles along the negative imaginary axis, and  
\begin{equation}
\mathfrak{b}=2 \Delta -2 \Delta  \log (\Delta )+2 a \Delta  \log (2 a \Delta )+(1-2 a \Delta ) \log (1-2 a \Delta ).
\end{equation}
The function $G_+$ also satisfies $G_+(+i\infty) = 1 +\mathcal{O}(\omega^{-1})$.

The equation for $u$ in \eqref{eq:u-eq-original} cannot be solved in closed analytic form, but it admits a recursive transseries solution. The function $f(\omega)$ entering the kernel of \eqref{eq:u-eq-original} has both a discontinuity along the positive real axis due to the exponentiated logarithm in \eqref{eq:G-template}, and inherits the poles of $r(-\omega)$ along the same half-line. One can deform the integral around the imaginary axis leading to a convergent integral from the discontinuity and a sum over the residues coming from the poles. Thus \eqref{eq:u-eq-original} can be rewritten as $(\omega'= i \xi')$
\begin{equation}
    u(\omega) = \frac{i}{\omega}+\frac{1}{2\pi i}\int_0^\infty \frac{e^{-2B\xi'}\mathcal{F}(i \xi') u(i\xi')}{i\xi'+\omega}d\xi'+\sum_k \frac{e^{-2B\xi_k}f_{\xi_k} u(i \xi_k) }{\omega+i\xi_k},
    \label{eq:u-int-two-pieces}
\end{equation}
where $\mathcal{F}$ is the discontinuity of $f(\omega)$ along the imaginary axis, $i\xi_k$ are the poles and $f_{\xi_k}$ is the residue of $f(\omega)$ at $i\xi_k$. 
As follows from \eqref{eq:G-template}, the discontinuity is of the form
\begin{equation}
  \mathcal{F}(i\xi) =   - 2\pi i e^{-2\Delta \xi\log\xi}\sum_{n\geq 1} d_n \xi^n.
  \label{eq:rho-disc}
\end{equation}
It is useful to introduce the auxiliary coupling $v$, defined as 
\begin{equation}
    \frac{1}{2v}- \Delta \log v \equiv  B,
    \label{eq:def-v}
\end{equation}
and the variable $\eta$, the rescaled imaginary component of $\omega$, defined as
\begin{equation}
\omega\equiv i v \eta\,.
\end{equation}
As we will shortly see, $v$ can be interpreted as a QFT coupling constant and, in terms of $v$ and $\eta$, $\mathcal{F}\sim {\cal O}(v)$.  
We can then write a iterative solution for $u$ as follows:
\begin{equation}
    u(i v \eta) = \frac{v^{-1}u_0}{1-v\mathfrak{I}} \equiv \sum_n v^{n-1} \mathfrak{I}^n u_0(\eta),
    \label{eq:u-pert}
\end{equation}
where $u_0$ is the seed solution
\begin{equation}
    u_0(\eta) = \frac{1}{\eta}-\sum_k \left(e^{-\frac{1}{v}}v^{2\Delta }\right)^{\xi_k}\frac{i v f_{\xi_k} u(i\xi_k)}{\xi_k+v\eta},
    \label{eq:u-seed}
\end{equation}
$\mathfrak{I} $ is a perturbative integral operator
\begin{equation}
    \mathfrak{I}f(\eta)= -\frac{1}{\pi}\sum_{n\geq 0} v^n \int_0^\infty \frac{e^{-\eta'} p_n(\eta',\log \eta')}{\eta+\eta'}f(\eta')d\eta',
\end{equation}
and $p_n(x,y)$ are polynomials constructed from the $d_n$ coefficients in \eqref{eq:rho-disc}. The solution is found at each perturbative and non-perturbative order by iterating the operator $\mathfrak{I}$ and solving recursively for the $u(i\xi_k)$. See \cite{Marino:2023epd} for more explicit details on the recursion. 

The condition \eqref{eq:u-bc} relates $v$ to $h$, giving
\begin{equation}
    e^{-\frac{1}{v}}v^{2\Delta} \propto \frac{m_a^2}{h^2}.
    \label{eq:v-massgap}
\end{equation}
One can use this relation to rewrite $v$ in terms of a QFT coupling. 
In terms of the coupling
\begin{equation}
\alpha \equiv 2\beta_0 g^2\,,
\end{equation}
where $g^2$ is the GN coupling in \eqref{eq:UV1} and $\beta_0 = 1/(4\pi \Delta)$ is its associated one-loop beta function coefficient, we can define
the ``BA scheme'', where $\alpha(\mu)$ is defined as
\begin{equation}\label{eq:TBAscheme}
    \frac{1}{\alpha}+\frac{\beta_1}{2\beta_0^2}\log\alpha = \log \frac{\mu}{m_a},
\end{equation}
with $\beta_1$ the two-loop beta function coefficient. Interestingly enough, $\Delta=- \beta_1/(2\beta_0^2)$ and hence, matching
\eqref{eq:v-massgap} with \eqref{eq:TBAscheme}, we find that the parameter $v$ can be interpreted as a QFT coupling at a scale of order $h$:
\begin{equation}
v \approx \frac 12 \alpha (\mu \propto  h)\,.
\end{equation}
Note that the full relation $\alpha(v)$ is itself a trans-series which must be inverted order by order to write $v$ as a function of $\alpha$.

With the recursive solution for $u$ and the relation $\alpha(v)$, we can use \eqref{eq:u-free-energy} to write the trans-series for the free energy in terms of the coupling $\alpha$. First, we note that the first term in \eqref{eq:u-free-energy}, proportional to the number 1 in square parenthesis, reproduces the free theory result \eqref{eq:Freegeq0}, as $G_+(0) = 1/\sqrt{a}$, and $u(i) = 1$ at leading order. A second, less trivial, universal contribution arises from the residue of the pole at $\omega=i$ coming from the integral in \eqref{eq:u-free-energy}:
\begin{equation}
\mathcal{F}_{\text{Res}\, \omega=i}(h) =  \frac{m_a^2}{8} \frac{\sin \pi \Delta}{\sin^2 a \pi \Delta}e^{\mp i \pi \Delta} = \frac{m^2}{8} (\cot (\pi  \Delta )\mp i) \,,
   \label{eq:ResomegaI}
\end{equation}
where we used \eqref{eq:u-bc} and the mass formula \eqref{eq:spectrum}.
As argued in \textit{e.g.} \cite{Fateev:1992tk, Zamolodchikov:1995xk}, the real part of this term can be identified with the subtraction of the vacuum free energy of the Gross--Neveu model (normalized with respect to the free fermion) that ensures $\mathcal{F}(h\rightarrow m_a)=0$. Because such a quantity is independent of the chemical potential, it should be independent of $a$, as we observe. In fact, it agrees with the prediction of \cite{Saleur:2009bf} and the $a=1$ calculation of \cite{Marino:2021dzn}.
It also matches the equivalent term in the  large $N$ analysis, see \eqref{eq:phi-m}.
The $\pm$ imaginary ambiguity depends on which branch of the log term in \eqref{eq:G-template} is taken for $G_-(\omega)$, 
and should cancel with  the leading ambiguity from Borel summation of perturbation theory.

The remaining, and more interesting, non-perturbative effects are dictated by the poles of $f$, which can be read from \eqref{eq:rDef-a}.
We have pole singularities at two sequences of points $i\xi_k$ and $i\xi'_k$:
\begin{equation}
\xi_k = \frac{2 k-1}{1- 2 a \Delta } = (2 k - 1) \frac{N-1}{N-a-1},\quad \xi'_k = \frac{k}{a \Delta } = 2 k \frac{N-1}{a},\quad k\in\mathbb{N}.
\label{res-a}
\end{equation}
The residues of $f$ at these points are given by
\begin{align}\label{eq:residues}
if_{\xi_k} &= 
e^{\mp i \pi  \Delta  \xi _k} \frac{2  (-1)^{k+1}}{\Gamma(k)^2} 
\left(\frac{ e^{\xi _k(\mathfrak{b} -2 \Delta  \log \left(\xi _k\right))}}{1-2 a\Delta }\right)
\frac{\Gamma \left(\frac{3}{2}+\frac{\xi _k}{2}\right) \Gamma \left(1+\Delta  \xi _k\right) \Gamma \left(1-a \Delta  \xi _k\right)}{\Gamma \left(\frac{3}{2}-\frac{\xi _k}{2}\right) \Gamma \left(1-\Delta  \xi _k\right) \Gamma \left(1+a \Delta  \xi _k\right)},
\\
if_{\xi'_k} &=
ie^{\mp i \pi  \Delta  \xi' _k} \frac{2(-1)^{k+1}}{k\Gamma(k)^2} 
\left(\frac{e^{\xi' _k \left(\mathfrak{b}-2 \Delta  \log \left(\xi' _k\right)\right)}}{2 a \Delta  }\right)
\frac{\Gamma \left(\frac{3}{2}+\frac{\xi' _k}{2}\right) \Gamma \left(1+\Delta  \xi' _k\right) \Gamma \left(\frac{1}{2}-\frac{1}{2} (1-2 a \Delta ) \xi'_k\right)}{\Gamma \left(\frac{3}{2}-\frac{\xi'_k}{2}\right) \Gamma \left(1-\Delta  \xi'_k\right) \Gamma \left(\frac{1}{2}+\frac{1}{2} (1-2 a \Delta ) \xi'_k\right)}. \nn
\end{align}
When the poles \eqref{eq:residues} overlap, the residue is given by a different formula, which can be derived from \eqref{eq:rDef-a}. 
Note that in that case we do {\it not} have a double pole for $f$ as another factor develops a zero.%
\footnote{This is easy to see. Given two positive integers $k$ and $k'$,  the relation $\xi_k = \xi'_{k'}$ is possible only when $a\Delta = k/(2k+2k'-1)$. Plugging this value of $a$ to $\xi_k$, we get $\xi_{k}(a(k,k')) = 2k+2k'-1$, value for which the first gamma function in the denominator of \eqref{eq:rDef-a} develops a simple pole. As a result, all the coefficients $a_{k,p}^{[m,n]}$ in eqn.~(4) of \cite{Benini:2025riz} vanish when $p\neq 0$, and no logarithms appear.}
As in \eqref{eq:ResomegaI}, the ambiguities coming from the first exponential factor appearing in \eqref{eq:residues} cancel those coming from the Borel summation of the free energy transseries.
We can identify these two sequences of poles with the non-perturbative scales $\Lambda_{\rm n}$ and $\Lambda_{\rm c}$ defined in \eqref{eq:lambda-12}, respectively. Concretely, we have
\begin{equation}
    \left(e^{-\frac{2}{\alpha}}\alpha^{2\Delta}\right)^{\xi_k}\propto \left(\frac{\Lambda_{\rm n}^2}{h^2}\right)^{2k-1}
    \,
    \qquad
    \left(e^{-\frac{2}{\alpha}}\alpha^{2\Delta}\right)^{\xi'_k}\propto \left(\frac{\Lambda_{\rm c}^2}{h^2}\right)^{k}.
\end{equation}
This is the key result of this section: from the BA we find that, even at finite $N$, non-perturbative effects are controlled by the two scales $\Lambda_{\rm n}$ and $\Lambda_{\rm c}$. 

The results obtained apply for any $a$ in the range $[1,N-2]$. 
For $a=1$, we note from \eqref{eq:rDef-a} that the scale $\Lambda_\text{c}$ does not exist, as two gamma functions simplify, removing the simple poles $\xi'_k$. This is in agreement with the QFT results in section~\ref{sec:pert theory}.
The non-integer powers found in \cite{Marino:2021dzn} for $a=1$ disappear once the surviving poles $\xi_k$ are expressed in powers of $\Lambda_\text{n}$ rather than $\Lambda$. For $a=N$, the kernel \eqref{eq:kernel-ab} should be replaced by the one for kink scattering. Aside from that, the Wiener-Hopf decomposition described still applies. The BA analysis of the $a=N$ case has been worked out in \cite{Melin:2024oee} at leading order in large $N$, but the transseries of the free energy has not been analyzed there. Starting from the kernel for kink-kink scattering, it is straightforward to verify that only one tower of residues appear, and these are associated precisely to the scale $\Lambda_c$ in \eqref{eq:lambda-12} for $a=N$. 
The only case left is $a=N-1$. The state with the highest charge to mass ratio is the kink, like for $a=N$. However, in contrast to the $a=N$ case, the kink- kink scattering kernel is expected to be given by a $2\times 2$ system, as both kink polarizations along the neutral direction are allowed.
We expect that the physics in this case is similar to the $a=N$ case, but it would be interesting to perform a detailed analysis.

Solving for the recursion \eqref{eq:u-pert} to higher and higher terms produces a transseries for the free energy, which has the form
\begin{equation}
\label{eq:ts-12-nologs}
\mathcal{F}(h) \sim h^2  \sum_{m,n\geq 0} \frac{\Lambda_\mathrm{n}^{2m} \Lambda_\mathrm{c}^{2n} }{ h^{2(m+n)}} \, \sum_{0 \leq k} \, a^{[m,n]}_{k} \, g^{2k} - c_0 \Lambda^2.
\end{equation}
For illustration, we report below the first few coefficient terms:
\begin{equation}
\label{eq:ts-terms}
\begin{aligned}
a_0^{[0,0]}&= - \frac{a}{2\pi},
\quad
a_1^{[0,0]}=  \frac{a}{(2\pi)^2},
\\
a_2^{[0,0]} &= -\frac{a}{\left(2 \pi\right)^3  } \left\{\frac{1}{\Delta }\log \left(\frac{2}{e} \Gamma (\Delta +1) \frac{\sin (\pi  a \Delta )}{\pi  a \Delta }\right)+\left(\frac{3}{2}-(1-2 a) \log (2)\right)\right\},
\\
a_{0}^{[1,0]}&= -\frac{iaf_{\xi_1}e^{-\mathfrak{b}\xi_1}}{4\pi} \frac{(2 e)^{2 \Delta \xi_1}}{\xi_1(\xi_1^2-1)},
\quad a_{0}^{[0,1]}= -\frac{iaf_{\xi'_1}e^{-\mathfrak{b}\xi'_1}}{4\pi} \frac{(2 e)^{2 \Delta \xi'_1}}{\xi'_1(\xi'_1\,\!^2-1)}, 
\\
c_0 &=  -\frac{(2 e)^{2 \Delta } e^{\mp i \pi  \Delta } \Gamma (\Delta )}{8 \pi  \Gamma (1-\Delta )},
\end{aligned}
\end{equation}
where we use $\mu = a h$ in \eqref{eq:TBAscheme}.%
\footnote{\label{foo: m/Lambda finite N}%
One relates $m_a$ in \eqref{eq:TBAscheme} to $\Lambda$ using the exact mass gap derived in \cite{fnw1} (with appropriate normalization of $\Lambda$):
\begin{equation}
    \frac{m}{\Lambda} = \frac{(2 e)^{\Delta }}{\Gamma (1-\Delta )} \,.
\end{equation}
This is the finite $N$ version of \eqref{eq:gap-largeN}.}
Note that we have arbitrary products of $\Lambda_{\rm n,c}$ (and not just those that appear at the poles) because in the process of solving the recursion and inverting the couplings, successive products of the intermediate trans-series are taken. For example, there is a term in $u(\eta)$ proportional to $(e^{-\frac{1}{v}}v^{2\Delta})^{\xi_1}u(i\xi_1)$. But after solving a few orders of the recursion, one finds that $u(i\xi_1)$ itself contains corrections of orders $(e^{-\frac{1}{v}}v^{2\Delta})^{\xi_1}$ and $(e^{-\frac{1}{v}}v^{2\Delta})^{\xi_2}$, leading to terms proportional to $(e^{-\frac{1}{v}}v^{2\Delta})^{2\xi_1}$ and $(e^{-\frac{1}{v}}v^{2\Delta})^{\xi_1+\xi_2}$ in the final answer.

It should be stressed that \eqref{eq:ts-12-nologs} provides a trans-series representation from which the actual function $\mathcal{F}(h)$ is found by Borel summation. However, due to singularities in the Borel transform, this summation requires the choice of a small deformation which can be phrased as giving an infinitesimal phase to $g$. The sign of this phase is important and should be coordinated with the $\pm$ signs in the coefficients in \eqref{eq:ts-terms}. The ``ambiguity'' from the deformed Borel summation and that of the coefficients cancel, leading to a well-defined unique function. For more details on this procedure in the $a=1$ case and similar integrable models, see \textit{e.g.} \cite{Reis:2022tni, Bajnok:2025mxi} and references therein. For reviews of resurgence and Borel summation see \cite{Sauzin:2014qzt, Dorigoni:2014hea, Serone:2024uwz, Aniceto:2018bis}.

The full or resummed function $\mathcal{F}(h)$ can be compared to the semi-classical result at large $N$. We do this analysis numerically in section~\ref{sec:num-largeN}. But one can also compare the large $N$ limit of the trans-series \eqref{eq:ts-12-nologs} (rephrased in terms of the 't~Hooft coupling) with the high density expansion of the large $N$ analytical result. This match is \textit{a priori} subtle because the order of limits (large $N$ and weak coupling/high density) is the inverse. Nonetheless, at leading order, one finds that the large $N$ limit of the coefficients in \eqref{eq:ts-terms} reproduces the expansion \eqref{grand potential high density y less 1}. Such agreement is a non-trivial consistency check of both results, and of the interaction between large $N$ expansion and weak coupling trans-series. That the two orders of expansion match was expected given similar analysis to higher order in the $a=1$ case, see \cite{Marino:2021dzn,Marino:2023epd,Reis:2022tni, DiPietro:2021yxb}.

A more sophisticated approach for calculating the trans-series from this type of integral equation was developed in \cite{Bajnok:2022xgx}. In that approach, the perturbative part of the free energy is calculated using the generalised Volin's method as in \cite{Marino:2019eym} and then the non-perturbative sectors are related to the perturbative one through ODEs in the parameter $v$. While more indirect, this method is very efficient at generating many terms in the trans-series and allows for resurgence relations to be extracted analytically to all orders. 

\subsection[Integrability at large \tps{N}{N}]{Integrability at large \tpsb{N}{N}}
\label{subsec:BAlargeN}

In this section we consider the BA equations in the large $a$, large $N$ limit, at fixed $y$.
At leading order in $N$, the mass and charges of the bound states and kink, $q_a$, $q_\text{k}$, $m_a$ and $m_\text{k}$, are given by
\begin{equation}
\begin{aligned}
q_a & = y N \equiv N q_y  \,, \qquad \qquad \qquad \qquad  q_\text{k} = N \frac{y}{2} \equiv N \hat q_\text{k}\,, \\
m_a & = N \frac{2m}{\pi} \sin\Big(\frac{\pi y}{2}\Big) \equiv N m_y \,, \qquad m_\text{k} = N \frac{m}{\pi} \equiv N \hat m_\text{k}\,.
\label{eq:chargmass}
\end{aligned}
\end{equation}
Here $m_a$ and $q_a$ are the mass and charge of the highest possible rank $a$ composite.
As previously explained, the particles which proliferate are expected to be the ones with the highest $q/m$ ratio. For any $y\neq 1$ these are rank $a$ composites, see appendix~\ref{app:anti}.
For $y=1$ these are kink-anti kink bound states. 

The large $N$ limit of the kernel \eqref{eq:kernel-ab} can be written as 
\be\label{eq:TBAy1}
\widetilde K_a(\omega) =  - \frac{N}{\pi} \bigg(  \frac{\tanh\frac{\pi \omega}{2}}{\omega} +   \frac{\sinh\frac{\pi \omega (2y-1)}{2}}{\omega \cosh\frac{\pi \omega}{2}}  \bigg) \equiv N \widetilde K_y(\omega)\,.
\ee
It is useful to take an inverse Fourier transform 
\be\label{eq:kink2}
K_\text{a}(\theta)  = \int_{-\infty}^\infty\!\frac{d\omega}{2\pi} e^{-i \omega \theta} \widetilde K_a(\omega)\,,
\ee
which gives
\be\label{eq:TBAy2}
K_a(\theta) =  - \frac{N}{2\pi^2} \bigg( \log \Big(1-\frac{2\cos (\pi y)}{\cos(\pi y) +\cosh \theta}\Big) + \log \coth^2\frac{\theta}{2}  \bigg) \equiv N K_y(\theta)\,.
\ee
The kernel $K_a$ for rank $a$ particle scales as $N$, so the BA equations \eqref{eq:FhPCF7} turn into singular integral equations in the large $N$ limit:
\be
\label{eq:intetwo}
 -\int_{-B}^{B} d\theta'\, K_y(\theta-\theta') \epsilon_y(\theta')= h q_y -m_y \cosh \theta\,.
\ee
For the kink case at $y=1$, we have the BA equation
\be\label{eq:kink8}
-\int_{-B}^B \!d\theta' \epsilon_\text{k}(\theta') \widehat K_\text{k}(\theta-\theta') = \frac{h}{2}-\frac{m}{\pi}\cosh(\theta) \,,
\ee
where
\be\label{eq:kink7}
\widehat K_\text{k}(\theta)  = -\frac{1}{4\pi^2} \log \coth^2\frac{\theta}{2} \equiv \frac{K_\text{k}(\theta)}{N}\,,
\ee
is the kink kernel.%
\footnote{In principle both a description in terms of kinks and in terms of rank $N-2$ states should apply. In fact, the free energy computed using one or the other description agree with each other. This is due to a factor of two in $\lim_{y\to 1}K_y(\theta) = 4K_\text{k}(\theta)$ which implies $\epsilon_\text{k}(\theta) = 2 \epsilon_{y=1}(\theta)$, compensating for the factor in the masses.}
We have not been able to find an analytic solution $\epsilon_y(\theta)$ of \eqref{eq:intetwo}, except for the special case $y=1/2$.
This is closely related to the kink case at $y=1$, which was studied in \cite{Melin:2024oee}. Since $y=1$ is the simpler of the two, we begin by reviewing the results of \cite{Melin:2024oee}, providing additional details that will be useful for the analysis of the $y=1/2$ case.
A more in depth study of the $y=1$ case can be found in \cite{Melin:2025eyw}, in which the connection with the semiclassical large $N$ analysis is studied in detail.

\subsubsection[Review of kink-antikink case (\tps{y=1}{y=1})]{Review of kink-antikink case (\tpsb{y=1}{y=1})}

Following the original derivation in \cite{Melin:2024oee}, the integral equation \eqref{eq:kink8} can be solved using the results reviewed in appendix~\ref{app:IntEq}. 
By taking a derivative with respect to $\theta$, \eqref{eq:intetwo} gives precisely \eqref{eq:sinh1}, with $f$ as in \eqref{eq:sinh1a}.
The resulting solution reads
\be
\epsilon_\text{k}(\theta) = 2 m \sqrt{\sinh^2 B - \sinh^2\theta}\,.
\label{eq:sinh5}
\ee
We still have to determine the value of $B$. This can be done by plugging the solution \eqref{eq:sinh5} in \eqref{eq:kink8} and demand that it is verified at a given $\theta$. We choose $\theta = 0$. The integral in \eqref{eq:kink8} can be computed by changing variables
\be\label{eq:kink9}
\cosh \theta' = \frac 1y
\ee
and expanding in $y$. We get
\be\label{eq:kink10}
\frac{\pi}{m} \int_{-B}^B \!d\theta' \epsilon_\text{k}(\theta') \widehat K_\text{k}(\theta') \equiv \sum_{n=0}^\infty\frac{F_n}{2n+1} = 1- \frac{h \pi}{2m}\,,
\ee
where
\be\label{eq:Bdef}
k =\frac{1}{\cosh B}\,,
\ee
and
\be
F_n = \frac{2}{\pi k} \int_{k}^1 \!dy \sqrt{\frac{y^2-k^2}{1-y^2}} y^{2n-1}\,.
\ee 
The explicit form of the $F_n$ terms is
\begin{align}\label{eq:kink12}
F_0 & = 1- \frac{1}{k}\,, \\
F_n  & = -\frac{ \Gamma\Big(n-\frac 12\Big) }{2k\sqrt{\pi}\Gamma(n+1)} \bigg( 2(n-1){}_2F_1\Big(-\frac 12,-n,\frac 32 - n; k^2\Big)+{}_2F_1\Big(\frac 12,-n,\frac 32 - n; k^2\Big)\bigg)\,, \quad n\geq 1\,. \nn
\end{align}
We can now further expand for small $k$ each $F_n$ and, for each order in $k$, perform the whole sum over $n$ in \eqref{eq:kink10}. In this way the expansion reproduces the one of the complete elliptic integral of the second kind $E(k^2)$:
\be\label{eq:kink13}
1 - \frac{\pi}{2k}+ \frac{k \pi}8 +\frac{3k^3\pi}{128} +\frac{5 k^5 \pi}{512} + \frac{175 k^7 \pi}{32768} + {\cal O}(k^9)  =  1-\frac{E(k^2)}{k}\,.
\ee
From \eqref{eq:kink10} we implicitly determine $B$ as
\be\label{eq:kink14}
\frac{E(k^2)}{k} = \frac{\pi h}{2m}\,.
\ee
The free energy is easily computed:
\be\label{eq:kink15}
\mathcal{F}_\text{k} = -\frac{m_\text{k}}{2\pi} \int_{-B}^B \! d\theta \, \cosh \theta \,\epsilon_\text{k}(\theta) = - \frac{N m^2}{2\pi}\sinh^2 B 
 =  \frac{Nm^2}{2\pi} - \frac{Nm^2}{2\pi k^2}\,.
\ee
For $h\to\infty$, we have $h=m\cosh B$ and hence 
\be
\lim_{h\to\infty} \mathcal{F}_\text{k} = - \frac{N h^2}{2\pi}\,,
\ee
in agreement with the free energy of $N$ species of free fermions.%
\footnote{For free fermions $\epsilon(\theta) = h - \cosh\theta$ and $\cosh B = h$.}

We now solve for the density of states. The formula is similar to \eqref{eq:kink8}. In the large $N$ limit we get
\be\label{eq:kink16}
-\int_{-B}^B \!d\theta' \chi_\text{k}(\theta') \widehat K_\text{k}(\theta-\theta') = \frac{m}{\pi}\cosh(\theta) \,.
\ee
Upon differentiating, the equation is identical (up to a sign) to the one discussed before. In contrast to $\epsilon$, $\chi$ is not required to be bounded at the end-points. So the general solution reads (see appendix~\ref{app:IntEq}) 
\be\label{eq:kink17}
\chi_\text{k}(\theta) = -  2m \sqrt{\sinh^2 B - \sinh^2\theta}+ \frac{mc}{2\sqrt{\sinh^2 B - \sinh^2\theta}}\,.
\ee
Following \cite{Melin:2024oee}, we parametrize the constant $c$ as
\be\label{eq:kink20}
c = \frac{2\sinh(2B)}{c_s} = \frac{4 k'}{c_s k^2}\,, \qquad k' = \sqrt{1-k^2} = \tanh B\,.
\ee
We fix $c_s$ by demanding that \eqref{eq:kink16} is satisfied for $\theta=0$. Proceeding as before, after some algebra, we 
get
\be\label{eq:kink24}
c_s = \frac{k' K(k^2)}{E(k^2)}\,,
\ee
where $K$ is the elliptic function of the first kind.
The particle and baryon density are respectively
\be
\label{eq:kink26}
\begin{split}
    \rho_\text{k}  & = \frac{1}{2\pi}\int_{-B}^B \! \!d\theta \,  \chi_\text{k}(\theta) = \frac{m} {k K(k^2)} = \frac{2\epsilon_+\epsilon_-}{\pi h c_s} \,, \\
    B_\text{k} & = q_\text{k}  \rho_\text{k}  = \frac{N} {2k K(k^2)}\,,
    \end{split}
\ee
where
\be
\label{eq:kink27}
\epsilon_+ = m\cosh B\,, \qquad \epsilon_- = m\sinh B \,.
\ee
The energy density reads
\begin{align}
   e_\text{k} & = \frac{m_\text{k}}{2\pi}\int_{-B}^B \! \! d\theta\, \chi_\text{k}(\theta)\cosh \theta = -\frac{Nm^2}{2\pi}\sinh^2B + \frac{Nm^2}{\pi}\frac{E(k^2)}{k^2 K(k^2)} \nn \\
 & = \mathcal{F}_\text{k} + \rho_\text{k} h_\text{k} = \mathcal{F}_\text{k} + B_\text{k} h \,,   \label{eq:kink29} 
\end{align}
where we used \eqref{eq:chargmass}, \eqref{eq:kink14}, \eqref{eq:kink15}, and \eqref{eq:kink26}. We reproduce the expected relation between free energy and energy density, related by a Legendre transform.

We now turn to dispersion relations, recalling \eqref{eq:chi-dpdt}, with $p$ the momentum of the excitation and $-\epsilon$ its energy.
The Fermi momentum $p_F$ of particles in the vacuum is determined by $\rho$ as in
\eqref{eq:pf-pirho}.
It is easier to determine $p = p(\epsilon)$, which reads
\be
\label{eq:kink34}
-c_s \frac{dp}{d\epsilon} = \frac{4 \epsilon_+ \epsilon_- -  c_s^2\epsilon^2}{\sqrt{(4\epsilon_+^2-\epsilon^2)(4\epsilon_-^2-\epsilon^2)}}\,.
\ee
The relation \eqref{eq:kink34} can be analytically integrated, using the boundary condition at $\theta=B$ to fix the integration constant:
$p(\epsilon(B))= p(0) = p_F$. We get
\be\label{eq:kink35}
p = P(\epsilon) + p_F\,,
\ee
where $P(0) = 0$ and $P(\epsilon)$ is a complicated expression in terms of elliptic functions which will not be reported.
For $p-p_F\ll 1$, we have $P(\epsilon) = \epsilon/c_s(1+ {\cal O}(\epsilon^2))$, from which we see that hole excitations close to the Fermi surface
are gapless with sound velocity given by $c_s$. It is reasonable to identify them as some sort of phonons of the crystal phase. This interpretation is confirmed by 
taking the limit of critical density $h\to h_c= 2m/\pi$, namely $B\to 0$ (strong coupling). Since $\epsilon < 2\epsilon_-$, in this limit
we can approximate $4\epsilon_+^2- \epsilon^2 \approx 4 \epsilon_+^2 \approx 4m^2$. The sound velocity goes to zero as $c_s \approx \epsilon_-/m$ and one has
\be
\label{eq:kink36}
\lim_{h\to h_c} -c_s \frac{dp}{d\epsilon} \approx  \frac{2 \epsilon_-}{\sqrt{4\epsilon_-^2-\epsilon^2}}\,.
\ee
Integrating, we have
\be\label{eq:kink37}
\lim_{h\to h_c} \epsilon(p) \approx 2 c_s \rho_{\rm k} \Big|\sin \Big( \frac{p-p_F}{2\rho_{\rm k}}\Big)\Big|\,.
\ee
Given that $c_s$ can be identified as a sound velocity, we can see its behaviour as a function of $h$.
For $h\to \infty$, we have $h\to \cosh B$, $\rho_{\rm k}\to (2/\pi) \cosh B$ and hence
\be
\lim_{h\to \infty} c_s = 1\,.
\ee
At the critical value $h\to h_c$, $B\to 0$, we get $\rho_{\rm k}\to -2m/\log\Big((1-k)/8\Big)$ from which
\be
\lim_{h\to h_c} c_s = B \log\frac{4}{B}\,.
\ee
Thus, phonons are relativistic in the weak coupling regime and slow down as we move toward $h_c$.

We can also consider fermion excitations. The kink-fermion kernel at large $N$ reads
\be
K_{1\text{k}}(\theta) = - \frac{1}{2\pi \cosh \theta}\,.
\ee
The fermion excitation $\epsilon_f$ is obtained by 
\be
\overline\epsilon_1(\theta) - \int_{-B}^B K_{1\text{k}}(\theta-\theta') \epsilon_\text{k}(\theta') = h - m\cosh \theta\,.
\ee
Integrating, we get
\be
\overline\epsilon_1(\theta) = h - m\sqrt{\sinh^2B + \cosh^2 \theta}\,.
\ee
Similarly, one gets
\be
\overline\chi_1 =  m\sqrt{\sinh^2B + \cosh^2 \theta} - \frac{k'}{c_s k^2} \frac{m}{ \sqrt{\sinh^2B + \cosh^2 \theta}}\,.
\ee
By simple manipulations, we get
\be
- c_s \frac{d p}{d\overline\epsilon_1} = \frac{c_s \epsilon^2- \epsilon_+ \epsilon_-}{\sqrt{(\epsilon^2-\epsilon_+^2)(\epsilon^2-\epsilon_-^2)}}\,,
\quad \epsilon \equiv h - \bar \epsilon_1\,.
\ee
We have a band gap equal to 
\be
\Delta =   m (\epsilon_+ - \epsilon_-) =  m e^{-B} \approx  \frac{m^2}{2h}\,.
\ee
The mass gap is obtained from \eqref{eq:mass-gap-eps}.
We get, for large $B$,
\be
M_1 = \min_{\theta\in [-B,B]} |\overline\epsilon_1(\theta)| = |\overline\epsilon_1(\theta=0)|  = m \frac{e^{-B}}{2}\approx \frac{m^2}{4h}\,,
\ee
which precisely matches the scale $\Lambda_\text{c}$ in \eqref{eq:lambda-12} in the limit $y\rightarrow 1$.

\subsubsection[A solvable bound state case (\tps{y=1/2}{y=1/2})]{A solvable bound state case (\tpsb{y=1/2}{y=1/2})}
\label{sec:y-half}

For $y=1/2$, the kernel \eqref{eq:TBAy2} becomes proportional to the kink one, 
\begin{equation}
\label{eq:nH3}
K_{\frac{1}{2}}(\theta) = 2 \widehat K_\text{k}(\theta)\,.
\end{equation}
We can then repeat verbatim the analysis reviewed for the kink, paying attention to match the two cases, given 
also the mass and charges in \eqref{eq:chargmass}. We have
\be\label{eq:nH5}\begin{split}
\epsilon_{\frac 12}(\theta) & = \frac{\epsilon_\text{k}(\theta)}{\sqrt{2}} =m\sqrt{2} \sqrt{\sinh^2 B - \sinh^2\theta}\,, \quad \frac 1k = \cosh B\,, \quad  
\frac{E(k^2)}{k} = \frac{\pi h}{2m\sqrt{2}}\,, \\
\chi_{\frac 12}(\theta) & = \frac{\chi_\text{k}(\theta)}{\sqrt{2}}\,,
\end{split}\ee
with $\chi_\text{k}$ as in \eqref{eq:kink17} and $c_s$ as in \eqref{eq:kink24}. 
The free energy is
\be\label{eq:nH6}
\mathcal{F}_{\frac N2} = -N \frac{m_{\frac 12}}{2\pi} \int_{-B}^B \! d\theta \, \cosh \theta \epsilon_{\frac 12}(\theta) = - \frac{Nm^2}{2\pi}\sinh^2 B \,.
\ee
For $h\to\infty$, we have $h=\sqrt{2}m \cosh B$ and 
\be\label{eq:nH7}
\lim_{h\to\infty} \mathcal{F}_{\frac N2} = - \frac{N h^2}{4\pi}\,,
\ee
in agreement with the free energy of $N/2$ species of free fermions.
The particle and energy densities read 
\be\label{eq:nH8}
\begin{split}
\rho_{\frac N2} & = \frac{m}{\sqrt{2}}\frac{1}{kK(k^2)} = \frac{2\epsilon_+ \epsilon_-}{\pi h c_s} \,, \qquad c_s = \frac{k'K(k^2)}{E(k^2)} \,, \\
 e_{\frac N2} & = -\frac{Nm^2}{2\pi} \sinh^2 B + \frac{Nm^2}{\pi} \frac{E(k^2)}{k K(k^2)} = \mathcal{F}_{\frac N2} + h_{\frac N2} \rho_{\frac N2} \,.
 \end{split}
\ee
The dispersion relation equals the one in \eqref{eq:kink34}. So, we get phonons with velocity $c_s$. Close to the critical fugacity 
$h_c =2 \sqrt{2}/\pi$, \eqref{eq:kink37} applies unchanged.
At high densities, 
\be\label{eq:nH9}
\lim_{h\to \infty} c_s = 1\,,
\ee
while at the critical value we get 
\be\label{eq:nH10}
\lim_{h\to h_c} c_s = B \log\frac{4}{B}\,,
\ee
exactly as in the kink case. Phonons are relativistic in the weak coupling regime and slow down as we move towards $h_c$.

We now consider charged fermion probes in the background of the $y=1/2$ bound states. The associated BA equation is
\be
\label{eq:nH11}
\overline\epsilon_1(\theta) -\int_{-B}^{B} d\theta'\, K_{1,\tfrac{1}{2}}(\theta-\theta') \epsilon_{\frac 12}(\theta')= h - m\cosh \theta\,,
\ee
where
\be\label{eq:nH12}
K_{1,\tfrac{1}{2}}(\theta) = -\frac{\sqrt{2}}{\pi}\frac{\cosh \theta}{\cosh 2\theta}
\ee
is the kernel for fermion - rank $N/2$ scattering and $\epsilon_{\frac{1}{2}}$ is given by \eqref{eq:nH5}. Integrating, we get
\be
\label{eq:nH12a}
\overline\epsilon_1(\theta) = h -m\cos \left(\gamma(\theta)\right)\sqrt{R(\theta)}\,,
\ee
where
\be
\label{eq:nH13}
\begin{split}
R(\theta) & = \sqrt{\cosh^2(2B)+\sinh^2(2\theta)} \,, \\
\gamma(\theta) & = \frac 12 \arctan\Big(\frac{\sinh (2\theta)}{\cosh (2B)} \Big)\,.
\end{split}
\ee
The density of states $\overline\chi_1(\theta)$ is similarly obtained as
\be
\label{eq:nH14}
\overline\chi_1 = \int_{-B}^{B} d\theta'\, K_{1,\tfrac{1}{2}}(\theta-\theta') \chi_{\frac 12}(\theta') + m\cosh \theta\,,
\ee
with  $\chi_{\frac{1}{2}}$ given by \eqref{eq:nH5}. Integrating, one has
\be
\label{eq:nH15}
\overline\chi_1 = \frac{m\cos \left(\gamma(\theta)\right)}{c_s \sqrt{R(\theta)}}\Big(c_s R(\theta) - \frac{2\epsilon_+ \epsilon_-}{m^2} \Big)\,,
\ee
with $\epsilon_\pm$ given in \eqref{eq:kink27}. Finding the dispersion relations from \eqref{eq:nH12a} and \eqref{eq:nH15} is complicated, but we can use again \eqref{eq:mass-gap-eps} to compute the gap, which gives us the scale of charged fermions $\Lambda_\text{c}$.
At high densities, $B\to \infty$, one has 
\be\label{eq:nH16}
M_1 = \min_{\theta\in [-B,B]} |\overline\epsilon_1(\theta)| = |\overline\epsilon_1(\theta=0)| \approx \frac{m e^{-3B}}{4\sqrt{2}} \approx \frac{1}{16}\frac{m^4}{h^3}\,,
\ee
which precisely matches the expected scaling.

With little effort, we can determine the mass gap for charged rank $b$ states, with $b$ finite. 
The kernel for charged rank $b$-charged $N/2$ scattering, at leading order in $1/N$, is given by
\be\label{eq:nH17}
K_{b,\tfrac{1}{2}}(\theta) = b K_{1,\tfrac{1}{2}}(\theta)\,.
\ee
 The right-hand side of the BA equation for charged rank $b$ states is $b$ times the one in \eqref{eq:nH11}, so
$\overline \epsilon_b(\theta) = b\overline \epsilon_1(\theta)$, and hence
\be\label{eq:nH18}
M_b = b M_1  \approx \frac{b}{16}\frac{m^4}{h^3}\,.
\ee
This is the expected result at leading order at large $N$ from QFT considerations.
The mass gap for charged $b$ particle bound states is detected by the large distance behaviour of the two-point function of the composite operator $\psi_1\ldots \psi_b$, where
$\psi_i$ are ``charged'' directions in flavour space. 
If
\be\label{eq:nH19}
\lim_{|x|\to \infty} \langle \psi_1(x) \psi_1(0) \rangle \sim e^{- M_1|x|} \,,
\ee
then, for $N\to \infty$, 
\be\label{eq:nH20}
\lim_{|x|\to \infty} \langle \psi_1\ldots \psi_b(x) \psi_1\ldots \psi_b(0) \rangle \sim 
\lim_{|x|\to \infty} \Big( \langle \psi_1(x) \psi_1(0) \rangle\Big)^b  \sim e^{- bM_1|x|} \sim e^{- M_b|x|}  \,,
\ee
in agreement with \eqref{eq:nH18}.

\subsection{Numerical analysis}
\label{sec:tba-num}

Another important feature of the integral equations found via integrability is that they are easily numerically solved, providing exact numerical results at finite chemical potential and finite $N$.

\subsubsection{Numerical method}

We follow the approach taken in \cite{Marino:2021dzn} of discretizing the interval.%
\footnote{An alternative approach which can also yield  very precise results is to project the integral equation in a basis of Chebyshev polynomials, as done in \cite{Abbott:2020qnl}.}
We need both integral equations for $\chi_a$ and $\epsilon_a$, \eqref{eq:FhPCF3} and \eqref{eq:FhPCF7}. The former is solved for fixed $B$, while the latter must be solved simultaneously with the boundary condition \eqref{eq:eps-bc}.

\begin{figure}[t!]
\centering
\begin{subfigure}{0.45\textwidth}
\includegraphics[width=\textwidth]{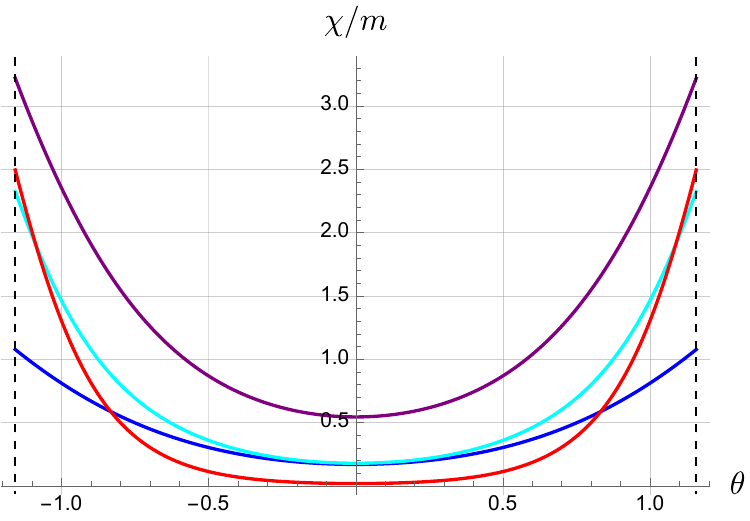}
\caption{$\chi_a$ and $\bar{\chi}_b$}
\end{subfigure}
\hfill
\begin{subfigure}{0.45\textwidth}
\includegraphics[width=\textwidth]{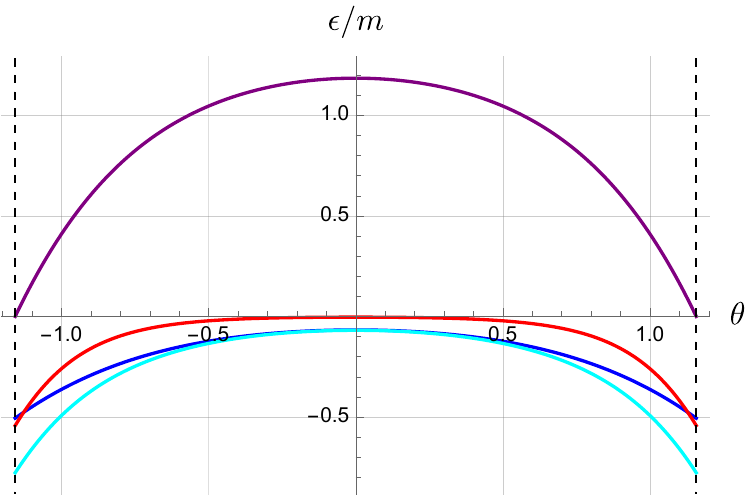}
\caption{$\epsilon_a$ and $\bar{\epsilon}_b$}
\end{subfigure}
\caption{Solutions to the integral equations for $N=6$ and $a=4$ at $h/m=\sqrt{e}$. The purple line correspond to the background solutions $\chi_a$ (left) and $\epsilon_a$ (right). The other full lines correspond to the probe solution: charged fermions $\overline{\chi}_1$/$\overline{\epsilon}_1$ (blue), charged hole $\overline{\chi}_{a-1}$/$\overline{\epsilon}_{a-1}$ (cyan) and neutral fermions $\overline{\chi}_{a+1}$/$\overline{\epsilon}_{a+1}$ (red). The vertical dashed lines are at $\pm B$.}
\label{fig:solutions}
\end{figure}

We discretize the system by introducing a lattice of points and weights $\{x_i,w_i\}$ such that
\begin{equation}
    \int_0^1 d x f(x)=\sum_i^M w_i f(x_i).
\end{equation}
We use the Gauss--Kronrod rule to generate the lattice. This rule over-samples near the edges of the interval, where the function is expected to vary faster.
The discretized equations are
\begin{align}
    \epsilon_a(2B(x_i-\tfrac{1}{2})) - 2B \sum_{j=1}^M w_j  K_{aa}(2B(x_i-x_j))\epsilon_a(2B(x_j-\tfrac{1}{2})) &= a h - m_a \cosh(2B(x_i-\tfrac{1}{2})),
   \nn  \\
    \chi_a(2B(x_i-\tfrac{1}{2})) - 2B \sum_{j=1}^M w_j  K_{aa}(2B(x_i-x_j))\chi_a(2B(x_j-\tfrac{1}{2})) &=  m_a \cosh(2B(x_i-\tfrac{1}{2})).
\end{align}
At fixed $B$ and $h$, the solution of the discrete integral equations amounts to a linear problem. However, to work at fixed $h$, one must take a trial $B$ (guided by large $h$ results, a good guess is $B\sim \log h$) and then do a binomial search for the value of $B$ that leads to \eqref{eq:eps-bc} by inspecting the sign of the edge values of $\epsilon_a$ at each $B$.

There are two main sources of calculation complexity in this problem. One is that the kernel in position space takes a long time to calculate due to the polygamma function, which one must evaluate $\mathcal{O}(N)$ times at each point due to the sum in \eqref{fnm-aux}. For intermediate calculations, it is convenient to save the kernel function evaluated in a dense and wide set of points and then construct an interpolation function. The other source of complexity is the binomial search for $B(h)$ due to the boundary conditions, which requires the resolution of the discretised equation around $10-20$ times.
Once the integral equations for the ``background'' are solved, it is simple to obtain the solutions for the probe equations \eqref{eq:chiProbe} and \eqref{eq:epsProbe} using the same method.

\begin{figure}[t!]
\centering
\begin{subfigure}{0.4\textwidth}
\includegraphics[width=\textwidth]{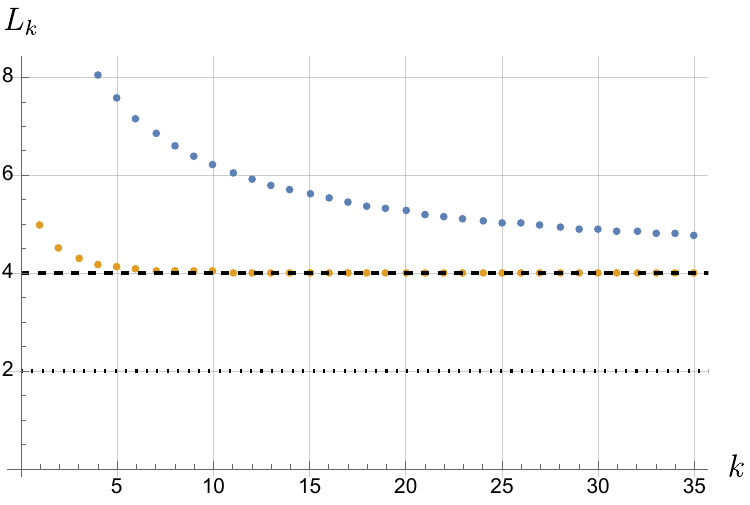}
\caption{$a=2$, $b=1$}
\end{subfigure}
\hfill
\begin{subfigure}{0.4\textwidth}
\includegraphics[width=\textwidth]{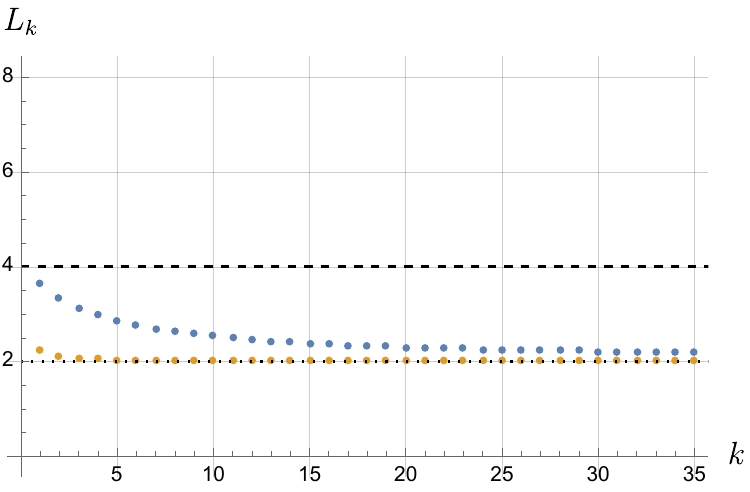}
\caption{$a=2$, $b=3$}
\end{subfigure}
\\
\begin{subfigure}{0.4\textwidth}
\includegraphics[width=\textwidth]{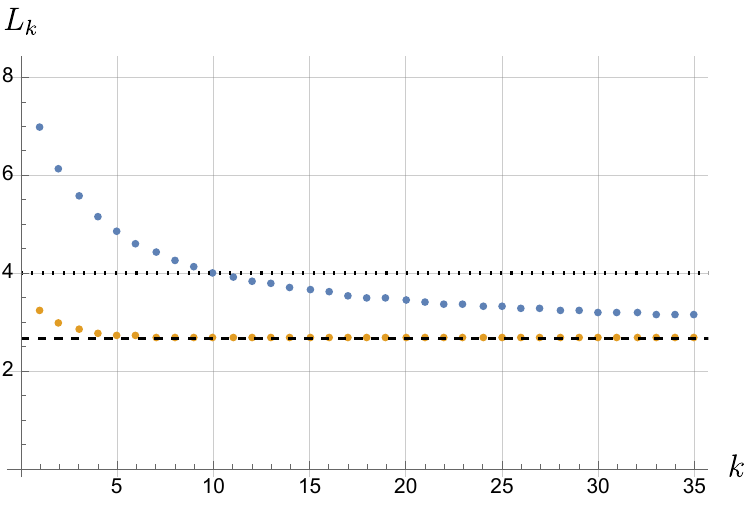}
\caption{$a=3$, $b=1$}
\end{subfigure}
\hfill
\begin{subfigure}{0.4\textwidth}
\includegraphics[width=\textwidth]{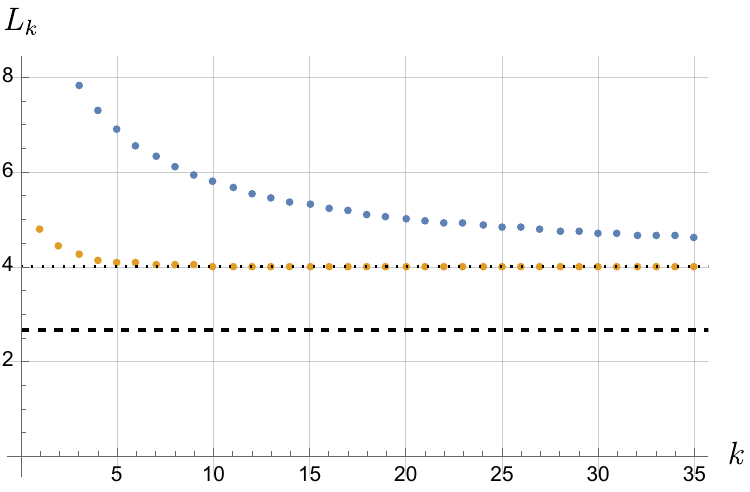}
\caption{$a=3$, $b=4$}
\end{subfigure}
\caption{Sequences $L_k^{(b)}$ for charged excitations ($b=1$) and neutral excitations ($b=a+1$) for $N=5$ and two values of $a$. The blue dots are the sequence and the orange dots are the third Richardson transform. We use $u_0=\sqrt{e}$ and $u=e^{1/8}$. The dashed line is $\tfrac{2N-2}{a}$ and the dotted line is $\frac{N-1}{N-1-a}$, verifying \eqref{eq:sequence-limit}.}
\label{fig:mass-gaps}
\end{figure}

\subsubsection[Finite \tps{N}{N} numerical analysis]{Finite \tpsb{N}{N} numerical analysis}

We can use the numerical analysis to inquire whether some of the physical results we found in the large $N$ analysis also hold at finite $N$, even if they are analytically unreachable.

In Figure~\ref{fig:solutions} we illustrate some numerical solutions of the integral equations, including both the background and probe solutions. We can see that for charged probes with small $\theta$, corresponding to momenta close to the Fermi surface, the solutions for charged fermion probes and charged hole probes overlap, as expected from effective relativistic invariance.

\paragraph{Scaling of the mass gaps.}

As explained in  \eqref{eq:mass-gap-eps}, the minimum value of probe energy distribution $\bar\epsilon_b$, which for probes occurs at $\theta=0$, gives the value of the mass gap. 
We can use this to validate the correspondence between the two dynamically generated scales and the masses of charged and neutral excitations also at finite $N$.

\begin{figure}[t!]
\centering
\includegraphics[width=0.6\textwidth]{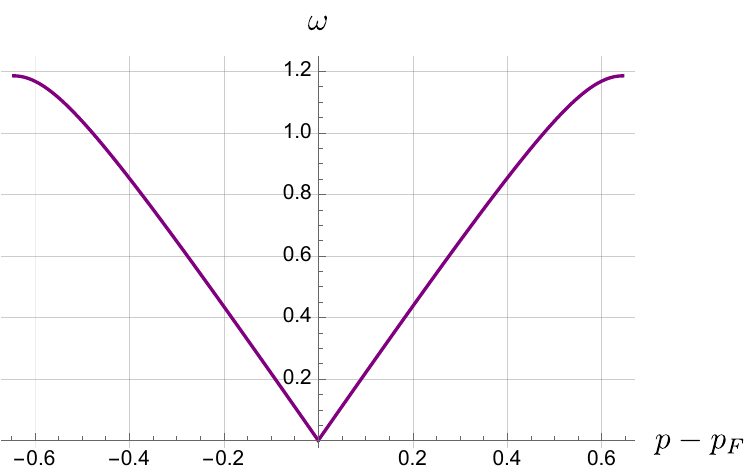}
\caption{Dispersion relation for background excitations for $N=6$ and $a=4$ at $h/m = \sqrt{e}$. The units of the plot are such that $m=1$.}
\label{fig:massless}
\end{figure}

We focus on three types of probes: charged fundamental fermions, charged fermion holes and neutral fundamental fermions. As explained in section~\ref{sec:tba-general}, these correspond to $b=1, a-1,a+1$, respectively. For such probes, the minimum of \eqref{eq:mass-gap-eps} always occurs at $\theta=0$. Therefore, in order to verify that $\Lambda_{\rm c}$ is proportional to the mass gaps of the charged excitations and $\Lambda_{\rm n}$ to the neutral ones, we should have the following leading order dependence on $h$ at fixed $\Lambda$:
\begin{equation}
  | \bar\epsilon_1(\theta=0) |\propto h^{1-\frac{2N-2}{a}},\quad
   | \bar\epsilon_{a-1}(\theta=0)| \propto h^{1-\frac{2N-2}{a}},\quad
  |  \bar\epsilon_{a+1}(\theta=0)| \propto h^{1-\frac{N-1}{N-1-a}}.
\end{equation}
In order to precisely test the scaling behavior of the gaps numerically, we construct auxiliary sequences that remove the overall unknown constant factor. Since, from integrability arguments, we expect corrections to the leading behaviour of $\bar\epsilon_b(\theta)$ to scale like $1/B\sim 1/\log(h/m)$, we build the sequences
\begin{equation}
    L_k^{(b)} = 1-\frac{\log\left|\bar\epsilon_b(\theta=0;h/m = u_0 u^{k-1})\right|}{(k-1)\log(u)+\log(u_0)},
    \label{eq:sequence-def}
\end{equation}
where $u$ and $u_0$ are some arbitrary finite values. Independently of these parameters, at large $k$ we expect
\begin{equation}
 L_k^{(1)} \sim  \frac{2N-2}{a} + \mathcal{O}\left(\frac{1}{k}\right),
 \quad 
 L_k^{(a-1)} \sim  \frac{2N-2}{a} + \mathcal{O}\left(\frac{1}{k}\right),
 \quad 
 L_k^{(a+1)} \sim  \frac{N-1}{N-1-a} + \mathcal{O}\left(\frac{1}{k}\right).
  \label{eq:sequence-limit}
\end{equation}
Since the corrections are proportional to $1/k$, these sequences are amenable to Richardson acceleration which improves convergence to higher power of $1/k$ \cite{Bender1978}. For the $r$-th  Richardson transform, one expects convergence of order $1/k^r$. However for higher $r$ more terms are required and there is greater sensitivity to numerical error. Thus we found it optimal to use between two to five transforms. This way, we could then check that the scaling is correct.
We plot some examples in Figure~\ref{fig:mass-gaps}, confirming \eqref{eq:sequence-limit}. 
We do not include the sequence $L_k^{(a-1)}$ corresponding to charged holes because it is identical to that of charged fermions (as is illustrated in Figure~\ref{fig:solutions}).
One could also extract the overall numerical factor at each $N$ and $a$.

\paragraph{Existence of massless modes.}

For probes of the same particle type that populates the ground state ($a$-particle bound states), the dispersion relation should be dictated by the solutions to the principal BA, $\epsilon_a$ and $\chi_a$. We can construct the curve for the derivative of momentum with respect to energy by taking
\begin{equation}
    \left(\omega, \frac{d p}{d \omega}\right) = \left(\epsilon_a(\theta),\frac{\chi}{\epsilon'_a}(\theta)\right),\quad \theta\in[-B,B].
    \label{eq:dispersion-curve}
\end{equation}
This curve can be integrated in $\omega$ to obtain $(\omega,p)$ and then inverted to obtain the dispersion relation such as in Figure~\ref{fig:massless}.
Integrating $dp/d\omega$ naturally generates an integration constant which we identify with $p_F$.
This constant can be explicitly calculated from the definition \eqref{eq:pf-pirho} using $\rho$ which is in turned obtained from $\chi$ through \eqref{eq:FhPCF5}. We can then numerically plot the resulting dispersion relation as we exemplify in Figure~\ref{fig:massless}. 

\begin{figure}[t!]
\begin{subfigure}{0.49\textwidth}
\includegraphics[height=0.625\textwidth]{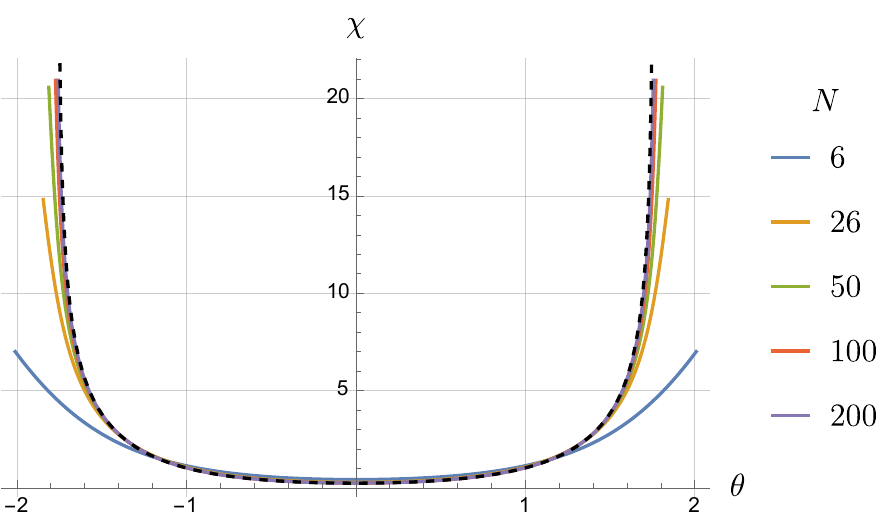}
\caption{$\chi_a$ for $a=N/2$}
\end{subfigure}
\hspace{0.04\textwidth}
\begin{subfigure}{0.49\textwidth}
\includegraphics[height=0.625\textwidth]{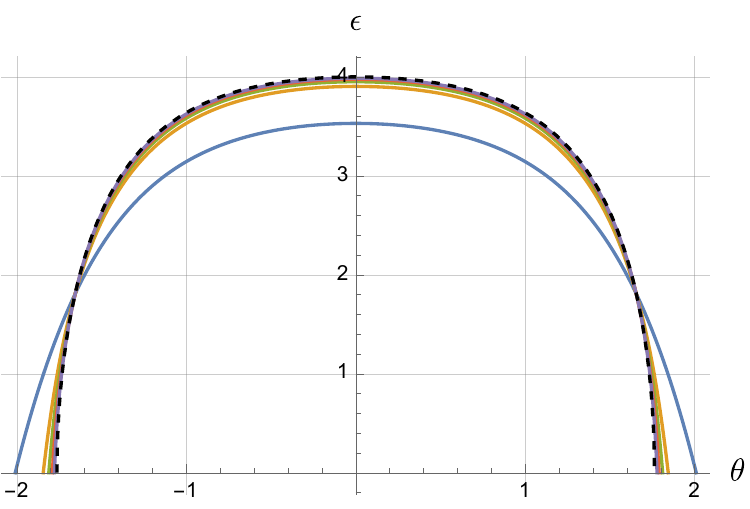}
\caption{$\epsilon_a$ for $a=N/2$}
\end{subfigure}
\caption{Plot of $\chi_a$ and $\epsilon_a$ for multiple values of $N$ with $a=N/2$. The analytical large $N$ solution is marked with a dashed line.}
\label{fig:solutions-half}
\end{figure}

While we only plot for a particular value of $N$ and $a$, the qualitative shape of the profile is fairly general since it follows from general properties of the dispersion.
We always find a massless mode, as anticipated below \eqref{eq:mass-gap-eps}, as a consequence of the boundary condition $\epsilon(\pm B)=0$. Moreover,  $\omega(p)$ is always linear near the origin, because $\frac{\epsilon'}{\chi}(B)$ is finite and non-vanishing. 
Finally, since $\epsilon'(0)=0$, we get a local maximum at the edges. Thus the profile of Figure~\ref{fig:massless}, with a massless mode accompanied by linear behavior in its vicinity while the velocity vanishes at the edges, is representative for any $N$ finite.
We find a similar profile from the analytical large $N$ analysis at $y=1/2$ and $y=1$.

\subsubsection[Numerical checks at large \tpsb{N}{N}]{Numerical checks at large \tpsb{N}{N}}
\label{sec:num-largeN}

Another interesting line of numerical tests is to take $N$ large but finite and compare them to the analytical infinite $N$ results. These show that the analytical results truly characterize the large $N$ limit and that it is smoothly connected to finite $N$. Furthermore, they help to guide the correspondence between the more semiclassical large $N$ results and the integrability framework, the latter of which can often be more obscure about its physics. 

\paragraph{Comparison with $y=1/2$ exact solution}

\begin{figure}[t!]
\centering
\begin{subfigure}{0.53\textwidth}
\includegraphics[height=4.8cm]{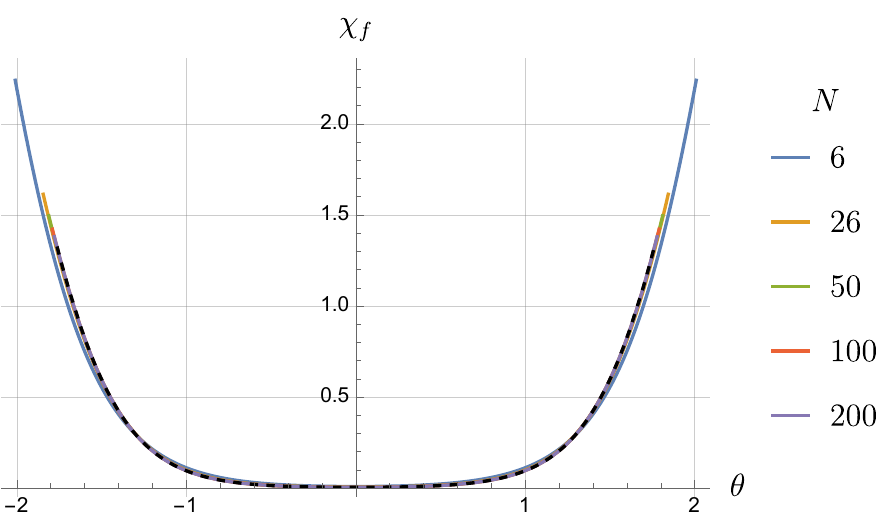}
\caption{$\overline{\chi}_1$ for $a=N/2$}
\end{subfigure}
\begin{subfigure}{0.46\textwidth}
\includegraphics[height=4.8cm]{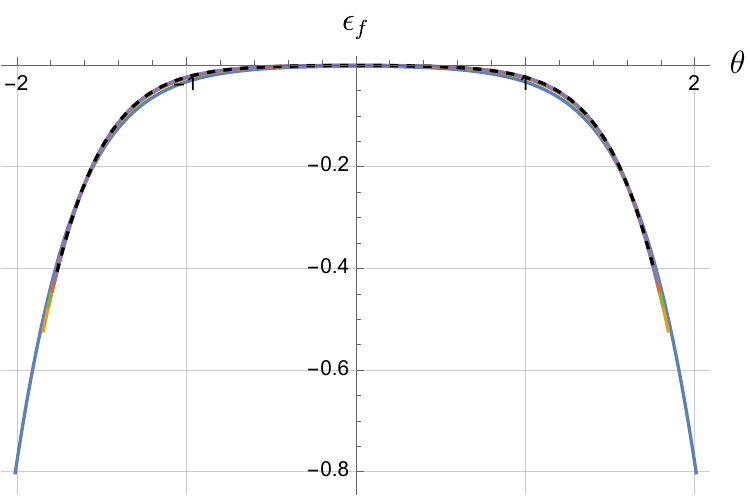}
\caption{$\overline{\epsilon}_1$ for $a=N/2$}
\end{subfigure}
\caption{\label{fig:solutions-probe-half}%
Plot of $\overline{\chi}_1$ and $\overline{\epsilon}_1$ for multiple values of $N$ with $a = N/2$. The analytical large $N$ solution is marked with a dashed line.}
\end{figure}

For $y=1/2$, we have access to the exact solution in the large $N$ limit from the results in section~\ref{sec:y-half}. In Figure~\ref{fig:solutions-half} we compare finite $N$ solutions for $\chi_a$ and $\epsilon_a$ with the analytical results \eqref{eq:nH5}. As can clearly be seen, convergence for the value of $B$ and the function itself is smooth and around $N\sim 100$ the finite $N$ answer is very close to the large $N$ result. Nonetheless, while $\chi_a$ is divergent at $\theta=B$ in the large $N$ solution, it is not at finite $N$, as has been derived from the edge limit analysis (see \cite{Volin:2009wr,Marino:2019eym}).
In Figure~\ref{fig:solutions-probe-half}, we do the same analysis for charged fermions, reporting $\bar \chi_1$ and $\bar \epsilon_1$ as a function of $\theta$. The convergence is much faster than in the background case discussed before, with $N=6$ already being almost  graphically indistinguishable from the large $N$ result.

\begin{figure}[t!]
\centering
\includegraphics[width=0.9\textwidth]{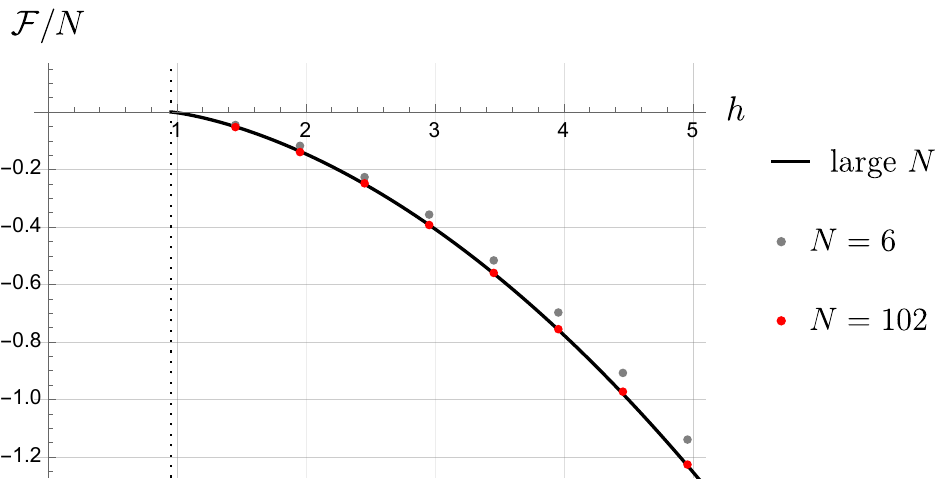}
\caption{\label{fig:free-energy}%
Comparison between the free energy computed using the BA for two values of $N$ and the semiclassical result \eqref{grand potential on-shell y<1}, with $y=1/3$.}
\end{figure}

\paragraph{Free energy at large $\boldsymbol{N}$.}
Given a discretized solution, we can obtain observables such as the free energy through
\begin{equation}
\mathcal{F}(h) = m_a\frac{ B}{\pi}\sum_{i=1}^M w_i \epsilon_a(2B(x_i-\tfrac{1}{2}))\cosh(2B(x_i-\tfrac{1}{2})).
\end{equation}
We can use this observable to compare the integrability approach to the large $N$ methods of section~\ref{sec: GN model part chem pot semiclassical}. As can be seen in Figure~\ref{fig:free-energy}, the function $\mathcal{F}(h)$ as obtained from \eqref{grand potential on-shell y<1}  matches closely the integrability results for a large but finite value of $N$. This is an indication that the semiclassical large $N$ analysis is not only correct but a good approximation of finite $N$ behavior.

\begin{figure}
\centering
\includegraphics[width=0.75\textwidth]{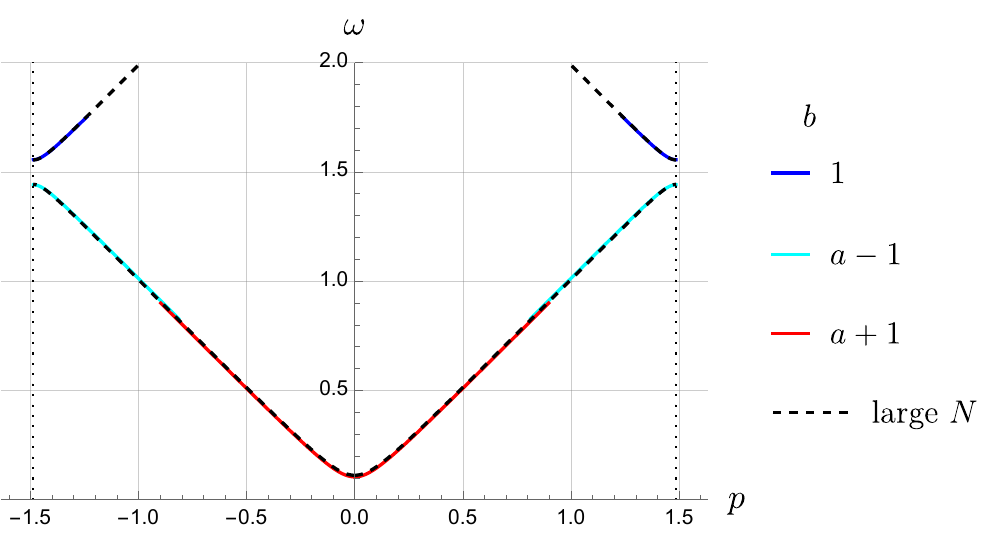}
\caption{Dispersion relation for probes of different $b$ overlayed with the analytical large $N$ result obtained from \eqref{density of states} in semiclassical analysis. 
We take $N=102$ and $a=68$ for the BA numerical data, and $y=2/3$ at large $N$. The difference between the numerical and the analytical 
value of $p_F$ is approximately $0.1\%$.}
\label{fig:disp-colors}
\end{figure}

\paragraph{Dispersion relation at large $\boldsymbol{N}$.}
A more detailed match between the semiclassical analysis and the integrability results is to match the analytical dispersion relation obtained by integrating \eqref{density of states} with the dispersion relation obtained by the probe equations. Here, we need to cobble together different types of probe in the integrability side. The energy of fermion excitations is given by $h-\overline{\epsilon}_1$, while for hole ones we have $h+\overline{\epsilon}_{a-1}$, and for neutral fermion simply $-\overline{\epsilon}_{a+1}$. We calculate $p$ analogously to \eqref{eq:dispersion-curve}, adding the integration constant $p_F$ (also numerically calculated) for the charged excitation and not adding it for the neutral ones. The curves are generated with rapidites in $[-B,B]$, as with the background excitations. The resulting dispersion curve matches the expected parts of the analytical result as can be seen in Figure~\ref{fig:disp-colors}. 

\begin{figure}
\centering\includegraphics[width=0.75\textwidth]{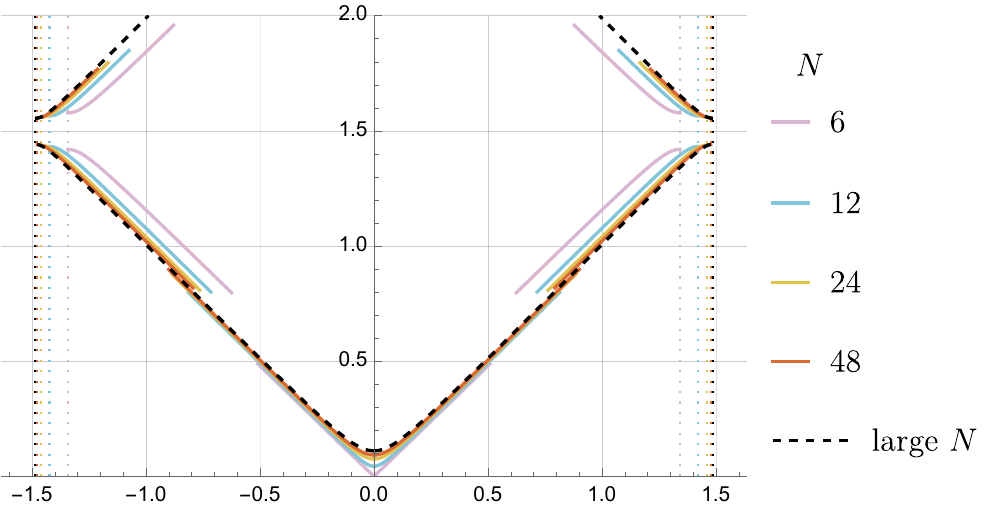}
\caption{Dispersion relation for probes at different values of $N$ with $a=2N/3$.
The three types of probe are colored identically for each value of $N$.
The dotted lines mark the Fermi momentum $p_F$ of the respective color.
The analytical results are obtained with $y=2/3$.}
\label{fig:disp-N}
\end{figure}

We can also inspect the convergence towards the large $N$ results, as exemplified in Figure~\ref{fig:disp-N}. Point wise the finite $N$ dispersion curve converges to the semiclassical result, as does the value of $p_F$. Note however that, at lower $N$, the dispersion of $a-1$ bound states and $a+1$ bound states (which we associated with charged hole and neutral fermions, respectively) are not continuously connected in the bottom band, as it occurs at large $N$.
Thus, in this particular aspect, we find a qualitative difference between the finite $N$ and the large $N$ results. This is related to the ``miracle'' at large $N$ discussed in \cite{Melin:2025eyw}.

\section{Outlook}
\label{sec:conclusions}

We have studied the $O(2N)$--symmetric Gross--Neveu model at finite density in the presence of a chemical potential $h$ for a generic number $a \leq N-2$ of fermion fields. By combining perturbative QFT, semiclassical large $N$, and Bethe ansatz methodologies, a coherent and detailed picture emerges.
At finite $N$, two new dynamically generated scales $\Lambda_\mathrm{n}$ and $\Lambda_\mathrm{c}$ -- governing the mass gap of neutral and charged fermions, respectively --  appear in the theory. For $h\geq h_\text{crit}$, bound states made of $a$-fermions condense and form an inhomogeneous configuration, which at infinite $N$ is a crystal spontaneously breaking translations. At large $h$, this crystal has mean $\Lambda_\mathrm{n}$ and spatial oscillations of amplitude $\Lambda_\mathrm{c}$. The two scales also control the nonperturbative corrections to the free energy, resolving a puzzle concerning fractional-power renormalons and predicting new ones.

There are several directions and open questions that would be interesting to explore in future studies. One may wonder whether additional dynamically generated scales, associated with the mass gaps of bound states of rank $a > 1$, could exist. At large $N$, factorization implies that all such masses are governed by suitable combinations of $\Lambda_\mathrm{n}$ and $\Lambda_\mathrm{c}$, with the precise relation depending on the $O(2N-2a)\times U(a)$ quantum numbers of the states. These two scales are also the only ones that appear in the inhomogeneous condensate.
Moreover, at finite $N$ and finite $a$, the Bethe ansatz computation of the free energy $\mathcal{F}(h)$ does not reveal any additional scale. It is therefore natural to conjecture that no further dynamically generated scales arise at finite $N$, although it would be desirable to establish this conclusion on a firmer footing.

It would be nice to extend our analysis for $T\neq 0$, generalizing for any $a$ the semiclassical analysis of \cite{Schnetz:2005ih}.
Bethe ansatz techniques become increasingly complicated for $T \neq 0$, as the medium is populated by an ensemble containing all particle species. However, quantum field–theoretical methods should still apply and may help clarify which features of the large-$N$ semiclassical analysis persist at finite $N$.
The appearance of inhomogeneous phases at finite $N$ has started to be analyzed using also numerical lattice methods, see e.g. \cite{Narayanan:2020uqt, Lenz:2020bxk}. It would be very interesting if lattice studies will be generalized to any $a$ in the future.

Do the crystal phases found in the GN model show up in other integrable gapped models?
In particular, renormalons exhibiting fractional powers when expressed in terms of the (vacuum) dynamically generated scale
were found in \cite{Marino:2021dzn} also for the principal chiral field model. Inspecting the kernel for the equation with the rank $a$ bound state, we see that the resulting fractional power also follows the pattern $N\rightarrow N-a$ in $\beta_0$ we found here. Also there, for $a>1$ there is apparently a second scale.
Thus, as in the Gross–Neveu case, we expect these fractional powers to become integral when expressed in terms of the (medium) dynamically generated scale(s), although it would be interesting to work out this mechanism in detail.

The transition at $h=h_\text{crit}$ between the homogeneous and the inhomogeneous phases is of second order and a critical
behaviour is expected there. What is the nature of the CFT governing this second order transition?

In \cite{Schnetz:2005ih} it was shown that the semiclassical analysis can be extended when UV fermion mass terms are added in the Lagrangian, 
and that crystal phases persist in this setting. 
This is an interesting result, as mass terms break integrability, and the latter enters crucially in the determination of the ansatz.
If this is indeed the case, it is reasonable to expect that our semiclassical analysis for any $a$ can be generalized in this context. 
Bethe ansatz is unavailable, but quantum field–theoretical methods may help clarify what happens at finite $N$.
One might wonder if crystal phases appear for more general non-integrable theories. 
There are also broader classes of ansatze, such as the one used in \cite{Basar:2008im, Basar:2009fg, Thies:2025mro} for the chiral Gross--Neveu model and the higher Lamé potentials, discussed for example in \cite{Dunne:1997ia, Schnetz:2004vr}. Typically these ansatz are characterised by more complicated gap structures.

QFT computations based on condensate calculus are a possible alternative to resurgence to obtain transseries perturbative expressions.
See e.g. \cite{Marino:2024uco, Marino:2025ido, Liu:2025bqq} for recent works in 2d models. At high densities, the inhomogeneous VEVs reduce to the simple expression \eqref{eq:intro-sigma-osc}.
Can we set-up a condensate calculus in this way? Setting up a condensate calculus for inhomogeneous VEVs would be important, as renormalon physics in QCD is often discussed in this context.

\section*{Acknowledgments}
We thank Marcos Mari\~no and Konstantin Zarembo for useful discussions and correspondence.
This work was supported by the ERC-COG grant NP-QFT No. 864583,
 ``Non-perturbative dynamics of quantum fields: from new deconfined phases of matter to quantum black holes'',
the MUR-FARE2020 grant No. R20E8NR3HX,
 ``The Emergence of Quantum Gravity from Strong Coupling Dynamics''
and the MUR-PRIN2022 grant No. 2022NY2MXY.
We are also supported by the INFN ``Iniziativa Specifica"  GAST and  ST\&FI.

\appendix

\section{Determining the maximum \tpsb{q/m}{q/m}}
\label{app:anti}

We determine which states populate the vacuum at fixed $a$ by looking at the $O(2)_i$ charges of the components of the various states present in the spectrum. Recall that for $O(2N)$ these are in the completely antisymmetric rank $r$ irreducible representations of $O(2N)$, with $r=1,\ldots, N-2$.
We are interested in the charge of the states under the $U(1)_V$, which is given by the sum over all the $O(2)_i$ charges. 
If we switch on equal fugacities for $a$ different $O(2)$ subgroups, as in \eqref{eq:UV1}, the fundamental representation $F$ of $O(2N)$ splits
into $2N-2a$ neutral states, and $a$ states in the $O(2)_V$ representation with charge 1. In terms of characters, we have
\be
\chi_F(x) = {\rm Tr} \, F = 2N-2a + a (x+x^{-1})\,,
\ee
where $x$ represents the $O(2)_V$ fugacity. The charges of states in completely antisymmetric rank $p$ representations (denoted for short rank $p$
in what follows) is determined in terms of the plethystic exponential for fermions:
\be\label{eq:ZSA}
\exp\Big(\sum_{p=1}^\infty \frac{(-1)^{p+1}}{p} \epsilon^p \chi_F(x^p) \Big) = {\rm det} (1+ \epsilon F) = (1+\epsilon)^{2N-2a} (1+ \epsilon x)^a (1+\epsilon x^{-1})^a  \,,
\ee 
where $\epsilon$ is a counting parameter. The charges of the states in the rank $p$ representations are given by the powers of $x$
in the $\epsilon^p$ terms appearing in \eqref{eq:ZSA}. It is manifest from this formula that, at fixed $a$, the rank $p$ representation of $O(2N)$ has states with maximum charge
\be\label{app:qrelMax}
\begin{split}
q_p & = p\,, \qquad \text{for} \; p<a\,, \\
q_p & = a \,, \qquad \text{for} \; p\geq a\,.
\end{split}
\ee
The kink states with maximum charge are given by
\be\label{app:qkinkMax}
q_\text{k} = \frac{a}{2}\,,
\ee
Using \eqref{app:qrelMax}, \eqref{app:qkinkMax}, as well as the mass formulas in the main text, \eqref{eq:spectrum} and \eqref{eq:skinkmass}, we 
easily derive the maximum charge to mass ratio for each value of $a$:
\be\begin{split}
\text{rank--a} & \qquad \text{for} \quad 1 \leq  a \leq N-2 \,,\\
\text{kinks} & \qquad \text{for} \quad   a = N-1, N \,.
\end{split}
\ee
Moreover, it is evident from \eqref{eq:ZSA} that for $a\leq N-2$ there is only a single state with maximal charge, 
and hence the S-matrix associated to its scattering is 1-dimensional.
The fact that the maximal charge is unique is easy to understand. Given a rank $a$ completely antisymmetric tensor of $O(2N)$
$V_{[A_1\ldots A_a]}$, its maximal charged state under $O(2)_V$ is obtained by taking all its indices in the $O(2a)$ directions 
with charge +1 under $O(2)_i$. There is only one component of this kind, namely $V_{+_1 \ldots +_a}$, where $\pm_a$ denote the 
two components of the ${\bf 2}_i$ representations of $O(2)_i$, $i=1,\ldots,a$.

\section{Free fermions with a chemical potential}
\label{app:freefermions}

In order to make the discussion self-contained, we review the fermion propagator and the Feynman rules for external states in the presence of a chemical potential $h$.

\subsection{The propagator}
\label{app:Canonical Quantization Propagator}

Let us canonically quantize a free fermion of mass $m$ in the presence of a chemical potential $h$. The Lagrangian is:
\begin{equation}
    \mathscr{L} = \bar\psi (i\parslash + h \gamma^0 - m) \psi \,.
\end{equation}
As is well known, when $h>m$ fermions populate the vacuum with increasing momentum $\bp$, up to the Fermi momentum $p_F = \sqrt{h^2 - m^2}$. In $d=2$ dimensions, the Fermi surface consists of just two points $\bp = p_F$ and $\bp = -p_F$.

We determine the time-ordered propagator, to be used in the perturbative expansion in Feynman diagrams around a Fermi surface. The equations of motion are
\begin{equation}
    \left(i\gamma^{\mu}\partial_{\mu} - m + h\gamma^0\right)\psi = 0 \,.
\end{equation}
By choosing $\gamma^{0} = \sigma^{1} = \begin{psmallmatrix} 0 & 1 \\ 1 & 0 \end{psmallmatrix}$, $\gamma^{1}  = i\sigma^{2} = \begin{psmallmatrix} 0 & 1 \\ -1 & 0 \end{psmallmatrix}$, and denoting $\bx = x^1$, $\bp = p_1 = -p^1$ for convenience, we find the two solutions:
\begin{subequations}
\begin{gather}
\label{eq:dirac plane waves0}
    \psi\left(x\right)
    = \begin{cases} e^{ih x^0} e^{-ip_\mu x^\mu} u\left(\bp\right) \\ e^{ih x^0}e^{ip_\mu x^\mu}v\left(\bp\right) \end{cases}
    \qquad \text{with} \qquad p_0 = E_\bp \equiv \sqrt{m^2 + \bp^2} > 0 \,, \\
    \label{eq:dirac plane waves}
    u^\alpha\left(\bp\right) = \begin{pmatrix} \sqrt{p_0 - p^1} \\ \sqrt{p_0 + p^1} \end{pmatrix}\bigg|_{p_0 = E_\bp}\,,  \qquad \qquad v^\alpha\left(\bp\right) = \begin{pmatrix} \sqrt{p_0 - p^1} \\ -\sqrt{p_0 + p^1} \end{pmatrix}\bigg|_{p_0 = E_\bp} \,.
\end{gather}
\end{subequations}
The two polarization vectors satisfy
\begin{equation}
(\slashed{p} - m) \, u(\bp) = 0 \;,\qquad\qquad (\slashed{p} + m) \, v(\bp) = 0 \;.
\end{equation}
Note that the dependence on $h$ can be absorbed by the time-dependent shift in \eqref{eq:dirac plane waves0}, so the fermion polarizations $u$ and $v$ are the same as in the case $h=0$. They satisfy:
\begin{equation}\begin{aligned}
        u^\dag(\bp) \, u(\bp) &= v^\dag(\bp) \, v(\bp) = 2E_\bp \,,\qquad & \bar u(\bp) \, u(\bp) &= -\bar v(\bp) \, v(\bp) = 2m \,, \\
        \bar u(\bp) \, v(\bp) &= \bar v(\bp) \, u(\bp) = 0 \,,& u^\dag(\bp) \, v(-\bp) &= v^\dag(\bp) \, u(-\bp) = 0 \,, \\
        u^\alpha(\bp) \, \bar u_\beta(\bp) &= (\slashed p + m)^\alpha{}_\beta \,,& v^\alpha(\bp) \, \bar v_\beta(\bp) &= ( \slashed{p} - m)^\alpha{}_\beta \,.
\end{aligned}\end{equation}

The canonical momentum is $\pi_\psi(x) = \frac{\partial \mathscr{L}}{\partial \, \partial_0\psi(x)} = i\psi^\dagger$ and the Hamiltonian is
\begin{equation}
    H = \int\! d\bx \; \psi^\dag(\bx) \, \cH \, \psi(\bx) \,, \qquad\qquad \cH \equiv m\gamma^0 - h - i\gamma^0\gamma^1\partial_1 \,.
\end{equation}
We can expand the fermionic field into modes:
\begin{equation}\begin{aligned}
\label{eq:mode expansion fermions}
    \psi(\bx) &= \int\! \frac{d\bp}{2\pi} \frac{1}{\sqrt{2E_\bp}} \, \Bigl( a_\bp \, u(\bp) \, e^{-i \, \bp \, \bx} + b_\bp^\dag \, v(\bp) \, e^{ i \, \bp \, \bx} \Bigr) \,, \\
    \pi_\psi(\bx) &= \int\! \frac{d\bp}{2\pi} \frac{i}{\sqrt{2E_\bp}} \, \Bigl( a_\bp^\dag \, u^\dag(\bp) \, e^{i \, \bp \, \bx} + b_\bp \, v^\dag(\bp) \, e^{-i \, \bp \, \bx} \Bigr) \,.
\end{aligned}\end{equation}
The system is quantized by imposing the anticommutation relations 
\begin{equation}
    \{a_\bp, a^\dagger_\bq\} = \{b_\bp, b^\dagger_\bq\} = (2\pi) \, \delta(\bp-\bq) \,,
\end{equation}
with all other anticommutators vanishing. These imply $\bigl\{ \psi^\alpha(\bx), \pi_{\psi \beta}(\bx') \big\} = i \delta^\alpha_\beta \delta(\bx - \bx')$ with all other anticommutators vanishing. When $h>0$, the Hamiltonian is conveniently rewritten as$\,$%
\footnote{For $h<0$, the role of the oscillators $a_\bp$ and $b_\bp$ (particle and antiparticle) is exchanged.}
\begin{equation}
\label{eq:free ferm hamiltonian}
    H = \int\! \frac{d\bp}{2\pi} \, \biggl[ \theta\bigl( |\bp| - p_F \bigr) \, (E_\bp - h) \, a^\dag_\bp a_\bp + \theta \bigl( p_F - |\bp| \bigr) \, (h-E_\bp) \, a_\bp a^\dagger_\bp + (E_\bp + h) \, b_\bp^\dag b_\bp \biggr]
\end{equation}
where all excitations have positive energy. The momentum operator reads
\begin{equation}
\label{eq:free ferm momentum}
    P = \int \frac{d\bp}{2\pi} \, \biggl[ \theta \bigl( |\bp| - p_F \bigr) \, a^\dag_\bp a_\bp -  \theta \bigl( p_F - |\bp| \bigr) \, a_\bp a^\dag_\bp + b_\bp^\dag b_\bp \biggr] \, \bp \;.
\end{equation}   
In \eqref{eq:free ferm hamiltonian} and \eqref{eq:free ferm momentum} we offset the infinite constant coming from the reordering of operators so that the vacuum defined below has zero energy and momentum.
The vacuum state is the Fermi surface $|h\rangle$, defined such that excitations have positive energy:
\begin{equation}
\label{eq:abpbbp}
    \begin{cases}
        a_\bp |h\rangle = b_\bp |h\rangle = 0 & \text{if } |\bp| \ge p_F \,, \\
        a^\dagger_\bp |h\rangle = b_\bp |h\rangle = 0 & \text{if }|\bp| < p_F \,.
    \end{cases}
\end{equation}
This is different from the vacuum $|h=0\rangle$ of the theory without chemical potential. We use normalization $\langle h | h \rangle = 1$. There are three types of elementary excitations: 
\begin{itemize}
\item $|\bp\rangle = a_\bp^\dag |h\rangle$ with $|\bp| \geq  p_F$: particle states with energy $\mathcal{E}_\bp = E_{\bp} - h$ and momentum $\bp$. 
\item $|\hat \bp\rangle = a_{-\bp} |h\rangle$ with $|\bp| < p_F$: hole states with energy $\mathcal{E}_\bp = h-E_{\bp}$ and momentum $\bp$.
\item $|\bp\rangle = b_\bp^\dag |h\rangle$ for all $\bp$: antiparticle states with energy $\mathcal{E}_\bp = E_{\bp} + h$ and momentum $\bp$.
\end{itemize}

The (Heisenberg-picture) field operators at spacetime coordinates $x^\mu = (t, \bx)$ are defined by evolving the operators \eqref{eq:mode expansion fermions} in time:
\begin{equation}\begin{aligned}
\label{eq:field operator Heisenberg}
    \psi(x) &= \int \frac{d\bp}{2\pi} \frac{1}{\sqrt{2E_\bp}} \, \Bigl( a_\bp \, u(\bp) \, e^{-i(p_\mu x^\mu - hx^0)} + b_\bp^\dag \, v(\bp) \, e^{i(p_\mu x^\mu + hx^0)} \Bigr) \Big|_{p_0 = E_\bp} \,, \\
    \bar\psi(x) &= \int \frac{d\bp}{2\pi} \frac{1}{\sqrt{2E_\bp}} \, \Bigl( a_\bp^\dag \, \bar u(\bp) \, e^{i(p_\mu x^\mu - hx^0)} + b_\bp \, \bar v(\bp) \, e^{-i(p_\mu x^\mu + hx^0)} \Bigr) \Big|_{p_0 = E_\bp} \,.
\end{aligned}\end{equation}
The Wightman functions are:
\begin{equation}\begin{aligned}
    S^{\alpha+}_{\;\;\beta}(x-y) &\equiv \langle h | \psi^\alpha(x) \bar \psi_\beta(y) |h\rangle = \int_{|\bp|\ge p_F} \frac{d\bp}{2\pi} \, \frac{(\slashed{p} + m)^\alpha{}_\beta}{2E_\bp} \, e^{-ip_\mu (x-y)^\mu + i h (x-y)^0 } \Big|_{p_0 = E_\bp} \,, \\ 
    S^{\alpha-}_{\;\;\beta}(x-y) &\equiv \langle h | \bar\psi_\beta(y) \psi^\alpha(x) |h\rangle = \int_{|\bp| \leq p_F} \frac{d\bp}{2\pi} \, \frac{(\slashed{p} + m)^\alpha{}_\beta}{2E_\bp} \, e^{-ip_\mu (x-y)^\mu + i h (x - y)^0} \Big|_{p_0 = E_\bp} \\ 
    & \hspace{3.8cm} + \int_\mathbb{R} \frac{d\bp}{2\pi} \, \frac{(\slashed{p} - m)^\alpha{}_\beta}{2E_\bp} \, e^{i p_\mu (x-y)^\mu + ih(x-y)^0} \Big|_{p_0 = E_\bp} \,.
\end{aligned}\end{equation}
The time-ordered (or Feynman) propagator on the Fermi surface is then
\begin{equation}
\label{Feynman propagator}
    S^{\alpha F}_{\;\;\beta}(x-y) \equiv \langle h| \cT \psi^\alpha(x) \bar\psi_\beta(y) |h\rangle = \theta \bigl(x^0 - y^0 \bigr) \, S^{\alpha +}_{\;\;\beta}(x-y) - \theta \bigl(y^0 - x^0 \bigr) \, S^{\alpha -}_{\;\;\beta}(x-y)
\end{equation}
where $\cT$ is time ordering. The propagator can be written as the following contour integral:
\begin{equation}
\label{eq:Fermion prop derivation}
    S^{\alpha F}_{\;\;\beta}(x-y) = \int_{\mathcal{C}} \! \frac{d^2p}{(2\pi)^2} \, \frac{i(\slashed{p} + m + h\gamma^0)^\alpha{}_\beta }{(p_0+h)^2 - \bp^2 - m^2} \, e^{-ip_\mu(x-y)^\mu} = \int_{\mathcal{C}} \! \frac{d^2p}{(2\pi)^2} \, \biggl( \frac{i \, e^{-ip_\mu(x-y)^\mu}}{\slashed{p} + h\gamma^0 - m} \biggr)_{\!\raisebox{0.7em}[0pt][0pt]{\scriptsize$\alpha$}\beta} .
\end{equation}
Here $d^2p = dp_0 \, d\bp$. The countour $\mathcal{C}$ in the complex $p_0$-plane depends on $\bp$ and it is as follows:
\begin{equation}
\label{eq:appContour}
    |\bp| < p_F: \;
\begin{tikzpicture}[baseline=-3]
	\draw (-2.7, 0) -- (1.5, 0);
	\draw (0, -1) -- (0, 1);
	\filldraw (-1.65, 0) circle [radius = 0.06] node [above] {\scriptsize$-E_\bp{-}h$};
	\filldraw (-0.55, 0) circle [radius = 0.06] node [above] {\scriptsize$E_\bp{-}h$};
	\draw [line width = 1] (-2.5, 0) -- (-2, 0);
	\draw [line width = 1] (-2, 0) arc [start angle = 180, end angle = 360, radius = 0.35];
	\draw [line width = 1] (-1.3, 0) -- (-0.9, 0);
	\draw [line width = 1] (-0.9, 0) arc [start angle = 180, end angle = 360, radius = 0.35];
	\draw [line width = 1] (-0.2, 0) -- (1.35, 0);
	\node at (1.25, 0.8) {\small$p_0$};
	\draw (1, 1) -- (1, 0.6) -- (1.5, 0.6);
\end{tikzpicture}
\qquad\;\;
|\bp| > p_F: \;
\begin{tikzpicture}[baseline=-3]
	\draw (-2.5, 0) -- (2.5, 0);
	\draw (0, -1) -- (0, 1);
	\filldraw (-0.85, 0) circle [radius = 0.06] node [above] {\scriptsize$-E_\bp - h$};
	\filldraw (1.55, 0) circle [radius = 0.06] node [below] {\scriptsize$E_\bp - h$};
	\draw [line width = 1] (-2.25, 0) -- (-1.2, 0);
	\draw [line width = 1] (-1.2, 0) arc [start angle = -180, end angle = 0, radius = 0.35];
	\draw [line width = 1] (-0.5, 0) -- (1.2, 0);
	\draw [line width = 1] (1.9, 0) arc [start angle = 0, end angle = 180, radius = 0.35];
	\draw [line width = 1] (1.9, 0) -- (2.25, 0);
	\node at (2.25, 0.8) {\small$p_0$};
	\draw (2, 1) -- (2, 0.6) -- (2.5, 0.6);
\end{tikzpicture} \,.
\end{equation}
When $x^0 > y^0$ we can close the contour in the lower half-plane (with clockwise orientation). In this way, we pick up minus the residue at $p_0 = E_\bp - h$ for $|\bp| > p_F$, and no residue at all for $|\bp| < p_F$. When $x^0 < y^0$ we can close the contour in the upper half-plane (with anticlockwise orientation). In this way, we pick up the residue at $p_0 = - E_\bp - h$ for all $\bp$, as well as the residue at $p_0 = E_\bp - h$ for $|\bp| < p_F$. This reproduces the expression in (\ref{Feynman propagator}). The location of the poles for $|\bp| < p_F$ reflects the fact that no particle excitations are present for such momenta, only antiparticles and holes.

The Lorentizian contour $\mathcal{C}$ is such that it can always be smoothly rotated to a Euclidean contour along the imaginary $p_0$ axis by means of an anticlockwise rotation, for any $\bp$. In other words, in Euclidean signature the integration in $p_0$ is always along the imaginary axis, irrespective of the position of the poles.

\subsection{Physical amplitudes}
\label{app:Physical amplitudes}

The momentum-space four-point correlators computed in the main text can be connect to scattering amplitudes. 
In Feynman diagrams, the external legs associated with the states described after \eqref{eq:abpbbp} are given by 
the following Wick contractions: \vspace{-0.7em}
\begin{align*} 
\text{incoming particle:} \quad \psi |\bp\rangle &= \quad \begin{tikzpicture}[baseline=-3]
    \draw[line width=0.6] (-1, 0) -- (0, 0); \draw [-{Latex[length=7]}] (-1, 0) -- (-0.3, 0);
    \filldraw (0, 0) circle [radius = 0.08];
    \draw [-{Stealth[length=5]}] (-0.8, 0.3) -- (-0.2, 0.3) node [above, pos=0.5] {\small$\bp$};
    \end{tikzpicture} \quad = u(\bp) \,, \\
\text{outgoing particle:} \quad \langle \bp| \bar\psi &= \quad \begin{tikzpicture}[baseline=-3]
    \draw[line width=0.6] (-1, 0) -- (0, 0); \draw [-{Latex[length=7, reversed]}] (-1, 0) -- (-0.4, 0);
    \filldraw (0, 0) circle [radius = 0.08];
    \draw [{Stealth[length=5]}-] (-0.8, 0.3) -- (-0.2, 0.3) node [above, pos=0.5] {\small$\bp$};
    \end{tikzpicture} \quad = \bar{u}(\bp) \,, \\
\text{incoming hole:} \quad \bar\psi |\hat\bp\rangle &= \quad \begin{tikzpicture}[baseline=-3]
    \draw[line width=0.6] (-1, 0) -- (0, 0); \draw [-{Latex[length=7, reversed]}] (-1, 0) -- (-0.4, 0);
    \filldraw (0, 0) circle [radius = 0.08];
    \draw [-{Stealth[length=5]}] (-0.8, 0.3) -- (-0.2, 0.3) node [above, pos=0.5] {\small$\bp$};
    \end{tikzpicture} \quad = \bar u(-\bp) \,, \\
\text{outgoing hole:} \quad \langle \hat\bp| \psi &= \quad \begin{tikzpicture}[baseline=-3]
    \draw[line width=0.6] (-1, 0) -- (0, 0); \draw [-{Latex[length=7]}] (-1, 0) -- (-0.3, 0);
    \filldraw (0, 0) circle [radius = 0.08];
    \draw [{Stealth[length=5]}-] (-0.8, 0.3) -- (-0.2, 0.3) node [above, pos=0.5] {\small$\bp$};
    \end{tikzpicture} \quad = u(-\bp) \,, \\
\text{incoming anti-fermion:} \quad \bar\psi |\bp\rangle &= \quad \begin{tikzpicture}[baseline=-3]
    \draw[line width=0.6] (-1, 0) -- (0, 0); \draw [-{Latex[length=7, reversed]}] (-1, 0) -- (-0.4, 0);
    \filldraw (0, 0) circle [radius = 0.08];
    \draw [-{Stealth[length=5]}] (-0.8, 0.3) -- (-0.2, 0.3) node [above, pos=0.5] {\small$\bp$};
    \end{tikzpicture} \quad = \bar{v}(\bp) \,, \\
\text{outgoing anti-fermion:} \quad \langle\bp| \psi &= \quad \begin{tikzpicture}[baseline=-3]
    \draw[line width=0.6] (-1, 0) -- (0, 0); \draw [-{Latex[length=7]}] (-1, 0) -- (-0.3, 0);
    \filldraw (0, 0) circle [radius = 0.08];
    \draw [{Stealth[length=5]}-] (-0.8, 0.3) -- (-0.2, 0.3) node [above, pos=0.5] {\small$\bp$};
    \end{tikzpicture} \quad = v(\bp) \,.
\end{align*}
The dots denote the rest of the diagram, the flow of spatial momentum is indicated, and the $p_0$ component of the external leg is the physical energy $\mathcal{E}_\bp$ of the relevant excitation defined after \eqref{eq:abpbbp}.

Let us report the polarization spinors \eqref{eq:dirac plane waves} for the case of interest of massless fermions:
\begin{equation}\begin{aligned}
\label{eq:massless spinors}
 \bp>0:\qquad    u^\alpha(\bp) &= \begin{psmallmatrix} 0 \\ \sqrt{2\bp} \end{psmallmatrix} \,,  &\bar u^\alpha(\bp) &= \epsilon^{\alpha\beta} \bar u_\beta(\bp) = \begin{psmallmatrix} 0 \\ \sqrt{2\bp} \end{psmallmatrix} \,, \\
\bp < 0: \qquad     u^\alpha(\bp) &= \begin{psmallmatrix} \sqrt{2|\bp|} \\ 0 \end{psmallmatrix} \,,  &\bar u^\alpha(\bp) &= \epsilon^{\alpha\beta} \bar u_\beta(\bp) = \begin{psmallmatrix} - \sqrt{2|\bp|} \\ 0 \end{psmallmatrix} \,, \\
 \bp>0:\qquad    v^\alpha(\bp) &= \begin{psmallmatrix} 0 \\ - \sqrt{2\bp} \end{psmallmatrix} \,,  &\bar v^\alpha(\bp) &= \epsilon^{\alpha\beta} \bar v_\beta(\bp) = \begin{psmallmatrix} 0 \\ -\sqrt{2\bp} \end{psmallmatrix} \,, \\
\bp < 0: \qquad     v^\alpha(\bp) &= \begin{psmallmatrix} \sqrt{2|\bp|} \\ 0 \end{psmallmatrix} \,,  &\bar v^\alpha(\bp) &= \epsilon^{\alpha\beta} \bar v_\beta(\bp) = \begin{psmallmatrix} -\sqrt{2|\bp|} \\ 0 \end{psmallmatrix} \,.
\end{aligned}\end{equation}
Scattering amplitudes of these excitations are related to the amputated correlators \eqref{eq:four point function explicit} computed in section~\ref{sec:pert theory}.
For neutral fermions, \eqref{eq:Gamma4Neut} can blow up in elastic interaction processes at very low energies and momenta:  $q \to (0,0)$. For example, for incoming particles of type $m$ with momentum $\bp_1 > 0$ and type $n$ with momentum $\bp_2 < 0$ scattering into outgoing particles of type $m$ with momentum $\bp_4 = \bp_2<0$ and type $n$ with momentum $\bp_3 = \bp_1>0$, we have: 
\begin{equation}
\label{eq:particle hole different flavors N}
    \langle \bp_3^n, \bp_4^m \, | \, \bp_1^m, \bp_2^n \rangle = \;
\raisebox{0pt}[2.2em][1.3em]{\begin{tikzpicture}[baseline=-3]
	\draw [line width=0.6] (-0.9, 0.9) -- (0.9, -0.9);
	\draw [line width=0.6] (-0.9, -0.9) -- (0.9, 0.9);
	\draw [-{Latex[length=7]}] (-0.9, 0.9) -- +(-45: 0.6);
	\draw [-{Latex[length=7]}] (0.9, -0.9) -- +(135: 0.6);
	\draw [-{Latex[length=7, reversed]}] (-0.9, -0.9) -- +(45: 0.55);
	\draw [-{Latex[length=7, reversed]}] (0.9, 0.9) -- +(-135: 0.55);
	\draw [-{Stealth[length=5]}] (-0.95, 0.55) node[shift={(-0.1, -0.25)}] {\small$\bp_1$} -- +(-45: 0.6);
	\draw [{Stealth[length=5]}-] (-0.95, -0.55) node[shift={(-0.1, 0.35)}] {\small$\bp_4$} -- +(45: 0.6);
	\draw [-{Stealth[length=5]}] (0.95, -0.55) node[shift={(0.1, 0.3)}] {\small$\bp_2$} -- +(135: 0.6);
	\draw [{Stealth[length=5]}-] (0.95, 0.55) node[shift={(0.1, -0.35)}] {\small$\bp_3$} -- +(-135: 0.6);
	\fill [white] (0,0) circle [radius = 0.35];
	\draw [pattern = north west lines, line width = 0.6] (0,0) circle [radius = 0.35];
\end{tikzpicture}}
    = u^\alpha(\bp_1) \, \bar u^\beta(\bp_4) \, u^\gamma(\bp_2) \, \bar u^\delta(\bp_3) \, \Gamma_{\alpha\beta\gamma\delta}^\text{\rm neut}(p_1,-p_4,p_2,-p_3) .
\end{equation}
The only contribution comes from $\Gamma^\text{\rm neut}_{+--+}$, where the indices $\pm$ refer to the top and bottom component of the 2-spinors in \eqref{eq:massless spinors}. Other processes relating particles and antiparticles are similarly derived.

For charged fermions, the amplitude can blow up for interaction processes near the Fermi surface:  $(q_0,q_1) \to (0,2h)$. For example, the scattering of a hole and a particle of type $m$ into a particle and a hole of type $n$ with momenta close to the positive part of the Fermi surface ($\bp_{1,2,3,4} > 0$) gives
\begin{equation}
\label{eq:particle hole different flavors}
    \langle \bp_3^n, \hat\bp_4^n \, | \, \bp_1^m, \hat\bp_2^m \rangle = \;
\raisebox{0pt}[2.2em][1.3em]{\begin{tikzpicture}[baseline=-3]
	\draw [line width=0.6] (-0.9, 0.9) -- (0.9, -0.9);
	\draw [line width=0.6] (-0.9, -0.9) -- (0.9, 0.9);
	\draw [-{Latex[length=7]}] (-0.9, 0.9) -- +(-45: 0.6);
	\draw [-{Latex[length=7]}] (0.9, -0.9) -- +(135: 0.6);
	\draw [-{Latex[length=7, reversed]}] (-0.9, -0.9) -- +(45: 0.55);
	\draw [-{Latex[length=7, reversed]}] (0.9, 0.9) -- +(-135: 0.55);
	\draw [-{Stealth[length=5]}] (-0.95, 0.55) node[shift={(-0.1, -0.25)}] {\small$\bp_1$} -- +(-45: 0.6);
	\draw [-{Stealth[length=5]}] (-0.95, -0.55) node[shift={(-0.1, 0.3)}] {\small$\bp_2$} -- +(45: 0.6);
	\draw [{Stealth[length=5]}-] (0.95, -0.55) node[shift={(0.1, 0.35)}] {\small$\bp_4$} -- +(135: 0.6);
	\draw [{Stealth[length=5]}-] (0.95, 0.55) node[shift={(0.1, -0.35)}] {\small$\bp_3$} -- +(-135: 0.6);
	\fill [white] (0,0) circle [radius = 0.35];
	\draw [pattern = north west lines, line width = 0.6] (0,0) circle [radius = 0.35];
\end{tikzpicture}}
    = \bar u^\alpha(\bp_1) \, u^\beta(-\bp_2) \, u^\gamma(-\bp_4) \, \bar u^\delta(\bp_3) \, \Gamma^{\rm ch}_{\alpha\beta\gamma\delta}(p_1,p_2,-p_4,-p_3)
\end{equation}
whose only contribution comes from $\Gamma^{\rm ch}_{+--+}$.

\section{Details about the 1-loop computation of \tpsb{\cI_{\alpha\beta\gamma\delta}(q,h_1,h_2)}{I(q,h1,h2)}}
\label{app:cI computation}

Let us calculate $\cI_{\alpha\beta\gamma\delta}(q,h_1,h_2)$ defined in \eqref{eq:cI def}, that can be rewritten as
\begin{equation}
    \cI_{\alpha\beta\gamma\delta}(q,h_1,h_2) = - \int\! \frac{d^2p}{(2\pi)^2} \, \frac{ \left( \slashed{p} + h_1 \gamma^0 \right)_{\alpha\beta} \, \left( \slashed{p} - \slashed{q} + h_2 \gamma^0 \right)_{\gamma\delta} }{ \left[ \left( p_0 + h_1 \right)^2 - p_1^2 \right] \left[ \left( p_0 - q_0 + h_2 \right)^2 - \left( p_1 - q_1 \right)^2 \right] } \,.
\end{equation}
We perform a Wick rotation of the contour $p_0 = ip_2$ with $p_2$ real, so that $p_0$ is along the imaginary axis. As discussed in app.~\ref{app:Canonical Quantization Propagator}, this can be done without crossing any pole and it sets the first propagator along the correct contour, but not the second propagator. Therefore, we also rotate the external momentum to Euclidean signature: $q_0 = i q_2$. Defining
\begin{equation}
\label{eq:composite gamma mats}
\gamma_a = (\gamma^0)_{\alpha\beta} (\gamma^0)_{\gamma\delta} \;,\quad \gamma_b = (\gamma^0)_{\alpha\beta} (\gamma^1)_{\gamma\delta} \;,\quad \gamma_c = (\gamma^1)_{\alpha\beta} (\gamma^0)_{\gamma\delta} \;,\quad\gamma_d = (\gamma^1)_{\alpha\beta} (\gamma^1)_{\gamma\delta} \;,
\end{equation}
the integral can be written as
\begin{equation}
\cI = i \int\! \frac{dp_1 dp_2}{(2\pi)^2} \, \frac{ \Omega }{ \left[ \left( p_2 -i h_1 \right)^2 + p_1^2 \right] \left[ \left( p_2 - q_2 - i h_2 \right)^2 + \left(p_1 - q_1 \right)^2 \right] }
\end{equation}
with
\begin{equation}
\Omega = (p_2 - i h_1)(p_2 - q_2 - i h_2) \gamma_a - i (p_2 - ih_1)(p_1 - q_1) \gamma_b - i p_1 (p_2 - q_2 - i h_2) \gamma_c - p_1(p_1 - q_1) \gamma_d \;.
\end{equation}
We integrate in $p_2$ by closing the contour in the upper complex half-plane and using the residue theorem, whereas we regulate the integral in $p_1$ by introducing a cutoff $|p_1| < M$ (assuming that $M$ is very large). There are four poles in the complex $p_2$-plane:
\begin{equation}
    p_2 = i \left(h_1 \pm p_1 \right) \,,\qquad\qquad p_2 = q_2 + i \bigl( h_2 \pm \left(p_1 -q_1 \right) \bigr) \,.
\end{equation}
We compute the residues in turn.
{\setlength{\leftmargini}{1.5em}
\begin{enumerate}
    \item The pole at $p_2 = i (h_1 + p_1 )$ is inside the contour for $p_1 > - h_1$, therefore the integral in $p_1$ is $\int_{-h_1}^M dp_1$, resulting in
    \begin{equation} \hspace{-1.2em}
        \frac{ i \, (M + h_1) (\gamma_a - \gamma_b - \gamma_c + \gamma_d ) }{ 8\pi \, (h_1 - h_2 + q_1 + iq_2 )} + \frac{i}{16\pi} \log \biggl[ \frac{2M + h_1 - h_2 - q_1 + iq_2 }{ iq_2 - q_1 - h_1 - h_2} \biggr] (\gamma_a + \gamma_b - \gamma_c - \gamma_d ) \,.
    \end{equation}
    
    \item The pole at $p_2 = i (h_1 - p_1 )$ similarly leads to the integral $\int_{-M}^{h_1} dp_1$, resulting in
    \begin{equation} \hspace{-1.5em}
        \frac{i \, (M + h_1) (\gamma_a + \gamma_b + \gamma_c + \gamma_d) }{ 8\pi \, (h_1 - h_2 - q_1 + iq_2 )} + \frac{i}{16\pi} \log\biggl[ \frac{2M + h_1 - h_2 + q_1 + iq_2 }{ q_1 + iq_2 - h_1 -h_2} \biggr] (\gamma_a - \gamma_b + \gamma_c - \gamma_d) \,.
    \end{equation}

    \item The pole at $p_2 = q_2 + i( h_2 + p_1 - q_1)$ is inside the contour for $p_1 > q_1 - h_2$, therefore the integral in $p_1$ is $\int_{q_1 - h_2}^M dp_1$, resulting in
    \begin{equation}
        \!\!\!\!\!\!\!\! \frac{i \, (M + h_2 - q_1) (\gamma_a - \gamma_b - \gamma_c + \gamma_d) }{ 8\pi \, (h_1 - h_2 + q_1 + iq_2)} + \frac{i}{16\pi}\log\biggl[ \frac{ 2M - h_1 + h_2 - q_1 - iq_2 }{ q_1 - iq_2 - h_1 - h_2} \biggr] (\gamma_a - \gamma_b + \gamma_c - \gamma_d ) \,.
    \end{equation}

    \item The pole at $p_2 = q_2 + i( h_2 - p_1 + q_1)$ similarly leads to the integral $\int_{-M}^{h_2+q_1} dp_1$, resulting in
    \begin{equation}
        \!\!\!\!\!\!\!\! \frac{i \, (M + h_2 + q_1) (\gamma_a + \gamma_b + \gamma_c + \gamma_d) }{ 8\pi \, (h_2 - h_1 + q_1 - iq_2)} + \frac{i}{16\pi} \log\biggl[ \frac{2M - h_1 + h_2 + q_1 - iq_2}{ -q_1 - iq_2 - h_1 - h_2} \biggr] (\gamma_a + \gamma_b - \gamma_c - \gamma_d) \,.
    \end{equation}
\end{enumerate}}\noindent
Summing over all contributions and taking the cutoff to infinity, $M\to\infty$, we are left with
\begin{equation}\begin{aligned}
\label{eq:cI temp}
    \cI_{\alpha\beta\gamma\delta}(q,h_1,h_2)
    &= \frac{\gamma_d - \gamma_a}{2} \, B(q,h_1,h_2) + \frac{\gamma_c - \gamma_b}{2} \, A(q,h_1,h_2) \\
    &\quad + \frac{ (\gamma_a + \gamma_d) \, i \, \bigl[ (h_1 - h_2)^2 - q_1^2 + i (h_1 - h_2)q_2 \bigr] + (\gamma_b + \gamma_c) \, q_1 q_2 }{ 4\pi \, (h_1 - h_2 + q_1 + i q_2 )( h_1 - h_2 - q_1 + i q_2) }
\end{aligned}\end{equation}
where
\begin{equation}\begin{aligned}
    A(q,h_1,h_2) &= \frac{i}{8\pi} \log \biggl[ \frac{ (h_1 + h_2 + q_1)^2 + q_2^2 }{ (h_1 + h_2 - q_1)^2 + q_2^2} \biggr] \,, \\
    B(q,h_1,h_2) &= \frac{i}{8\pi} \log \bigg[ \frac{ \bigl( (h_1 + h_2 - q_1)^2 + q_2^2 \bigr) \bigl( (h_1 + h_2 + q_1)^2 + q_2^2 \bigr) }{ 16M^4} \biggr] \,.
\end{aligned}\end{equation}
The dependence on $M$ is logarithmic, as expected for a marginal coupling in a renormalizable theory. Using $\slashed{q} = q_0 \gamma^0 + q_1 \gamma^1 = iq_2 \gamma^0 + q_1 \gamma^1$ and noticing that
\begin{equation}
\slashed{q}_{\alpha\beta} \, \slashed{q}_{\gamma\delta} = - q_2^2 \, \gamma_a + i q_1 q_2 \, (\gamma_b + \gamma_c) + q_1^2 \, \gamma_d \;,
\end{equation}
we can simplify the expression in \eqref{eq:cI temp} to
\begin{equation}\begin{aligned}
\cI &= \frac{\gamma_d - \gamma_a}2 \, B(q, h_1, h_2) + \frac{\gamma_c - \gamma_b}2 \, A(q, h_1, h_2) \\
&\quad + \frac{ -i \, \slashed{q}_{\alpha\beta} \, \slashed{q}_{\gamma\delta} + (h_1 - h_2) \bigl[ q_2 (\gamma_a - \gamma_d) + i (h_1 - h_2) \gamma_d \bigr] }{ 4\pi \, (h_1 - h_2 + q_1 + i q_2 )( h_1 - h_2 - q_1 + i q_2) } + \frac{i \, \gamma_a}{4\pi} \;.
\end{aligned}\end{equation}
Note that when $h_1 = h_2 = 0$ we find $\cI_{\alpha\beta\gamma\delta} = \frac{i}{4\pi} \frac{ \slashed{q}_{\alpha\beta} \slashed{q}_{\gamma\delta} }{ q^2} + \frac{ i \gamma^0_{\alpha\beta} \gamma^0_{\gamma\delta} }{ 4\pi}$. This would have been Lorentz covariant if not for the last term. The violation is due to the choice of regulator, momentum cutoff, which explicitly breaks Lorentz invariance. But these terms cancel between the different diagrams, and the physical amplitude remains Lorentz covariant.

\subsection{The 1-loop diagrams}
\label{app:1-loop diagrams}
The 1-loop diagrams appearing in \autoref{fig:1-loop-correlator} contribute with a sign that depends on the pattern of contractions between the four external fermions and the two vertices. Each crossing of contractions and each contraction where $\psi$ is to the right of the contracted $\bar\psi$ give a minus sign. For each diagram, the contractions and the contribution of the diagram are (implicitly summing over spinorial indices that appear twice): 
\begin{enumerate}
    \item[(\ref{fig:1-loop-1})] 
            $\wick{\c3{\bar{\psi}_i^\alpha} \c2{\psi_i^\beta} \quad ( \c2{\bar{\psi}_i} \c3{\psi_i} )} 
            \wick{( \c2{\bar{\psi}_k} \c1{\psi_k}) \quad ( \c1{\bar{\psi}_k} \c2{\psi_k})} \wick{(\c2{\bar{\psi}_j} \c1{\psi_j} ) \quad \c1{\bar{\psi}_j^\gamma} \c2{\psi_j^\delta}}$, sign: $-$.
            \begin{equation}
                \label{eq:1-loop-1}
                - (ig^2)^2 \, \epsilon_{\alpha\beta} \epsilon_{\gamma\delta}  \sum_{k=1}^N \cI^{\lambda\phantom{\sigma}\sigma}_{\phantom{\lambda}\sigma\phantom{\sigma}\lambda}(q,h_k,h_k) \,,
            \end{equation}
    \item[(\ref{fig:1-loop-2})] 
            $\wick{\c5{\bar{\psi}_i^\alpha} \c2{\psi_i^\beta} \quad (\c2{\bar{\psi}_i} \c3{\psi_i}) (\c4{\bar{\psi}_i} \c5{\psi_i}) \quad (\c3{\bar{\psi}_i} \c4{\psi_i})} \wick{(\c2{\bar{\psi}_j} \c1{\psi_j}) \quad \c1{\bar{\psi}_j^\gamma}\c2{\psi_j^\delta}}$, sign: $+$.
            \begin{equation}
                (ig^2)^2 \, \cI^{\phantom{\beta}\phantom{\lambda}\lambda}_{\beta\lambda\phantom{\lambda}\alpha}(-q,h_i,h_i)   \epsilon_{\delta\gamma} \,,
            \end{equation}
    \item[(\ref{fig:1-loop-3})] 
        $\wick{\c3{\bar{\psi}_i^\alpha} \c2{\psi_i^\beta} \quad (\c2{\bar{\psi}_i} \c3{\psi_i})} \wick{(\c3{\bar{\psi}_i} \c2{\psi_i}) \quad (\c4{\bar{\psi}_i} \c3{\psi_i}) (\c2{\bar{\psi}_j}  \c1{\psi_j}) \quad \c1{\bar{\psi}_j^\gamma} \c4{\psi_j^\delta}}$, sign: $+$.
        \begin{equation}
                (ig^2)^2 \, \epsilon_{\beta\alpha} \cI^{\phantom{\delta}\phantom{\lambda}\lambda}_{\delta\lambda\phantom{\lambda}\gamma}(q,h_j,h_j) \,,
            \end{equation}
    \item[(\ref{fig:1-loop-4})] 
        $\wick{\c5{\bar{\psi}_i^\alpha} \c2{\psi_i^\beta} \quad (\c2{\bar{\psi}_i} \c3{\psi_i}) (\c6{\bar{\psi}_j} \c4{\psi_j}) \quad (\c3{\bar{\psi}_i} \c5{\psi_i}) (\c4{\bar{\psi}_j}  \c2{\psi_j}) \quad \c2{\bar{\psi}_j^\gamma} \c6{\psi_j^\delta}}$, sign: $+$.
        \begin{equation}
                (ig^2)^2 \, \bigl(-\cI_{\beta\alpha\delta\gamma}(q_{13},h_i,-h_j)\bigr) \,,
            \end{equation}
    \item[(\ref{fig:1-loop-5})] 
        $\wick{\c5{\bar{\psi}_i^\alpha} \c2{\psi_i^\beta} \quad (\c2{\bar{\psi}_i} \c3{\psi_i}) (\c4{\bar{\psi}_j} \c6{\psi_j}) \quad (\c3{\bar{\psi}_i} \c5{\psi_i}) (\c2{\bar{\psi}_j}  \c4{\psi_j}) \quad \c6{\bar{\psi}_j^\gamma} \c2{\psi_j^\delta}}$, sign: $+$.
        \begin{equation}
                (ig^2)^2 \,  \cI_{\beta\alpha\delta\gamma}(q_{14},h_i,h_j) \,.
        \end{equation}
\end{enumerate}

\subsection{Scattering of identical fermions}
\label{app:Identical Fermions}
Here we compute the amplitude \eqref{eq:corr def} when $i=j$. In that case there's another set of diagrams, identical to those in Figure~\ref{fig:1-loop-correlator} up to exchanging $\left(2\leftrightarrow4,\beta\leftrightarrow\delta\right)$, which also introduces an overall minus sign (because of exchanging two fermions). The amplitude is now 
\begin{equation}\label{eq:Gamma4App}
    \begin{aligned}
        \Gamma{}_{\alpha\beta\gamma\delta}^{ii}&(q_{1},q_{2},q_{3},q_{4}) = g^{4}\epsilon_{\alpha\beta}\gamma_{\gamma\delta}^{*}A(q,h_{i}) - g^{4}\gamma_{\alpha\beta}^{*}\epsilon_{\gamma\delta}A(q,h_{i}) \\  
        &  + \epsilon_{\alpha\beta}\epsilon_{\gamma\delta}\biggl[ig^{2} + i\delta_{g} - 2(N - a)g^{4}B(q,0) - 2ag^{4}B(q,h) + g^{4}B(q,h_{i}) + g^{4}B(q,h_{j})\biggr] \\  
        &  + g^{4}\eta_{\mu\nu}\gamma_{\beta\alpha}^{\mu}\gamma_{\delta\gamma}^{\nu}\bigl[B(q_{14},h_{i}) - B(q_{13},h_{i}, - h_{i})\bigr] + g^{4}\epsilon_{\mu\nu}\gamma_{\beta\alpha}^{\mu}\gamma_{\delta\gamma}^{\nu}\bigl[A(q_{14},h_{i}) - A(q_{13},h_{i}, - h_{i})\bigr] \\  
        &  - g^{4}\epsilon_{\alpha\delta}\gamma_{\gamma\beta}^{*}A(q_{14},h_{i}) + g^{4}\gamma_{\alpha\delta}^{*}\epsilon_{\gamma\beta}A(q_{14},h_{i}) \\  
        &  - \epsilon_{\alpha\delta}\epsilon_{\gamma\beta}\biggl[ig^{2} + i\delta_{g} - 2(N - a)g^{4}B(q_{14},0) - 2ag^{4}B(q_{14},h) + g^{4}B(q_{14},h_{i}) + g^{4}B(q_{14},h_{i})\biggr] \\  
        &  - g^{4}\eta_{\mu\nu}\gamma_{\delta\alpha}^{\mu}\gamma_{\beta\gamma}^{\nu}\bigl[B(q,h_{i}) - B(q_{13},h_{i}, - h_{i})\bigr] - g^{4}\epsilon_{\mu\nu}\gamma_{\delta\alpha}^{\mu}\gamma_{\beta\gamma}^{\nu}\bigl[A(q,h_{i}) - A(q_{13},h_{i}, - h_{i})\bigr] \\  
        &  + \text{non logarithmic terms.}
    \end{aligned}
\end{equation}
where the last four lines are those that arose from the new diagrams. We now use the identities
\begin{equation}
    \eta_{\mu\nu}\gamma_{\delta\alpha}^{\mu}\gamma_{\beta\gamma}^{\nu} = \epsilon_{\delta\gamma}\epsilon_{\beta\alpha} -\gamma_{\delta\gamma}^{*}\gamma_{\beta\alpha}^{*}\,, \qquad \epsilon_{\mu\nu}\gamma_{\delta\alpha}^{\mu}\gamma_{\beta\gamma}^{\nu} = \epsilon_{\delta\gamma}\gamma_{\beta\alpha}^{*} -\gamma_{\delta\gamma}^{*}\epsilon_{\beta\alpha} \,,
\end{equation}
to simplify the third and sixth lines.

Let's start with the neutral scattering. The new terms contributing to the $\epsilon_{\alpha\beta}\epsilon_{\gamma\delta}$ component just cancel each other (and there would be another identical term proportional to $-\epsilon_{\alpha\delta}\epsilon_{\gamma\beta}$), resulting in the same value for the counter term as when $i \neq j$, and so the same beta function.  

For scattering of charged fermions, the renormalization scale is set to the upper part of the Fermi surface. The renormalization conditions is as in \eqref{eq:renorm cond kinematics} with $\bar\mu=(2h,0)$. The terms that will contribute to the logarithmic divergence for the momenta we chose are those that contain $B\left(q,h\right)$ and $B\left(q_{13},h,-h\right)$, while those containing $B\left(q_{14},h,h\right)$ will not (we remind the reader that our notation is $q_{ij}=q_{i}+q_{j}$). These terms are
\begin{multline}
    g^{4}\bigl(\epsilon_{\alpha\beta}\gamma^*_{\gamma\delta} -\gamma_{\alpha\beta}^{*}\epsilon_{\gamma\delta}\bigr) A(q,h_{i}) + \epsilon_{\alpha\beta}\epsilon_{\gamma\delta}\biggl[ig^{2}+i\delta_{g}-2\left(a-1\right)g^{4}B(q,h)\biggr] \\ - g^{4}\left(\epsilon_{\alpha\beta}\epsilon_{\gamma\delta}-\gamma_{\alpha\beta}^{*}\gamma_{\gamma\delta}^{*}\right)\left(B(q,h)-B\left(q_{13},h,-h\right)\right) - g^{4}\left(\epsilon_{\alpha\beta}\gamma_{\gamma\delta}^{*}-\gamma_{\alpha\beta}^{*}\epsilon_{\gamma\delta}\right)A(q,h) \,,
\end{multline}
where the first line is the same one that appeared before, and the second is the new contribution coming from the diagrams (\ref{fig:1-loop-4}), (\ref{fig:1-loop-5}) after exchanging $\left(2\leftrightarrow4\right)$. Up to possible finite terms, which do not matter for the computation of the 1-loop beta function, near the Fermi surface the amplitudes are
\begin{align}
    \Gamma_{-+-+}^{{\rm ch}} & = ig^{2}+i\delta_{g}-2\left(a-1\right)g^{4}B(q,h)\,, \\ 
    \Gamma_{+-+-}^{{\rm ch}} & = ig^{2}+i\delta_{g}-2\left(a-1\right)g^{4}B(q,h)\,, \\ 
    \Gamma_{-++-}^{{\rm ch}} & = -ig^{2}-i\delta_{g}+2g^{4}\left(aB(q,h)-B\left(q_{13},h,-h\right)\right) \,, \\ 
    \Gamma_{+--+}^{{\rm ch}} & = -ig^{2}-i\delta_{g}+2g^{4}\left(aB(q,h)-B\left(q_{13},h,-h\right)\right) \,.
\end{align}
where we have used $\gamma_{\alpha\beta}^{*} = \begin{psmallmatrix}0 & -1\\
-1 & 0
\end{psmallmatrix}$. Our renormalization condition is such that $B\left(q,h\right)=B\left(q_{13},h,-h\right)$ up to a momentum independent shift, and so ultimately
\begin{equation}
    \Gamma_{+--+}^{{\rm ch}} = \Gamma_{-++-}^{{\rm ch}} = -ig^{2}-i\delta_{g}+2g^{4}\left(a-1\right)B(q,h) \,.
\end{equation}
The beta function near the Fermi surface, using this renormalization condition on four identical fermions, is thus $\beta = - \frac{a-1}{2\pi}g^{4}$.

\section{Elliptic functions and useful integrals}
\label{app:ellipticFun}

\subsection{Some basics of elliptic functions}
\label{app:ellipticFunbasics}

The incomplete elliptic integral of the first kind $F(\varphi|\mathfrak{m})$ is defined as
\be
\label{eq:Ellipt1}
F(\varphi|\mathfrak{m}) \,\equiv\, \int_0^\varphi \! \frac{d\theta}{\sqrt{1-\mathfrak{m}\sin^2\theta}} = \int_0^{\sin\varphi} \! \frac{ dt}{ \sqrt{1-t^2} \, \sqrt{1- \mathfrak{m} \, t^2}} \;.
\ee
Usually $\mathfrak{m}$ is called the parameter. In the basic definition, $0 \leq \mathfrak{m} \leq 1$ and $0 \leq \varphi \leq \frac\pi2$ but they can be extended. The inverse of $F$ is called the amplitude:
\be
F(\varphi|\mathfrak{m}) = u \qquad\Leftrightarrow\qquad \varphi = \operatorname{am}(u|\mathfrak{m}) \;.
\ee
The Jacobi elliptic functions are then defined as
\be
\sn u = \sin\varphi \;,\qquad \cn u = \cos\varphi \;,\qquad \dn u = \sqrt{1 - \mathfrak{m} \sin^2 \varphi} \;,
\ee
where we left the parameter $\mathfrak{m}$ implicit. Note that $0\leq \dn u \leq 1$. Three basic identities follow:
\begin{equation}\begin{aligned}
\sn(u| \mathfrak{m})^2 + \cn(u| \mathfrak{m})^2 = 1 \;,\qquad\qquad \dn(u| \mathfrak{m})^2 + \mathfrak{m} \sn(u| \mathfrak{m})^2  &= 1 \;, \\
\dn(u|\mathfrak{m})^2 - \mathfrak{m} \cn(u| \mathfrak{m})^2 & = 1-\mathfrak{m} \;.
\end{aligned}\end{equation}
We also have, by definition:
\be
\label{eq:Ellipt5}
\varphi = \operatorname{am}(u|\mathfrak{m}) = \arcsin \bigl( \sn(u|\mathfrak{m}) \bigr) \;.
\ee
We see that $F\bigl( \arcsin(z) |\mathfrak{m}\bigr) = u$ is the inverse function to $\sn(u|\mathfrak{m}) = z$. When evaluated at $\varphi = \frac\pi2$, $F(\varphi|\mathfrak{m})$ gives the complete elliptic integral of the first kind:
\be
\label{eq:Kdefinition}
K(\mathfrak{m}) = F \bigl( \tfrac\pi2 \big| \mathfrak{m} \bigr) = \int_0^{\frac\pi2} \frac{d\theta}{\sqrt{1-\mathfrak{m} \sin^2\theta}} \;.
\ee
The incomplete elliptic integral of the first kind has useful properties. For parameter $\mathfrak{m}$ bigger than 1 one can use:
\be
F(\varphi|\mathfrak{m}) = \frac1{\sqrt\mathfrak{m}} \, F \bigl( \theta \big| \mathfrak{m}^{-1} \bigr) \qquad\text{with}\qquad \sin\theta = \mathfrak{m}^{1/2} \sin\varphi \;.
\ee
This can be rewritten as
\be
F\bigl( \arcsin \sqrt{z} \big| \mathfrak{m} \bigr) = \frac1{\sqrt{\mathfrak{m}}} \, F\bigl( \arcsin \sqrt{\mathfrak{m}\, z} \big| \mathfrak{m}^{-1} \bigr) \;.
\ee
For imaginary arguments one can use:%
\footnote{To prove it: $F(i\varphi|\mathfrak{m}) = \int_0^{i\sinh \varphi} \bigl( (1-t^2)(1-\mathfrak{m} \, t^2) \bigr)^{-1/2} \, dt = \int_0^{\sin\theta} \bigl( (1-w^2)(1-\mathfrak{m}' \, w^2) \bigr)^{-1/2} \, dw$ with the change of variable $t = iw / \sqrt{1-w^2}$ and $w = -i t / \sqrt{1-t^2}$ for $\mathfrak{m}' = 1-\mathfrak{m}$ and $\sinh\varphi = \tan \theta$.}
\be
F(i \varphi | \mathfrak{m}) = i \, F (\theta, 1-\mathfrak{m}) \qquad\text{with}\qquad \begin{aligned} \tan\theta &= \sinh\varphi \\ \sin\theta &= \tanh \varphi \end{aligned} \;.
\ee
This can be rewritten as
\be
F\bigl( i \operatorname{arcsinh}\sqrt{w} \big| \mathfrak{m} \bigr) = i \, F\biggl( \arcsin \sqrt{\frac{w}{w+1}} \bigg| 1- \mathfrak{m} \biggr) \;.
\ee
One also has$\,$%
\footnote{To prove it, use the change of variable $t = \bigl( 1 - (1-\mathfrak{m}) w^2 \bigr)^{-1/2}$ that is $w = (1-\mathfrak{m})^{-\frac12} \sqrt{ 1-1/t^2 }$.}
\be
F \bigl( \tfrac\pi2 + i \varphi \big| \mathfrak{m} \bigr) = K(\mathfrak{m}) + i \, F(\theta | 1-\mathfrak{m}) \qquad\text{with}\qquad \sin\theta = (1 - \mathfrak{m})^{-\frac12} \tanh\varphi \;.
\ee
For small real $\varphi$ one finds that $\theta$ is real, but for $|\varphi| > \operatorname{arctanh} \sqrt{1-\mathfrak{m}}$ one finds that $\theta$ is complex. In that case one can use
\be
    F \bigl( \tfrac\pi2 + i \varphi \big| \mathfrak{m} \bigr) = \begin{cases}
        F(\theta|\mathfrak{m}) + 2K + i K' & \text{with } \sin\theta = -\dfrac1{\sqrt\mathfrak{m}\, \cosh\varphi} \text{ for } \varphi \geq \operatorname{arccosh} \frac1{\sqrt{m}} \\
        F(\theta|\mathfrak{m}) - i K' & \text{with } \sin\theta = \dfrac1{\sqrt\mathfrak{m}\, \cosh\varphi} \text{ for } \varphi \leq - \operatorname{arccosh} \frac1{\sqrt{m}}
    \end{cases}
\ee
Here $K \equiv K(\mathfrak{m})$ and $K' \equiv K(1-\mathfrak{m})$.
Another relation is
\be
\label{doubly sing relation for F}
    F(\varphi|\mathfrak{m}) + F(\psi|\mathfrak{m}) = K(\mathfrak{m}) \qquad\text{if}\qquad \sqrt{1-\mathfrak{m}} \; \tan(\varphi) \tan(\psi) = 1 \;.
\ee

The incomplete elliptic integral of the second kind $E(\varphi | \mathfrak{m})$ is defined as
\be
E(\varphi|\mathfrak{m}) \equiv \int_0^\varphi \sqrt{1 - \mathfrak{m} \sin^2\theta} \, d\theta = \int_0^{\sin\varphi} \sqrt{ \frac{1-\mathfrak{m}\, t^2}{1-t^2} } \, dt \;.
\ee
The complete elliptic integral of the second kind is
\be\label{eq:Edefinition}
E(\mathfrak{m}) = E\bigl( \tfrac\pi2 \big| \mathfrak{m} \bigr) = \int_0^{\frac\pi2} \sqrt{ 1 - \mathfrak{m} \sin^2\theta} \; d\theta \;.
\ee
There is a similar function, called Jacobi epsilon function $\mathcal{E}(u|\mathfrak{m})$ but sometimes indicated with the letter $E$ and sometimes called the incomplete elliptic integral of the second kind, defined such that
\be
E(\varphi | \mathfrak{m}) = \mathcal{E}\bigl( F(\varphi| \mathfrak{m}) \big| \mathfrak{m} \bigr) \qquad\Leftrightarrow\qquad \mathcal{E}(u|\mathfrak{m}) = E \bigl( \operatorname{am}(u|\mathfrak{m}) \big| \mathfrak{m} \bigr) \;.
\ee
The function can be defined by
\be
\mathcal{E}(u| \mathfrak{m}) = \int_0^u \dn^2(w | \mathfrak{m}) \, dw \;.
\ee
The Jacobi zeta function can be defined as
\be
\label{eq:jacobi zeta}
\zn(u| \mathfrak{m}) = \mathcal{E}(u| \mathfrak{m}) - \frac{E(\mathfrak{m})}{K(\mathfrak{m})} \, u \;.
\ee
Some derivatives are:
\begin{equation}\begin{aligned}
\frac{d}{du} \sn(u| \mathfrak{m}) &= \cn(u| \mathfrak{m}) \dn(u| \mathfrak{m}) \;,\qquad&
\frac{d}{du} \dn(u| \mathfrak{m}) &= - \mathfrak{m} \sn(u| \mathfrak{m}) \cn(u| \mathfrak{m}) \;,\\
\frac{d}{du} \cn(u| \mathfrak{m}) &= - \sn(u| \mathfrak{m}) \dn(u| \mathfrak{m}) \;,\qquad&
\frac{d}{du} \, \mathcal{E}(u| \mathfrak{m}) &= \dn(u| \mathfrak{m})^2 \;.
\end{aligned}\end{equation}

The incomplete elliptic integral of the third kind $\Pi(\varphi | \mathfrak{n}, \mathfrak{m})$ is defined as$\,$%
\footnote{In \texttt{Mathematica} the order of the arguments is $\mathfrak{n}, \varphi, \mathfrak{m}$.}
\be
\Pi(\varphi | \mathfrak{n}, \mathfrak{m}) = \int_0^\varphi \frac{d\theta}{ (1-\mathfrak{n} \sin^2\theta) \sqrt{ 1 - \mathfrak{m} \sin^2\theta} } = \int_0^{\sin\varphi} \frac{ dt}{ (1-\mathfrak{n} \, t^2) \sqrt{ (1-t^2)(1-\mathfrak{m} \, t^2) }} \;.
\ee
In the basic definition, $0 \leq \mathfrak{n},\mathfrak{m} \leq 1$ and $0 \leq \varphi \leq \frac\pi2$. Note that $\Pi(\varphi| 0,\mathfrak{m}) = F(\varphi | \mathfrak{m})$. The complete elliptic integral of the third kind is
\be
\Pi(\mathfrak{n},\mathfrak{m}) = \Pi \bigl( \tfrac\pi2 \big| \mathfrak{n}, \mathfrak{m} \bigr) = \int_0^{\frac\pi2} \frac{d\theta}{ (1-\mathfrak{n} \sin^2\theta) \sqrt{ 1 - \mathfrak{m} \sin^2\theta} } \;.
\ee
When the parameter is bigger than 1, one can use
\be
\Pi(\varphi | \mathfrak{n}, \mathfrak{m}) = \frac1{\sqrt\mathfrak{m}} \, \Pi\bigl( \theta \big| \mathfrak{m}^{-1} \mathfrak{n} , \mathfrak{m}^{-1} \bigr) \qquad\text{with}\qquad \sin\theta = \mathfrak{m}^{1/2} \sin\varphi \;.
\ee
An identity for complete elliptic integrals of the third kind is:
\be
\label{sum of Pi's}
(\mathfrak{m} - \alpha^2) \, \Pi(\alpha^2, \mathfrak{m}) + (\mathfrak{m} - \beta^2) \, \Pi (\beta^2, \mathfrak{m}) = \mathfrak{m} \, K(\mathfrak{m}) \qquad\text{if}\quad (1-\alpha^2)(1-\beta^2) = 1-\mathfrak{m} \;.
\ee

\subsubsection{Series expansions}
\label{app: series expansions}

For $\mathfrak{m} \to 1$, $K(\mathfrak{m})$ diverges. To derive the asymptotic behavior we can proceed as follows. Write
\be
K(1-p) = \int_0^{\frac\pi2} \! \frac{d\theta}{ \sqrt{ \sin^2\theta + p \cos^2\theta}} = \int_0^1 \! \frac{ dt}{ \sqrt{ (1-t^2) [ (1-p) t^2 + p ] }} \;.
\ee
In the first equality we took the standard definition but mapped $\theta \to \frac\pi2 - \theta$; in the second equality we changed $\sin\theta = t$. Now the divergence as $p\to 0$ is at $t=0$. We decompose
\be
\label{expansion of K}
K(1-p) = \int_0^1 \! \frac{ dt}{ \sqrt{ (1-p) t^2 + p }} + \int_0^1 \biggl[ \frac1{\sqrt{1-t^2}} - 1 \biggr] \frac{ dt}{ \sqrt{ (1-p) t^2 + p }} \;.
\ee
The first integral can be computed exactly, and we can extract the divergent and finite parts as $p\to0$. The second integral is finite at $p=0$. We obtain:
\be
\label{eq:K-1-p}
K(1-p) = \frac{ \operatorname{arcsinh} \sqrt{ \frac1p - 1}}{\sqrt{1-p}} + \Bigl( \log 2 + \mathcal{O}(p\log p) \Bigr) = \frac12 \log \frac{16}{p} + \mathcal{O}(p \log p) \;.
\ee
Similarly, we find:
\begin{equation}
\label{eq:E-1-p}
E(1-p) = 1 + \frac p4 \Bigl( \log \frac{16}p - 1 \Bigr) + \mathcal{O}(p^2\log p).
\end{equation}

Computing the subleading terms is more complicated, due to the fact that we are not dealing with a standard Taylor expansion. Indeed, consider a series
\be
f(p) = a_0 + \sum\nolimits_{n=1}^\infty \bigl( a_n + b_n \log p \bigr) \, p^n \;.
\ee
The finite part is $\lim_{p\to0} f(p) = a_0$. In order to extract $a_1$ and $b_1$, consider the derivative
\be
f'(p) = \sum\nolimits_{n=1}^\infty \Bigl( n\, a_n + b_n + n \, b_n \log p \Bigr) \, p^{n-1} = (a_1 + b_1) + b_1 \log p + \mathcal{O}(p \log p) \;.
\ee
We see that we cannot simply take $\lim_{p\to 0} f'(p) = \infty$, but rather we should compute the asymptotic expansion of $f'(p)$ around $p=0$, up to vanishing terms $\mathcal{O}(p\log p)$. If we take as $f(p)$ the second integral in (\ref{expansion of K}), we find:
\begin{equation}\begin{aligned}
f'(p) &= - \int_0^1 \biggl[ \frac1{\sqrt{1-t^2}} - 1 \biggr] \frac{1-t^2}{ 2 \bigl[ (1-p)t^2 + p \bigr]^\frac32} \\
&= \int_0^1 \frac{-t^2/4}{ \bigl[ (1-p)t^2 + p \bigr]^\frac32 } + \int_0^1 \biggl[ - \biggl( \frac1{\sqrt{1-t^2}} - 1 \biggr) \frac{1-t^2}2 + \frac{t^2}4\biggr] \frac1{ \bigl[ (1-p)t^2 + p \bigr]^\frac32 } \;.
\end{aligned}\end{equation}
The first integral can be computed exactly and then expanded in $p$: it produces the $\log p$ divergent term. The second integral can be computed at $p=0$. We get
\be
f'(p) = \frac18 \Bigl[ 1 + \log p \Bigr] + \mathcal{O}(p\log p) \qquad\Rightarrow\qquad a_1 = 0 \;,\quad b_1 = \frac18 \;.
\ee
This gives us the first correction:
\be
K(1-p) = \frac12 \log \frac{16}p + \frac p8 \biggl( -2 + \log \frac{16}p \biggr) + \mathcal{O}(p^2\log p) \;.
\ee

We can apply the same technique to $F$ by writing
\be
\label{expansion of F}
F\Bigl( \arcsin\sqrt{1-x^2} \Bigm| 1-p \Bigr) = \int_{\arcsin x}^{\frac\pi2} \frac{d\theta}{ \sqrt{\sin^2\theta + p \cos^2\theta}} = \int_x^1 \! \frac{ dt}{ \sqrt{ (1-t^2) [ (1-p) t^2 + p ] }} \;.
\ee
We perform the same decomposition as above. The first integral gives
\be
\int_x^1 \frac{dt}{\sqrt{(1-p)t^2 + p}} = \frac{\operatorname{arcsinh} \sqrt{ \frac1p -1} - \operatorname{arcsinh} \Bigl( x \sqrt{ \frac1p -1} \, \Bigr) }{ \sqrt{1-p}} \;.
\ee
The asymptotic behavior of the second arcsinh depends on the behavior of its argument. The second integral is finite at $p=0$ and/or $x=0$:
\be
\int_x^1 \biggl[ \frac1{\sqrt{1-t^2}} - 1 \biggr] \frac{ dt}{ \sqrt{ (1-p) t^2 + p }} = \log \Bigl( 1 + \sqrt{1-x^2} \Bigr) + \mathcal{O}(p\log p) \;.
\ee
If we assume that $p\to 0$ and $p/x^2 \to 0$ (for instance, by keeping $x$ fixed), the expansion of $F$ becomes very simple, because we can perform a Taylor expansion of the integrand in (\ref{expansion of F}) in powers of $p$ and integrate term by term. We obtain:
\be
F\Bigl( \arcsin\sqrt{1-x^2} \Bigm| 1-p \Bigr) = \log \frac{ 1 + \sqrt{1{-}x^2} }{x} + \frac p4 \biggl( \log \frac{ 1 + \sqrt{1{-}x^2} }{x} - \frac{\sqrt{1{-}x^2}}{x^2} \biggr) + \mathcal{O}(p^2) .
\ee
Notice that we cannot substitute $x=0$ in this formula and connect with the asymptotic expansion of $K(1-p)$, because $x=0$ would violate our assumption.
In the case that $x\to 0$ (still assuming that $p/x^2 \to 0$), we find
\be
\label{eq:F-expasion}
F\Bigl( \arcsin\sqrt{1-x^2} \Bigm| 1-p \Bigr) = \log \frac2x - \frac{x^2}4 + \biggl( -\frac14 + \mathcal{O}(x^2\log x) \biggr) \frac{p}{x^2} + \mathcal{O} \Bigl(x^4, \frac{p^2}{x^4} \Bigr) \;.
\ee

The function $\Pi$ has more parameters and so, once again, the asymptotic expansion depends on the relative behavior of those parameters. Consider $\Pi(1-\mathfrak{m}, 1-p)$ in the limit $p\to 0$, $\mathfrak{m} \to 0$. We assume that $p/\mathfrak{m} \to 0$, which gives the following asymptotic expansion:
\be
\label{eq:Pi-exp}
\Pi(1-\mathfrak{m}, 1-p) = \frac1{2\mathfrak{m}} \biggl[ \log \frac{4\mathfrak{m}}p + \mathcal{O} \Bigl( \frac{p}{\mathfrak{m}} \log \frac{p}{\mathfrak{m}} \Bigr) + \mathcal{O}(\mathfrak{m} \log \mathfrak{m}) \biggr] \;.
\ee
Notice that this asymptotic expansion, even if pushed, cannot fix the finite terms because $\frac{1}{\mathfrak{m}} \bigl( \frac {p}{\mathfrak{m}} \bigr)^\#$ could go to zero or diverge. For the incomplete integral, we are interested in
\be
\label{eq:incomp-Pi-exp}
\Pi \biggl( \arcsin \sqrt{ \frac{1-\mathfrak{m}}{1-p}} \,\bigg|\, 1-\mathfrak{m}, 1-p \biggr) = \frac{1}{2\mathfrak{m}} \biggl[ \log 2 + \mathcal{O}\Bigl( \frac {p}{\mathfrak{m}} \Bigr) + \mathcal{O}(\mathfrak{m} \log \mathfrak{m}) \biggr] \;.
\ee
This expansion is computed assuming that $p/\mathfrak{m} \to 0$, although it could do so very slowly and thus this expansion cannot fix the finite terms.

The function $E$ is defined as 
\be
\label{expansion of E}
E\Bigl( \arcsin\sqrt{1-x^2} \Bigm| 1-p \Bigr) = \int_{\arcsin x}^{\frac\pi2} \! \sqrt{\sin^2\theta + p \cos^2\theta} \; d\theta = \int_x^1 \! \sqrt{ \frac{ (1-p) t^2 + p }{ 1-t^2} } \; dt \;.
\ee
The function is finite at $p=x=0$ and in particular $E\bigl( \frac\pi 2 \big| 1 \bigr) \equiv E(1) = 1$. To compute the first corrections, we assume that $p\to0$ and $p/x^2 \to 0$, and expand in $p$:
\be
\frac d{dp} \, E\big( \arcsin\sqrt{1-x^2} \big| 1-p \bigr) = \frac{ F \bigl( \arcsin \sqrt{1-x^2} \big| 1-p \bigr) - E\bigl( \arcsin\sqrt{1-x^2} \big| 1-p \bigr) }{2(1-p)} \;.
\ee
The derivative has to be evaluated at $p=0$ and expanded in $x$ (the latter is compatible with our previous analysis):
\begin{equation}\begin{aligned}
F\bigl( \arcsin \sqrt{1-x^2} \big| 1 \bigr) &= \operatorname{arctanh}\sqrt{1-x^2} &&= \log \frac2x - \frac{x^2}4 - \frac{3x^4}{32} + \mathcal{O}(x^6) \;, \\
E\bigl( \arcsin \sqrt{1-x^2} \big| 1 \bigr) &= \sqrt{1-x^2} &&= 1 - \frac{x^2}2 - \frac{x^4}8 + \mathcal{O}(x^6) \;.
\end{aligned}\end{equation}
We obtain:
\be
\label{eq:incomp-E-exp}
E = 1 - \frac{x^2}2 - \frac{x^4}8 + \biggl[ \frac{x^2}2 \biggl( \log \frac2x - 1 \biggr) + \frac{x^4}8 + \mathcal{O}(x^6) \biggr] \frac p{x^2} + \mathcal{O}\Bigl( x^6 \,,\, \frac{p^2}{x^4} \Bigr) \;.
\ee
A useful application of this formula is to derive
\be
\label{eq:incomp-E-parms}
E\Bigl( \arcsin \bigl( \dn ( b | \mathfrak{m} ) \bigr) \Bigm| q^2 \Bigr) = 1 - \frac{\mathfrak{m}}{2} - \frac{\mathfrak{m}^2}8 + \mathcal{O}(\mathfrak{m}^3,\, \nu^2 \, \mathfrak{m} \log \mathfrak{m}) \;. 
\ee
This is used in section~\ref{sec: high-density limit}, where the parameter $\nu$ is defined.

\subsection{Some useful integrals}
\label{sec: useful integrals}

We define, following the notation of \cite{Thies:2005wv}:
\begin{equation}\begin{aligned}
\eta &= \frac1{\sn^2 b} -1 - \mathfrak{m} \;,\qquad&
\chi &= \sqrt{1+\eta} = \sqrt{\frac1{\sn^2 b} - \mathfrak{m}} = \frac{\dn b}{\sn b} \;, \\
q &= \sqrt{(1-\mathfrak{m}) \, n} = \frac{\sqrt{1-\mathfrak{m}}}{\dn b} \;,\qquad&
n &= \frac{1+\mathfrak{m}+\eta}{1 + \eta} = \frac1{1 - \mathfrak{m} \sn^2 b} = \frac1{\dn^2 b} \;.
\end{aligned}\end{equation}
We assume $0 < \mathfrak{m} < 1$ and use $0 < b \leq K(\mathfrak{m})$ (although we do not need the latter restriction). Then $\eta > - \mathfrak{m}$, $\chi > 0$, $n \geq 1$ and it has a maximum in $b$ for every $\mathfrak{m}$, $0<q\leq 1$. Besides $\sn^2 b \in [0,1]$. We also define
\be
p_+(\varepsilon) = \arcsin \sqrt{ \frac1n \, \frac{ \varepsilon - 1 - \mathfrak{m}}{\varepsilon - 1} } \;,\qquad\quad p_-(\varepsilon) = \arcsin \sqrt{ \frac{ 1 - \varepsilon}{(1-\mathfrak{m})(1+\mathfrak{m} - \varepsilon)} } \;.
\ee
These are the variables that we will use in the primitives below. Finally
\be
\check p = \arcsin \frac1{\sqrt{n}} = \arcsin(\dn b) \;,\quad q^2 = \frac{(1-\mathfrak{m})(1+\mathfrak{m} + \eta)}{1+\eta} = \frac{1-\mathfrak{m}}{1-\mathfrak{m} \sn^2 b} = \frac{1-\mathfrak{m}}{\dn^2 b}
\ee
will be useful. In the integrals that follow, $b$ never appears explicitly: they should be thought of in terms of $\mathfrak{m}, \eta$ with $0<\mathfrak{m}<1$ and $\eta > -\mathfrak{m}$.

The first integral is $\int \bigl[ ( \varepsilon +\eta ) (\varepsilon - \mathfrak{m}) (\varepsilon -1) (\varepsilon - 1 - \mathfrak{m}) \big]^{-\frac12} d\varepsilon$. The integrand is real positive for $\varepsilon < -\eta$, for $\mathfrak{m} < \varepsilon < 1$, and for $1+\mathfrak{m} < \varepsilon$. The definite integral is convergent both at the singularities as well as at infinity. For $\varepsilon > 1 + \mathfrak{m}$ we find:%
\footnote{Note that $\operatorname{arccot}\sqrt{w} = \arcsin \sqrt{ \frac1{1+w}}$ and $\arctan\sqrt{w} = \arcsin \sqrt{ \frac{w}{w+1}}$ for $w>0$. Besides:
\be
\arcsin\sqrt{w} = \frac\pi2 - i \operatorname{arccosh}\sqrt{w} = \frac\pi2 - i \operatorname{arcsinh}\sqrt{w-1} \;.
\ee
These are useful if one has to compute the arcsin of a number bigger than 1.}
\begin{equation}\begin{aligned}
\label{inverse quadratic integral}
\int \! \frac{d\varepsilon}{\sqrt{ ( \varepsilon +\eta ) (\varepsilon - \mathfrak{m}) (\varepsilon -1) (\varepsilon - 1 - \mathfrak{m}) }} &= \frac2\chi \, F \bigl( p_+ \,\big|\, q^2 \bigr) \;, \\
\int_{1+\mathfrak{m}}^\infty \frac{d\varepsilon}{\sqrt{ ( \varepsilon +\eta ) (\varepsilon - \mathfrak{m}) (\varepsilon -1) (\varepsilon - 1 - \mathfrak{m}) }} &= \frac2\chi \, F \bigl( \check p \,\big|\, q^2 \bigr) \;.
\end{aligned}\end{equation}
For $\mathfrak{m} < \varepsilon < 1$ we find:
\begin{align}
\int \! \frac{d\varepsilon}{\sqrt{ ( \varepsilon +\eta ) (\varepsilon - \mathfrak{m}) (\varepsilon -1) (\varepsilon - 1 - \mathfrak{m}) }} &= - \frac2\chi \, F \bigl( p_- \,\big|\, q^2 \bigr)\,, \nn\\
\int_\mathfrak{m}^1 \frac{d\varepsilon}{\sqrt{ ( \varepsilon +\eta ) (\varepsilon - \mathfrak{m}) (\varepsilon -1) (\varepsilon - 1 - \mathfrak{m}) }} &= \frac2\chi \, K(q^2) \;.
\end{align}

The second integral is $\int \bigl[ ( \varepsilon +\eta ) / (\varepsilon - \mathfrak{m}) (\varepsilon -1) (\varepsilon - 1 - \mathfrak{m}) \big]^{\frac12} d\varepsilon$. The integrand is real positive in the same regions as before.
For $1+\mathfrak{m} < \varepsilon$ the indefinite integral is
\be
\int \! \frac{ \sqrt{\varepsilon +\eta} \;\; d\varepsilon}{\sqrt{ (\varepsilon - \mathfrak{m}) (\varepsilon -1) (\varepsilon - 1 - \mathfrak{m}) }} = \frac{2\mathfrak{m}}\chi \, \Pi \bigl( p_+ \,\big|\, n, q^2 \bigr) + 2\chi \, F \bigl( p_+ \,\big|\, q^2 \bigr) \;.
\ee
Notice that the second parameter $n>1$, however the range of the variable $p_+$ is such that the singularity in the integrand is hit only for $\varepsilon \to \infty$. In the limit, one can use the following asymptotic expansion, quoted in eqn.~(24) of \cite{Thies:2005wv}:
\begin{align}
\label{logarithmic divergence from Pi}
\Pi\biggl( \arcsin \biggl(\, \frac1{\sqrt{\mathfrak{n}}} - \epsilon \biggr) \bigg| \mathfrak{n}, \mathfrak{m} \biggr)
&= \frac12 \sqrt{ \frac{\mathfrak{n}}{(\mathfrak{n}-1)(\mathfrak{n}-\mathfrak{m})}} \, \log \frac{ 2(\mathfrak{n} - \mathfrak{m})(\mathfrak{n}-1)}{ \sqrt{\mathfrak{n}}\, (\mathfrak{n}^2 - \mathfrak{m}) \, \epsilon} \\
&\quad {} + F\Bigl( \arcsin \bigl(\mathfrak{n}^{-\frac12} \bigr) \Big| \mathfrak{m} \Bigr) - \Pi \Bigl( \arcsin \bigl( \mathfrak{n}^{-\frac12} \bigr) \Big| \frac{\mathfrak{m}}{\mathfrak{n}} , \mathfrak{m} \Bigr) + \mathcal{O}(\epsilon) \;. \nn
\end{align}
Since in our case $\epsilon = \mathfrak{m} / 2\sqrt{n}\, \varepsilon_\text{max}$, we find:
\begin{align}\label{eq:UsefulIntD}
& \int_{1+\mathfrak{m}}^{\varepsilon_\text{max}} \!\!\! \frac{ \sqrt{\varepsilon +\eta} \;\; d\varepsilon}{\sqrt{ (\varepsilon - \mathfrak{m}) (\varepsilon -1) (\varepsilon - 1 - \mathfrak{m}) }}  \\
&\hspace{2cm} = \log \biggl( \frac{4 \, \varepsilon_\text{max}}{ 2 + \eta} \biggr) + \frac{2}{\chi \sn^2(b)} \, F\bigl( \check p \,\big|\, q^2 \bigr) - \frac{2\mathfrak{m}}\chi \, \Pi \bigl( \check p \,\big|\, 1-\mathfrak{m} , q^2 \bigr) + \mathcal{O}\bigl( \varepsilon_\text{max}^{-1} \bigr) . \nn
\end{align}

For $\mathfrak{m} < \varepsilon < 1$ the definite integral is convergent:
\begin{align}
\int \! \frac{ \sqrt{\varepsilon +\eta} \;\; d\varepsilon}{\sqrt{ (\varepsilon - \mathfrak{m}) (\varepsilon -1) (\varepsilon - 1 - \mathfrak{m}) }} &= 
 \frac{2(\mathfrak{m} + \eta)}\chi\, \Pi \Bigl( \arcsin \sqrt{ \frac{1+\eta}{1-\mathfrak{m}} \, \frac{\varepsilon - \mathfrak{m}}{ \varepsilon + \eta} } \,\Big|\, \sn^2(b) \, q^2 , q^2  \Bigr)\,, \nn\\
\int_\mathfrak{m}^1 \frac{ \sqrt{\varepsilon +\eta} \;\; d\varepsilon}{\sqrt{ (\varepsilon - \mathfrak{m}) (\varepsilon -1) (\varepsilon - 1 - \mathfrak{m}) }} &= 
 \frac{2(\mathfrak{m} + \eta)}\chi\, \Pi \bigl( \sn^2(b) \, q^2, q^2 \bigr) \;. \label{eq:UsefulIntD2}
\end{align}
The prefactor can be written as $2(\mathfrak{m}+\eta)/\sqrt{1+\eta}$, while $\sn^2(b)\, q^2 = (1-\mathfrak{m})/(1+\eta)$. There is an alternative expression, still for $\mathfrak{m} < \varepsilon < 1$:
\begin{align}
& \int \! \frac{ \sqrt{\varepsilon +\eta} \;\; d\varepsilon}{\sqrt{ (\varepsilon - \mathfrak{m}) (\varepsilon -1) (\varepsilon - 1 - \mathfrak{m}) }} = \frac{2\mathfrak{m}}\chi \, \Pi \bigl( p_- \,\big|\, 1-\mathfrak{m} , q^2 \bigr) - 2 \, \frac{\eta + 1 + \mathfrak{m}}\chi \, F \bigl( p_- \,\big|\, q^2 \bigr)\,, \nn \\
& \int_\mathfrak{m}^1 \frac{ \sqrt{\varepsilon +\eta} \;\; d\varepsilon}{\sqrt{ (\varepsilon - \mathfrak{m}) (\varepsilon -1) (\varepsilon - 1 - \mathfrak{m}) }}
= \frac{2}{\chi \sn^2(b)} \, K(q^2) - \frac{2\mathfrak{m}}\chi \, \Pi \bigl( 1-\mathfrak{m}, q^2\bigr) \;.
\end{align}
The two expressions are equivalent, and in particular the following identity holds:
\be
\label{identity on Pi}
K(q^2) - \mathfrak{m} \sn^2(b) \, \Pi(1-\mathfrak{m}, q^2) = \cn^2(b) \; \Pi \biggl( \frac{\sn^2(b)}{\dn^2(b)} \, (1-\mathfrak{m}) \,,\, q^2 \biggr) \;.
\ee
This is a special case of (\ref{sum of Pi's}).
One has to be careful when taking the limit $b \to K(\mathfrak{m})$: the functions $K$ and $\Pi$ diverge; the RHS remains finite because $\cn^2(b) \to 0$; the LHS remains finite because the two divergences cancel out.

The third indefinite integral, for $1 + \mathfrak{m} < \varepsilon$, is:
\begin{align}
& \int\! \frac{ (\varepsilon - \mathfrak{m} - E/K) \; d\varepsilon }{ \sqrt{ (\varepsilon-\mathfrak{m}) (\varepsilon - 1) (\varepsilon - 1- \mathfrak{m}) }} = 2 \sqrt{ \frac{ (\varepsilon - \mathfrak{m}) (\varepsilon -1 -\mathfrak{m}) }{ \varepsilon -1} } \\
&\hspace{1cm} + 2 \, \Bigl( 1 - \tfrac EK \Bigr) \, F \Bigl( \arcsin \sqrt{ \frac{\varepsilon - 1 - \mathfrak{m}}{\varepsilon -1}} \,\Big|\, 1-\mathfrak{m} \Bigr) - 2 E \Bigl( \arcsin \sqrt{ \frac{\varepsilon - 1 - \mathfrak{m}}{\varepsilon -1}} \,\Big|\, 1-\mathfrak{m} \Bigr) \;. \nn
\end{align}
Notice that this integral does not depend on $\eta$ (and thus on $b$). The definite integral has a divergence as the square root of $\varepsilon_\text{max}$, while it vanishes at the lower end. We get $2\varepsilon_\text{max}^{1/2} + \mathcal{O}(\varepsilon_\text{max}^{-1/2})$ from the square root. The other terms give $2\bigl( K(1-\mathfrak{m}) - E(\mathfrak{m}) \, K(1-\mathfrak{m}) / K(\mathfrak{m}) - E(1-\mathfrak{m}) \bigr)$ in the limit $\varepsilon_\text{max} \to \infty$, and we can write
\be\label{eq:usefulD4}
\int_{1+\mathfrak{m}}^{\varepsilon_\text{max}} \!\!\! \frac{ (\varepsilon - \mathfrak{m} - E/K) \; d\varepsilon }{ \sqrt{ (\varepsilon-\mathfrak{m}) (\varepsilon - 1) (\varepsilon - 1- \mathfrak{m}) }}
= 2 \, \varepsilon_\text{max}^{1/2} - \frac\pi{K(\mathfrak{m})} + \mathcal{O} \bigl( \varepsilon_\text{max}^{-1/2} \bigr)
\ee
using Legendre's relation $EK' + E'K - KK' = \frac\pi2$.

The fourth indefinite integral, for $1+\mathfrak{m}<\varepsilon$, is:
\begin{align}
& \int \! \frac{ (\varepsilon - \mathfrak{m} - E/K) \, \sqrt{\varepsilon + \eta} \; d\varepsilon }{ \sqrt{ (\varepsilon - \mathfrak{m}) (\varepsilon - 1) (\varepsilon - 1 - \mathfrak{m}) }}
= \sqrt{ \frac{ (\varepsilon + \eta)(\varepsilon - \mathfrak{m})(\varepsilon - 1 - \mathfrak{m}) }{ \varepsilon-1} } \\
&\hspace{1.7cm} + \chi \, \Bigl( 2 - \mathfrak{m} - 2\tfrac EK \Bigr) \, F\bigl( p_+ \,\big|\, q^2 \bigr) - \chi \, E \bigl( p_+ \,\big|\, q^2 \bigr) + \frac{\mathfrak{m}}{\chi} \, \Bigl( 2 + \eta - 2\tfrac EK \Bigr) \, \Pi \bigl(p_+ \,\big|\, n, q^2 \bigr) \;. \nn
\end{align}
The definite integral has a divergence linear in $\varepsilon_\text{max}$. We get $\varepsilon_\text{max} + \frac{\eta - 2\mathfrak{m}}2 + \mathcal{O}(\varepsilon_\text{max}^{-1})$ from the square root, $F$ and $E$ remain finite, while $\Pi$ gives a logarithmic divergence as in (\ref{logarithmic divergence from Pi}). We obtain:
\begin{align} \label{eq:usedulD5}
& \int_{1+\mathfrak{m}}^{\varepsilon_\text{max}} \! \frac{ (\varepsilon - \mathfrak{m} - E/K) \, \sqrt{\varepsilon + \eta} \; d\varepsilon }{ \sqrt{ (\varepsilon - \mathfrak{m}) (\varepsilon - 1) (\varepsilon - 1 - \mathfrak{m}) }} = \varepsilon_\text{max} + \frac12 \Bigl( 2 + \eta - 2\tfrac EK \Bigr) \log\biggl( \frac{ 4 \varepsilon_\text{max} }{ 2+\eta} \biggr) + \biggl( \frac\eta2 - \mathfrak{m} \biggr) \nn \\
&\quad + \chi \, \Bigl( 2 - \mathfrak{m} - 2\tfrac EK \Bigr) \, F\bigl( \check p \,\big|\, q^2 \bigr) - \chi \, E \bigl( \check p \,\big|\, q^2 \bigr) \nn \\
&\quad + \frac{\mathfrak{m}}{\chi} \, \Bigl( 2 + \eta - 2\tfrac EK \Bigr) \Bigl( F \bigl( \check p \,\big|\, q^2 \bigr) - \Pi \bigl( \check p \,\big|\, 1-\mathfrak{m}, q^2 \bigr) \Bigr) + \mathcal{O}\bigl( \varepsilon_\text{max}^{-1} \bigr) \;.
\end{align}
For $\mathfrak{m} < \varepsilon < 1$ we have:
\begin{align}
& \int \! \frac{ \bigl\lvert \varepsilon - \mathfrak{m} - E/K \bigr\rvert \, \sqrt{\varepsilon + \eta} \; d\varepsilon }{ \sqrt{ (\varepsilon - \mathfrak{m}) (\varepsilon - 1) (\varepsilon - 1 - \mathfrak{m}) }}
= \sqrt{ \frac{ (\varepsilon + \eta) ( \varepsilon - \mathfrak{m}) (1- \varepsilon) }{ 1+ \mathfrak{m} - \varepsilon} } - \chi \, E \bigl( p_- \,\big|\, q^2 \bigr) \\
&\hspace{0.7cm} + \frac{\mathfrak{m}}{\chi} \, \Bigl( 2 + \eta - 2\tfrac EK \Bigr) \, \Bigl( F \bigl( p_- \,\big|\, q^2 \bigr) - \Pi \bigl( p_- \,\big|\, 1-\mathfrak{m}, q^2 \bigr) \Bigr)
+ \chi \, \Bigl( 2 - \mathfrak{m} - 2 \tfrac EK \Bigr) \, F \bigl( p_- \,\big|\, q^2 \bigr) \;. \nn
\end{align}
We included an absolute value in the integrand because its argument is negative on the lower band. The definite integral gives:
\begin{align}\label{eq:usedulD7}
& \int_\mathfrak{m}^1 \! \frac{ \bigl\lvert \varepsilon - \mathfrak{m} - E/K \bigr\rvert \, \sqrt{\varepsilon + \eta} \; d\varepsilon }{ \sqrt{ (\varepsilon - \mathfrak{m}) (\varepsilon - 1) (\varepsilon - 1 - \mathfrak{m}) }} \\
&\quad = \frac{\mathfrak{m}}{\chi} \, \Bigl( 2 + \eta - 2\tfrac EK \Bigr) \, \Bigl( \Pi \bigl( 1-\mathfrak{m}, q^2 \bigr) - K(q^2) \Bigr) + \chi \, E(q^2) - \chi \, \Bigl( 2 - \mathfrak{m} - 2 \tfrac EK \Bigr) \, K(q^2) \;. \nn
\end{align}

\section{Properties of the spectral problem}
\label{app:eigenfuncs}

Here we review some details of the spectrum of the operator $\mathcal{D}$ defined in \eqref{eq:DopeDefi}, following mostly \cite{Schnetz:2005ih}. Using the ansatz \eqref{eq:sigma-prof}, the ``potentials'' in (\ref{eqn for eigenfunctions 2nd order}) are
\be
\label{potentials u+-}
u_+(x) \,\equiv\, \sigma(x)^2 - \sigma'(x) = A^2 \biggl( 2 \mathfrak{m} \sn(Ax| \mathfrak{m})^2 + \frac1{\sn(b| \mathfrak{m})^2} - 1 - \mathfrak{m} \biggr)
\ee
and $u_-(x) \equiv \sigma^2 + \sigma' = u_+(x+b/A)$ which is just shifted. Then, the function $\phi_+(x) = \tilde\phi_+(Ax)$ solves \eqref{eqn for eigenfunctions 2nd order} if and only if $\tilde \phi_+(\xi)$ solves
\be
\label{Lame equation}
\biggl( \parfrac{^2}{\xi^2} + \varepsilon(\omega) - 2 \mathfrak{m} \sn(\xi| \mathfrak{m})^2 \biggr) \tilde\phi_+(\xi) = 0 \quad\text{with}\quad \varepsilon(\omega) = \frac{\omega^2}{A^2} + 1 + \mathfrak{m} - \frac1{\sn(b| \mathfrak{m})^2} \;.
\ee
This second-order equation is called Lam\'e equation. The function $\sigma(x)$ and the potentials $u_\pm(x)$ have periods $2K(\mathfrak{m})/A$ in $x$, while the potential in (\ref{Lame equation}) has period $2K(\mathfrak{m})$ in $\xi$. It follows, according to Bloch's theorem, that we look for solutions that are periodic up to a phase. Such solutions are known (see \textit{e.g.} \cite{Kusnezov:2000}). They can be parametrized by $\alpha$:
\be
\tilde\phi_+(\xi) = \mathcal{N} \; \frac{\theta_1\bigl( \frac{\pi(\xi+\alpha)}{2K(\mathfrak{m})} \big| q \bigr) }{ \theta_4 \bigl( \frac{\pi \xi}{2K(\mathfrak{m})} \big| q \bigr) } \; e^{-\xi \zn(\alpha| \mathfrak{m})}
\qquad\text{where}\qquad q = \operatorname{nome}(\mathfrak{m}) = e^{- \pi \, \frac{K(1-\mathfrak{m})}{K(\mathfrak{m})}} \;.
\ee
The elliptic theta functions are defined as $\theta_1(z,q) = 2 q^{1/4} \sum_{n=0}^\infty (-1)^n q^{n(n+1)} \sin(2n+1)z $ and $\theta_4(z,q) = 1 + 2 \sum_{n=0}^\infty (-1)^n q^{n^2} \cos 2nz$. $K(\mathfrak{m})$ is the elliptic integral of the first kind, and finally:
\be
\zn(\alpha|\mathfrak{m}) = \parfrac{}{\alpha} \log \theta_4 \bigl( \tfrac{\pi\alpha}{2K(\mathfrak{m})} \big| q \bigr) \;.
\ee
These solutions give
\be
\varepsilon(\alpha) = 1 + \mathfrak{m} \cn(\alpha| \mathfrak{m})^2 = \mathfrak{m} + \dn(\alpha| \mathfrak{m})^2 \;.
\ee
Solutions periodic up to a phase are obtained by imposing $\zn(\alpha| \mathfrak{m}) \in i\mathbb{R}$. There are two bands:
\be
\begin{array}{llllclll}
\text{lower:} & \alpha = K + iK' - i \gamma \;, & \gamma & \text{from } 0& \to & \varepsilon = \mathfrak{m} & \tilde p = 0 & \frac\omega A= \sqrt{ \frac1{\sn^2 b} - 1} \\
&&& \text{to } K' & \to & \varepsilon = 1 & \tilde p = \frac{\pi}{2K} & \frac\omega A = \sqrt{ \frac1{\sn^2 b} - \mathfrak{m}} \\[1em]
\text{upper:} & \alpha = i\gamma \;, & \gamma & \text{from } 0 & \to & \varepsilon = 1+ \mathfrak{m} & \tilde p = \frac{\pi}{2K} & \frac\omega A = \frac1{\sn b} \\[0.5em]
&&& \text{to } K' & \to & \varepsilon = +\infty & \tilde p = +\infty & \frac\omega A = + \infty
\end{array}
\ee
Here $K \equiv K(\mathfrak{m})$ and $K' \equiv K(1-\mathfrak{m})$. The energy can be written as
\be
\omega = A \sqrt{ \dn(\alpha|\mathfrak{m})^2 + \sn(b|\mathfrak{m})^{-2} -1} \;.
\ee
The rescaled Bloch momentum is:%
\footnote{The rescaled Bloch momentum is defined by $\tilde\phi_+(\xi+2K) = e^{-i \tilde p \, 2K} \, \tilde\phi_+(\xi)$ because the period is $2K$. The rescaled momentum is defined modulo $\pi/K$, and we can take it inside the Brillouin zone $\bigl( - \frac\pi{2K}, \frac\pi{2K} \bigr]$.}
\be
\tilde p(\alpha) = - i \zn(\alpha| \mathfrak{m}) + \frac\pi{2K( \mathfrak{m})} \;.
\ee
Expressed in the $\omega$ variable, it leads to \eqref{eq:disp-analytic}. Solutions with $\alpha \to -\alpha$ give the same $\varepsilon$ (and complete the two-dimensional vector space of solutions) and opposite $\tilde p$. The physical momentum is $p = A \, \tilde p$. Using the relation $(\varepsilon - \mathfrak{m})(\varepsilon -1)(\varepsilon - 1 - \mathfrak{m}) = - \mathfrak{m}^2 \sn^2(\alpha) \cn^2(\alpha) \dn^2(\alpha)$ one proves \eqref{density of states}.

The argument of the square root in \eqref{density of states} is always positive (for $\varepsilon$ within the bands), whereas the argument of the absolute value is positive on the upper band and negative on the lower band. The bands are then those given in \eqref{eq:bands}.
From the density of states \eqref{density of states} at large energies, one finds
\be
\label{eq:eps-p-large}
\varepsilon = \frac{p^2}{A^2} + 2 \, \biggl( 1 - \frac{E(\mathfrak{m})}{K(\mathfrak{m})} \biggr) + \mathcal{O}\bigl( p^{-2} \bigr)
\ee
as well as $\omega = p + \mathcal{O}(p^{-1})$.
The dispersion relation $\tilde\omega(\tilde p)$ is hard to write in closed form (here $\tilde\omega$ is the rescaled energy, so that $\omega = A \, \tilde\omega$). However $\tilde p(\tilde\omega)$ can be found and analytically continued to \eqref{eq:disp-analytic}.

The positive constant $\mathcal{N}$ is chosen in such a way that $\bigl\lvert \tilde\phi_+(\xi) \bigr\rvert{}^2$ has average $1/2$:
\be
\frac1{2K} \int_0^{2K} \! d\xi \, \bigl\lvert \tilde\phi_+(\xi) \bigr\rvert^2 = \frac{A}{2K} \int_0^{\frac{2K}A} \! dx \, \bigl\lvert \phi_+(x) \bigr\rvert^2 = \frac12 \;.
\ee
One finds
\be
\mathcal{N}^{\,2} = \frac{K(\mathfrak{m}) \, \sqrt{\mathfrak{m} (1-\mathfrak{m})} }{ \pi \, \theta_4^2\bigl( \frac{\pi\alpha}{2K(\mathfrak{m})} \big| q \bigr) \, \Bigl\lvert \dn(\alpha| \mathfrak{m})^2 - \frac{E(\mathfrak{m})}{K(\mathfrak{m})} \Bigr\rvert } \;.
\ee
The absolute value at denominator is necessary because its argument is negative on the lower band (while it is positive on the upper band). Following the discussion between (37)--(46) in \cite{Schnetz:2005ih}, $\phi_-$ is fixed by $\phi_-(x) = - \frac1\omega \bigl( \parfrac{}{x} + \sigma(x) \bigr) \phi_+(x)$. On the other hand, the equation for $\phi_-(x)$ in (\ref{eqn for eigenfunctions 2nd order}) is just a shift of the one for $\phi_+(x)$, therefore $\tilde\phi_-(\xi) = e^{i\varphi} \, \tilde\phi_+(\xi+b)$ for some phase $e^{i\varphi}$ that could be computed (but is not necessary). It follows that
\be
\langle \psi^\dag_\omega \psi_\omega \rangle = \bigl\langle |\tilde\phi_+|^2 + |\tilde\phi_-|^2 \bigr\rangle = 1 \qquad\text{for each $\omega$} \;,
\ee
where $\langle \;\rangle$ is the spatial average. This confirms that the eigenfunctions of the Dirac operator are normalized. Using the definitions \eqref{eq:sigma-tilde} and \eqref{eq:stu} we can write the spectral quantities
\be
|\tilde\phi_+|^2 = \frac12 \, \frac{\varepsilon - \mathfrak{m} - \dn^2(\xi|\mathfrak{m}) }{ \varepsilon - \mathfrak{m} - E/K} \;,\qquad
\psi^\dag \psi = \frac{(\omega/A)^2 + \bigl( \tilde\sigma(\xi)^2 - 3s + 1 + \mathfrak{m} \bigr)/2 }{ (\omega/A)^2 -s + u} \;.
\ee
We used
\be
\varepsilon(\omega) = \omega^2 / A^2 - \eta \;,\qquad \eta = \sn(b| \mathfrak{m})^{-2} -1 - \mathfrak{m} \;.
\ee
Then, to compute $\bar\psi\psi$ one uses $\bar\psi\psi = - \phi_+^* \phi_- - \phi_-^* \phi_+ = \frac1\omega \bigl( \partial_x + 2\sigma(x) \bigr) |\phi_+(x)|^2$ and obtains \eqref{psi-bar psi}.

\section{Homogeneous limit of inhomogeneous configurations}
\label{app:homLimit}

In this appendix we rederive the (density of) grand canonical potential for homogeneous configurations $\sigma(x) = \sigma_0$ by starting from the inhomogeneous ansatz and taking the homogeneous limit. 
We will see that this agrees with the more direct approach of section~\ref{sec:hom-instability}. 
We can consider homogeneous ansatze using the previous formalism by setting
\be
\mathfrak{m} =1 \;,\qquad\qquad A = \sigma_0 \tanh(b) \;,
\ee
for any $b$. Note that:
\be
\sn(b|1) = \tanh(b) \;,\qquad \dn(b|1) = \cn(b|1) = \frac1{\cosh(b)} \;,\qquad \eta+2 = \tanh^{-2}(b) \;.
\ee
In this case the lower band disappears, while the upper band has $\varepsilon \in [2, \infty)$ and $\omega \in [\sigma_0, \infty)$ with density of states and dispersion relation:
\be
\frac{d\tilde p}{d\varepsilon} = \frac1{2\sqrt{\varepsilon -2}} \;,\qquad\qquad \omega = A \sqrt{ \varepsilon + \eta} \;,\qquad\qquad \omega^2 = p^2 + \sigma_0^2 \;.
\ee
This is just a relativistic fermion with mass $\sigma_0$.

To compute the grand canonical potential we use (\ref{grand potential integrals}). For $h \leq \sigma_0$ it is essentially the computation we already did, because $h$ is below the band:
\begin{align}
\label{integral for mu < sigma0}
E_1 &= - \frac{A^2}{2\pi} \int_2^{\varepsilon_\text{max}} \sqrt{ \frac{ \varepsilon+\eta}{ \varepsilon -2} } \, d\varepsilon = - \frac{A^2}{2\pi} \biggl[ \sqrt{ (\varepsilon_\text{max} + \eta)(\varepsilon_\text{max} - 2)} + (2+\eta) \operatorname{arcsinh} \sqrt{ \frac{ \varepsilon_\text{max} - 2}{ 2 + \eta} } \biggr] \nn\\
&= - \frac{M^2}{2\pi} - \frac{A^2 (2+\eta) }{2\pi} \biggl( \frac12 + \log \frac{ 2M}{ A \sqrt{2+\eta}} \biggr) + \mathcal{O}(M^{-2})
\end{align}
and
\be
E_2 = \frac{A^2}{2\lambda} \, \tanh^{-2}(b) = \frac1{2\pi} \log \Bigl( \frac{2M}{\Lambda} \Bigr) \, \sigma_0^2 \;,
\ee
while $E_3=0$ and $\varrho = 0$. This gives
\be
\Phi = \frac{\sigma_0^2}{2\pi} \biggl( \log \frac{\sigma_0}{\Lambda} - \frac12 \biggr)+ \frac{\Lambda^2}{4\pi} \qquad\text{for } h \leq \sigma_0 \;.
\ee
For $h>\sigma_0$, the chemical potential is inside the band. We can reorganize $E_1 + E_3$ in such a way to have an integral over the whole band, plus an integral only up to $h$ weighted by $y$. The latter contribution is as on the first line of (\ref{integral for mu < sigma0}), but with upper limit $\varepsilon_\text{F} = \frac{h^2}{A^2} - \eta$. Notice that $\varepsilon_\text{F} + \eta = h^2 / \sigma_0^2 \tanh^2(b)$ and $\varepsilon_\text{F} - 2 = (h^2 / \sigma_0^2 -1)/ \tanh^2(b)$. We also get a contribution from the density $\varrho$:%
\footnote{Indeed the fermion current operator does not contain derivatives (as opposed to a scalar current operator), thus there can be a density of charge even with a constant condensate.}
the integral on the whole band gives a pure divergence linear in $M$, but when $h$ cuts inside the band we get minus the integral up to $\varepsilon_F$. In total, the extra contribution is
\begin{align}
E_4 &= \int_0^h \! d\omega \, (\omega - h) = \frac{A}{2\pi} \int_2^{\varepsilon_\text{F}} \! \frac{ A \sqrt{\varepsilon + \eta} - h }{\sqrt{\varepsilon - 2}} \, d\varepsilon \\
&= \frac{A^2}{2\pi} \biggl[ \sqrt{(\varepsilon_\text{F} + \eta)(\varepsilon_\text{F} - 2)} + (2+\eta) {\rm arcsinh} \sqrt{ \frac{\varepsilon_\text{F} - 2}{ 2+\eta}} \, \biggr] - \frac{Ah}{2\pi} \, 2 \sqrt{ \varepsilon_\text{F}-2} \nn \\
&= \frac{\sigma_0^2}{2\pi} \biggl[ \log \biggl( \frac{h}{\sigma_0} + \sqrt{ \frac{h^2}{\sigma_0^2} - 1} \,\biggr) - \frac{h}{\sigma_0} \sqrt{ \frac{h^2}{\sigma_0^2} - 1} \,\biggr] \;, \nn
\end{align}
weighted by $y$. Summing the two contributions we obtain:
\be
\Phi = \frac{\sigma_0^2}{2\pi} \, \Biggl\{ \log \frac{\sigma_0}{\Lambda} - \frac12 + \Theta(h-\sigma_0) \, y \, \Biggl[  \log \biggl( \frac{h}{\sigma_0} + \sqrt{ \frac{h^2}{\sigma_0^2} - 1 } \, \biggr) - \frac{h}{\sigma_0} \sqrt{ \frac{h^2}{\sigma_0^2} - 1 } \,\Biggr] \Biggr\}  + \frac{\Lambda^2}{4\pi}\;.
\ee
This matches \eqref{eq:Phi-hom}.
Note that such a $\Phi$ is $\mathcal{C}^1$: only the second derivative is discontinuous. 

\section{Integral equations}
\label{app:IntEq}

An important class of singular integral equations which admit an analytic solution is
\be
\text{P} \int_{B_1}^{B_2}\! \! K(x-y) \epsilon(y) dy = f(x)\,.
\label{eq:AELSM1}
\ee
Here $x,y\in \mathbb{R}$, P is the principal part of the integral, $f\in L^p$ on the interval $[B_1,B_2]$, with $p>4/3$, and $K$ is the Cauchy kernel 
\be
K(x) = \frac 1x\,.
\ee
The unique solution in $L^p$, with $p>2$, reads (see e.g. chapter IV of \cite{tricomi1985})
\be
\epsilon_\text{B}(x) =  \frac{\sqrt{(B_2-x)(x-B_1)}}{\pi^2} \int_{B_1}^{B_2}\frac{dt}{t-x}\frac{f(t)}{\sqrt{(t-B_1)(B_2-t)}}\,, 
\label{eq:1xGen2}
\ee
under the condition that
\be
 \int_{B_1}^{B_2} \! dt  \frac{f(t)}{\sqrt{(t-B_1)(B_2-t)}}  = 0\,.
\label{eq:1xGen3}
\ee
If \eqref{eq:1xGen3} is relaxed, there are no solutions in $L^{p>2}$.
More general solutions are allowed on $L^p$, with $p<4/3$, and reads
\be
\epsilon(x) = \epsilon_\text{B}(x) + \frac{c}{\sqrt{(B_2-x)(x-B_1)}}\,, 
\label{eq:1xGen2a}
\ee
where $c$ is an arbitrary constant. Note that the second term is the solution of \eqref{eq:AELSM1} with $f=0$ and is $L^p$, with $p<2$.

Singular integral equations of the form
\be
\text{P} \int_{-B}^{B} \frac{\epsilon(\theta')}{\sinh(\theta-\theta')}  d\theta' = f(\theta)\,,
\label{eq:sinh1}
\ee
where
\be\label{eq:sinh1a}
f(\theta) = 2\pi \sinh \theta\,,
\ee
can be recast in the form \eqref{eq:AELSM1} by a simple change of variables. 
Let $x=\exp(2\theta)$, so that
\be
\sinh(\theta-\theta') = \frac 12 \frac{x- x'}{\sqrt{x x'}}\,,
\ee
and \eqref{eq:sinh1} turns into
\be
\text{P} \int_{a_1}^{a_2} \frac{\widetilde \epsilon(y)}{x-y}  dy = f(x)\,,
\ee
where
\be\label{eq:sinh2a}
 \widetilde \epsilon (x) = \frac{\epsilon(\theta(x))}{\sqrt{x}}\,, \qquad f(x) = \pi \Big(1-\frac 1 x\Big)\,, \qquad a_1 = e^{-2B}, \qquad a_2 = e^{2B}\,.
\ee
A bounded solution is given by:%
\footnote{Note that the condition $a_1 = 1/a_2$ is necessary to solve \eqref{eq:1xGen3}.} 
\be
    g(x) = \pi\,, \qquad \widetilde \epsilon(x) =  \frac{ \sqrt{(x-a_1)(a_2-x)}}{x} \,.
\ee
In terms of the original variables, this reproduces \eqref{eq:sinh5} in the main text.

\bibliographystyle{JHEP}
\bibliography{condensates_long.bib}

@article{Abbott:2020qnl,
    author = "Abbott, Michael C. and Bajnok, Zolt{\'a}n and Balog, J{\'a}nos and Heged{\'{u}}s, {\'A}rp{\'a}d and Sadeghian, Saeedeh",
    title = "{Resurgence in the O(4) sigma model}",
    eprint = "2011.12254",
    archivePrefix = "arXiv",
    primaryClass = "hep-th",
    doi = "10.1007/JHEP05(2021)253",
    journal = "JHEP",
    volume = "05",
    pages = "253",
    year = "2021"
}

@article{Alberte:2020eil,
    author = "Alberte, Lasma and Nicolis, Alberto",
    title = "{Spontaneously broken boosts and the Goldstone continuum}",
    eprint = "2001.06024",
    archivePrefix = "arXiv",
    primaryClass = "hep-th",
    reportNumber = "Imperial/TP/2020/LA/01",
    doi = "10.1007/JHEP07(2020)076",
    journal = "JHEP",
    volume = "07",
    pages = "076",
    year = "2020"
}

@article{Aniceto:2018bis,
    author = "Aniceto, In\^es and Ba\c{s}ar, Gokce and Schiappa, Ricardo",
    title = "{A Primer on Resurgent Transseries and Their Asymptotics}",
    eprint = "1802.10441",
    archivePrefix = "arXiv",
    primaryClass = "hep-th",
    doi = "10.1016/j.physrep.2019.02.003",
    journal = "Phys. Rept.",
    volume = "809",
    pages = "1--135",
    year = "2019"
}

@article{Azaria:2016mqb,
    author = "Azaria, P. and Konik, R. M. and Lecheminant, Ph. and Palmai, T. and Takacs, G. and Tsvelik, A. M.",
    title = "{Particle Formation and Ordering in Strongly Correlated Fermionic Systems: Solving a Model of Quantum Chromodynamics}",
    eprint = "1601.02979",
    archivePrefix = "arXiv",
    primaryClass = "hep-th",
    doi = "10.1103/PhysRevD.94.045003",
    journal = "Phys. Rev. D",
    volume = "94",
    number = "4",
    pages = "045003",
    year = "2016"
}

@article{Bajnok:2022xgx,
    author = "Bajnok, Zoltan and Balog, Janos and Vona, Istvan",
    title = "{The full analytic trans-series in integrable field theories}",
    eprint = "2212.09416",
    archivePrefix = "arXiv",
    primaryClass = "hep-th",
    doi = "10.1016/j.physletb.2023.138075",
    journal = "Phys. Lett. B",
    volume = "844",
    pages = "138075",
    year = "2023"
}

@article{Bajnok:2025mxi,
    author = "Bajnok, Zolt{\'a}n and Balog, J{\'a}nos and Vona, Istv{\'a}n",
    title = "{The complete trans-series for conserved charges in integrable field theories}",
    eprint = "2501.16435",
    archivePrefix = "arXiv",
    primaryClass = "hep-th",
    doi = "10.1016/j.aop.2025.170152",
    journal = "Annals Phys.",
    volume = "481",
    pages = "170152",
    year = "2025"
}

@article{Barducci:1994cb,
    author = "Barducci, A. and Casalbuoni, R. and Modugno, M. and Pettini, Giulio and Gatto, Raoul",
    title = "{Thermodynamics of the massive Gross-Neveu model}",
    eprint = "hep-th/9406117",
    archivePrefix = "arXiv",
    reportNumber = "UGVA-DPT-1994-06-854",
    doi = "10.1103/PhysRevD.51.3042",
    journal = "Phys. Rev. D",
    volume = "51",
    pages = "3042--3060",
    year = "1995"
}

@article{Basar:2008im,
    author = "Basar, Gokce and Dunne, Gerald V.",
    title = "{Self-consistent crystalline condensate in chiral Gross-Neveu and Bogoliubov-de Gennes systems}",
    eprint = "0803.1501",
    archivePrefix = "arXiv",
    primaryClass = "hep-th",
    doi = "10.1103/PhysRevLett.100.200404",
    journal = "Phys. Rev. Lett.",
    volume = "100",
    pages = "200404",
    year = "2008"
}

@article{Basar:2009fg,
    author = "Ba\c{s}ar, Gokce and Dunne, Gerald V. and Thies, Michael",
    title = "{Inhomogeneous condensates in the thermodynamics of the chiral NJL$_2$ model}",
    eprint = "0903.1868",
    archivePrefix = "arXiv",
    primaryClass = "hep-th",
    doi = "10.1103/PhysRevD.79.105012",
    journal = "Phys. Rev. D",
    volume = "79",
    pages = "105012",
    year = "2009"
}

@book{Bender1978,
    author = "Bender, Carl M. and Orszag, Steven A.",
    title = "{Advanced mathematical methods for scientists and engineers}",
    publisher = "{McGraw-Hill, NY, USA}",
    year = "1978",
    doi = "10.1007/978-1-4757-3069-2",
    note = "page 375"
}

@article{Benfatto:1990zz,
    author = "Benfatto, G. and Gallavotti, G.",
    title = "{Renormalization-group approach to the theory of the Fermi surface}",
    doi = "10.1103/PhysRevB.42.9967",
    journal = "Phys. Rev. B",
    volume = "42",
    pages = "9967--9972",
    year = "1990"
}

@article{Benini:2025riz,
    author = "Benini, Francesco and Mamroud, Ohad and Reis, Tomas and Serone, Marco",
    title = "{Condensates, Crystals, and Renormalons in the Gross-Neveu Model at Finite Density}",
    eprint = "2505.23388",
    archivePrefix = "arXiv",
    primaryClass = "hep-th",
    doi = "10.1103/trj9-r9j8",
    journal = "Phys. Rev. Lett.",
    volume = "135",
    number = "14",
    pages = "141601",
    year = "2025"
}

@article{Campbell:1981dc,
    author = "Campbell, D. K. and Bishop, A. R.",
    title = "{Soliton excitations in polyacetylene and relativistic field theory models}",
    doi = "10.1016/0550-3213(82)90089-X",
    journal = "Nucl. Phys. B",
    volume = "200",
    pages = "297--328",
    year = "1982"
}

@article{Ciccone:2022zkg,
    author = "Ciccone, Riccardo and Di Pietro, Lorenzo and Serone, Marco",
    title = "{Inhomogeneous phase of the Chiral Gross\textendash{}Neveu model}",
    eprint = "2203.07451",
    archivePrefix = "arXiv",
    primaryClass = "hep-th",
    doi = "10.1103/PhysRevLett.129.071603",
    journal = "Phys. Rev. Lett.",
    volume = "129",
    pages = "071603",
    year = "2022"
}

@article{Ciccone:2023pdk,
    author = "Ciccone, Riccardo and Di Pietro, Lorenzo and Serone, Marco",
    title = "{Anomalies and persistent order in the chiral Gross\textendash{}Neveu model}",
    eprint = "2312.13756",
    archivePrefix = "arXiv",
    primaryClass = "hep-th",
    doi = "10.1007/JHEP02(2024)211",
    journal = "JHEP",
    volume = "02",
    pages = "211",
    year = "2024"
}

@article{Coleman1973,
    author = "Coleman, Sidney R.",
    doi = "10.1007/BF01646487",
    journal = "Commun. Math. Phys.",
    pages = "259--264",
    title = "{There are no Goldstone bosons in two-dimensions}",
    volume = "31",
    year = "1973"
}

@article{Dashen:1975xh,
    author = "Dashen, Roger F. and Hasslacher, Brosl and Neveu, Andre",
    title = "{Semiclassical bound states in an asymptotically free theory}",
    doi = "10.1103/PhysRevD.12.2443",
    journal = "Phys. Rev. D",
    volume = "12",
    pages = "2443",
    year = "1975"
}

@article{DiPietro:2021yxb,
    author = "Di Pietro, Lorenzo and Mari\~no, Marcos and Sberveglieri, Giacomo and Serone, Marco",
    title = "{Resurgence and $1/N$ expansion in integrable field theories}",
    eprint = "2108.02647",
    archivePrefix = "arXiv",
    primaryClass = "hep-th",
    doi = "10.1007/JHEP10(2021)166",
    journal = "JHEP",
    volume = "10",
    pages = "166",
    year = "2021"
}

@article{Dorigoni:2014hea,
    author = "Dorigoni, Daniele",
    title = "{An Introduction to Resurgence, Trans-Series and Alien Calculus}",
    eprint = "1411.3585",
    archivePrefix = "arXiv",
    primaryClass = "hep-th",
    reportNumber = "DAMTP-2014-44",
    doi = "10.1016/j.aop.2019.167914",
    journal = "Annals Phys.",
    volume = "409",
    pages = "167914",
    year = "2019"
}

@article{Dunne:1997ia,
    author = "Dunne, Gerald V. and Feinberg, Joshua",
    title = "{Self isospectral periodic potentials and supersymmetric quantum mechanics}",
    eprint = "hep-th/9706012",
    archivePrefix = "arXiv",
    reportNumber = "UCONN-97-10, NSF-ITP-97-054",
    doi = "10.1103/PhysRevD.57.1271",
    journal = "Phys. Rev. D",
    volume = "57",
    pages = "1271--1276",
    year = "1998"
}

@article{Evans:1998ek,
    author = "Evans, Nick J. and Hsu, Stephen D. H. and Schwetz, Myckola",
    title = "{An Effective field theory approach to color superconductivity at high quark density}",
    eprint = "hep-ph/9808444",
    archivePrefix = "arXiv",
    reportNumber = "OITS-657, RU-98-37, BUHEP-98-22",
    doi = "10.1016/S0550-3213(99)00175-3",
    journal = "Nucl. Phys. B",
    volume = "551",
    pages = "275--289",
    year = "1999"
}

@article{Fateev:1992tk,
    author = "Fateev, V. A. and Onofri, E. and Zamolodchikov, Alexei B.",
    title = "{Integrable deformations of the $O(3)$ sigma model. The sausage model}",
    reportNumber = "PAR-LPTHE-92-46, LPTHE-92-46",
    doi = "10.1016/0550-3213(93)90001-6",
    journal = "Nucl. Phys. B",
    volume = "406",
    pages = "521--565",
    year = "1993"
}

@article{Feinberg:2003qz,
    author = "Feinberg, Joshua",
    title = "{All about the static fermion bags in the Gross-Neveu model}",
    eprint = "hep-th/0305240",
    archivePrefix = "arXiv",
    doi = "10.1016/j.aop.2003.08.004",
    journal = "Annals Phys.",
    volume = "309",
    pages = "166--231",
    year = "2004"
}

@article{fendley-tba,
    author = "Fendley, Paul",
    title = "{Integrable sigma models and perturbed coset models}",
    eprint = "hep-th/0101034",
    archivePrefix = "arXiv",
    doi = "10.1088/1126-6708/2001/05/050",
    journal = "JHEP",
    volume = "05",
    pages = "050",
    year = "2001"
}

@article{fnw1,
	author = {Forgacs, P. and Niedermayer, F. and Weisz, P.},
	doi = {10.1016/0550-3213(91)90044-X},
	journal = {Nucl. Phys.},
	pages = {123-143},
	reportnumber = {MPI-PH-91-37},
	slaccitation = {%%CITATION = NUPHA,B367,123;%%},
	title = {{The Exact mass gap of the Gross-Neveu model. 1. The Thermodynamic Bethe ansatz}},
	volume = {B367},
	year = {1991}}

@article{fnw2,
	author = {Forgacs, P. and Niedermayer, F. and Weisz, P.},
	doi = {10.1016/0550-3213(91)90045-Y},
	journal = {Nucl. Phys.},
	pages = {144-157},
	reportnumber = {MPI-PH-91-38},
	slaccitation = {%%CITATION = NUPHA,B367,144;%%},
	title = {{The Exact mass gap of the Gross-Neveu model. 2. The 1/N expansion}},
	volume = {B367},
	year = {1991}}

@article{gn-inst,
    author = "Koenigstein, Adrian and Pannullo, Laurin and Rechenberger, Stefan and Steil, Martin J. and Winstel, Marc",
    title = "{Detecting inhomogeneous chiral condensation from the bosonic two-point function in the (1 + 1)-dimensional Gross\textendash{}Neveu model in the mean-field approximation*}",
    eprint = "2112.07024",
    archivePrefix = "arXiv",
    primaryClass = "hep-ph",
    doi = "10.1088/1751-8121/ac820a",
    journal = "J. Phys. A",
    volume = "55",
    number = "37",
    pages = "375402",
    year = "2022"
}

@article{Gross:1974jv,
    author = "Gross, David J. and Neveu, Andre",
    title = "{Dynamical Symmetry Breaking in Asymptotically Free Field Theories}",
    doi = "10.1103/PhysRevD.10.3235",
    journal = "Phys. Rev. D",
    volume = "10",
    pages = "3235",
    year = "1974"
}

@article{hmn,
	author = {Hasenfratz, P. and Maggiore, M. and Niedermayer, F.},
	doi = {10.1016/0370-2693(90)90685-Y},
	journal = {Phys. Lett.},
	pages = {522-528},
	reportnumber = {BUTP-90/12-BERN},
	slaccitation = {%%CITATION = PHLTA,B245,522;%%},
	title = {{The Exact mass gap of the O(3) and O(4) nonlinear sigma models in d = 2}},
	volume = {B245},
	year = {1990}}

@article{Karowski:1980kq,
    author = "Karowski, M. and Thun, H. J.",
    title = "{Complete $S$-matrix of the $O(2N)$ Gross\textendash{}Neveu model}",
    doi = "10.1016/0550-3213(81)90484-3",
    journal = "Nucl. Phys. B",
    volume = "190",
    pages = "61--92",
    year = "1981"
}

@article{Kusnezov:2000,
doi = {10.1088/0305-4470/33/36/310},
url = {https://doi.org/10.1088/0305-4470/33/36/310},
year = {2000},
month = {sep},
publisher = {},
volume = {33},
number = {36},
pages = {6413},
author = {Hui Li and Dimitri Kusnezov and Francesco Iachello},
title = {Group theoretical properties and band structure of the
Lamé Hamiltonian},
journal = {Journal of Physics A: Mathematical and General}
}

@article{Lajer:2021kcz,
    author = "Lajer, Marton and Konik, Robert M. and Pisarski, Robert D. and Tsvelik, Alexei M.",
    title = "{When cold, dense quarks in 1+1 and 3+1 dimensions are not a Fermi liquid}",
    eprint = "2112.10238",
    archivePrefix = "arXiv",
    primaryClass = "hep-th",
    doi = "10.1103/PhysRevD.105.054035",
    journal = "Phys. Rev. D",
    volume = "105",
    number = "5",
    pages = "054035",
    year = "2022"
}

@article{Lenz:2020bxk,
    author = {Lenz, Julian and Pannullo, Laurin and Wagner, Marc and Wellegehausen, Bj\"orn and Wipf, Andreas},
    title = "{Inhomogeneous phases in the Gross\textendash{}Neveu model in $1{+}1$ dimensions at finite number of flavors}",
    eprint = "2004.00295",
    archivePrefix = "arXiv",
    primaryClass = "hep-lat",
    doi = "10.1103/PhysRevD.101.094512",
    journal = "Phys. Rev. D",
    volume = "101",
    pages = "094512",
    year = "2020"
}

@article{Liu:2025bqq,
    author = "Liu, Yizhuang and Mari{\~n}o, Marcos",
    title = "{Trans-series from condensates in the non-linear sigma model}",
    eprint = "2507.02605",
    archivePrefix = "arXiv",
    primaryClass = "hep-th",
    month = "7",
    year = "2025"
}

@article{Marino:2019eym,
    author = "Mari\~no, Marcos and Reis, Tom\'as",
    title = "{Renormalons in integrable field theories}",
    eprint = "1909.12134",
    archivePrefix = "arXiv",
    primaryClass = "hep-th",
    doi = "10.1007/JHEP04(2020)160",
    journal = "JHEP",
    volume = "04",
    pages = "160",
    year = "2020"
}

@article{Marino:2021dzn,
    author = "Mari\~no, Marcos and Miravitllas, Ramon and Reis, Tomas",
    title = "{New renormalons from analytic trans-series}",
    eprint = "2111.11951",
    archivePrefix = "arXiv",
    primaryClass = "hep-th",
    doi = "10.1007/JHEP08(2022)279",
    journal = "JHEP",
    volume = "08",
    pages = "279",
    year = "2022"
}

@article{Marino:2023epd,
    author = "Marino, Marcos and Miravitllas, Ramon and Reis, Tom{\'a}s",
    title = "{On the Structure of Trans-Series in Quantum Field Theory}",
    eprint = "2302.08363",
    archivePrefix = "arXiv",
    primaryClass = "hep-th",
    doi = "10.3842/SIGMA.2025.065",
    journal = "SIGMA",
    volume = "21",
    pages = "065",
    year = "2025"
}

@article{Marino:2024uco,
    author = "Mari\~no, Marcos and Miravitllas, Ramon",
    title = "{Trans-series from condensates}",
    journal = "SciPost Phys.",
    volume = "18",
    pages = "101",
    year = "2025",
    doi = "10.21468/SciPostPhys.18.3.101",
    eprint = "2402.19356",
    archivePrefix = "arXiv",
    primaryClass = "hep-th",
}

@article{Marino:2025ido,
    author = "Marino, Marcos",
    title = "{Anatomy of the simplest renormalon}",
    eprint = "2504.12044",
    archivePrefix = "arXiv",
    primaryClass = "hep-th",
    month = "4",
    year = "2025"
}

@misc{melin2023integrability,
  title={Integrability and Thermodynamics of the Gross-Neveu Model},
  author={Melin, Valdemar},
  year={2023}
}

@article{Melin:2024oee,
    author = "Melin, Valdemar and Sekiguchi, Yuta and Wiegmann, Paul and Zarembo, Konstantin",
    title = "{Peierls transition in Gross\textendash{}Neveu model from Bethe ansatz}",
    eprint = "2404.07307",
    archivePrefix = "arXiv",
    primaryClass = "hep-th",
    doi = "10.1103/PhysRevLett.133.101601",
    journal = "Phys. Rev. Lett.",
    volume = "133",
    pages = "101601",
    year = "2024"
}

@article{Melin:2025eyw,
    author = "Melin, Valdemar and Wiegmann, Paul and Zarembo, Konstantin",
    title = "{Finite-gap potentials as a semiclassical limit of the thermodynamic Bethe Ansatz}",
    eprint = "2512.19655",
    archivePrefix = "arXiv",
    primaryClass = "hep-th",
    month = "12",
    year = "2025"
}

@article{Mermin1966,
    author = "Mermin, N. D. and Wagner, H.",
    doi = "10.1103/PhysRevLett.17.1133",
    journal = "Phys. Rev. Lett.",
    pages = "1133--1136",
    title = "{Absence of ferromagnetism or antiferromagnetism in one-dimensional or two-dimensional isotropic Heisenberg models}",
    volume = "17",
    year = "1966"
}

@article{Narayanan:2020uqt,
    author = "Narayanan, Rajamani",
    title = "{Phase diagram of the large $N$ Gross\textendash{}Neveu model in a finite periodic box}",
    eprint = "2001.09200",
    archivePrefix = "arXiv",
    primaryClass = "hep-th",
    doi = "10.1103/PhysRevD.101.096001",
    journal = "Phys. Rev. D",
    volume = "101",
    pages = "096001",
    year = "2020"
}

@article{PhysRevA.44.R2251,
  title = {Multipolaron solutions of the Gross-Neveu field theory: Toda potential and doped polymers},
  author = {Saxena, Avadh and Bishop, A. R.},
  journal = {Phys. Rev. A},
  volume = {44},
  issue = {4},
  pages = {R2251--R2254},
  numpages = {0},
  year = {1991},
  month = {Aug},
  publisher = {American Physical Society},
  doi = {10.1103/PhysRevA.44.R2251},
  url = {https://link.aps.org/doi/10.1103/PhysRevA.44.R2251}
}

@inproceedings{Polchinski:1992ed,
    author = "Polchinski, Joseph",
    title = "{Effective field theory and the Fermi surface}",
    booktitle = "{Theoretical Advanced Study Institute (TASI 92): From Black Holes and Strings to Particles}",
    eprint = "hep-th/9210046",
    archivePrefix = "arXiv",
    reportNumber = "NSF-ITP-92-132, UTTG-20-92",
    pages = "0235--276",
    month = "6",
    year = "1992"
}

@article{Polyakov:1983tt,
    author = "Polyakov, Alexander M. and Wiegmann, P. B.",
    title = "{Theory of nonabelian Goldstone bosons in two dimensions}",
    doi = "10.1016/0370-2693(83)91104-8",
    journal = "Phys. Lett. B",
    volume = "131",
    pages = "121--126",
    year = "1983"
}

@phdthesis{Reis:2022tni,
    author = "Reis, Tomas",
    title = "{On the resurgence of renormalons in integrable theories}",
    eprint = "2209.15386",
    archivePrefix = "arXiv",
    primaryClass = "hep-th",
    school = "U. of Geneva",
    year = "2022"
}

@article{Saleur:2009bf,
    author = "Saleur, Hubert and Pozsgay, Balazs",
    title = "{Scattering and duality in the 2 dimensional OSP(2|2) Gross Neveu and sigma models}",
    eprint = "0910.0637",
    archivePrefix = "arXiv",
    primaryClass = "hep-th",
    doi = "10.1007/JHEP02(2010)008",
    journal = "JHEP",
    volume = "02",
    pages = "008",
    year = "2010"
}

@article{Sauzin:2014qzt,
    author = "Sauzin, David",
    title = "{Introduction to 1-summability and resurgence}",
    eprint = "1405.0356",
    archivePrefix = "arXiv",
    primaryClass = "math.DS",
    month = "5",
    year = "2014"
}

@article{Schafer:1998na,
    author = {Sch{\"a}fer, Thomas and Wilczek, Frank},
    title = "{High density quark matter and the renormalization group in QCD with two and three flavors}",
    eprint = "hep-ph/9810509",
    archivePrefix = "arXiv",
    reportNumber = "IASSNS-HEP-98-90",
    doi = "10.1016/S0370-2693(99)00162-8",
    journal = "Phys. Lett. B",
    volume = "450",
    pages = "325--331",
    year = "1999"
}

@article{Schnetz:2004vr,
    author = "Schnetz, Oliver and Thies, Michael and Urlichs, Konrad",
    title = "{Phase diagram of the Gross\textendash{}Neveu model: Exact results and condensed matter precursors}",
    eprint = "hep-th/0402014",
    archivePrefix = "arXiv",
    doi = "10.1016/j.aop.2004.06.009",
    journal = "Annals Phys.",
    volume = "314",
    pages = "425--447",
    year = "2004"
}

@article{Schnetz:2005ih,
    author = "Schnetz, Oliver and Thies, Michael and Urlichs, Konrad",
    title = "{Full phase diagram of the massive Gross\textendash{}Neveu model}",
    eprint = "hep-th/0511206",
    archivePrefix = "arXiv",
    doi = "10.1016/j.aop.2005.12.007",
    journal = "Annals Phys.",
    volume = "321",
    pages = "2604--2637",
    year = "2006"
}

@article{Schnetz:2005vh,
    author = "Schnetz, Oliver and Thies, Michael and Urlichs, Konrad",
    title = "{The Phase diagram of the massive Gross-Neveu model, revisited}",
    eprint = "hep-th/0507120",
    archivePrefix = "arXiv",
    reportNumber = "FAU-TP3-05-5",
    month = "7",
    year = "2005"
}

@article{Schon:2000he,
    author = "Sch{\"{o}}n, Verena and Thies, Michael",
    title = "{Emergence of Skyrme crystal in Gross\textendash{}Neveu and 't~Hooft models at finite density}",
    eprint = "hep-th/0003195",
    archivePrefix = "arXiv",
    doi = "10.1103/PhysRevD.62.096002",
    journal = "Phys. Rev. D",
    volume = "62",
    pages = "096002",
    year = "2000"
}

@article{Serone:2024uwz,
    author = "Serone, Marco",
    title = "{Lectures on Resurgence in Integrable Field Theories}",
    eprint = "2405.02224",
    archivePrefix = "arXiv",
    primaryClass = "hep-th",
    month = "5",
    year = "2024"
}

@article{Shifman:1978bx,
    author = "Shifman, Mikhail A. and Vainshtein, A. I. and Zakharov, Valentin I.",
    title = "{QCD and Resonance Physics. Theoretical Foundations}",
    doi = "10.1016/0550-3213(79)90022-1",
    journal = "Nucl. Phys. B",
    volume = "147",
    pages = "385--447",
    year = "1979"
}

@article{Thies:2003br,
    author = "Thies, Michael",
    title = "{Analytical solution of the Gross\textendash{}Neveu model at finite density}",
    eprint = "hep-th/0308164",
    archivePrefix = "arXiv",
    doi = "10.1103/PhysRevD.69.067703",
    journal = "Phys. Rev. D",
    volume = "69",
    pages = "067703",
    year = "2004"
}

@article{Thies:2003kk,
    author = "Thies, Michael and Urlichs, Konrad",
    title = "{Revised phase diagram of the Gross\textendash{}Neveu model}",
    eprint = "hep-th/0302092",
    archivePrefix = "arXiv",
    doi = "10.1103/PhysRevD.67.125015",
    journal = "Phys. Rev. D",
    volume = "67",
    pages = "125015",
    year = "2003"
}

@article{Thies:2005wv,
    author = "Thies, Michael and Urlichs, Konrad",
    title = "{From non-degenerate conducting polymers to dense matter in the massive Gross\textendash{}Neveu model}",
    eprint = "hep-th/0505024",
    archivePrefix = "arXiv",
    doi = "10.1103/PhysRevD.72.105008",
    journal = "Phys. Rev. D",
    volume = "72",
    pages = "105008",
    year = "2005"
}

@article{Thies:2006ti,
    author = "Thies, Michael",
    title = "{From relativistic quantum fields to condensed matter and back again: Updating the Gross\textendash{}Neveu phase diagram}",
    eprint = "hep-th/0601049",
    archivePrefix = "arXiv",
    doi = "10.1088/0305-4470/39/41/S04",
    journal = "J. Phys. A",
    volume = "39",
    pages = "12707--12734",
    year = "2006"
}

@article{Thies:2024anc,
    author = "Thies, Michael",
    title = "{Nonperturbative phase boundaries in the Gross-Neveu model from a stability analysis}",
    eprint = "2408.09803",
    archivePrefix = "arXiv",
    primaryClass = "hep-th",
    doi = "10.1103/PhysRevD.110.096012",
    journal = "Phys. Rev. D",
    volume = "110",
    number = "9",
    pages = "096012",
    year = "2024"
}

@article{Thies:2025mro,
    author = "Thies, Michael",
    title = "{Peierls instability for systems with several Fermi surfaces: An example from the chiral Gross-Neveu model}",
    eprint = "2511.04212",
    archivePrefix = "arXiv",
    primaryClass = "hep-th",
    doi = "10.1103/rgbk-f6gt",
    journal = "Phys. Rev. D",
    volume = "113",
    number = "3",
    pages = "036016",
    year = "2026"
}

@book{tricomi1985,
  title={Integral Equations},
  author={Tricomi, F.G.},
  isbn={9780486648286},
  lccn={84025917},
  series={(Pure and applied mathematics, v. 5)},
  url={https://books.google.it/books?id=FYs8ua1X6xIC},
  year={1985},
  publisher={Dover Publications}
}

@article{Volin:2009wr,
    author = "Volin, Dmytro",
    title = "{From the mass gap in $O(N)$ to the non-Borel-summability in $O(3)$ and $O(4)$ sigma-models}",
    eprint = "0904.2744",
    archivePrefix = "arXiv",
    primaryClass = "hep-th",
    doi = "10.1103/PhysRevD.81.105008",
    journal = "Phys. Rev. D",
    volume = "81",
    pages = "105008",
    year = "2010"
}

@article{Wolff:1985av,
    author = "Wolff, U.",
    title = "{The phase diagram of the infinite N Gross-Neveu model at finite temperature and chemical potential}",
    doi = "10.1016/0370-2693(85)90671-9",
    journal = "Phys. Lett. B",
    volume = "157",
    pages = "303--308",
    year = "1985"
}

@article{Zamolodchikov:1978xm,
    author = "Zamolodchikov, Alexander B. and Zamolodchikov, Alexei B.",
    title = "{Factorized $S$-matrices in two-dimensions as the exact solutions of certain relativistic quantum field theory models}",
    doi = "10.1016/0003-4916(79)90391-9",
    journal = "Annals Phys.",
    volume = "120",
    pages = "253--291",
    year = "1979"
}

@article{Zamolodchikov:1995xk,
    author = "Zamolodchikov, Alexei B.",
    title = "{Mass scale in the sine-Gordon model and its reductions}",
    doi = "10.1142/S0217751X9500053X",
    journal = "Int. J. Mod. Phys. A",
    volume = "10",
    pages = "1125--1150",
    year = "1995"
}

\end{document}